\RequirePackage{amsmath} 
\documentclass[longauth,traditabstract]{aa}

\usepackage[nonamebreak]{natbib}
\usepackage[stable]{footmisc}
\usepackage{amsmath,amsfonts,amssymb}
\usepackage{txfonts}
\usepackage{placeins}
\usepackage{natbib}
\bibpunct{(}{)}{;}{a}{}{,} %
\usepackage{graphicx}
\usepackage[table,usenames,dvipsnames]{xcolor}
\usepackage{ifthen}
\definecolor{linkcolor}{rgb}{0.6,0,0}
\definecolor{citecolor}{rgb}{0,0,0.75}
\definecolor{urlcolor}{rgb}{0.12,0.46,0.7}
\usepackage[breaklinks, colorlinks, urlcolor=urlcolor, linkcolor=linkcolor,citecolor=citecolor,pdfencoding=auto]{hyperref}
\hypersetup{linktocpage}
\usepackage{overpic}
\usepackage{rotating}
\usepackage[utf8]{inputenc}
\def\setsymbol#1#2{\expandafter\def\csname #1\endcsname{#2}}
\def\getsymbol#1{\csname #1\endcsname}

\def\Planck{\textit{Planck}}

\newbox\tablebox    \newdimen\tablewidth
\def\leaderfil{\leaders\hbox to 5pt{\hss.\hss}\hfil}
\def\endPlancktable{\tablewidth=\columnwidth 
    $$\hss\copy\tablebox\hss$$
    \vskip-\lastskip\vskip -2pt}
\def\endPlancktablewide{\tablewidth=\textwidth 
    $$\hss\copy\tablebox\hss$$
    \vskip-\lastskip\vskip -2pt}
\def\tablenote#1 #2\par{\begingroup \parindent=0.8em
    \abovedisplayshortskip=0pt\belowdisplayshortskip=0pt
    \noindent
    $$\hss\vbox{\hsize\tablewidth \hangindent=\parindent \hangafter=1 \noindent
    \hbox to \parindent{$^#1$\hss}\strut#2\strut\par}\hss$$
    \endgroup}
\def\doubleline{\vskip 3pt\hrule \vskip 1.5pt \hrule \vskip 5pt}

\def\L2{\ifmmode L_2\else $L_2$\fi}

\def\DeltaT{\ifmmode \Delta T\else $\Delta T$\fi}
\def\deltat{\ifmmode \Delta t\else $\Delta t$\fi}
\def\fknee{\ifmmode f_{\rm knee}\else $f_{\rm knee}$\fi}
\def\Fmax{\ifmmode F_{\rm max}\else $F_{\rm max}$\fi}
\def\solar{\ifmmode{\rm M}_{\mathord\odot}\else${\rm M}_{\mathord\odot}$\fi}
\def\Msolar{\ifmmode{\rm M}_{\mathord\odot}\else${\rm M}_{\mathord\odot}$\fi}
\def\Lsolar{\ifmmode{\rm L}_{\mathord\odot}\else${\rm L}_{\mathord\odot}$\fi}
\def\inv{\ifmmode^{-1}\else$^{-1}$\fi}
\def\mo{\ifmmode^{-1}\else$^{-1}$\fi}
\def\sup#1{\ifmmode ^{\rm #1}\else $^{\rm #1}$\fi}
\def\expo#1{\ifmmode \times 10^{#1}\else $\times 10^{#1}$\fi}
\def\,{\thinspace}
\def\lsim{\mathrel{\raise .4ex\hbox{\rlap{$<$}\lower 1.2ex\hbox{$\sim$}}}}
\def\gsim{\mathrel{\raise .4ex\hbox{\rlap{$>$}\lower 1.2ex\hbox{$\sim$}}}}

\def\simprop{\mathrel{\raise .4ex\hbox{\rlap{$\propto$}\lower 1.2ex\hbox{$\sim$}}}}
\def\deg{\ifmmode^\circ\else$^\circ$\fi}
\def\pdeg{\ifmmode $\setbox0=\hbox{$^{\circ}$}\rlap{\hskip.11\wd0 .}$^{\circ}
          \else \setbox0=\hbox{$^{\circ}$}\rlap{\hskip.11\wd0 .}$^{\circ}$\fi}
\def\arcs{\ifmmode {^{\scriptstyle\prime\prime}}
          \else $^{\scriptstyle\prime\prime}$\fi}
\def\arcm{\ifmmode {^{\scriptstyle\prime}}
          \else $^{\scriptstyle\prime}$\fi}
\newdimen\sa  \newdimen\sb
\def\parcs{\sa=.07em \sb=.03em
     \ifmmode \hbox{\rlap{.}}^{\scriptstyle\prime\kern -\sb\prime}\hbox{\kern -\sa}
     \else \rlap{.}$^{\scriptstyle\prime\kern -\sb\prime}$\kern -\sa\fi}
\def\parcm{\sa=.08em \sb=.03em
     \ifmmode \hbox{\rlap{.}\kern\sa}^{\scriptstyle\prime}\hbox{\kern-\sb}
     \else \rlap{.}\kern\sa$^{\scriptstyle\prime}$\kern-\sb\fi}
\def\ra[#1 #2 #3.#4]{#1\sup{h}#2\sup{m}#3\sup{s}\llap.#4}
\def\dec[#1 #2 #3.#4]{#1\deg#2\arcm#3\arcs\llap.#4}
\def\deco[#1 #2 #3]{#1\deg#2\arcm#3\arcs}
\def\rra[#1 #2]{#1\sup{h}#2\sup{m}}

\def\dots{\relax\ifmmode \ldots\else $\ldots$\fi}
\def\WHzsr{\ifmmode $W\,Hz\mo\,sr\mo$\else W\,Hz\mo\,sr\mo\fi}
\def\mHz{\ifmmode $\,mHz$\else \,mHz\fi}
\def\GHz{\ifmmode $\,GHz$\else \,GHz\fi}
\def\mKs{\ifmmode $\,mK\,s$^{1/2}\else \,mK\,s$^{1/2}$\fi}
\def\muKs{\ifmmode \,\mu$K\,s$^{1/2}\else \,$\mu$K\,s$^{1/2}$\fi}
\def\muKRJs{\ifmmode \,\mu$K$_{\rm RJ}$\,s$^{1/2}\else \,$\mu$K$_{\rm RJ}$\,s$^{1/2}$\fi}
\def\muKHz{\ifmmode \,\mu$K\,Hz$^{-1/2}\else \,$\mu$K\,Hz$^{-1/2}$\fi}
\def\MJysr{\ifmmode \,$MJy\,sr\mo$\else \,MJy\,sr\mo\fi}
\def\MJysrmK{\ifmmode \,$MJy\,sr\mo$\,mK$_{\rm CMB}\mo\else \,MJy\,sr\mo\,mK$_{\rm CMB}\mo$\fi}
\def\microns{\ifmmode \,\mu$m$\else \,$\mu$m\fi}

\def\muK{\ifmmode \,\mu$K$\else \,$\mu$\hbox{K}\fi}
\def\microK{\ifmmode \,\mu$K$\else \,$\mu$\hbox{K}\fi}
\def\muW{\ifmmode \,\mu$W$\else \,$\mu$\hbox{W}\fi}
\def\kms{\ifmmode $\,km\,s$^{-1}\else \,km\,s$^{-1}$\fi}
\def\kmsMpc{\ifmmode $\,\kms\,Mpc\mo$\else \,\kms\,Mpc\mo\fi}
\providecommand{\sorthelp}[1]{}

\newcommand{\planck}{\Planck}
\newcommand{\WMAP}{\text{WMAP}}  
\newcommand{\COBE}{\text{COBE}}

\newcommand{\muKcmb}{\ifmmode \,\mu$K$_{\rm CMB}$\else \,$\mu$K$_{\rm CMB}$\fi}
 
\newcommand{\NHUNIT}{\ifmmode {\rm \,cm^{-2}} \else $\rm \,cm^{-2}$ \fi} 

\newcommand{\SMICA}{\texttt{SMICA}}
\newcommand{\NILC}{\texttt{NILC}}
\newcommand{\SEVEM}{\texttt{SEVEM}}
\newcommand{\commander}{\texttt{Commander}}
\newcommand{\simall}{\texttt{Simall}}

\newcommand{\mksym}[1]{\ifmmode {\rm #1}\else #1\fi}

\def\LCDM{$\Lambda$CDM}

\newcommand{\ns}{n_{\rm s}}

\newcommand{\Aphiphi}{A_{\rm L}^{\phi\phi}}

\newcommand{\OmegaM}{\ifmmode\Omega_{\rm M}\else $\Omega_{\rm M}$\fi}

\newcommand{\Neff}{\ifmmode N_{\rm eff} \else $N_{\rm eff}$\fi}

\newcommand{\rdrag}{r_{\rm drag}}

\newcommand{\rstar}{r_{\ast}}

\newcommand{\hn}{\hat{\boldsymbol{n^{\,}}}\mkern-1.5mu}

\def\aap{{A\&A}}

\def\apjs{{ApJ Supp.}}

\defcitealias{planck2014-a10}{LowEll2016}
\defcitealias{planck2014-a09}{Maps2015}
\defcitealias{planck2016-ES}{Explanatory Supplement}

\newcommand{\sroll}{{\tt SRoll}}

\begin{document}

\title{\vglue -10mm\Planck\ 2018 results. I.\\
 Overview and the cosmological legacy of \Planck}
\titlerunning{The cosmological legacy of \Planck}
\authorrunning{Planck Collaboration}

\author{\small
Planck Collaboration: N.~Aghanim\inst{56}
\and
Y.~Akrami\inst{59, 61}
\and
F.~Arroja\inst{63}
\and
M.~Ashdown\inst{69, 5}
\and
J.~Aumont\inst{98}
\and
C.~Baccigalupi\inst{79}
\and
M.~Ballardini\inst{22, 42}
\and
A.~J.~Banday\inst{98, 8}
\and
R.~B.~Barreiro\inst{64}
\and
N.~Bartolo\inst{31, 65}
% ...
}
\institute{\small
AIM, CEA, CNRS, Universit\'{e} Paris-Saclay, Universit\'{e} Paris-Diderot, Sorbonne Paris Cit\'{e}, F-91191 Gif-sur-Yvette, France\goodbreak
\and
APC, AstroParticule et Cosmologie, Universit\'{e} Paris Diderot, CNRS/IN2P3, CEA/lrfu, Observatoire de Paris, Sorbonne Paris Cit\'{e}, 10, rue Alice Domon et L\'{e}onie Duquet, 75205 Paris Cedex 13, France\goodbreak
\and
African Institute for Mathematical Sciences, 6-8 Melrose Road, Muizenberg, Cape Town, South Africa\goodbreak
\and
Aix Marseille Univ, CNRS, CNES, LAM, Marseille, France\goodbreak
\and
Astrophysics Group, Cavendish Laboratory, University of Cambridge, J J Thomson Avenue, Cambridge CB3 0HE, U.K.\goodbreak
\and
Astrophysics \& Cosmology Research Unit, School of Mathematics, Statistics \& Computer Science, University of KwaZulu-Natal, Westville Campus, Private Bag X54001, Durban 4000, South Africa\goodbreak
\and
CITA, University of Toronto, 60 St. George St., Toronto, ON M5S 3H8, Canada\goodbreak
\and
CNRS, IRAP, 9 Av. colonel Roche, BP 44346, F-31028 Toulouse cedex 4, France\goodbreak
\and
Cahill Center for Astronomy and Astrophysics, California Institute of Technology, Pasadena CA,  91125, USA\goodbreak
\and
California Institute of Technology, Pasadena, California, U.S.A.\goodbreak
\and
Centre for Theoretical Cosmology, DAMTP, University of Cambridge, Wilberforce Road, Cambridge CB3 0WA, U.K.\goodbreak
\and
Computational Cosmology Center, Lawrence Berkeley National Laboratory, Berkeley, California, U.S.A.\goodbreak
\and
DTU Space, National Space Institute, Technical University of Denmark, Elektrovej 327, DK-2800 Kgs. Lyngby, Denmark\goodbreak
\and
D\'{e}partement de Physique Th\'{e}orique, Universit\'{e} de Gen\`{e}ve, 24, Quai E. Ansermet,1211 Gen\`{e}ve 4, Switzerland\goodbreak
\and
Departamento de Astrof\'{i}sica, Universidad de La Laguna (ULL), E-38206 La Laguna, Tenerife, Spain\goodbreak
\and
Departamento de F\'{\i}sica, Universidad de Oviedo, C/ Federico Garc\'{\i}a Lorca, 18 , Oviedo, Spain\goodbreak
\and
Departamento de F\'{i}sica Matematica, Instituto de F\'{i}sica, Universidade de S\~{a}o Paulo, Rua do Mat\~{a}o 1371, S\~{a}o Paulo, Brazil\goodbreak
\and
Department of Astrophysics/IMAPP, Radboud University, P.O. Box 9010, 6500 GL Nijmegen, The Netherlands\goodbreak
\and
Department of General Education, National Institute of Technology, Kagawa College, 355 Chokushi-cho, Takamatsu, Kagawa 761-8058, Japan\goodbreak
\and
Department of Mathematics, University of Stellenbosch, Stellenbosch 7602, South Africa\goodbreak
\and
Department of Physics \& Astronomy, University of British Columbia, 6224 Agricultural Road, Vancouver, British Columbia, Canada\goodbreak
\and
Department of Physics \& Astronomy, University of the Western Cape, Cape Town 7535, South Africa\goodbreak
\and
Department of Physics and Astronomy, University College London, London WC1E 6BT, U.K.\goodbreak
\and
Department of Physics and Astronomy, University of Sussex, Brighton BN1 9QH, U.K.\goodbreak
\and
Department of Physics, Gustaf H\"{a}llstr\"{o}min katu 2a, University of Helsinki, Helsinki, Finland\goodbreak
\and
Department of Physics, Princeton University, Princeton, New Jersey, U.S.A.\goodbreak
\and
Department of Physics, University of California, Berkeley, California, U.S.A.\goodbreak
\and
Department of Physics, University of California, One Shields Avenue, Davis, California, U.S.A.\goodbreak
\and
Department of Physics, University of California, Santa Barbara, California, U.S.A.\goodbreak
\and
Department of Physics, University of Illinois at Urbana-Champaign, 1110 West Green Street, Urbana, Illinois, U.S.A.\goodbreak
\and
Dipartimento di Fisica e Astronomia G. Galilei, Universit\`{a} degli Studi di Padova, via Marzolo 8, 35131 Padova, Italy\goodbreak
\and
Dipartimento di Fisica e Scienze della Terra, Universit\`{a} di Ferrara, Via Saragat 1, 44122 Ferrara, Italy\goodbreak
\and
Dipartimento di Fisica, Universit\`{a} La Sapienza, P. le A. Moro 2, Roma, Italy\goodbreak
\and
Dipartimento di Fisica, Universit\`{a} degli Studi di Milano, Via Celoria, 16, Milano, Italy\goodbreak
\and
Dipartimento di Fisica, Universit\`{a} degli Studi di Trieste, via A. Valerio 2, Trieste, Italy\goodbreak
\and
Dipartimento di Fisica, Universit\`{a} di Roma Tor Vergata, Via della Ricerca Scientifica, 1, Roma, Italy\goodbreak
\and
European Space Agency, ESAC, Planck Science Office, Camino bajo del Castillo, s/n, Urbanizaci\'{o}n Villafranca del Castillo, Villanueva de la Ca\~{n}ada, Madrid, Spain\goodbreak
\and
European Space Agency, ESTEC, Keplerlaan 1, 2201 AZ Noordwijk, The Netherlands\goodbreak
\and
Gran Sasso Science Institute, INFN, viale F. Crispi 7, 67100 L'Aquila, Italy\goodbreak
\and
Haverford College Astronomy Department, 370 Lancaster Avenue, Haverford, Pennsylvania, U.S.A.\goodbreak
\and
Helsinki Institute of Physics, Gustaf H\"{a}llstr\"{o}min katu 2, University of Helsinki, Helsinki, Finland\goodbreak
\and
INAF - OAS Bologna, Istituto Nazionale di Astrofisica - Osservatorio di Astrofisica e Scienza dello Spazio di Bologna, Area della Ricerca del CNR, Via Gobetti 101, 40129, Bologna, Italy\goodbreak
\and
INAF - Osservatorio Astronomico di Padova, Vicolo dell'Osservatorio 5, Padova, Italy\goodbreak
\and
INAF - Osservatorio Astronomico di Trieste, Via G.B. Tiepolo 11, Trieste, Italy\goodbreak
\and
INAF, Istituto di Radioastronomia, Via Piero Gobetti 101, I-40129 Bologna, Italy\goodbreak
\and
INAF/IASF Milano, Via E. Bassini 15, Milano, Italy\goodbreak
\and
INFN - CNAF, viale Berti Pichat 6/2, 40127 Bologna, Italy\goodbreak
\and
INFN, Sezione di Bologna, viale Berti Pichat 6/2, 40127 Bologna, Italy\goodbreak
\and
INFN, Sezione di Ferrara, Via Saragat 1, 44122 Ferrara, Italy\goodbreak
\and
INFN, Sezione di Milano, Via Celoria 16, Milano, Italy\goodbreak
\and
INFN, Sezione di Roma 1, Universit\`{a} di Roma Sapienza, Piazzale Aldo Moro 2, 00185, Roma, Italy\goodbreak
\and
INFN, Sezione di Roma 2, Universit\`{a} di Roma Tor Vergata, Via della Ricerca Scientifica, 1, Roma, Italy\goodbreak
\and
IUCAA, Post Bag 4, Ganeshkhind, Pune University Campus, Pune 411 007, India\goodbreak
\and
Imperial College London, Astrophysics group, Blackett Laboratory, Prince Consort Road, London, SW7 2AZ, U.K.\goodbreak
\and
Infrared Processing and Analysis Center, California Institute of Technology, Pasadena, CA 91125, U.S.A.\goodbreak
\and
Institut d'Astrophysique Spatiale, CNRS, Univ. Paris-Sud, Universit\'{e} Paris-Saclay, B\^{a}t. 121, 91405 Orsay cedex, France\goodbreak
\and
Institut d'Astrophysique de Paris, CNRS (UMR7095), 98 bis Boulevard Arago, F-75014, Paris, France\goodbreak
\and
Institut f\"{u}r Theoretische Teilchenphysik und Kosmologie, RWTH Aachen University, D-52056 Aachen, Germany\goodbreak
\and
Institute Lorentz, Leiden University, PO Box 9506, Leiden 2300 RA, The Netherlands\goodbreak
\and
Institute of Astronomy, University of Cambridge, Madingley Road, Cambridge CB3 0HA, U.K.\goodbreak
\and
Institute of Theoretical Astrophysics, University of Oslo, Blindern, Oslo, Norway\goodbreak
\and
Instituto de Astrof\'{\i}sica de Canarias, C/V\'{\i}a L\'{a}ctea s/n, La Laguna, Tenerife, Spain\goodbreak
\and
Instituto de Astrof\'{\i}sica e Ci\^{e}ncias do Espa\c{c}o, Faculdade de Ci\^{e}ncias da Universidade de Lisboa, Campo Grande, PT1749-016 Lisboa, Portugal\goodbreak
\and
Instituto de F\'{\i}sica de Cantabria (CSIC-Universidad de Cantabria), Avda. de los Castros s/n, Santander, Spain\goodbreak
\and
Istituto Nazionale di Fisica Nucleare, Sezione di Padova, via Marzolo 8, I-35131 Padova, Italy\goodbreak
\and
Jet Propulsion Laboratory, California Institute of Technology, 4800 Oak Grove Drive, Pasadena, California, U.S.A.\goodbreak
\and
Jodrell Bank Centre for Astrophysics, Alan Turing Building, School of Physics and Astronomy, The University of Manchester, Oxford Road, Manchester, M13 9PL, U.K.\goodbreak
\and
Kavli Institute for Cosmological Physics, University of Chicago, Chicago, IL 60637, USA\goodbreak
\and
Kavli Institute for Cosmology Cambridge, Madingley Road, Cambridge, CB3 0HA, U.K.\goodbreak
\and
Kavli Institute for the Physics and Mathematics of the Universe (Kavli IPMU, WPI), UTIAS, The University of Tokyo, Chiba, 277- 8583, Japan\goodbreak
\and
Laboratoire de Physique Subatomique et Cosmologie, Universit\'{e} Grenoble-Alpes, CNRS/IN2P3, 53, rue des Martyrs, 38026 Grenoble Cedex, France\goodbreak
\and
Laboratoire de Physique Th\'{e}orique, Universit\'{e} Paris-Sud 11 \& CNRS, B\^{a}timent 210, 91405 Orsay, France\goodbreak
\and
Lawrence Berkeley National Laboratory, Berkeley, California, U.S.A.\goodbreak
\and
Low Temperature Laboratory, Department of Applied Physics, Aalto University, Espoo, FI-00076 AALTO, Finland\goodbreak
\and
Max-Planck-Institut f\"{u}r Astrophysik, Karl-Schwarzschild-Str. 1, 85741 Garching, Germany\goodbreak
\and
Mullard Space Science Laboratory, University College London, Surrey RH5 6NT, U.K.\goodbreak
\and
NAOC-UKZN Computational Astrophysics Centre (NUCAC), University of KwaZulu-Natal, Durban 4000, South Africa\goodbreak
\and
Nicolaus Copernicus Astronomical Center, Polish Academy of Sciences, Bartycka 18, 00-716 Warsaw, Poland\goodbreak
\and
SISSA, Astrophysics Sector, via Bonomea 265, 34136, Trieste, Italy\goodbreak
\and
San Diego Supercomputer Center, University of California, San Diego, 9500 Gilman Drive, La Jolla, CA 92093, USA\goodbreak
\and
School of Chemistry and Physics, University of KwaZulu-Natal, Westville Campus, Private Bag X54001, Durban, 4000, South Africa\goodbreak
\and
School of Physical Sciences, National Institute of Science Education and Research, HBNI, Jatni-752050, Odissa, India\goodbreak
\and
School of Physics and Astronomy, Cardiff University, Queens Buildings, The Parade, Cardiff, CF24 3AA, U.K.\goodbreak
\and
School of Physics and Astronomy, Sun Yat-sen University, 2 Daxue Rd, Tangjia, Zhuhai, China\goodbreak
\and
School of Physics and Astronomy, University of Nottingham, Nottingham NG7 2RD, U.K.\goodbreak
\and
School of Physics, Indian Institute of Science Education and Research Thiruvananthapuram, Maruthamala PO, Vithura, Thiruvananthapuram 695551, Kerala, India\goodbreak
\and
School of Physics, The University of New South Wales, Sydney NSW 2052, Australia\goodbreak
\and
Simon Fraser University, Department of Physics, 8888 University Drive, Burnaby BC, Canada\goodbreak
\and
Sorbonne Universit\'{e}, Institut Lagrange de Paris (ILP), 98 bis Boulevard Arago, 75014 Paris, France\goodbreak
\and
Sorbonne Universit\'{e}, Observatoire de Paris, Universit\'{e} PSL, \'{E}cole normale sup\'{e}rieure, CNRS, LERMA, F-75005, Paris, France\goodbreak
\and
Sorbonne Universit\'{e}-UPMC, UMR7095, Institut d'Astrophysique de Paris, 98 bis Boulevard Arago, F-75014, Paris, France\goodbreak
\and
Space Research Institute (IKI), Russian Academy of Sciences, Profsoyuznaya Str, 84/32, Moscow, 117997, Russia\goodbreak
\and
Space Science Data Center - Agenzia Spaziale Italiana, Via del Politecnico snc, 00133, Roma, Italy\goodbreak
\and
Space Sciences Laboratory, University of California, Berkeley, California, U.S.A.\goodbreak
\and
The Oskar Klein Centre for Cosmoparticle Physics, Department of Physics, Stockholm University, AlbaNova, SE-106 91 Stockholm, Sweden\goodbreak
\and
UPMC Univ Paris 06, UMR7095, 98 bis Boulevard Arago, F-75014, Paris, France\goodbreak
\and
Univ. Grenoble Alpes, CNRS, IPAG, 38000 Grenoble, France\goodbreak
\and
Universit\'{e} de Toulouse, UPS-OMP, IRAP, F-31028 Toulouse cedex 4, France\goodbreak
% ...
}

\date{\vglue -1.5mm \today\vglue -5mm}

\abstract{\vglue -3mm 
The European Space Agency's \Planck\ satellite, which was dedicated to
studying the early Universe and its subsequent evolution, was launched on
14 May 2009. It scanned the microwave and submillimetre sky continuously
between 12 August 2009 and 23 October 2013, producing deep, high-resolution,
all-sky maps in nine frequency bands from 30 to 857\,GHz.  This paper
presents the cosmological legacy of \Planck, which currently provides our
strongest constraints on the parameters of the standard cosmological model
and some of the tightest limits available on deviations from that model.
The 6-parameter $\Lambda$CDM model continues to provide an excellent fit to
the cosmic microwave background data at high and low redshift,
describing the cosmological information
in over a billion map pixels with just six parameters.
With 18 peaks in the temperature and polarization angular power spectra
constrained well, \Planck\ measures five of the six parameters to better than
1\,\% (simultaneously), with the best-determined parameter ($\theta_\ast$) now
known to 0.03\,\%.  We describe the multi-component sky as seen by \Planck,
the success of the \LCDM\ model, and the connection to lower-redshift probes
of structure formation.  We also give a comprehensive summary of the major
changes introduced in this 2018 release.
The \Planck\ data, alone and in combination with other probes, provide
stringent constraints on our models of the early Universe and the large-scale
structure within which all astrophysical objects form and evolve.
We discuss some lessons learned from the \Planck\ mission, and highlight
areas ripe for further experimental advances.
}
   
\keywords{Cosmology: observations -- Cosmology: theory --
cosmic background radiation -- Surveys}

\maketitle


\tableofcontents

 







\section{Introduction} \label{sec:intro}

This paper, one of a set associated with the 2018 release of data
from the \Planck\footnote{\Planck\ (\url{http://www.esa.int/Planck}) is a
project of the European Space Agency (ESA) with instruments provided
by two scientific consortia funded by ESA member states
(in particular the lead countries France and Italy),
with contributions from NASA (USA), and telescope reflectors provided by a
collaboration between ESA and a scientific consortium led and funded
by Denmark.} mission, presents the cosmological legacy of \Planck.
\Planck\ was dedicated to studying the early Universe and its subsequent
evolution by mapping the anisotropies in the cosmic microwave background
(CMB) radiation.

The CMB, discovered in 1965 \citep{PenWil65,DPRW65}, has been a pillar of
our cosmological world view since it was determined to be of cosmological
origin.
The CMB spectrum is the best-measured blackbody in nature \citep{fixsen2009},
and the absence of spectral distortions places very strong constraints on
the amount of energy that could have been injected into the Universe at
epochs later than $z\simeq 2\times10^6$
\citep[e.g.,][]{Fixsen96,ChlubaSunyaev12}.
This limits the properties of decaying or annihilating particles, primordial
black holes, topological defects, primordial magnetic fields, and other exotic
physics.
Perhaps its largest impact, however, has come from CMB anisotropies,
the small deviations in intensity and polarization from point to point
on the sky.

The anisotropies in the CMB, first detected by the Cosmic Background Explorer
(\COBE) satellite
\citep{Smoot92}, provide numerous, strong tests of the cosmological paradigm
and the current best measurements on most of the parameters of our
cosmological model \citep{planck2013-p11,planck2014-a15,planck2016-l06}.
The \COBE\ detection cemented the gravitational instability paradigm within
a cold dark matter (CDM) model \citep{EBW92}.
Ground-based and balloon-borne experiments
\citep[e.g.,][]{deBernardis00,Balbi00,Miller02,macias2007} established
that the Universe has no significant spatial curvature
\citep{KnoxPage00,Pierpaoli00}.
The Wilkinson Microwave Anisotropy Probe (\WMAP) showed that the
fluctuations are predominantly adiabatic
(\citealt{kogut2003}; from the phasing of the peaks and polarization)
and provided multiple, simultaneous, tight constraints on cosmological
parameters \citep{bennett2003a} -- a legacy that the \Planck\ mission
has continued and enriched (Sect.~\ref{sec:lcdm_params}).

\Planck\ was the third-generation space mission dedicated to measurements of
CMB anisotropies.
It was a tremendous technical success, operating in a challenging environment
without interruption over three times the initially planned mission duration,
with performance exceeding expectations.
Currently our best measurements of the anisotropy spectra on the scales most
relevant for cosmology come from \Planck.

\begin{table}[tb]
\caption{Important milestones in the \Planck\ mission.} 
\label{tab:milestones}
\vskip -6mm
\footnotesize
\newdimen\tblskip \tblskip=5pt
\setbox\tablebox=\vbox{
 \newdimen\digitwidth
 \setbox0=\hbox{\rm 0}
 \digitwidth=\wd0
 \catcode`*=\active
 \def*{\kern\digitwidth}
  \newdimen\dpwidth
  \setbox0=\hbox{.}
  \dpwidth=\wd0
  \catcode`!=\active
  \def!{\kern\dpwidth}
\halign{\hbox to 0.9in{#\leaderfil}\tabskip 0.5em&
#\hfil\tabskip=0pt\cr
\noalign{\doubleline}
\omit\hfil Date\hfil& \omit\hfil Milestone \hfil\cr
\noalign{\vskip 5pt\hrule\vskip 5pt}
 Nov 1992& ESA call for M3 (of Horizon 2000 programme)\cr
 May 1993& Proposals for COBRAS and SAMBA submitted\cr
 Sep 1993& Selection of COBRAS and SAMBA for assessment\cr
 Dec 1994& Selection of COBRAS and SAMBA for Phase A\cr
 Jul 1996& (Combined) \textbf{Project selection as M3}\cr
 May 1998& Pre-selection of the instrument consortia\cr
 Feb 1999& Final approval of scientific payload and consortia\cr
 Jan 2001& First meeting of the full Planck Collaboration\cr
 Apr 2001& Prime contractor selected. \textbf{Start of phase B}\cr
 Jun 2001& \WMAP\ blazes the way for \Planck\cr
\noalign{\vskip 5pt\hrule\vskip 3pt}
 Sep 2001&System requirements review\cr
 Jul--Oct 2002& Preliminary design review\cr
 Dec 2002& Science ground segment (SGS) review\cr
 Apr--Oct 2004&  Critical design review\cr
 Jan 2005&  Delivery of HFI cryo-qualification model to ESA\cr
 Aug 2006& Calibration of flight instruments at Orsay and Laben\cr      
 Sep 2006& Delivery of instrument flight models to ESA\cr
 Nov 2006& HFI and LFI mating at Thales in Cannes\cr
 Jan 2007& Integration completed\cr
 Mar 2007& SGS implementation review\cr
 Feb--Apr 2007& Qualification review\cr
 Jun--Aug 2007& Final global test at Centre Spatial de Li{\'e}ges\cr
 Nov 2008& Ground segment readiness review\cr
 Jan 2009&  Flight acceptance review passed\cr
 19 Feb 2009& \Planck\ flies to French Guyana\cr
 14 May 2009& \textbf{Launch}\cr
\noalign{\vskip 5pt\hrule\vskip 3pt}
02 Jul 2009& Injection into $L_2$ orbit\cr
20 May 2009& Commissioning begins\cr
13 Aug 2009& Commissioning ends\cr
27 Aug 2009& End of ``First light survey''\cr
14 Feb 2010& Start of second all-sky survey\cr
05 Jul 2010& First all-sky image released\cr
14 Aug 2010& Start of third all-sky survey\cr
27 Nov 2010& \textbf{End of nominal mission}, start of extended mission\cr
14 Feb 2011& Start of fourth all-sky survey\cr
29 Jul 2011& Start of fifth all-sky survey\cr
14 Jan 2012& End of cryogenic mission, start of warm phase\cr
30 Jan 2012& LFI starts sixth all-sky survey\cr
08 Feb 2012& \Planck\ completes 1000 days in space\cr
14 Aug 2013& Departure manoeuvre executed\cr
04 Oct 2013& Start of end-of-life operations\cr
09 Oct 2013& De-orbiting from $L_2$\cr
09 Oct 2013& HFI, LFI, and SCS commanded off\cr
23 Oct 2013& \textbf{Last command}\cr
\noalign{\doubleline}
Feb 1996& Publication of the "Redbook" of \Planck\ science\cr
Jan 2005& Bluebook: The Scientific Programme of \Planck\cr
Sep 2009& First light survey press release\cr 
Mar 2010&  First (of 15) internal data releases\cr
Sep 2010& Pre-launch papers, special issue of A\&A, Vol.~520\cr
Jan 2011& Early release (compact source catalogue)\cr 
Dec 2011& Early results papers, special issue of A\&A, Vol.~536\cr
Mar 2013& \textbf{Nominal mission data release} (temperature, PR1)\cr 
Nov 2014& 2013 results papers, special issue of A\&A, Vol.~571\cr
Feb--Aug 2015& Extended mission data release (PR2)\cr
Sep 2016& 2015 results papers, special issue of A\&A, Vol.~594\cr
2018 & This \textbf{Legacy data release (PR3)}\cr
\noalign{\vskip 5pt\hrule\vskip 3pt}}}
\endPlancktable
\end{table}

Some milestones in the \Planck\ mission are listed in
Table~\ref{tab:milestones}.
A set of 13 pre-launch papers was published in a special issue of
Astronomy and Astrophysics (Vol.\ 520, 2010; see \citealt{tauber2010a}).
For an overview of the scientific operations of the \Planck\ mission see
\citet{planck2013-p01} and the Explanatory Supplement
\citep{planck2014-ES,planck2016-ES}.
The first set of scientific data, the Early Release Compact Source
Catalogue (ERCSC; \citealt{planck2011-1.10}), was released in January 2011.
A set of 26 papers related to astrophysical foregrounds was published
in another special issue of Astronomy and Astrophysics (Vol.\ 536, 2011;
see \citealt{planck2011-1.1}).
The first cosmological results from \Planck, based mainly on temperature
maps of the whole sky acquired during the nominal mission duration of
15.5~months, were reported in 2013 and the data products made available
(as ``PR1'') on the Planck Legacy Archive
(PLA\footnote{\url{http://pla.esac.esa.int}}).  These cosmological results
were published as a series along with further data-processing and astrophysics
papers in 2014 (A\&A Vol.~571, 2014; see \citealt{planck2013-p01}).
The first results from the full mission, including some polarization data,
were presented in 2015; for a summary see \citet{planck2014-a01}.
The raw time-ordered observations were released to the public in their
entirety in February 2015, as part of this second \Planck\ data release
(``PR2''), together with associated frequency and component sky maps and
higher-level science derivatives.

This paper is part of a final series of papers from the \Planck\ collaboration,
released along with the final data (``PR3'').
It presents an overview of the \Planck\ mission and the numerous contributions
\Planck\ has made to our understanding of cosmology, that is, we consider
the cosmological legacy of \Planck.  After a broad
overview of the useful products derived from \Planck\ data, from the maps at
nine frequencies to astrophysical components and their broad characterization
(specifics of this release are detailed in Appendix~\ref{sec:therelease}),
we discuss the CMB anisotropies, which were the main focus of the
\Planck\ mission.  We then turn to a comparison of our results to theoretical
models, and the way in which the \Planck\ data confirm and inform those
models, before comparing to a wider range of astrophysical and cosmological
data.  A discussion of how \Planck\ has placed constraints on models of the
early and late Universe and the relationship of the \Planck\ data to other
cosmological probes precedes a discussion of the post-\Planck\ landscape, and
finally our conclusions.  In appendices, we include some details of this
release, and a more detailed discussion of improvements in the data processing
between the 2015 and 2018 releases.

\section{The sky according to \Planck} \label{sec:thesky}

\begin{figure*}[hbtp]
\begin{center}
\resizebox{!}{9.1in}{\includegraphics{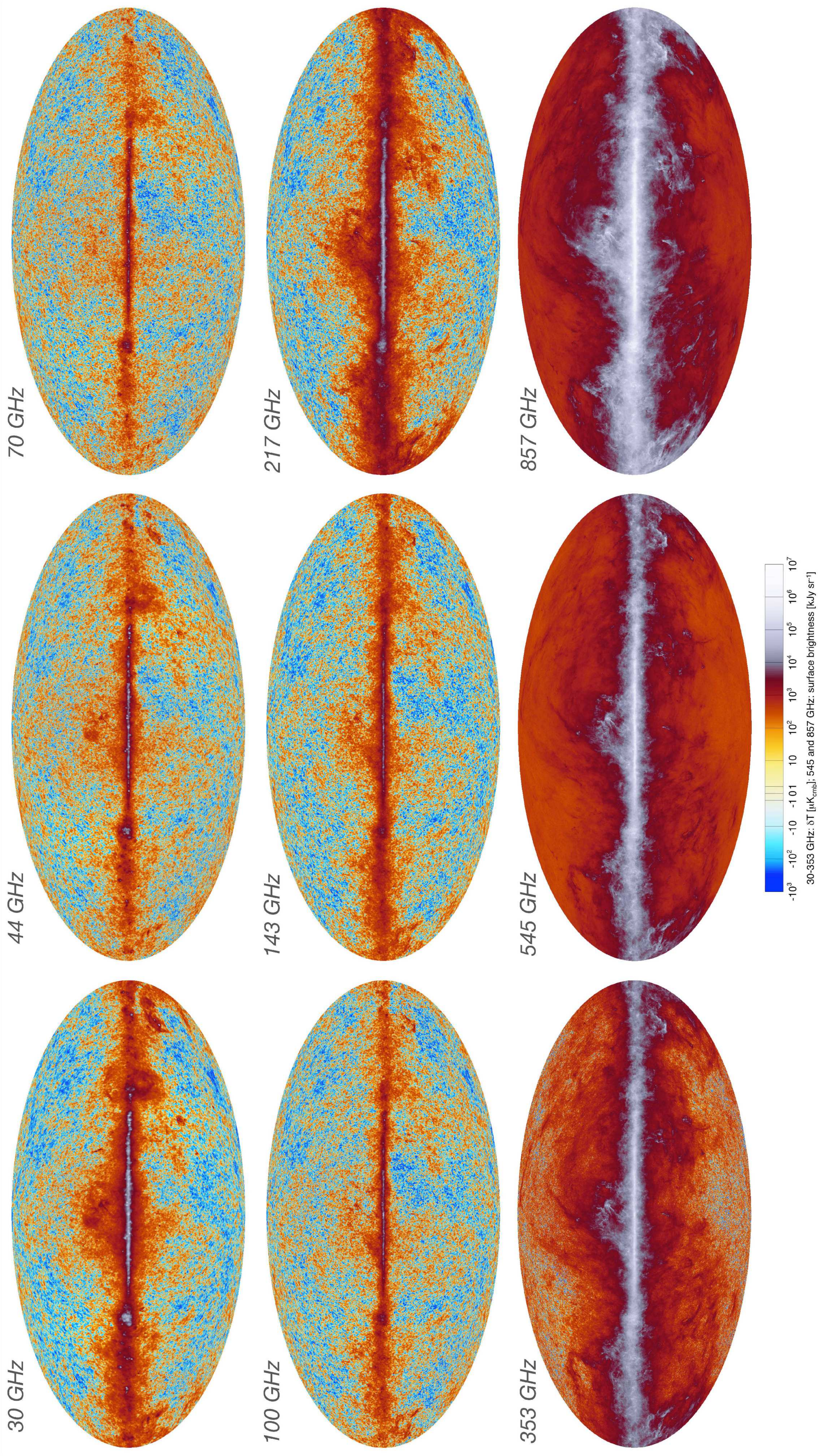}}
\end{center}
\caption{Fluctuations of sky emission in each of nine \Planck\ frequency bands, after removal of a common dipole component. The fluctuations are expressed as equivalent temperature variations at each of the seven lowest frequencies, so that fluctuations with a thermal spectrum will appear the same in each map (except for the effects of the varying resolution of the maps). The highest frequencies, which monitor the dust emission, are expressed in more conventional units.  
\label{fig:Ifreqmap} }
\end{figure*}

\begin{figure*}[!htbp]
\begin{center}
\resizebox{0.93\textwidth}{!}{\includegraphics{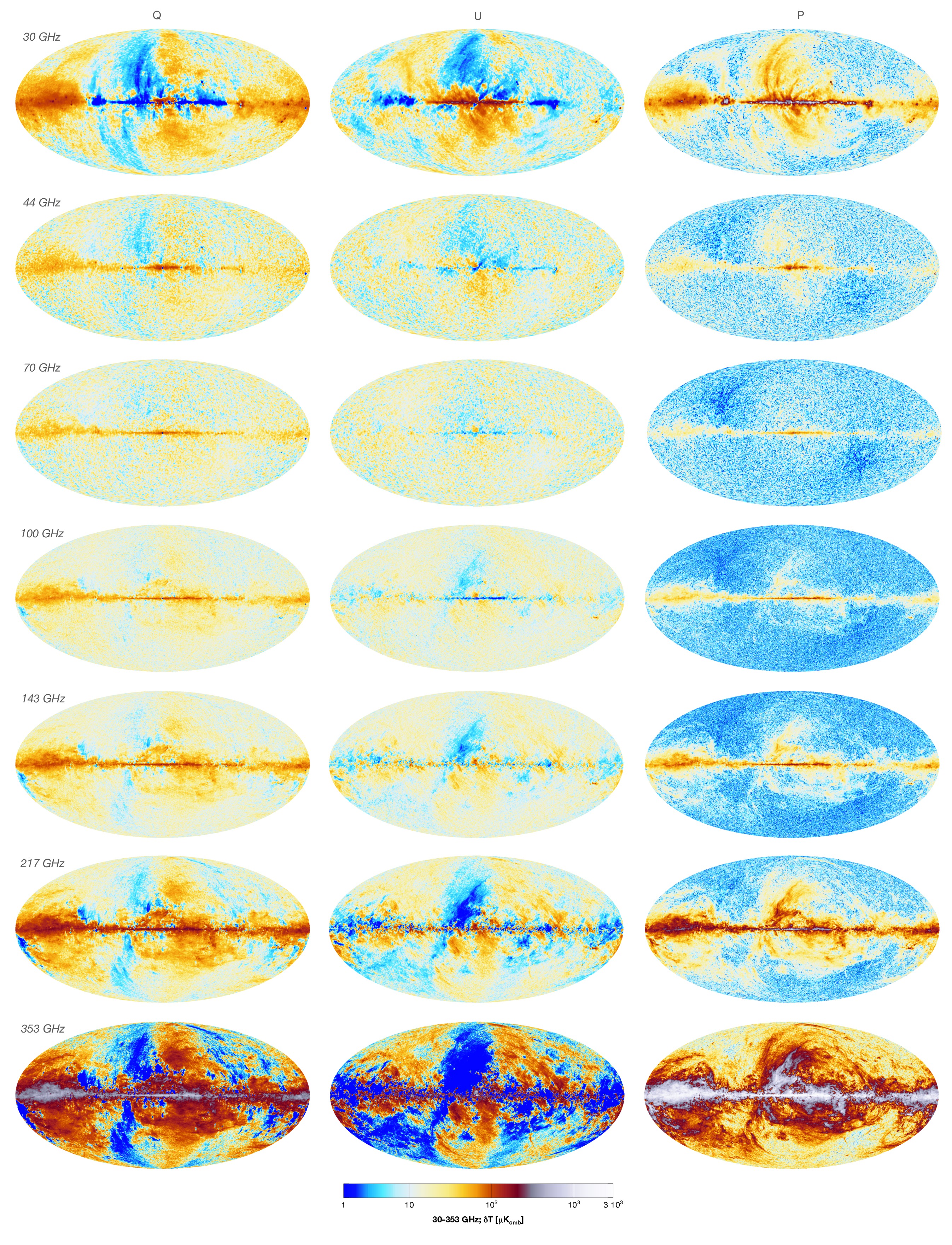}}
\end{center}
\caption{Sky polarization in seven polarized frequency bands of \Planck. The first two columns show the $Q$ and $U$ Stokes parameters; the last column indicates the polarized intensity, $P=\sqrt{Q^2+U^2}$ (although this emphasizes the strength of polarization in noisy regions). In addition to the rich science that they enable on their own, these maps set the baseline for all future CMB polarization experiments, for example by defining the most cosmologically challenged areas.  
\label{fig:Pfreqmap} }
\end{figure*}

Details about the \Planck\ mission and its scientific payload and performance 
have been discussed in previous publications
\citep[][and references therein]{planck2013-p01,planck2014-a01}.
\Planck\ was the first submillimetre mission to map the entire sky to
sub-Jansky sensitivity with angular resolution better than $10^\prime$.
In this section we describe the calibration and main properties of the
frequency maps (Figs.~\ref{fig:Ifreqmap} and \ref{fig:Pfreqmap}), and the methods used to separate the sky emission into
different components.  We briefly describe the main foreground components
before discussing the CMB anisotropies, whose characterization was the main
goal of the \Planck\ mission.

\subsection{The Solar dipole}

\begin{table*}[hbtp]
\newdimen\tblskip \tblskip=5pt
\caption{\COBE, \WMAP, LFI, HFI, and combined \Planck\ measurements of the Solar dipole.  The uncertainties are dominated by systematic effects, whose assessment is discussed in \cite{planck2016-l02} and \cite{planck2016-l03}.}
\label{tab:dipole}
\vskip -4mm
\footnotesize
\setbox\tablebox=\vbox{
 \newdimen\digitwidth
 \setbox0=\hbox{\rm 0}
 \digitwidth=\wd0
 \catcode`*=\active
 \def*{\kern\digitwidth}
  \newdimen\dpwidth
  \setbox0=\hbox{.}
  \dpwidth=\wd0
  \catcode`!=\active
  \def!{\kern\dpwidth}
\halign{\hbox to 4.5cm{#\leaderfil}\tabskip 2em&
    \hfil$#$\hfil \tabskip 2em&
    \hfil$#$\hfil \tabskip 2em&
    \hfil$#$\hfil \tabskip 0em\cr
\noalign{\doubleline}
\omit&&\multispan2\hfil\sc Galactic coordinates\hfil\cr
\noalign{\vskip -3pt}
\omit&\omit&\multispan2\hrulefill\cr
\noalign{\vskip 3pt} 
\omit&\omit\hfil\sc Amplitude\hfil&l&b\cr
\omit\hfil\sc Experiment\hfil&[\muK_{\rm
CMB}]&\omit\hfil[deg]\hfil&\omit\hfil[deg]\hfil\cr
\noalign{\vskip 3pt\hrule\vskip 5pt}
\COBE \rlap{$^{\rm a}$}&                  3358!**\pm24!**&     264.31*\pm0.20*&
     48.05*\pm0.11*\cr
\WMAP \rlap{$^{\rm b}$}&                  3355!**\pm*8!**&     263.99*\pm0.14*&
     48.26*\pm0.03*\cr
\Planck\ 2015 nominal \rlap{$^{\rm c}$}& 3364.5*\pm*2.0*&     264.00*\pm0.03*&
     48.24*\pm0.02*\cr
\noalign{\vskip 3pt}
LFI 2018 \rlap{$^{\rm d}$}&              3364.4*\pm*3.1*&     263.998\pm0.051&
     48.265\pm0.015\cr
HFI 2018 \rlap{$^{\rm d}$}&              3362.08\pm*0.99&     264.021\pm0.011&
     48.253\pm0.005\cr
\noalign{\vskip 3pt}
\bf \Planck\ 2018 \rlap{$^{\rm e}$}& \bf 3362.08\pm*0.99& \bf 264.021\pm0.011&
 \bf 48.253\pm0.005\cr
\noalign{\vskip 5pt\hrule\vskip 5pt}}}
\endPlancktablewide
\tablenote {{\rm a}} \citet{kogut1993,lineweaver1996}; we have added statistical and systematic uncertainty estimates linearly.\par
\tablenote {{\rm b}} \citet{hinshaw2009}.\par
\tablenote {{\rm c}} The 2015 \Planck\ ``nominal'' Solar dipole was chosen as a
plausible combination of the LFI and HFI 2015 measurements to subtract
the dipole from the 2018 frequency maps.  The difference compared with the
final determination of the dipole is very small for most purposes.\par
\tablenote {{\rm d}} Uncertainties include an estimate of systematic errors.
In the case of HFI, we have added statistical and systematic
errors linearly.\par
\tablenote {{\rm e}} The current best \Planck\ determination of the dipole is
that of HFI \citep{planck2016-l03}.  The central value for the direction
corresponds to ${\rm RA}=167\pdeg942\pm0\pdeg007$,
${\rm Dec}=-6\pdeg944\pm0\pdeg007$ (J2000). The uncertainties are the (linear) sum of the statistical and systematic uncertainties detailed in \citet{planck2016-l03}.  The uncertainty on the amplitude does not include the 0.02\% uncertainty on the temperature of the CMB monopole.\par
\end{table*}

We distinguish between two dipoles related to motion with respect to the
CMB rest frame.  The first is the ``Solar dipole,'' induced by the motion
of the Solar System barycentre with respect to the CMB.  The second is the
``orbital dipole,'' that is, the modulation of the Solar dipole induced by the
orbital motion of the satellite around the Solar System barycentre.
The orbital velocity is known exquisitely well, and hence the induced dipole
in $\Delta T/T$ units; this means that the accuracy of the predicted orbital
dipole is ultimately limited by the accuracy with which we know the
temperature of the CMB.
In the 2015 data release, photometric calibration from 30 to 353\,GHz was
based on the ``orbital dipole''.  This allowed us to measure the
amplitude and direction of the ``Solar dipole'' on the calibrated maps of
individual detectors, at frequencies where the CMB is the dominant signal
(70 to 353\,GHz).
The dipole parameters measured in 2015 were significantly more accurate than
the previous best measurements provided by \WMAP\ (see Table~\ref{tab:dipole}).
However, comparison of individual detector determinations showed clear
indications of the presence of small residual systematic effects
\citep{planck2014-a03,planck2014-a09}. 
The dipole amplitude and direction showed shifts with position in the focal
plane for LFI; for HFI the shifts were associated with frequency, as well as
with the Galactic mask and the component-separation method used, indicating
the presence of dipolar and quadrupolar residuals after removal of the dust
and CMB anisotropies.

In 2018, both instruments have achieved a significant reduction in the levels
of residual systematic effects (especially at the largest angular scales where
the dipole signals are present) and in the case of HFI also in the accuracy of
photometric calibration.
Furthermore, the HFI dust foreground effect was identified with large-scale
(mostly quadrupolar) spectral energy distribution changes.
Correcting these brought full consistency between frequencies, as well as
for detectors within each frequency band.
This has resulted in dramatic improvement in the determination of the 2018
Solar dipole parameters, which are presented in Table~\ref{tab:dipole}.
The independent LFI and HFI measurements are fully consistent with each other
and with those of \WMAP, and, as described in \citet{planck2016-l02} and
\citet{planck2016-l03}, they are no longer significantly affected by
systematic effects (in the sense that the results are consistent between
frequencies, sky fractions, and component-separation methods used, although the uncertainties are not purely statistical).
Considering that the uncertainties in the HFI determination are much lower than
those of LFI, we recommend that users adopt the HFI determination of the
Solar dipole as the most accurate one available from \Planck.

In the 2018 maps, the 2015 ``nominal'' Solar dipole, which is slightly different than the final best dipole, has been subtracted.
(The induced quadrupole has also been subtracted from the maps.)
This was done in order to produce a consistent data set that is independent
of the best determination of the dipole parameters, which was made at a later time
separately at each individual frequency.  This implies that a very small,
residual Solar dipole is present in all released maps.  This can be
removed if desired using the procedure described in \citet{planck2016-l03}.

The Solar dipole can still be measured with high
signal-to-noise ratio at 545\,GHz. The 545-GHz data were not calibrated on the
orbital dipole, however, but instead on observations of Uranus and Neptune
\citep{planck2016-l03}.
Therefore the photometric accuracy of this calibration is
limited by that of the physical emission model of the planets, to a level
of approximately 5\,\%. However, the dispersion of the Solar dipole amplitude
measured in individual 545-GHz detector maps is within 1\,\% of that
at lower frequencies. This implies that, in actual fact, the planet model can
be calibrated on this measurement more precisely than has been assumed so far
\citep{planck2016-LII}.
It also means that an improved model can be extended to recalibrate the
857\,GHz channel.  These improvements have not been implemented in the 2018
release.

The amplitude of the dipole provides a constraint for building a picture of the
local large-scale structure, through the expected convergence of bulk-flow
measurements for galaxies \citep[e.g.,][]{Scrimgeour2016}.
The new best-fit dipole amplitude is known more precisely than the CMB
monopole, and even when we fold in an estimate of systematic uncertainties it
is now known to about 0.025\,\% (essentially the same as the monopole).
The dipole amplitude corresponds to
$\beta\equiv v/c = (1.23357\pm 0.00036)\times10^{-3}$ or $v=(369.82\pm0.11)
\,{\rm km}\,{\rm s}^{-1}$, where we have added in the systematic
uncertainties linearly.  When giving the amplitude of the dipole in temperature units,
one should also include the uncertainty in $T_0$.

The Solar dipole direction lies just inside the
little-known constellation of Crater (near the boundary with Leo).
The error ellipse of \Planck's dipole direction (a few arcsec in radius, or
around 30\arcs\ including systematic uncertainties) is so
small that it is empty in most published astronomical catalogues.
We discuss the cosmological implications of the dipole in
Sect.~\ref{sec:dipole}.

The Sun's motion in the CMB frame is not the only relative velocity of
interest, and indeed from a cosmological perspective more relevant would be
the motion of the centre of our Galaxy relative to the CMB or the motion of
our group of galaxies relative to the CMB.  The
peculiar motion of the Local Group is well known to have a larger
speed than that of the Sun relative to the CMB, due to the roughly anti-coincident
direction of our rotation around the Galaxy.  It is this larger peculiar
velocity that has been the focus of studies to explain the origin of the
motion in the context of structures in our extragalactic neighbourhood
\citep[e.g.,][]{lyndenbell1988,tully2008}.
Estimates of the corrections required to obtain the motion of
the Galactic centre relative to the CMB and the motion of the centre of mass
of the Local Group relative to the CMB were given
by \citet{kogut1993}, and have seldom been revisited since then.  We summarize
more modern determinations in Table~\ref{tab:frames}.

Firstly, we take the estimate of the Sun's motion relative
to the Local Standard of Rest from \citet{schonrich2010}, which uses nearby
stars, and the estimate of the motion of the LSR around the centre of the Milky
Way from \cite{mcmillan2011}, which combines studies of larger-scale Galactic
dynamics.  These can be subtracted from the Solar dipole to give the velocity
of the Galactic centre relative to the CMB, as presented in the fourth line
of Table~\ref{tab:frames}.

Secondly, we take the estimate of the Sun's velocity relative to the
centre of the Local Group from \citet{diaz2014}, found by averaging velocities
of members galaxies \citep[as also performed by several other
studies, e.g.,][]{yahil1977,courteau1999,mikulizky2015}.
This vector can be subtracted from the Solar dipole velocity to derive the
velocity of the Local Group relative to the CMB.  The value is
$(620\pm15)\,{\rm km}\,{\rm s}^{-1}$ in a direction (known to about a couple
of degrees) that lies about $30\deg$ above the Galactic plane and is nearly
opposite in latitude to the direction of Galactic
rotation.  The uncertainty in the Local Group's
speed relative to the CMB is almost entirely due to the uncertainty in the 
speed of the Sun relative to the centre-of-mass of the Local Group.

\begin{table}[hbtp!]
\newdimen\tblskip \tblskip=5pt
\caption{Relative velocities involving the CMB frame, the Galactic centre, and
the Local Group.}
\label{tab:frames}
\vskip -4mm
\footnotesize
\setbox\tablebox=\vbox{
 \newdimen\digitwidth
 \setbox0=\hbox{\rm 0}
 \digitwidth=\wd0
 \catcode`*=\active
 \def*{\kern\digitwidth}
 \newdimen\signwidth
 \setbox0=\hbox{+}
 \signwidth=\wd0
 \catcode`!=\active
 \def!{\kern\signwidth}
 \newdimen\pointwidth
 \setbox0=\hbox{\rm .}
 \pointwidth=\wd0
 \catcode`?=\active
 \def?{\kern\pointwidth}
\halign{\hbox to 2.5cm{#\leaderfil}\tabskip 0.1em&
 \hfil#\hfil\tabskip 1.0em&
 \hfil#\hfil&\tabskip 0pt&
 \hfil#\hfil\tabskip 0pt\cr
\noalign{\doubleline}
\omit\hfil Relative\hfil& **Speed& $**l$& $*b$\cr
\omit\hfil velocity\hfil& **[${\rm km}\,{\rm s}^{-1}$]& *[deg]& **[deg]\cr
\noalign{\vskip 3pt\hrule\vskip 5pt}
Sun--CMB \rlap{$^{\rm a}$}& $369.82\pm0.11$& $264.021\pm0.011$& $48.253\pm0.005$\cr
\noalign{\vskip 8pt}
Sun--LSR \rlap{$^{\rm b}$}& $*17.9\pm2.0$& $*48\pm7$& $23\pm4$\cr
LSR--GC \rlap{$^{\rm c}$}& $239\pm5$& **90& *0\cr
GC--CMB \rlap{$^{\rm d}$}& $565\pm5$& $265.76\pm0.20$& $28.38\pm0.28$\cr
\noalign{\vskip 8pt}
Sun--LG \rlap{$^{\rm e}$}& $*299\pm15$& $*98.4\pm3.6$& $*-5.9\pm3.0!$\cr
LG--CMB \rlap{$^{\rm d}$}& $*620\pm15$& $271.9\pm2.0$& $*29.6\pm1.4*$\cr
\noalign{\vskip 5pt\hrule\vskip 5pt}}}
\endPlancktable
\tablenote {{\rm a}} Velocity of the Sun relative to the CMB; \Planck\ 2018.\par
\tablenote {{\rm b}} Velocity of the Sun relative to the Local Standard
of Rest from \citet{schonrich2010}, adding statistical and systematic
uncertainties.\par
\tablenote {{\rm c}} Rotational velocity of the LSR from
\cite{mcmillan2011}.\par
\tablenote {{\rm d}} Resulting velocity, using non-relativistic velocity
addition and assuming uncorrelated errors.\par
\tablenote {{\rm e}} Velocity of the Sun relative to the Local Group from
\citet{diaz2014}.\par
\end{table}

\subsection{Frequency maps and their properties}\label{sec:numaps}

\begin{table*}[htbp]
\caption{Main characteristics of \Planck\ frequency maps. 
\label{tab:planckmapsummary} }
\begingroup
\newdimen\tblskip \tblskip=5pt
\nointerlineskip
\vskip -3mm
\tiny
\setbox\tablebox=\vbox{
   \newdimen\digitwidth 
   \setbox0=\hbox{\rm 0} 
   \digitwidth=\wd0 
   \catcode`*=\active 
   \def*{\kern\digitwidth}
   \newdimen\signwidth 
   \setbox0=\hbox{+} 
   \signwidth=\wd0 
   \catcode`!=\active 
   \def!{\kern\signwidth}
\halign{
\hbox to 6.5 cm{#\leaderfil}\tabskip 1.0em&
\hfil #\hfil&
\hfil #\hfil&
\hfil #\hfil&
\hfil #\hfil&
\hfil #\hfil&
\hfil #\hfil&
\hfil #\hfil&
\hfil #\hfil&
\hfil #\hfil\tabskip 0pt\cr
\noalign{\doubleline}
\omit&\multispan9\hfil Frequency [GHz]\hfil\cr
\noalign{\vskip -2pt}
\omit&\multispan9\hrulefill\cr
\omit\hfil Property\hfil&
\omit\hfil 30\hfil&
\omit\hfil 44\hfil&
\omit\hfil 70\hfil&
\omit\hfil 100\hfil&
\omit\hfil 143\hfil&
\omit\hfil 217\hfil&
\omit\hfil 353\hfil&
\omit\hfil 545\hfil&
\omit\hfil 857\hfil\cr
\noalign{\vskip 3pt\hrule\vskip 5pt}
Frequency [GHz]\tablefootmark{\rm a}& 28.4& 44.1& 70.4 & 100& 143& 217& 353& 545& 857\cr
Effective beam FWHM [arcmin]\tablefootmark{\rm b}& 32.29& 27.94& 13.08 & 9.66& 7.22& 4.90& 4.92& 4.67& 4.22\cr
Temperature noise level [$\mu\mathrm{K_{CMB}}$\,deg]\tablefootmark{\rm c}& 2.5 & 2.7 & 3.5 & 1.29& 0.55& 0.78& 2.56& & \cr
\phantom{Temperature Sensitivity} [kJy sr$^{-1}$\,deg]\tablefootmark{\rm c}& & & & & & & & 0.78& 0.72\cr
Polarization noise level [$\mu\mathrm{K_{CMB}}$\,deg]\tablefootmark{\rm c} & 3.5& 4.0& 5.0& 1.96& 1.17& 1.75& 7.31& & \cr

Dipole-based calibration uncertainty [$\%$]\tablefootmark{\rm d}& 0.17& 0.12 & 0.20& 0.008& 0.021& 0.028& 0.024& $\sim$1& \cr
Planet submm inter-calibration accuracy [$\%$]\tablefootmark{\rm e}&\dots&\dots&\dots&\dots&\dots&\dots&\dots& \dots&$\sim$3\cr
Temperature transfer function uncertainty [$\%$]\tablefootmark{\rm f}& 0.25& 0.11 &Ref. & Ref. & 0.12& 0.36& 0.78& 4.3& \cr
Polarization calibration uncertainty [$\%$]\tablefootmark{\rm g}&$<0.01\,\%$ & $<0.01\,\%$ & $<0.01\,\%$ & 1.0& 1.0 & 1.0&\dots &\dots & \dots\cr
Zodiacal emission monopole level [$\mu\mathrm{K_{CMB}}$]\tablefootmark{\rm h} & 0& 0& 0&0.43& 0.94& 3.8& 34.0& \dots& \dots\cr
\phantom{Zodiacal emission monopole level} [MJy\,sr\mo]\tablefootmark{\rm h}&\dots&\dots&\dots&\dots&\dots&\dots&\dots& 0.04& 0.12\cr
LFI zero level uncertainty [$\mu\mathrm{K}_\mathrm{CMB}$]\tablefootmark{\rm i}& $\pm 0.7$& $\pm 0.7$& $\pm 0.6$&\dots&\dots&\dots&\dots&\dots&\dots\cr
HFI Galactic emission zero level uncertainty [MJy\,sr\mo]\tablefootmark{\rm j}&\dots&\dots&\dots& $\pm0.0008$& $\pm0.0010$& $\pm0.0024$& $\pm0.0067$& $\pm0.0165$& $\pm0.0147$\cr
HFI CIB monopole assumption [MJy\,sr\mo]\tablefootmark{\rm k}& & & &$\hphantom{\pm}0.0030$& $\hphantom{\pm}0.0079$&!0.033*&!0.13**& !0.35**& !0.64**\cr
HFI CIB zero level uncertainty [MJy\,sr\mo]\tablefootmark{\rm l}&\dots&\dots&\dots& $\pm0.0031$& $\pm0.0057$& $\pm0.016$*& $\pm0.038$*& $\pm0.066$*& $\pm0.077$*\cr
\noalign{\vskip 5pt\hrule\vskip 3pt}}}
\endPlancktablewide
\tablenote {{\rm a}} For LFI channels (30--70\,GHz), this is the centre frequency. For HFI channels (100--857\,GHz), it is a reference (identifier) frequency. \par
\tablenote {{\rm b}} Mean FWHM of the elliptical Gaussian fit of the effective beam.  \par
\tablenote {{\rm c}} Estimates of noise in intensity and polarization scaled to $1\deg$ assuming that the noise is white. These levels are unchanged from 2015. \par
\tablenote {{\rm d}} Absolute calibration accuracy obtained using the measurement of the Solar dipole at $\ell=1$. \par
\tablenote {{\rm e}} The 857-GHz channel retains the 2015 planet calibration, and the accuracy is calculated a posteriori using a model of planet emission
\citep{2017A&A...607A.122P} and the 545-GHz data.\par
\tablenote {{\rm f}} For LFI this is the ratio of 30- and 44-GHz half-ring cross-spectra in the range $\ell\simeq50$--850 to that of the 70-GHz cross-spectrum. For HFI it is the upper limit derived from the levels of the first three CMB acoustic peaks ($\ell\simeq15$--1000), relative to the 100\,GHz channel.\par
\tablenote {{\rm g}}  Additional calibration uncertainty applicable to $Q$ and $U$ only. For LFI, the additional uncertainty (based on simulations) is negligible. For HFI, the dominant inaccuracy is the knowledge of the polarization efficiency, which is currently derived from the relative levels of the first three CMB acoustic peaks ($\ell \simeq 15$--1000), in combination with a prediction of the best-fit $TT$-based cosmology. The best estimates \citep{planck2016-l03} indicate that a bias should be applied to the maps of 0.7, $-$1.7, and 1.9\,\%,
at 100, 143, and 217\,GHz, respectively, with an uncertainty as indicated in this table. \par
\tablenote {{\rm h}} Average contribution of the zodiacal emission to the monopole. As the level of this emission is dependent on the time of observation, it has been removed from the frequency maps during processing. \par
\tablenote {{\rm i}} Estimated uncertainty in the zero levels associated with Galactic emission. The zero levels were set by fitting a model of Galactic emission that varies as the cosecant of the latitude to the maps after CMB subtraction. The levels subtracted were 11.9, $-$15.4, and $-$35.7 $\mu\mathrm{K_{CMB}}$ at 30, 44, and 70\,GHz, respectively. \par
\tablenote {{\rm j}} The zero levels of the HFI maps are set by correlating the Galactic emission component to a map of the diffuse \ion{H}{i} column density, as in \citet{planck2013-p03f}. The uncertainties in the estimated zero levels are unchanged since 2013.  \par
\tablenote {{\rm k}} Once the Galactic zero level has been set, the monopole of the \citet{Bethermin_2012} CIB model has been added to the frequency maps. \par
\tablenote {{\rm l}} The estimated uncertainty of the CIB monopole that has been added to the maps.\par
\endgroup
\end{table*}

The Low and High Frequency Instruments together contained an array of
74 detectors in nine bands, covering frequencies between 25 and $1000\,$GHz,
imaging the whole sky twice per year with angular resolution between
$33^\prime$ and $5^\prime$.
Table~\ref{tab:planckmapsummary} gives the main characteristics of
the \Planck\ frequency maps, including angular resolution and sensitivity.

An extensive series of null tests for the consistency of the maps is
provided in \citet{planck2013-p01a}, \citet{planck2014-a01},
\citet{planck2016-l02}, and \citet{planck2016-l03}.
We find impressive consistency between the maps.
Consistency of absolute calibration across the nine frequency channels
is discussed extensively in the same papers, and we discuss inter-instrument
consistency in Appendix~\ref{sec:inst-consistency}. Some considerations about the principles followed in the \Planck\ analysis (including a discussion of
blinding) are given in Appendix~\ref{sec:blinding}.
For the main CMB channels (70--$217\,$GHz) the inter-calibration is
at the level of 0.2\,\% \citep{planck2014-a01}.
At $100\,$GHz, the absolute photometric calibration on large scales is
an astounding $0.008\,\%$.
For the HFI polarization maps, the largest source of uncertainty is the
polarization efficiency (Table~\ref{tab:planckmapsummary}).

The beams are estimated from planetary observations, and the polarized
beam models are combined with the specific scanning strategy to
generate ``effective beams,'' which describe the relation of maps to the
sky \citep[see][]{planck2014-a05,planck2014-a08}.  
The response in harmonic space is known as a window function, and both
the mean windows and the major error eigenmodes are provided in 
the PLA.
Typical uncertainties are well below 0.1\,\% for the main CMB channels.

Figures~\ref{fig:Ifreqmap} and \ref{fig:Pfreqmap} show views of the sky
as seen by \Planck\ in intensity and polarization.
\Planck\ uses {\tt HEALPix} \citep{gorski2005} as its pixelization
scheme, with resolution labelled by the $N_{\rm side}$ value.
In {\tt HEALPix} the sphere is divided into $12\,N_{\rm side}^2$ pixels.
At $N_{\rm side}\,{=}\,2048$, typical of \Planck\ maps, their mean spacing is
1\parcm7.
Each panel in Fig.~\ref{fig:Ifreqmap} shows the intensity in one of
\Planck's nine frequency channels, in Galactic coordinates.
In all cases the figures are unable to convey both the angular resolution
and the dynamic range of the \Planck\ data.
However, they serve to show the major features of the maps and the numerous
astrophysical components that contribute to the signal.
Similarly, Fig.~\ref{fig:Pfreqmap} shows the polarization properties measured
by \Planck\ at seven frequencies.  The CMB component of the maps has a
6\% linear polarization, though the foregrounds exhibit differing polarization
levels as a function of frequency.

The most prominent feature in the maps is the Galactic plane, steadily
brightening to both higher (where Galactic dust dominates the emission) and lower (where synchrotron and free-free emission
dominate) frequencies.  At high Galactic latitudes, and over much of the sky between
70 and 217\,GHz, the signal is dominated by the ``primary'' CMB
anisotropies, which were frozen in at the surface of last scattering
and provide the main information constraining our cosmological model.

To be more quantitative, it is useful to introduce two-point statistics, in the form of a two-point angular correlation function, or its harmonic-space counterpart, the angular power spectrum. We follow the usual convention and perform an harmonic decomposition of the sky maps.  
If $T$, $Q$, and $U$ represent the intensity and
polarization\footnote{\Planck\ uses the ``COSMO'' convention for
polarization (corresponding to the FITS keyword ``POLCCONV''),
which differs from the IAU convention often adopted for astrophysical
data sets \citep{planck2016-ES}.}
Stokes parameters (in thermodynamic temperature units), then we define
\begin{eqnarray}
  a_{\ell m} &=& \int d\hn\
  Y_{\ell m}^\ast(\hn)\,T(\hn), \\
  a^E_{\ell m}\pm ia^B_{\ell m} &=&
  \int d\hn\ {*}_{\pm 2}Y_{\ell m}^\ast(\hn)
  \,\left(Q \pm iU\right)(\hn),
\end{eqnarray}
where ${}_{\pm 2}Y_{\ell m}$ are the spin-spherical harmonics, which
are proportional to Wigner $\mathcal{D}$-functions\footnote{See e.g., \href{https://en.wikipedia.org/wiki/Wigner_D-matrix}{Wikipedia}.}.
The polarization is defined through the scalar $E$ and pseudo-scalar $B$
fields, which are non-local, linear combinations of $Q$ and $U$
\citep{ZalSel97,Kam97,HuWhi97,Dodelson03}.
For small patches of sky, $E$ and $B$ are simply $Q$ and $U$ in the
coordinate system defined by the Fourier transform coordinate $\vec{\ell}$
\citep{Sel97}.
Alternatively, near a maximum of the polarization the direction of greatest
change for an $E$ mode is parallel or perpendicular to the polarization
direction (see Fig.~\ref{fig:CMB-TPmap}).

When statistical isotropy may be assumed, it demands that $\langle a_{\ell m}^\ast a_{\ell' m'}\rangle$
be diagonal and depend only on $\ell$.  We write
\begin{equation}
  \left\langle a_{\ell m}^{T\ast}\, a_{\ell' m'}^T\right\rangle =
  C_\ell^{TT}\ \delta_{\ell'\ell}\,\delta_{m'm}\,,
\end{equation}
and similarly for $TE$, $EE$, $BB$, etc.
We find it convenient to define
\begin{equation}
  \mathcal{D}_\ell^{XY} = \frac{\ell(\ell+1)C_\ell^{XY}}{2\pi},
\end{equation}
which we will often refer to as the angular power spectrum.
An auto-spectrum, $\mathcal{D}_\ell^{XX}$ indicates
the approximate contribution per logarithmic interval of multipoles centred on
$\ell$ to the variance of the fluctuation, that is, the 2-point correlation
function at zero lag.  It thus captures the relative importance of various
contributions to the signal as a function of scale.  

\begin{figure*}[htpb]
\begin{center}
\resizebox{0.9\textwidth}{0.395\textwidth}{\includegraphics{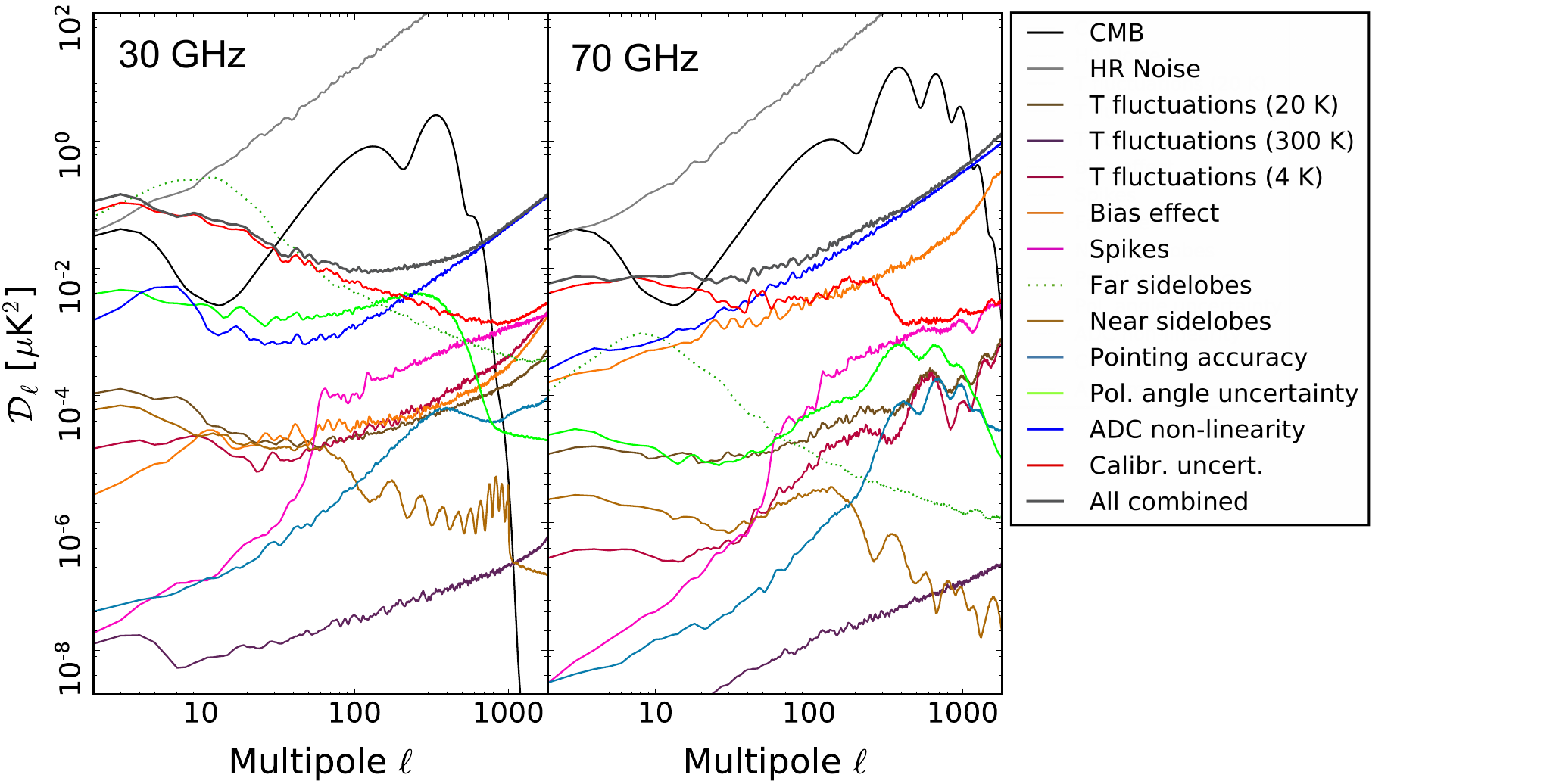}}
\resizebox{0.9\textwidth}{0.395\textwidth}{\includegraphics{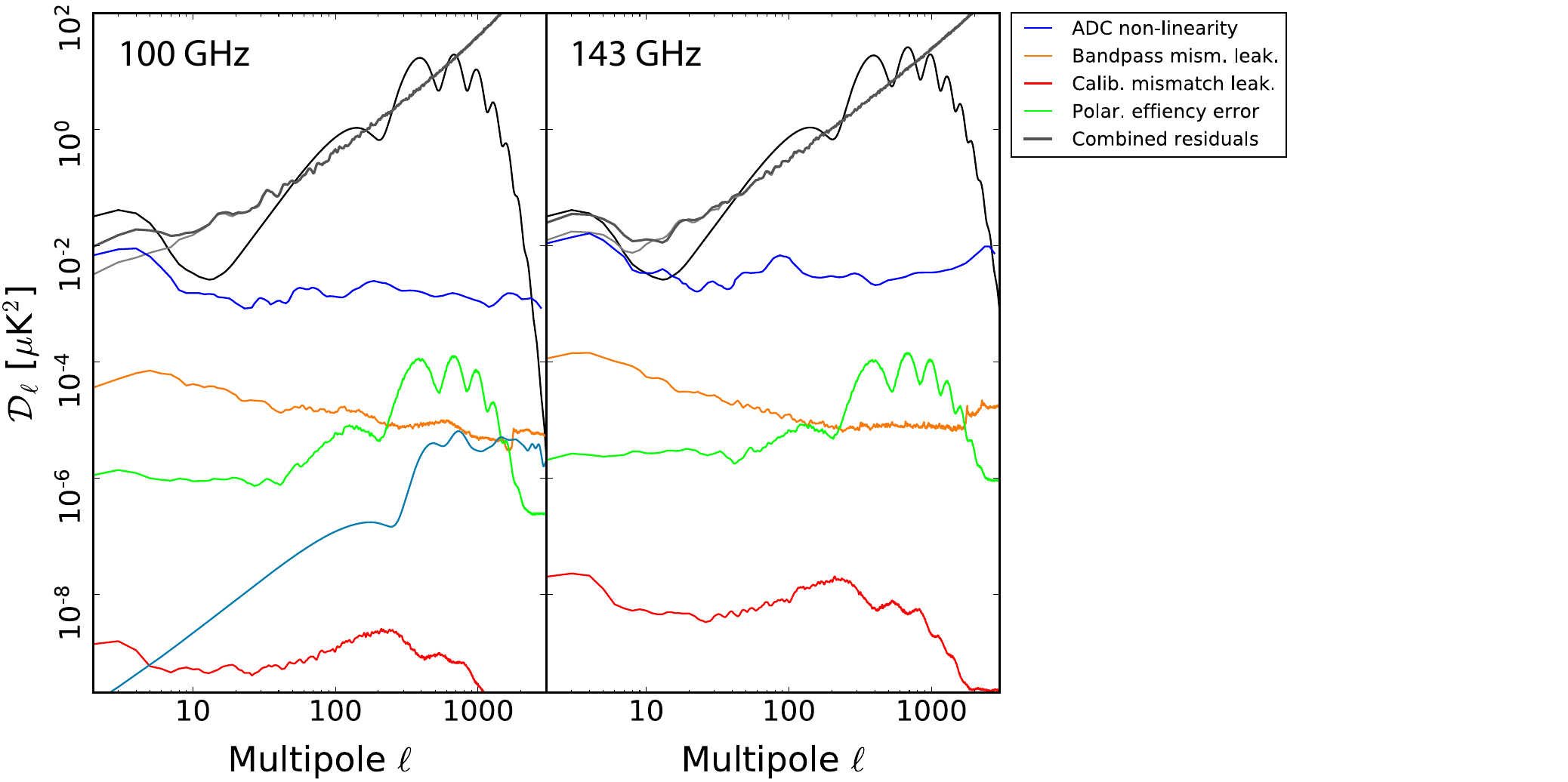}}
\resizebox{0.9\textwidth}{0.395\textwidth}{\includegraphics{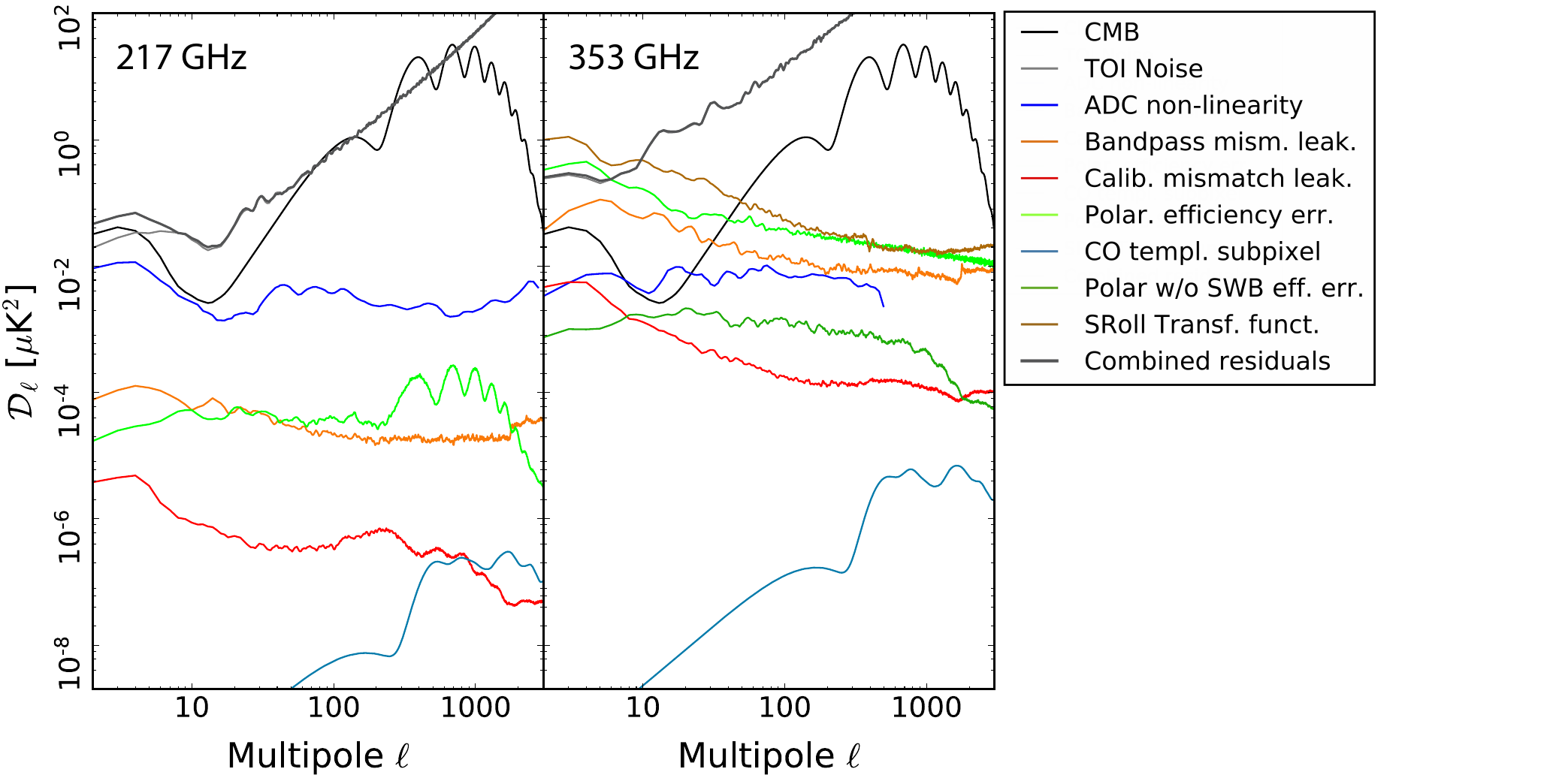}}
\end{center}
\caption{Estimates of the residual polarization systematic effects and noise across the core CMB channels at 70--217\,GHz and two adjacent foreground-monitoring channels at 30 and 353\,GHz. The residual systematics $E$-mode auto-power spectra are compared to that of the CMB signal after convolution with the beam window function at that frequency (noting that the CMB contribution to the total signal is small in the foreground-monitoring channels). The top panel displays the 30- and 70-GHz channels of the LFI instrument, with specific systematic effects colour-coded in the accompanying legend. The middle and lower panels show the HFI estimates at 100 and 143\,GHz, and at 217 and 353\,GHz, respectively.
\label{fig:syst_sims} }
\end{figure*}

Figure~\ref{fig:syst_sims} shows the estimated levels of CMB and residual systematics in frequency maps as a function of scale. The plots show 
the $E$-mode power spectrum, $\mathcal{D}_\ell^{EE}$, for all core CMB channels at 70, 100, 143, and 217\,\GHz, and at the adjacent 30- and 350-GHz channels, which are of particular use for understanding foregrounds. At the largest scales, the residual systematics are comparable to the noise level, which is itself close to the low level of the reionization bump determined by \Planck\ (see Sect.~\ref{sec:reionization}). This points to the great challenge of this measurement.  At small scales, residual systematic effects are significantly smaller than the signal and the noise in the main CMB channels.  This figure summarizes most of the data-processing work by the \Planck\ collaboration, in the sense that it embodies the final quantitative understanding of the measurements and their processing. This determines what has to be included in faithful end-to-end simulations.  

The all-sky, fully calibrated maps of sky intensity and polarization, shown in
Figs.~\ref{fig:Ifreqmap} and \ref{fig:Pfreqmap}, together with their detailed instrumental characterization and simulations, are the main legacy of the \Planck\ mission and will be a resource to multiple communities for addressing numerous science questions in decades to come.
In the next few sections, we discuss the separation of the maps into their
physical components and then the cosmological consequences that can be
derived from the CMB anisotropies.

\subsection{Component separation} \label{sec:components}

In addition to the primary anisotropies that are the main focus of
the \Planck\ mission, the sky emission contains many other astrophysical
components, which differ by their dependence on frequency as well as
their spatial properties.
By making measurements at multiple frequencies, spanning the peak of the
CMB blackbody spectrum, we are able to characterize the foregrounds and
reduce their contamination of the primary CMB anisotropies to unprecedented
levels.  

In order to separate the maps into their contributing signals and to clean
the CMB map from foregrounds, we have used four different approaches, as we
did in earlier releases \citep{planck2013-p06,planck2014-a11}.
The four approaches were initially selected as a representative of a
particular class of algorithm (blind, non-blind, configuration-space, 
and harmonic-space methods).  They were also checked with a common series of map
simulations, the last test being blind (and actually used to select a baseline). 
Combined, they represent most of the methods proposed in the literature.
They are:
\begin{itemize}
\item \commander, a pixel-based parameter and template fitting procedure
\citep{eriksen2008,planck2014-a12};
\item \NILC, a needlet-based internal linear combination approach
\citep{BasDel13};
\item \SEVEM, which employs template fitting \citep{leach2008,fernandez2012};
and
\item \SMICA, which uses an independent component analysis of power spectra
\citep{planck2016-l04}.
\end{itemize}
In addition we employ the {\tt GNILC} algorithm \citep{GNILC}
to extract high (electromagnetic) frequency foregrounds.

Each method produces: CMB maps in Stokes $I$, $Q$, and $U$;
confidence maps (i.e., masks); an effective beam; and a noise estimate map,
together characterizing the CMB.
The differences between the four maps can be used as an estimate of the
uncertainty in the recovery of the CMB, and is reassuringly small
\citep{planck2016-l04}.
These CMB maps and accompanying simulations are the basic input for all
analyses of homogeneity, stationarity, and Gaussianity of the CMB fields
\citep{planck2016-l07,planck2016-l08,planck2016-l09}.

For this release, the primary objective of the component-separation process
was to obtain the best possible polarization maps.
The steps taken to ensure high-fidelity polarization maps (especially at
100--353\,GHz) are described in detail in \citet{planck2016-l03};
see also Appendix~\ref{sec:changes}.
Some of the choices made for the sake of polarization compromised to some
extent the accuracy of the temperature maps; advice on how to use the
temperature maps is contained within \citet{planck2016-l03}.
The \Planck\ 2018 data release does not include new foreground reconstructions
in intensity, since the improved HFI processing regarding bandpass leakage 
requires new methodological developments in other areas that are not yet available
(see Appendices~\ref{sec:improvements} and \ref{sec:mapimprovements}).

Even with these compromises, the foreground maps produced by \Planck\ in
this and the 2015 release are a treasure trove for many areas of astrophysics,
including the study of the Galactic interstellar medium
\citep[see, e.g.,][]{planck2016-l11A,planck2016-l11B}, the cosmic
infrared background \citep[CIB;][]{planck2013-pip56},
and the Sunyaev-Zel'dovich (SZ) effect \citep{SunZel72,SunZel80}.
SZ-related science results from \Planck\ are reported in, for example
\citet{planck2014-a28} and \citet{planck2014-a30}.

\begin{figure*}[hbtp]
\resizebox{0.49\textwidth}{!}{\includegraphics{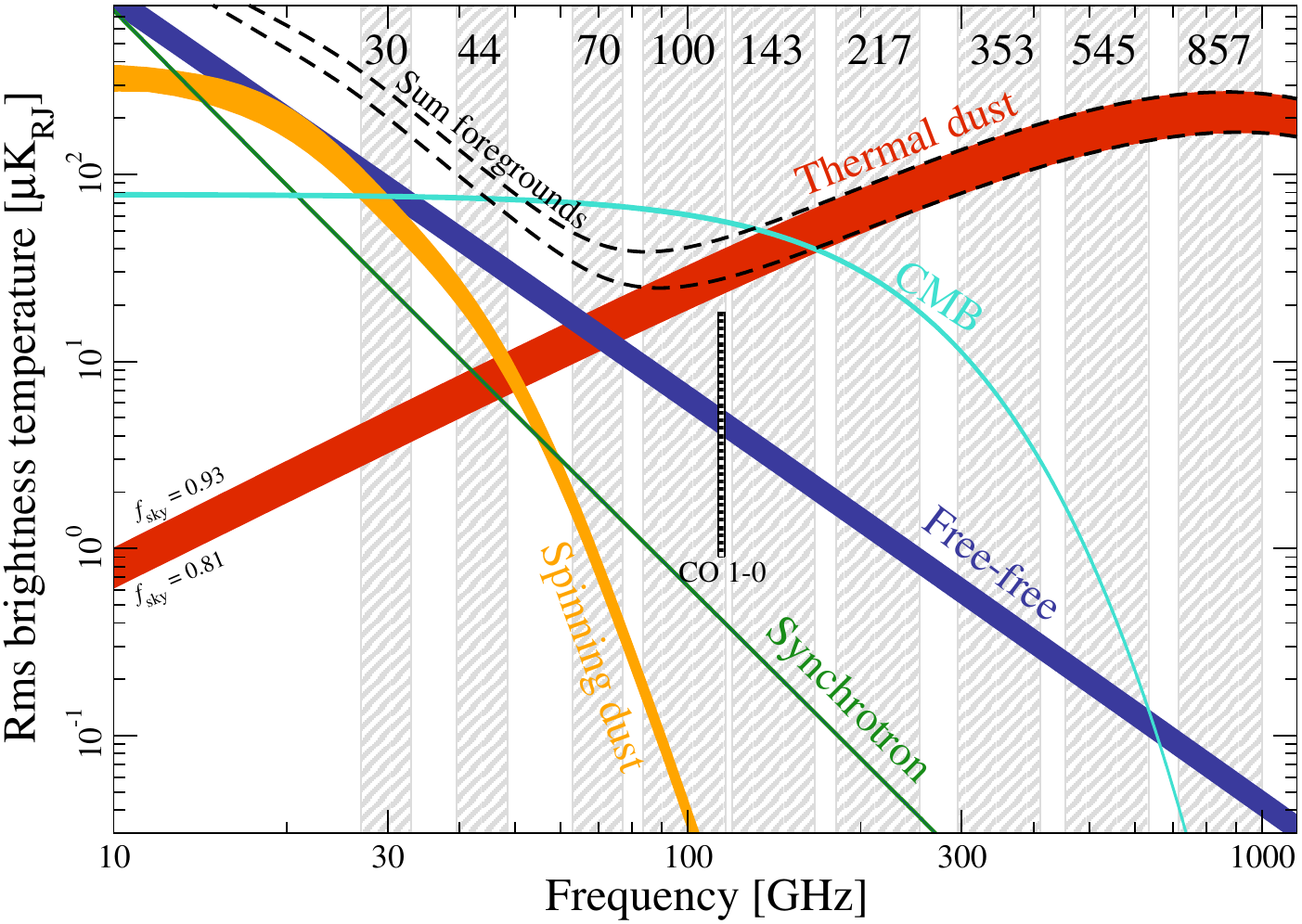}}
\resizebox{0.49\textwidth}{!}{\includegraphics{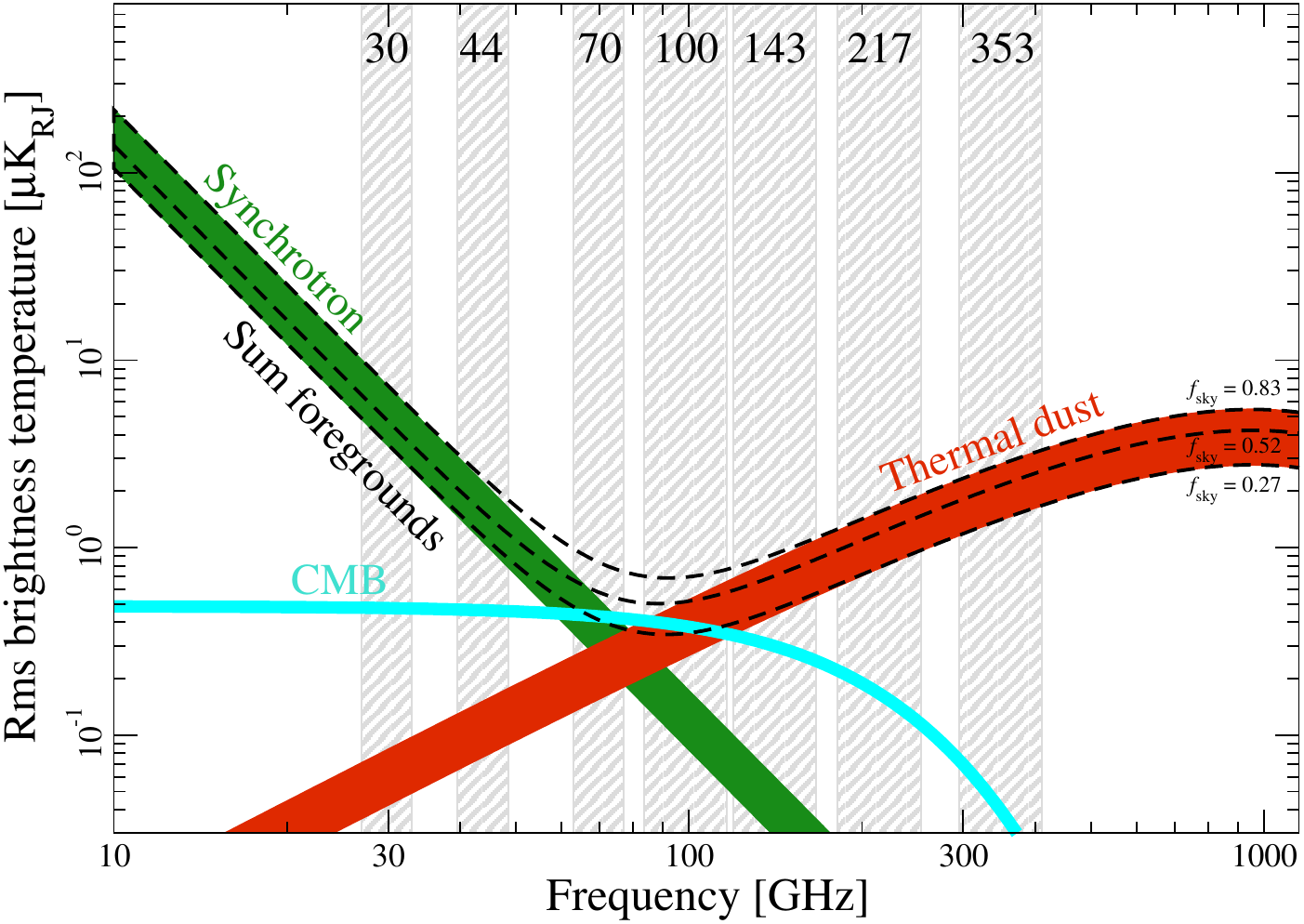}}
\caption{Frequency dependence of the main components of the submillimetre sky
in temperature (left) and polarization (right; shown as $P=\sqrt{Q^2+U^2}$).
The (vertical) grey bands show the \Planck\ channels, with the coloured bands
indicating the major signal and foreground components.
For temperature the components are smoothed to $1^\circ$ and the widths of
the bands show the range for masks with 81--$93\,\%$ sky coverage.
For polarization the smoothing is $40^\prime$ and the range is 73--$93\,\%$.
For steep spectra, the rms shown here is dominated by
the largest angular scales.  But as shown by Fig.~\ref{fig:foregrounds2},
on much smaller angular scales in regions far form the Galactic plane,
the foreground signals fall far below the cosmological signal
(except at the lowest $\ell$, in polarization).
\label{fig:foregrounds} }
\end{figure*}

\begin{figure*}[hbtp]
\includegraphics[width=\columnwidth]{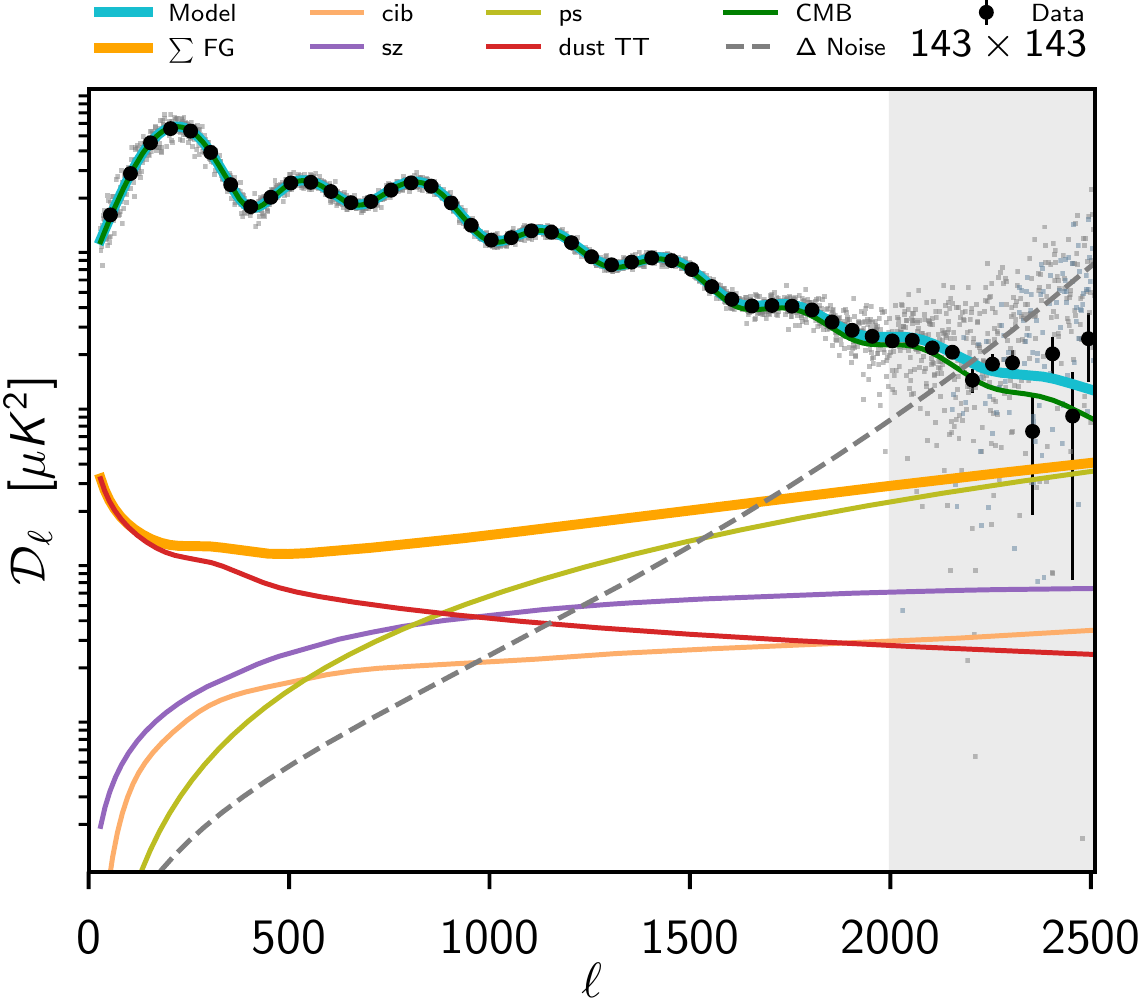}
\includegraphics[width=\columnwidth]{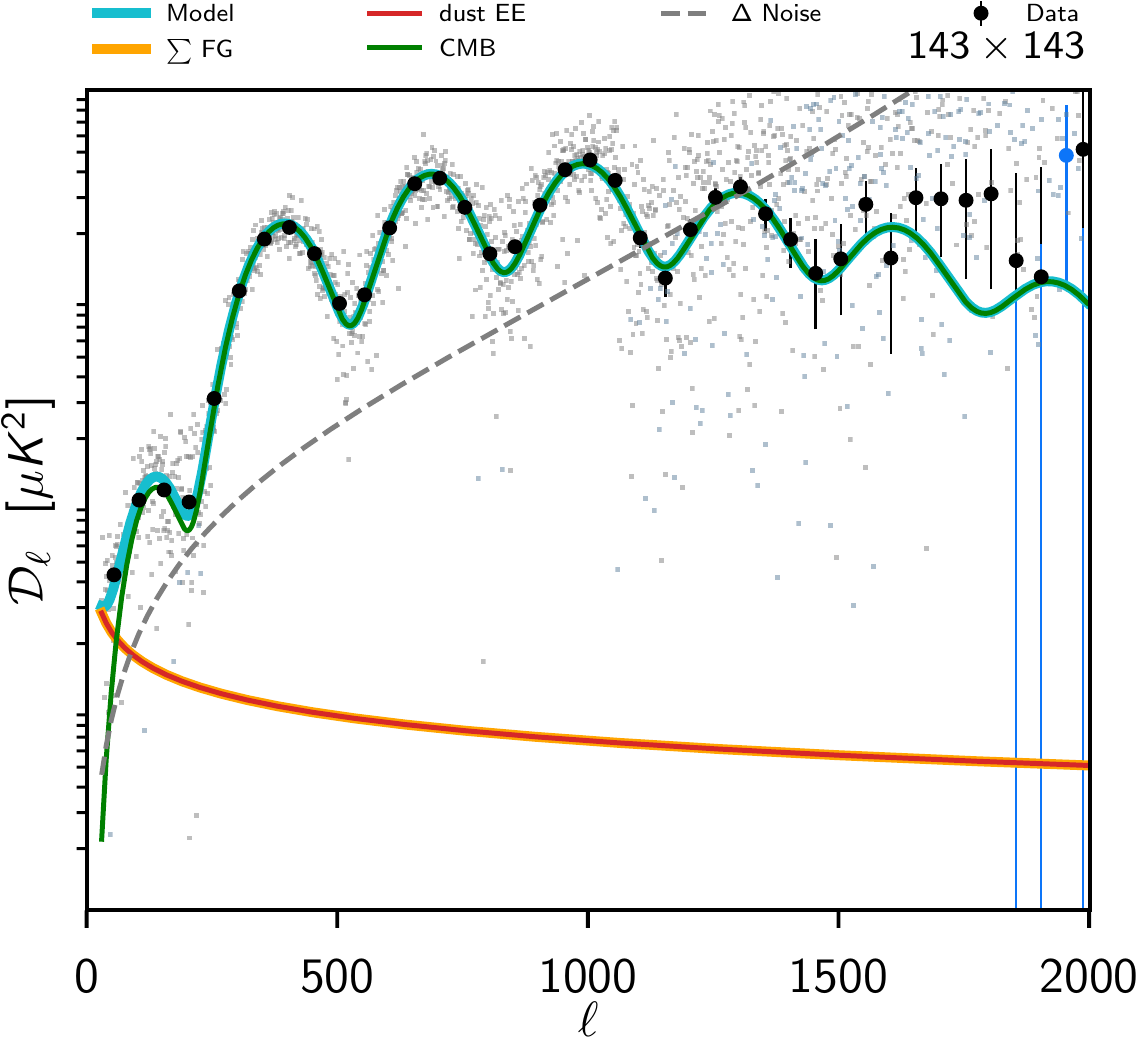}
\caption{Angular scale dependence of the main components of the submillimetre sky at 143\,GHz in temperature (left) and $E$-type polarization (right). These power spectra, ${\cal D}_{\ell} = \ell ( \ell + 1 )\, C_\ell/(2\pi)$, give approximately the contribution per logarithmic interval to the variance of the sky fluctuations. They are computed within the sky regions retained for the cosmological analysis (57\,\% of the 143\,GHz sky for the temperature and 50\,\% for polarization, in order to mask the resolved point sources and decrease the Galactic contributions). The grey dots are the values at individual multipoles, and the large black circles with error bars give their averages and dispersions in bands. The data (corrected for systematic effects) are very well fit by a model (cyan curves) that is largely dominated by the CMB fluctuation spectra (light blue curves, mostly inside the model), with a superposition of foreground emission (orange curves) dominated by dust at large scales (red curve), together with a noise contribution (dotted line).  We note, however, that foreground emission actually dominates the ``reionization bump'' at the lowest polarization multipoles. The grey shaded area shows the area in temperature which is not used for cosmology.
\label{fig:foregrounds2} }
\end{figure*}

\subsection{Foregrounds} \label{sec:foregrounds}

\Planck's unprecedented sensitivity and frequency coverage have 
enabled dramatic advances in component separation, reducing the
frequency maps into their astrophysical components, as described above.
These component products, which should be thought of as phenomenological
rather than being based on ab initio models, include maps in both
temperature and polarization of: the CMB; the thermal SZ effect;
thermal dust and the cosmic-infrared background; carbon monoxide;
synchrotron; free-free; and anomalous microwave emission.
They also effectively give rise to catalogues of compact Galactic and
extragalactic sources, including polarization information.
The maps and catalogues have a wide range of astrophysical uses that we
shall not attempt to survey here \citep[but see appendix~A of][for a guide to
the \Planck\ papers dealing with polarized thermal emission from
dust]{planck2016-l11B}.

An overview of the frequency dependence of the major components
(free-free emission, synchrotron, and dust)
is given in Fig.~\ref{fig:foregrounds}.
We first look at the angular power spectra of these contaminants, since this
allows us to better judge the foreground contributions at different angular
scales in regions actually used for the cosmology analysis.
Figure~\ref{fig:foregrounds2} shows the angular power spectra of the sky at
143\,GHz, compared to that of the primary CMB.
Out to $\ell\simeq 2500$ the latter dominates for the key cosmology channels.
This shows that the Galaxy is fortunately more transparent to the CMB over
most angular scales than one might fear based on the examination of
Fig.~\ref{fig:foregrounds}.
The full angular spectra at all frequencies, including the $TE$
cross-spectra, can be found in \citet{planck2016-l05}.

The foregrounds can be usefully characterized as Galactic or extragalactic, and
diffuse or compact.
Compact sources have been obtained by identifying locations with
a significantly high signal in a narrow band-pass spatial filter.
The Second \Planck\ Catalogue of Compact Sources, presented in
\citet{planck2014-a35}, lists compact sources over the entire sky in each of
the nine \Planck\ frequencies, including polarization information.  The \Planck\ 
catalogues of sources above $100\,$GHz represent the first such samples
ever, while the catalogues at $100\,$GHz and below represent a significant advance
over the previous state of the art.  The Galactic sources include
cold cores, \ion{H}{ii} regions, and young star-forming regions.
The extragalactic sources \citep{planck2012-VII,planck2014-a35} can be
characterized in frequency as radio
\cite[primarily quasars, blazars, and radio galaxies][]{planck2011-6.1,planck2011-6.2,planck2011-6.3a,planck2016-XLV,planck2018-LIV}
and infrared sources 
\cite[primarily dusty, star-forming galaxies][]{planck2011-6.3a,planck2011-6.4a,planck2014-XXVII},
with a special sub-population detected via the SZ effect
\citep{planck2014-a36}.
The \Planck\ SZ catalogue contained 1653 detections, of which 1203 were
confirmed, massive galaxy clusters with identified counterparts in external
data sets.
It was the first SZ catalogue with more than 1000 confirmed clusters.
Maps of the diffuse SZ effect have also been obtained, which are somewhat
sensitive to the outskirts of clusters \citep{planck2015-XXXVII}.
Finally, \citet{planck2015-XXXIX} provides a list of 2\,151 high-frequency
sources \citep[called the ``PHz'' catalogue, see also][]{planck2014-XXVII},
selected over 26\,\% of the sky using a combination of submillimetre colours.
These are likely to lie at high redshift ($z\ga2$), the majority being
over-densities of star-forming galaxies (including a population of
proto-clusters), with a small fraction representing some of the brightest
submillimetre gravitational lenses discovered so far \citep{Canameras2015}.
A discussion of how sources are treated (e.g., masked or modelled) for the
main cosmology analysis can be found in
\citet{planck2016-l04,planck2016-l05,planck2016-l06,planck2016-l08}.

\Planck\ detects many types of diffuse foregrounds, which must be
modelled or removed in order to study the primary CMB anisotropies.
The separation of the diffuse emission into component maps is described
in \citet{planck2014-a11}.
At frequencies below 50\,GHz, the total intensity is dominated by
free-free (bremsstrahlung from electron-ion collisions), synchrotron, and
spinning dust emission, while polarization is dominated by synchrotron
emission from relativistic cosmic ray electrons spiralling in the Galactic
magnetic field \citep[e.g.,][]{planck2014-a31}.
At higher frequencies (above 100\,GHz) the total intensity is dominated by
thermal dust emission from our Galaxy (extending to high Galactic latitudes
and sometimes referred to as ``cirrus'') at low $\ell$, and the cosmic
infrared background (CIB; primarily unresolved, dusty, star-forming galaxies)
at high $\ell$ \citep{planck2013-pip56}.
Only the former contribution is significantly polarized.
There is also a small contribution from free-free and synchrotron emission 
near $100\,$GHz.

Above $70\,$GHz, polarized thermal emission from diffuse, interstellar,
Galactic dust is the main foreground for CMB polarization.  Grain sizes
are thought to range from microns to that of large molecules, with the
grains made primarily of carbon, silicon, and oxygen.
The dust is made up of different components with different polarization
properties, and has a complex morphology.

\Planck\ has already determined that there are no dust-free windows on the
sky at the level relevant for future CMB experiments, so measuring and
understanding this important foreground signal will be a major component
of all future CMB polarization experiments.
The \Planck\ results show that pre-\Planck\ dust models were too simplistic,
and suggest that more accurate models, which include the insights from
\Planck, will take many years to fully develop.
However, \Planck\ observations already provide us with unprecedented data to
describe, at least on a statistical basis, the turbulent component of
the Galactic magnetic field and its interplay with the structure of
interstellar matter on scales ranging from a fraction of a parsec to
$100\,$pc \citep{planck2014-XIX}.
The data show that the interstellar magnetic fields have a coherent
orientation with respect to density structures, aligned with filamentary
structures in the diffuse interstellar medium, and mainly perpendicular in
star forming molecular clouds \citep{planck2014-XXXII,planck2015-XXXV}.
This result is far from being clearly understood, but it may signal the
importance of magnetic fields in the formation of structures in the
interstellar medium.

The polarization power spectra of dust are well described by power laws,
with $C_\ell^{EE,BB}\,{\propto}\,\ell^{-2.42\pm0.02}$, and frequency
dependence given by a modified blackbody (similar to that for the total
intensity, namely an emissivity index of about $1.55$ and a temperature of
about $20\,$K).
The power spectrum analyses presented in \citet{planck2014-XXX} led to three
unexpected results: a positive $TE$ correlation; 
$C_\ell^{BB}\,{\simeq}\,0.5\,C_\ell^{EE}$ for $40\,{<}\,\ell\,{<}\,600$;
and a non-negative $TB$ signal from Galactic dust emission.
Several studies \citep{Clark2015,planck2015-XXXVIII,Ghosh2017} have shown
that both the observed $TE$ correlation and the asymmetry between $E$- and
$B$-mode amplitudes for dust polarization can be accounted for by the preferred
alignment between the filamentary structure of the diffuse ISM and the
orientation of the magnetic field inferred from the polarization angle
(while the non-zero $TB$ correlation is also related to the fact that the
Milky Way is not parity invariant).
\citet{planck2016-L} further demonstrated that the frequency spectral index
of the emission varies across the sky.
We discuss this important foreground component further in
\citet{planck2016-l04}, \citet{planck2016-l11A}, and \citet{planck2016-l11B}.

\Planck\ produced the first well-calibrated, all-sky maps across the
frequencies relevant for CMB anisotropies.  The dramatic increase in
our understanding of the submillimetre sky has wide legacy value.  For
cosmology, perhaps the most important lesson is the realization that there
are no ``holes'' in which one can see $B$ modes comparable to the
signal from a tensor-to-scalar ratio $r\,{\sim}\,10^{-2}$ without component
separation.  At this level, foreground contamination comes from both low
frequencies (synchrotron) and high frequencies (dust), with neither component
being negligible.
In this component-separation-dominated regime, wide frequency coverage,
such as attained by \Planck, will be essential.

\subsection{CMB anisotropy maps} \label{sec:cmbmap}

\begin{figure*}[htbp]
\begin{center}
\resizebox{0.68\textwidth}{!}{\includegraphics[angle=90]{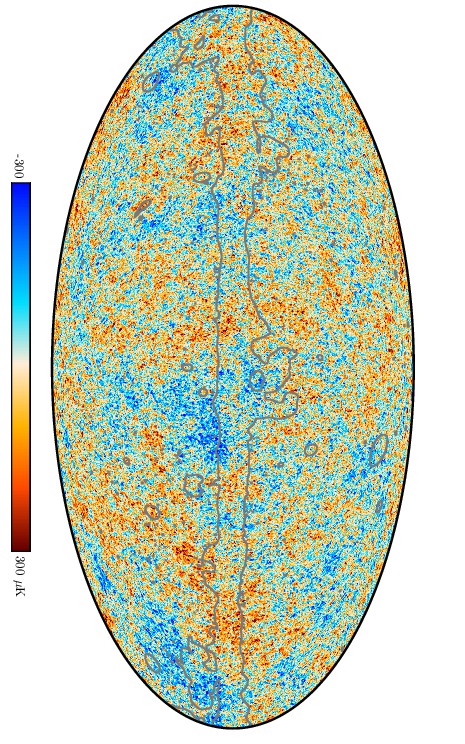}}\par 
\vspace{0.1truecm}
\resizebox{0.68\textwidth}{!}{\includegraphics[angle=90]{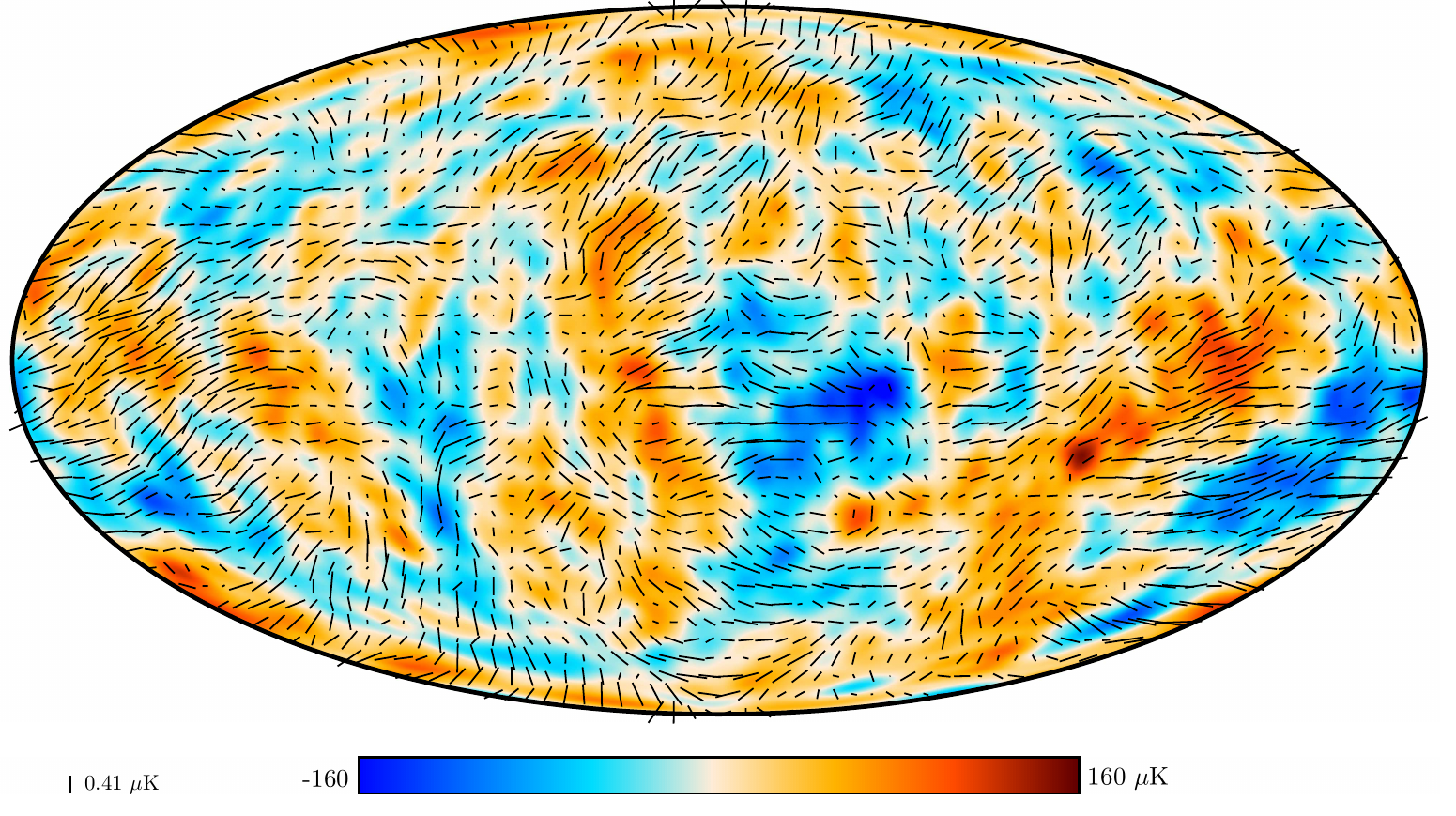}}\par
\vspace{0.05truecm}
\resizebox{0.68\textwidth}{!}{\includegraphics{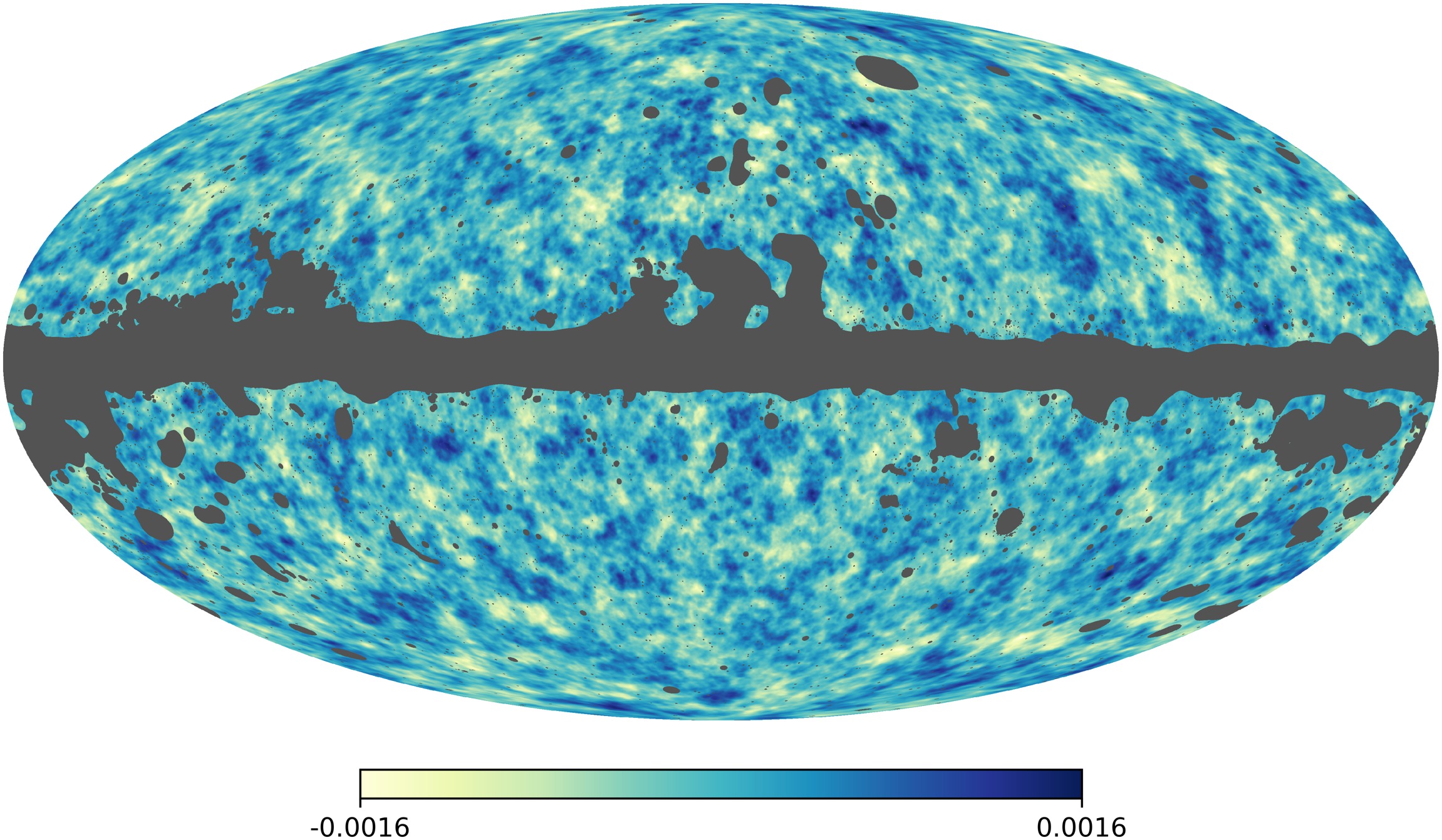}}\par
\vspace{-0.45truecm}
\end{center}
\caption{\Planck\ CMB sky. The top panel shows the 2018 \SMICA\ temperature map. The middle panel shows the polarization field as rods of varying length superimposed on the temperature map, with both smoothed to 5\deg. This smoothing is done for visibility purposes; the enlarged region presented in Fig.~\ref{fig:CMB-TPmap} shows that the \Planck\ polarization map is still dominated by signal at much smaller scales. Both CMB maps have been masked and inpainted in regions where residuals from foreground emission are expected to be substantial. This mask, mostly around the Galactic plane, is delineated by a grey line in the full resolution temperature map. The bottom panel shows the \Planck\ lensing map (derived from $\nabla\phi$, that is, the $E$ mode of the lensing deflection angle), specifically a minimum variance, Wiener filtered, map obtained from both temperature and polarization information; the unmasked area covers 80.7\,\% of the sky, which is larger than that used for cosmology.
\label{fig:sigmaps} }
\end{figure*}

\begin{figure}[htbp]
\begin{center}\vbox{
\includegraphics[width=\columnwidth]{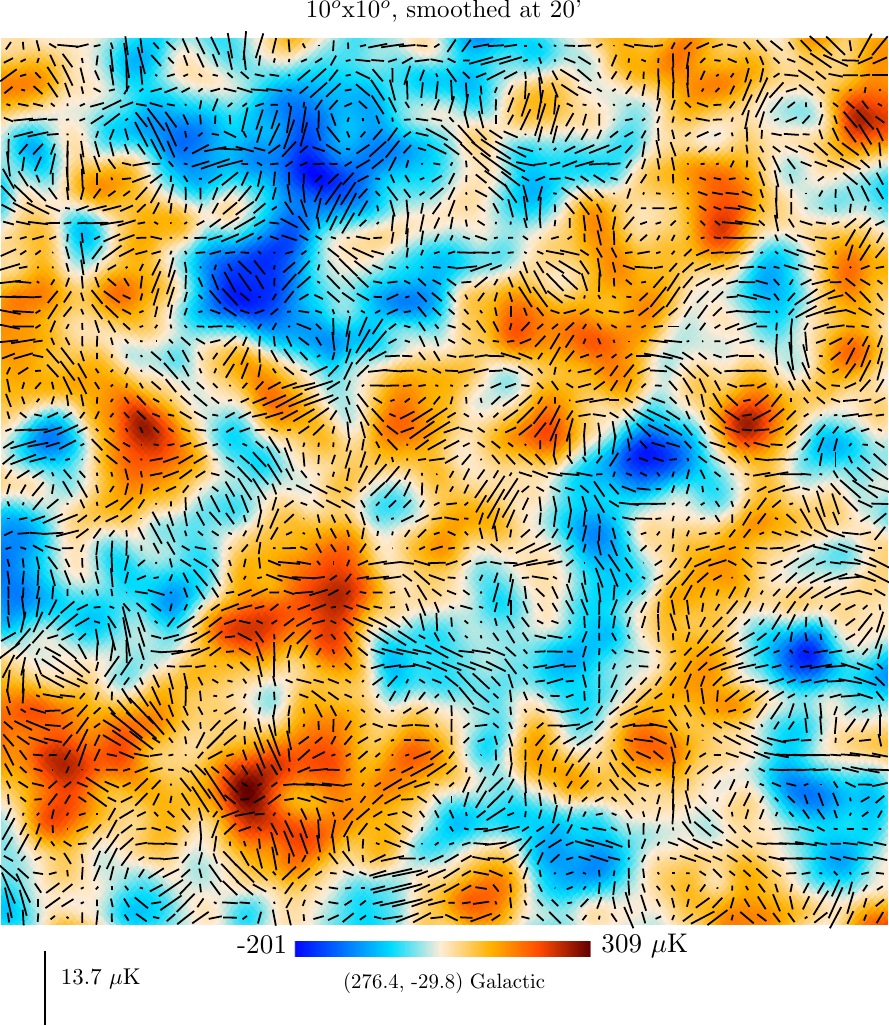} \vspace{-0.2cm}
\includegraphics[width=\columnwidth]{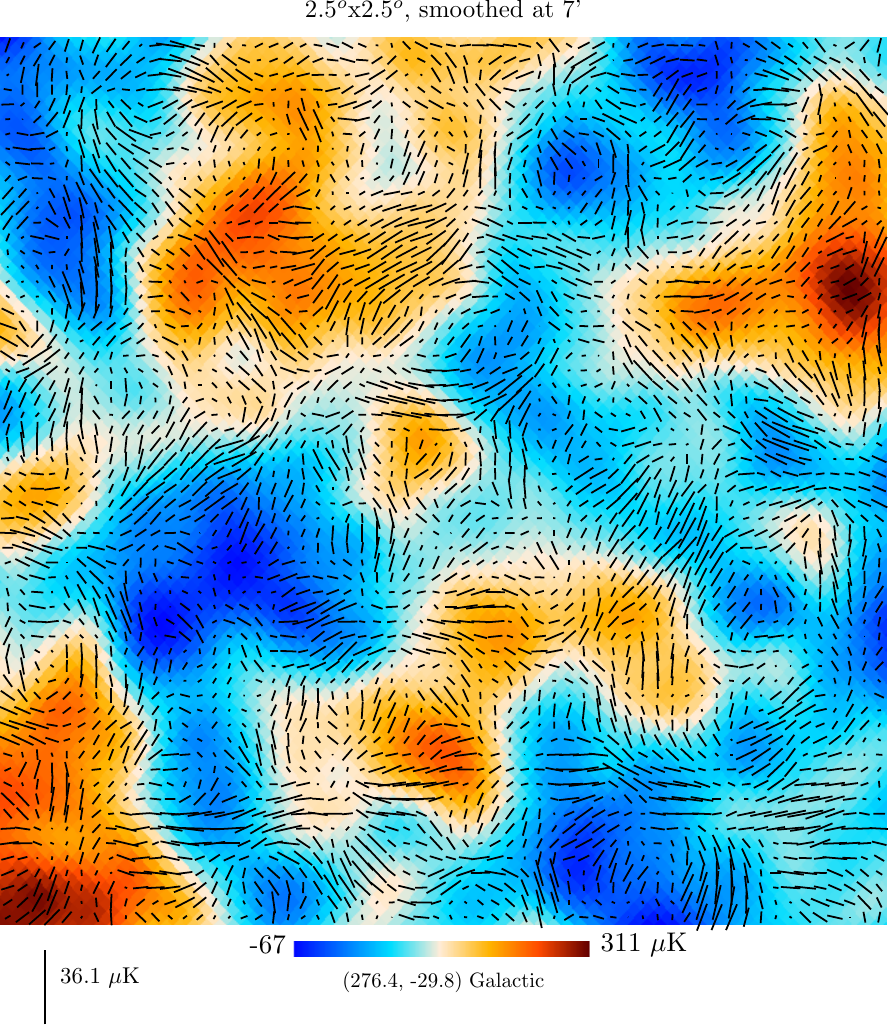}}
\end{center} 
\caption{Enlargement of part of the \planck\ 2018 CMB polarization map. The coloured background shows the temperature anisotropy field smoothed to the same scale as the polarization field, enabling us to visualize the correlation between the two fields. The top map shows a 10\deg\ $\times$ 10\deg\ patch centred on the south ecliptic pole, smoothed with a 20\arcm\ FWHM Gaussian (the data are natively at 5\arcm\ resolution). The bottom panel is a further expansion of a 2.5\deg $\times$ 2.5\deg region in the same direction.
\label{fig:CMB-TPmap} }
\end{figure}

\begin{figure}[htbp]
\centering
\includegraphics[width=\columnwidth]{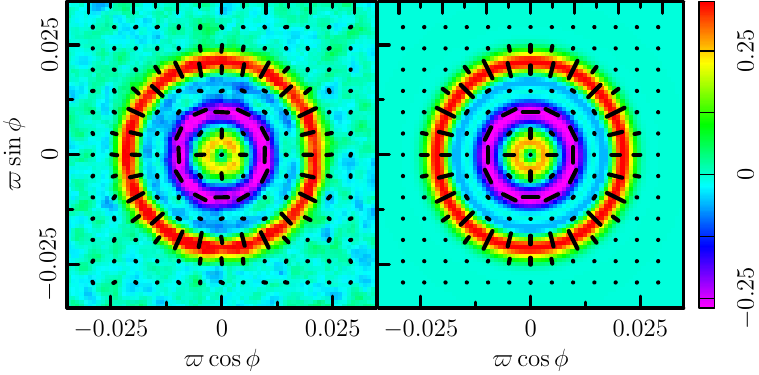}
\caption{Stacked $Q_\mathrm{r}$ image around temperature hot spots selected
  above the null threshold ($\nu = 0$) in the \SMICA\ sky map.  The quantity
  $Q_\mathrm{r}$ \citep[and its partner $U_\mathrm{r}$, introduced
  in][]{Kam97} is a transformed version of the Stokes parameters $Q$ and $U$,
  where $Q_\mathrm{r}$ measures the tangential-radial component of the
  polarization relative to the centre and $U_\mathrm{r}$ measures
  the polarization at $\pm45\deg$ relative to a radial vector.
  The left panel corresponds to the observed data, and the right panel
  shows the ensemble average of CMB-only maps for the fiducial cosmology.
  The axes are in degrees, and the image units are \microK.
  The black solid lines show the polarization directions
  for stacked $Q$ and $U$, with lengths proportional to
  the polarization amplitude $P$.
  {}From \cite{planck2014-a18}.}
\label{fig:stackings}
\end{figure}

Figure~\ref{fig:sigmaps} shows the maps of CMB anisotropies on which
we base our analyses of the statistical character of these 
fluctuations.\footnote{Galactic and extragalactic foregrounds have been removed from the maps.  Cosmological parameter constraints are mostly based on a likelihood analysis of the angular (cross-)power spectra of the frequency
maps, which are analysed with a model of the foreground spectra whose
parameters are treated as nuisance parameters, together with other parameters
characterizing uncertainties of instrumental origin.}
The component with the highest signal-to-noise ratio (S/N) is the temperature
anisotropy.
As shown later, \Planck\ has measured more than a million harmonic modes
of the temperature map with a signal-to-noise greater than unity.

The (linear) polarization signal is shown in the middle panel with a relatively low angular resolution of 5\deg\ to increase legibility. The polarization signal, shown by rods of varying length and orientation,\footnote{The orientation is computed in the local tangent plane with respect to the local meridian, and then rotated so that the meridian would be vertical, i.e., the rods are shown in the plane of the page with the north pointing to the top of the page.} 
is smaller in amplitude than the temperature signal. It is dominated by $E$ modes generated by Thomson scattering in the last-scattering surface of the anisotropic temperature field.  Unlike the temperature, \Planck's measurement of the polarization is limited by noise. The small-scale polarization pattern and its relationship to temperature anisotropies can be appreciated in Fig.~\ref{fig:CMB-TPmap}, which displays a $10\deg\times10\deg$ patch in the vicinity of the south ecliptic pole and a zoom into the central $2.5\deg\times2.5\deg$ patch. In these figures, the polarization is superimposed on the temperature anisotropies (shown in the background). It is clear that the two fields are correlated, as expected in the standard model (Sect.~\ref{sec:thecmbsky}).
This is directly visualized in Fig.~\ref{fig:stackings} by stacking the polarization pattern around hot spots of the temperature anisotropy map. It reveals that the pattern is mirror-symmetric, that is, it is predominantly $E$ modes, as expected. This trace of the dynamics of acoustic perturbations at the last scattering surface behaves precisely accordingly to \LCDM\ predictions (simulated in the right panel).

Most of the signal seen in the first two maps of Fig.~\ref{fig:sigmaps} is
dominated by processes occurring at $z\,{\simeq}\,10^3$.
However, the deflection of CMB photons by the gravitational potentials
associated with large-scale structure subtly modifies the signals
\Planck\ observes.
By measuring the impact of this CMB lensing on such wide-area but
high-angular-resolution sky maps, \Planck\ is able to measure the
lensing potential over much of the sky
\citep{planck2013-p12,planck2014-a17,planck2016-l08}.
This is shown in the bottom panel of Fig.~\ref{fig:sigmaps} and provides sensitivity to the lower-redshift Universe and a powerful test of the gravitational instability paradigm.

The primary use of CMB maps is to study their statistical properties.
It turns out that the primary CMB anisotropies (formed at the last-scattering
epoch) are extremely close to Gaussian-distributed
\citep{planck2016-l07,planck2016-l09},
although there are a number of potential deviations (or ``anomalies'')
to which we shall return in Sect.~\ref{sec:anomalies}.
This is in accord with the predictions of the simplest models of inflation,
and indeed provides strong constraints on many inflationary models
(see Sect.~\ref{sec:physics} and \citealt{planck2016-l10}). 
Such models also imply that the information content in the CMB comes from
its statistical properties, rather than the precise locations of individual
features, and that those properties are statistically isotropic.
Since a Gaussian field can be entirely described by its mean and correlation
function, and since the mean is zero by definition for the anisotropies,
essentially all of the cosmologically-relevant information in the CMB
anisotropies resides in their correlation functions or power spectra.
This allows a huge compression, with concomitant increase in S/N:
the 1.16 billion pixels in the 23 maps can be compressed to $10^6$ high-S/N
multipoles.
As we will see later, the $\Lambda$CDM model allows even more dramatic
compression: only six numbers describe around $10^3\,\sigma$ worth of
power spectrum detection.

\subsection{CMB angular power spectra} \label{sec:spectra}

\begin{figure*}[htbp]
\begin{center}
\resizebox{\textwidth}{0.83\textheight}{\includegraphics{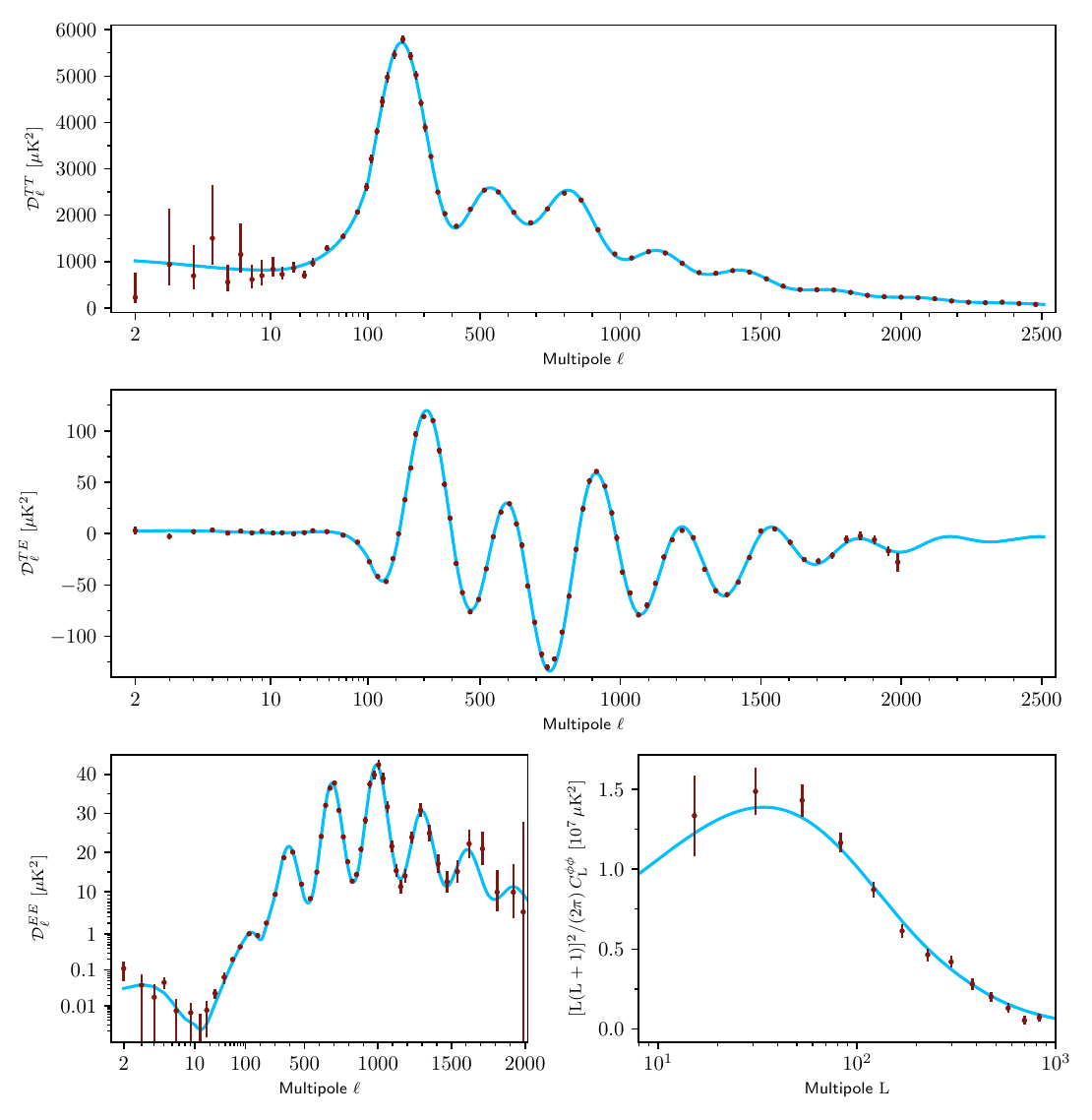}}
\end{center}
\caption{\Planck\ CMB power spectra. These are foreground-subtracted,
frequency-averaged, cross-half-mission angular power spectra for temperature
(top), the temperature-polarization cross-spectrum (middle), the $E$ mode of
polarization (bottom left), and the lensing potential (bottom right). 
Within \LCDM\ these spectra contain the majority of the cosmological
information available from \Planck, and the blue lines show the best-fitting
model.  The uncertainties of the $TT$ spectrum are dominated by sampling
variance, rather than by noise or foreground residuals, at all scales
below about $\ell=1800$ -- a scale at which the CMB information is essentially
exhausted within the framework of the \LCDM\ model.
The $TE$ spectrum is about as constraining as the $TT$ one, while the $EE$
spectrum still has a sizeable contribution from noise.
The lensing spectrum represents the highest signal-to-noise ratio detection of
CMB lensing to date, exceeding $40\,\sigma$.  The anisotropy power spectra use
a standard binning scheme (which changes abruptly at $\ell=30$), but are
plotted here with a multipole axis that goes smoothly from logarithmic at
low $\ell$ to linear at high $\ell$.  In all panels, the blue line is the best-fit \Planck\ 2018 model, based on the combination of $TT$, $TE$, and $EE$.}
\label{fig:cl}
\end{figure*}

\subsubsection{CMB intensity and polarization spectra} \label{sec:CMBspectra}

The foreground-subtracted, frequency-averaged, cross-half-mission $TT$, $TE$,
and $EE$ spectra are plotted in Fig.~\ref{fig:cl}, together with the
{\tt Commander} power spectrum at multipoles $\ell<30$.
The figure also shows the best-fit base-$\Lambda$CDM theoretical spectrum
fitted to the combined temperature, polarization, and lensing data.

Figure~\ref{fig:cl} clearly illustrates that \Planck\ has determined the
angular power spectrum of the primary temperature anisotropies
to high precision across all the physically relevant scales.  In
this sense, \Planck\ brings to an end an era in CMB studies that was opened
by the first detection of these anisotropies by \COBE\ in 1992 \citep{Smoot92}.
At the same time, \Planck\ has made important measurements of the
polarization power spectra and maps of the effects of gravitational lensing.
Improvements in these measurements will be the focus of the field in coming
years.

The impressive agreement between the $\Lambda$CDM model and the
\Planck\ data will be the subject of later sections.  For now let us focus on
a number of ways of characterizing the information obtained in the spectra
of Fig.~\ref{fig:cl}.

One way of assessing the constraining power contained in a particular
measurement of CMB anisotropies is to determine the effective number of
$a_{\ell m}$ modes that have been measured.
This is equivalent to estimating 2 times the square of the total S/N in
the power spectra, a measure that contains all the available
cosmological information \citep{ScottCNM} if we assume that the anisotropies
are purely Gaussian
(and hence ignore all non-Gaussian information coming from lensing,
the CIB, cross-correlations with other probes, etc.).
Translating this S/N into inferences about cosmology or particular parameters
is not straightforward, since it needs to take into account how the spectra
respond to changes in parameters and in particular to degeneracies; however, 
the raw numbers are still instructive.
For the \Planck\ 2013 $TT$ power spectrum, the number was 826\,000
(rounded to the nearest 1\,000, including the effects of instrumental noise,
cosmic variance, and masking).
The 2015 $TT$ data increased this value to 1\,114\,000, with $TE$ and
$EE$ adding a further 60\,000 and 96\,000 modes, respectively (where these
were from the basic likelihood, with a conservative sky fraction).
Based on the 2018 data the numbers are now 1\,430\,000 for $TT$,
64\,000 for $TE$, 109\,000 for $EE$, and also 3\,000 for $\phi\phi$
(the lensing spectrum).
For comparison, the equivalent number of modes from the final \WMAP\ $TT$
power spectrum is 150\,000.

\Planck\ thus represents a 900$\,\sigma$ detection of power
(for the sake of simplicity, we do not include the correlations of the
covariance in this calculation; doing so would increase these numbers by
about 10--20\,\%).
This increases even further if one is less conservative and includes more
sky, along with more complicated foreground modelling.

The acoustic peaks in the $\mathcal{D}_\ell$s reveal the underlying
physics of oscillating sound waves in the coupled photon-baryon fluid,
driven by gravitational potential perturbations.
One can easily see the fundamental mode (which reaches a density and
temperature maximum as the Universe recombines) at $\ell\,{\simeq}\,220$, and
then the first harmonic, the second harmonic, and so on.
It is natural to treat the positions of the individual peaks
in the power spectra as empirical information that becomes part of the
canon of facts now known about our Universe.

Fitting for the positions and amplitudes of features in the band powers
is a topic with a long history, with approaches becoming more sophisticated
as the fidelity of the data has improved
(e.g., \citealt{Scott94}, \citealt{Hancock97}, \citealt{Knox00},
 \citealt{deBernardis02}, \citealt{Bond03}, \citealt{page2003b},
 \citealt{Durrer03}, \citealt{Readhead04}, \citealt{Jones06},
 \citealt{hinshaw2007}, \citealt{Corasaniti08}, \citealt{Pryke09}).
We follow the approach (with small differences) described in \citet{planck2014-a01}, fitting Gaussians
to the peaks in $TT$ and $EE$, but parabolas to the peaks in $TE$. For $TT$ we
remove a featureless damping tail (using extreme lensing) in order to fit the
higher-$\ell$ region (starting with trough 3).\footnote{In
\citet{planck2014-a01}, peak 4 did not have this feature removed before fitting,
which explains the large discrepancy between our values here. Furthermore we
find that the marginal detection of peak~8 in \citet{planck2014-a01} has become
slightly poorer (even although in general the constraints have improved).} We also fit the first peak in $C^{EE}_{\ell}$
with a Gaussian directly.
Our numerical values, presented in in Table~\ref{tab:peaks}, are
consistent with previous estimates, but with increased precision.
\Planck\ detects 18 peaks
(with still only marginal detection of the eighth $TT$ peak) and 17 troughs,
for a total of 35 power spectra extrema (36 if the $C_\ell^{\phi\phi}$ peak
is included).

We shall use the rich structure of the anisotropy spectra, described above,
to constrain cosmological models in later sections.

\begin{table}[htbp]
\newdimen\tblskip \tblskip=5pt
\caption{The peaks of the CMB angular power spectra, $\mathcal{D}_\ell$,
as determined by \Planck.}
\label{tab:peaks}
\vskip -9mm
\footnotesize
\setbox\tablebox=\vbox{
 \newdimen\digitwidth
 \setbox0=\hbox{\rm 0}
 \digitwidth=\wd0
 \catcode`*=\active
 \def*{\kern\digitwidth}
 \newdimen\signwidth
 \setbox0=\hbox{+}
 \signwidth=\wd0
 \catcode`!=\active
 \def!{\kern\signwidth}
 \newdimen\pointwidth
 \setbox0=\hbox{\rm .}
 \pointwidth=\wd0
 \catcode`?=\active
 \def?{\kern\pointwidth}
 \halign{\tabskip=0pt\hbox to 1.0in{#\leaderfil}\tabskip=2em&
       \hfil$#$\hfil&
       \hfil$#$\hfil\tabskip=0pt\cr
\noalign{\doubleline}
\noalign{\vskip 3pt}
\omit\hfil Extremum\hfil&\omit\hfil Multipole\hfil& \omit \hfil
 Amplitude\ $*[\mu{\rm K}^2]$\hfil\cr
\noalign{\vskip 4pt\hrule\vskip 6pt}
\omit{\boldmath{$TT$}} \bf power spectrum\hfil\cr
\noalign{\vskip 5pt}
Peak 1&                *220.6\pm*0.6& !5733?*\pm39?**\cr
\hglue 1.0em Trough 1& *416.3\pm*1.1& !1713?*\pm20?**\cr
Peak 2&                *538.1\pm*1.3& !2586?*\pm23?**\cr
\hglue 1.0em Trough 2& *675.5\pm*1.2& !1799?*\pm14?**\cr
Peak 3&                *809.8\pm*1.0& !2518?*\pm17?**\cr
\hglue 1.0em Trough 3& 1001.1\pm*1.8& !1049?*\pm*9?**\cr
Peak 4&                1147.8\pm*2.3& !1227?*\pm*9?**\cr
\hglue 1.0em Trough 4& 1290.0\pm*1.8& !*747?*\pm*5?**\cr
Peak 5&                1446.8\pm*1.6& !*799?*\pm*5?**\cr
\hglue 1.0em Trough 5& 1623.8\pm*2.1& !*399?*\pm*3?**\cr
Peak 6&                1779?*\pm*3?*& !*378?*\pm*3?**\cr
\hglue 1.0em Trough 6& 1919?*\pm*4?*& !*249?*\pm*3?**\cr
Peak 7&                2075?*\pm*8?*& !*227?*\pm*6?**\cr
\hglue 1.0em Trough 7& 2241?*\pm24?*& !*120?*\pm*6?**\cr
\noalign{\vskip 4pt\hrule\vskip 6pt}
\omit{\boldmath{$TE$}} \bf power spectrum\hfil\cr
\noalign{\vskip 5pt}
Trough 1&              *146.0\pm*1.1& **-46.7\pm*1.4*\cr
\hglue 1.0em Peak 1&   *308.2\pm*0.8& *!117.1\pm*1.9*\cr
Trough 2&              *471.1\pm*0.7& **-74.1\pm*1.5*\cr
\hglue 1.0em Peak 2&   *595.8\pm*0.9& *!*27.8\pm*1.6*\cr
Trough 3&              *747.2\pm*0.8& *-128.0\pm*1.5*\cr
\hglue 1.0em Peak 3&   *917.1\pm*0.8& *!*59.0\pm*1.6*\cr
Trough 4&              1072.5\pm*1.2& **-79.1\pm*1.6*\cr
\hglue 1.0em Peak 4&   1221.3\pm*1.7& *!**3.5\pm*1.7*\cr
Trough 5&              1372.7\pm*2.8& **-60.0\pm*1.9*\cr
\hglue 1.0em Peak 5&   1532.1\pm*2.2& *!**8.9\pm*1.5*\cr
Trough 6&              1697.4\pm*5.9& **-27.2\pm*2.3*\cr
\hglue 1.0em Peak 6&   1859.7\pm*6.2& ***-1.0\pm*2.4*\cr
\noalign{\vskip 4pt\hrule\vskip 6pt}
\omit{\boldmath{$EE$}} \bf power spectrum\hfil\cr
\noalign{\vskip 5pt}
Peak 1&                *145?*\pm*3?*& !**1.11\pm*0.04\cr
\hglue 1.0em Trough 1& *195.0\pm*5.4& !**0.79\pm*0.08\cr
Peak 2&                *398.3\pm*1.0& !*21.45\pm*0.31\cr
\hglue 1.0em Trough 2& *522.0\pm*1.1& !**7.18\pm*0.29\cr
Peak 3&                *690.4\pm*1.2& !*38.1*\pm*0.6*\cr
\hglue 1.0em Trough 3& *831.8\pm*1.1& !*12.6*\pm*0.4*\cr
Peak 4&                *993.1\pm*1.8& !*42.6*\pm*0.8*\cr
\hglue 1.0em Trough 4& 1158.8\pm*2.6& !*12.0*\pm*1.1*\cr
Peak 5&                1296.4\pm*4.3& !*31.5*\pm*1.3*\cr
\noalign{\vskip 4pt\hrule\vskip 6pt}
}}
\endPlancktable
\end{table}

\subsubsection{CMB lensing spectrum} \label{sec:lensingspectra}

The photons that we see as the cosmic microwave background
must traverse almost the entire observable
Universe on their way to us.  During this journey they have their wavelengths
stretched by the cosmological expansion and their paths deflected by the
gravitational potentials associated with inhomogeneities in the Universe
\citep{1987A&A...184....1B}.
The lensing-induced deflections are of order 2\arcm\ to 3\arcm,
coherent over patches 2\deg\ to 3\deg\ across,
and arise primarily from structures at redshifts of 0.5--10.
Since each photon undergoes multiple deflections during its travel, this
``secondary'' anisotropy is enhanced over naive expectations and turns out
to be one of the most important secondary signals we measure.

This ``gravitational lensing'' of CMB photons by large-scale structures along
their path has several effects
\citep[see e.g.,][for reviews]{LewisChallinor06,Hanson10}.
One is to slightly smooth the peak and trough structure of the CMB power
spectra and the damping tail (this is fully accounted for by the numerical
codes when deriving the parameter constraints on a model;
\citealt{Seljak:1995ve}).
Another effect is to transform some of the polarization $E$ modes into $B$
modes, adding to the potentially pre-existing $B$-mode contribution from
primordial tensor fluctuations \citep{Zaldarriaga:1998ar}.
These distortions couple adjacent $\ell$ modes, which would otherwise be
uncorrelated if the initial fluctuations were statistically homogeneous.
This can then be used to obtain an estimator of the lensing potential by
cross-correlating CMB maps ($T$, $E$, $B$) and their derivatives, with
appropriate weightings \citep{Hu:2001kj,Hirata:2002jy}.
These lensing measurements provide sensitivity to parameters that affect
the late-time expansion, the geometry, or the clustering of matter, and can be
cross-correlated with large-scale structure surveys in a variety of ways
(see Sect.~\ref{sec:lsslensing}).

The lensing deflections are usually written as the gradient\footnote{The CMB
literature and the galaxy lensing literature differ in the sign of $\alpha$
and of $\psi$.  We follow the CMB convention here.} of a ``lensing potential,''
$\vec{\alpha}=\vec{\nabla}\psi(\hn)$, which is a measure of the integrated
mass distribution back to the surface of last scattering:
\begin{equation}
  \psi(\hn) = -2\int_0^{\chi_\ast} d\chi
  \ \left(\frac{\chi_\ast-\chi}{\chi_\ast\,\chi}\right)\,\Psi_W(\chi\hn),
\end{equation}
where $\chi_\ast$ is the comoving distance to the surface of last scattering
and $\Psi_{\rm W}$ is the (Weyl) potential, which probes the matter through
Poisson's equation.
For this reason the nearly all-sky lensing map shown in Fig.~\ref{fig:sigmaps}
provides a remarkable view of (essentially) all of the matter in the Universe,
as traced by photons travelling through $13.8\,$Gyr of cosmic history.
At $>40\,\sigma$, this is the largest area, and highest significance,
detection of weak lensing to date and constrains the amplitude of large-scale
structure power to $3.5\,\%$ \citep{planck2016-l08}. 
The highest S/N per mode is achieved near $L\,{=}\,60$, where the
signal-to-noise ratio per $L$ is close to unity (we follow the standard
convention and use $L$ rather than $\ell$ for lensing multipoles).
This drops by about a factor of 2 by $L\,{=}\,200$, though there is still some
power out beyond $L\,{=}\,1000$.

\Planck\ was the first experiment to measure the lensing power spectrum
to higher accuracy than it could be theoretically predicted from measurements
of the anisotropies.  This represents a turning point, where lensing
measurements start to meaningfully impact parameter constraints.
In the future, lensing will play an increasingly important role in CMB
experiments -- a future that \Planck\ has ushered in.

\begin{figure*}[htbp]
\centering
\includegraphics[width=\textwidth]{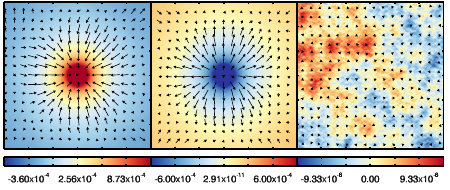}
\caption{Lensing stacking on CIB peaks. This figure first appeared as figure~4 of \cite{planck2013-p13}, a paper devoted to studying the gravitational lensing-infrared background correlation. It shows temperature maps of size $1\,{\rm deg}^2$ at 545\,GHz stacked on the 20\,000 brightest peaks (left column), troughs (centre column), and random map locations (right column). The stacked (averaged) temperature maps are in K. The arrows indicate the lensing deflection angle deduced from the gradient of the band-pass filtered lensing potential map stacked on the same peaks. The longest arrow corresponds to a deflection of 6\parcs3, which is only a fraction of the total deflection angle because of our filtering. This stacking allows us to visualize in real space the lensing of the CMB by the potential wells traced by galaxies that generate the CIB. This vividly demonstrates that our lensing map, albeit noisy, is well correlated with a high-$z$ tracer; it also warrants using the CIB as a lensing potential tracer at smaller scales (and in other experiments).
\label{fig:cibstackings}
}
\end{figure*}

In addition to enhancements of data processing into maps, the final data release includes several improvements in the lensing pipeline over the 2013 and 2015 analyses \citep{planck2016-l08}, 
including a polarization-only lensing reconstruction, as a demonstration
of consistency and a cross-check on the paradigm.
In addition to the lensing measured from the CMB channels,
\citet{planck2016-l08} also presents a joint analysis of lensing reconstruction
and the CIB, as probed by our highest frequency channels.
The CIB is a high-$z$ tracer of the density field that is around 80\,\%
correlated with the CMB lensing potential.
Figure~\ref{fig:cibstackings} shows the lensing deflection inferred from our
lensing maps, stacked on the 20\,000 brightest peaks and deepest troughs in
the CIB.  One can clearly see the high degree of correlation and the expected
convergence and divergence patterns around over and under-densities.
Having a high signal-to-noise ratio, the CIB map provides a good estimate
of the lensing modes on small scales and the best picture we have at present
of the lensing potential.
Finally, \citet{planck2016-l08} demonstrates that the smoothing effect of
lensing on the CMB acoustic peaks can be corrected, with ``delensing'' removing
approximately 50\,\% of the effect.  The ability to delens the CMB will grow
in importance as we move into a future of low-noise observations where
lensing-induced power becomes dominant.

The lensing potential power spectrum provides additional information on the
low-$z$ Universe, and thus an alternative route to constraining cosmological
parameters and a means of breaking degeneracies that affect the primary
anisotropies.
The reduction in the uncertainty of the effects of reionization afforded
by the new low-$\ell$ polarization data (see Sect.~\ref{sec:lcdm_params})
leads to a reduction in the uncertainty on the power spectrum normalization
when using primary anisotropies alone.
The constraints on the amplitude from the primary anisotropies are thus
tighter, and this reduces the impact of the lensing upon parameter shifts.
However, lensing still plays an important role and provides a consistency
check on the overall picture.  For example, the best-determined combination
of parameters from CMB lensing is $\sigma_8\Omega_{\rm m}^{0.25}$, which is
now determined to 3.5\,\% ($0.589\pm 0.020$; 68\,\% CL).
Combining this with the baryon density from big-bang nucleosynthesis (BBN)
and distance measurements from baryon acoustic oscillations (BAOs)
allows us to place competitive constraints on $\sigma_8$, $\Omega_{\rm m}$,
and $H_0$ individually \citep{planck2016-l08}.

Our baseline lensing likelihood is based on an $f_{\rm sky} \simeq70$\,\%,
foreground-cleaned combination of the high-frequency maps.
The usable range of multipoles extends from $L\,{=}\,8$ to $L\,{=}\,400$.
Multipoles below this are adversely affected by a large and uncertain
mean-field correction \citep{planck2016-l08}.
Although the lensing maps are provided to $L\,{=}\,4096$, the data above
$L\,{=}\,400$ do not pass some null tests \citep{planck2016-l08}
and thus are regarded as less reliable.  Multipoles $L\,{\gg}\,60$ become
increasingly noise dominated, but some residual signal is present even at
very high $L$, which can be of use in cross-correlation or stacking analyses.

In addition to the power-spectrum band powers and covariance, we have released
temperature-based, polarization-based, and joint temperature- and
polarization-based convergence maps, plus the simulations, response
functions, and masks necessary to use them for cosmological science.
We also release the joint CIB map, the likelihood, and parameter chains.

\section{The \texorpdfstring{\LCDM}{LCDM} model} \label{sec:lcdm}

Probably the most striking characteristic to emerge from the last few decades
of cosmological research is the almost unreasonable effectiveness of
the minimal 6-parameter $\Lambda$CDM model in accounting for cosmological
observations over many decades in length scale and across more than
$10\,$Gyr of cosmic time.
Though many of the ingredients of the model remain highly mysterious from
a fundamental physics point of view, $\Lambda$CDM is one of our most
successful phenomenological models.
As we will discuss later, it provides a stunning fit to an ensemble
of cosmological observations on scales ranging from Mpc to the Hubble scale,
and from the present day to the epoch of last scattering.

The $\Lambda$CDM model rests upon a number of assumptions, many of which
can be directly tested with \Planck\ data.
With the model tested and the basic framework established,
\Planck\ provides the strongest constraints on the
six parameters that specify the model (Tables~\ref{tab:LCDMbest_fit} and
\ref{tab:LCDMparams}).
Indeed, of these six parameters all but one -- the optical depth -- are now
known to sub-percent precision (for $n_{\rm s}$ this claim depends upon
the conventional choice that $n_{\rm s}\,{=}\,1$ represents scale-invariance).

\begin{table}[htbp] 
\begin{center}
\caption{The 6-parameter $\Lambda$CDM model that best fits the combination
of data from \Planck\ CMB temperature and polarization power spectra (including
lensing reconstruction), with and without BAO data (see text). A number of
convenient derived parameters are also given in the lower part of the table.
These best fits can differ by small amounts from the central values
of the confidence limits in Table \ref{tab:LCDMparams}.
\label{tab:LCDMbest_fit}}
\begingroup
\openup 4pt
\newdimen\tblskip \tblskip=5pt
\nointerlineskip
\vskip -5mm
\footnotesize
\setbox\tablebox=\vbox{
    \newdimen\digitwidth
    \setbox0=\hbox{\rm 0}
    \digitwidth=\wd0
    \catcode`*=\active
    \def*{\kern\digitwidth}
    \newdimen\signwidth
    \setbox0=\hbox{+}
    \signwidth=\wd0
    \catcode`!=\active
    \def!{\kern\signwidth}
\halign{
\hbox to 1.2in{$#$\leaderfil}\tabskip=1.0em&
        \hfil$#$\hfil&\hfil$#$\hfil\tabskip=0.5pt\cr
\noalign{\doubleline}
\omit\hfil Parameter\hfil&\omit\hfil \Planck\ alone\hfil
&\hfil\Planck + {\rm BAO}\hfil\cr
\noalign{\vskip 5pt\hrule\vskip 5pt}
\Omega_{\mathrm{b}} h^2&   0.022383& 0.022447\cr
\Omega_{\mathrm{c}} h^2&    0.12011& 0.11923\cr
100\theta_{\mathrm{MC}}&   1.040909& 1.041010\cr
\tau                   &     0.0543& 0.0568\cr
\ln(10^{10} A_\mathrm{s})&   3.0448& 3.0480\cr
n_\mathrm{s}           &    0.96605& 0.96824\cr
\noalign{\vskip 5pt\hrule\vskip 5pt}
H_0\,[{\rm km}\,{\rm s}^{-1}{\rm Mpc}^{-1}]& 67.32& 67.70\cr
\Omega_\Lambda         &   0.6842& 0.6894\cr
\Omega_{\rm m}         &   0.3158& 0.3106\cr
\Omega_{\rm m}h^2      &   0.1431& 0.1424\cr
\Omega_{\rm m}h^3      &   0.0964& 0.0964\cr
\sigma_8               &   0.8120& 0.8110\cr
\sigma_8(\Omega_{\rm m}/0.3)^{0.5}& 0.8331& 0.8253\cr
z_{\rm re}             &     7.68& 7.90\cr
{\rm Age}\,[{\rm Gyr}] &  13.7971& 13.7839\cr
\noalign{\vskip 4pt\hrule\vskip 3pt}
} 
} 
\endPlancktable
\endgroup
\end{center}
\end{table}

\begin{table}[htbp] 
\begin{center}
\caption{Parameter confidence limits from \Planck\ CMB temperature,
polarization, and lensing power spectra, and with
the inclusion of BAO data.
The first set of rows gives 68\,\% limits for the base-\LCDM\ model, while
the second set gives 68\,\% constraints on a number of derived parameters
(as obtained from the constraints on the parameters used to specify the
base-\LCDM\ model). The third set below the double line gives 95\,\% limits
for some 1-parameter extensions to the \LCDM\ model.
More details can be found in \citet{planck2016-l06}.
\label{tab:LCDMparams}}
\begingroup
\openup 4pt
\newdimen\tblskip \tblskip=5pt
\nointerlineskip
\vskip -5mm
\footnotesize
\setbox\tablebox=\vbox{
    \newdimen\digitwidth
    \setbox0=\hbox{\rm 0}
    \digitwidth=\wd0
    \catcode`*=\active
    \def*{\kern\digitwidth}
    \newdimen\signwidth
    \setbox0=\hbox{+}
    \signwidth=\wd0
    \catcode`!=\active
    \def!{\kern\signwidth}
\halign{
\hbox to 1.0in{$#$\leaderfil}\tabskip=1.0em&
        \hfil$#$\hfil&
        \hfil$#$\hfil\tabskip=0.5pt\cr
\noalign{\doubleline}
\omit\hfil Parameter\hfil&\omit\hfil \Planck\ alone\hfil&\omit\hfil \Planck\ + BAO\hfil\cr
\noalign{\vskip 5pt\hrule\vskip 5pt}
\Omega_{\mathrm{b}} h^2&0.02237\pm 0.00015&0.02242\pm 0.00014\cr
\Omega_{\mathrm{c}} h^2&0.1200\pm 0.0012&0.11933\pm 0.00091\cr
100\theta_{\mathrm{MC}}&1.04092\pm 0.00031&1.04101\pm 0.00029\cr
\tau&0.0544\pm 0.0073&0.0561\pm 0.0071\cr
\ln(10^{10} A_\mathrm{s})&3.044\pm 0.014&3.047\pm 0.014\cr
n_\mathrm{s}&0.9649\pm 0.0042&0.9665\pm 0.0038\cr
\noalign{\vskip 5pt\hrule\vskip 3pt}
H_0&67.36\pm 0.54*&67.66\pm 0.42*\cr
\Omega_\Lambda&0.6847\pm 0.0073&0.6889\pm 0.0056\cr
\Omega_{\mathrm{m}}&0.3153\pm 0.0073&0.3111\pm 0.0056\cr
\Omega_{\mathrm{m}} h^2&0.1430\pm 0.0011&0.14240\pm 0.00087\cr
\Omega_{\mathrm{m}} h^3&0.09633\pm 0.00030&0.09635\pm 0.00030\cr
\sigma_8&0.8111\pm 0.0060&0.8102\pm 0.0060\cr
\sigma_8(\Omega_{\mathrm{m}}/0.3)^{0.5}&0.832\pm 0.013&0.825\pm 0.011\cr
z_{\mathrm{re}}&7.67\pm0.73&7.82\pm0.71\cr
\mathrm{Age}[\mathrm{Gyr}]&13.797\pm 0.023*&13.787\pm 0.020*\cr
r_\ast[\mathrm{Mpc}]&144.43\pm 0.26**&144.57\pm 0.22**\cr
100\theta_\ast&1.04110\pm 0.00031&1.04119\pm 0.00029\cr
r_{\mathrm{drag}}[\mathrm{Mpc}]&147.09\pm 0.26**&147.57\pm 0.22**\cr
z_{\mathrm{eq}}&3402\pm 26**&3387\pm 21**\cr
k_{\rm{eq}}[\mathrm{Mpc}^{-1}]&0.010384\pm 0.000081&0.010339\pm 0.000063\cr
\noalign{\doubleline}
\Omega_K& -0.0096\pm 0.0061 & 0.0007\pm 0.0019\cr
\Sigma m_\nu\,[\mathrm{eV}]& < 0.241& < 0.120\cr
N_{\mathrm{eff}}& 2.89^{+0.36}_{-0.38}& 2.99^{+0.34}_{-0.33}\cr
r_{0.002}& < 0.101& < 0.106\cr
\noalign{\vskip 4pt\hrule\vskip 3pt}
} 
} 
\endPlancktable
\endgroup
\end{center}
\end{table}

\subsection{Assumptions underlying \texorpdfstring{$\Lambda$CDM}{LCDM}}
\label{sec:assumptions}

A complete list of the assumptions underlying the $\Lambda$CDM model is
not the goal of this section, but below we list several of the major
assumptions.

\begin{enumerate}
\item[{\sf A1}] Physics is the same throughout the observable Universe.
\item[{\sf A2}] General Relativity (GR) is an adequate description of gravity.
\item[{\sf A3}] On large scales the Universe is statistically the same
everywhere (initially an assumption, or ``principle,'' but now strongly implied
by the near isotropy of the CMB).
\item[{\sf A4}] The Universe was once much hotter and denser and has been
expanding since early times.
\item[{\sf A5}] There are five basic cosmological constituents:
    \begin{enumerate}
    \item[({\sf a})] Dark energy that behaves just like the energy density of
    the vacuum.
    \item[({\sf b})] Dark matter that is pressureless
          (for the purposes of forming structure),
          stable, and interacts with normal matter only gravitationally.
    \item[({\sf c})] Regular atomic matter that behaves just like it does on
          Earth.
    \item[({\sf d})] The photons we observe as the CMB.
    \item[({\sf e})] Neutrinos that are almost massless (again for structure
          formation) and stream like non-interacting, relativistic particles
          at the time of recombination.
    \end{enumerate}
\item[{\sf A6}] The curvature of space is very small.
\item[{\sf A7}] Variations in density were laid down everywhere at early times,
and are Gaussian, adiabatic, and nearly scale invariant
(i.e., proportionally in all constituents and with similar amplitudes
 as a function of scale)
as predicted by inflation.
\item[{\sf A8}] The observable Universe has ``trivial'' topology (i.e., like
$\mathbb{R}^3$).  In particular it is not periodic or multiply connected.
\end{enumerate}

With these assumptions it is possible to predict a wide range of observations
with a very small number of parameters.  The observed fact that the
fluctuations in temperature and polarization in the CMB are small makes
the calculation of CMB observables an exercise in linear perturbation theory
(see \citealt{Peacock99}, \citealt{Dodelson03}, \citealt{2005pfc..book.....M}, \citealt{Peter:1208401} and \citealt{LL09} for textbook treatments, and
\citealt{Partridge95} and \citealt{Peebles2009} for historical discussions).
The evolution of the perturbations in each species can be computed to high
accuracy using a ``Boltzmann code'' once the initial conditions, constituents,
and ionization history are specified.
The initial conditions are part of our assumptions.
The high-$z$ part of the ionization history can be computed to high
accuracy given the assumptions above
(see, e.g., extensive discussion and references in \citealt{planck2014-a25}).
Thus one needs to specify only the values of the constituents and the low-$z$
part of the ionization history.

\subsection{\Planck's constraints on \texorpdfstring{$\Lambda$CDM}{LCDM} parameters}
\label{sec:lcdm_params}

To fully prescribe the $\Lambda$CDM model we need to specify its parameters.
Adopting the convention that the Hubble parameter today is
$H_0=100\,h\,{\rm km}\,{\rm s}^{-1}{\rm Mpc}^{-1}$, we take these to be:
the density of cold dark matter, $\omega_{\rm c}=\Omega_{\rm c}h^2$;
the density of baryons, $\omega_{\rm b}=\Omega_{\rm b}h^2$
(consisting of hydrogen, and helium with mass fraction $Y_{\rm P}$
obtained from standard BBN); the amplitude, $A_{\rm s}$, and spectral index,
$\ns$, of a power-law spectrum of adiabatic perturbations;
a proxy ($\theta_\mathrm{MC}$; Eq.~6 of \citealt{planck2013-p11})
for the angular scale of the acoustic oscillations, $\theta_\ast$; and
the optical depth to Thomson scattering from reionization, $\tau$.
The best-fit model and constraints on these parameters are given in
Tables~\ref{tab:LCDMbest_fit} and \ref{tab:LCDMparams}.

We assume that the radiation is made up of photons (as a blackbody
with $T=2.7260\,$K, \citealt{fixsen2009}) and neutrinos with
$\rho_\nu=N_{\rm eff}(7/8)(4/11)^{4/3}\rho_\gamma$ and\footnote{A newer
evaluation gives $N_{\rm eff}=3.045$ \citep{SalPas16}.  The difference is
negligible for our purposes, so we keep the older number for consistency
with previous results.} $N_{\rm eff}=3.046$ \citep{Man02}.
The neutrinos are assumed to have very low masses, which we approximate
as a single eigenstate with $m_\nu=0.06\,$eV.
Other parameters can be derived from these and the assumptions that we
already spelled out.  For example, since
$\left|\Omega_K\right|\ll 1$, we have $\Omega_\Lambda=1-\Omega_{\rm m}$, and the
redshift of equality can be found from
$\rho_\gamma+\rho_\nu=\rho_{\rm c}+\rho_{\rm b}$ (assuming neutrinos are
relativistic at $z>10^3$, as required by the current data).
A list of derived parameters and their relation to the base parameters
can be found in \citet{planck2014-a16} or Tables~\ref{tab:LCDMbest_fit}
and \ref{tab:LCDMparams}.
Further discussion of how the parameters affect the anisotropy spectra can
be found in the aforementioned textbooks or in
\citet{planck2014-a16} and \citet{planck2016-LI}.

Figure~\ref{fig:cl} shows the measured angular power spectra from \Planck,
with the blue line representing the best-fit $\Lambda$CDM model.
Beginning with the $TT$ spectrum, one can see three regions, separated by
two characteristic scales.
On scales larger than the Hubble scale at last scattering (low $\ell$) the
almost scale-invariant spectrum is a pristine imprint of the initial conditions.
On degree angular scales the almost harmonic sequence of power maxima
represents the peaks and troughs in density and temperature of the
baryon-photon fluid as it oscillates in the gravitational potentials
prior to recombination.
On scales smaller than the geometric mean\footnote{The diffusion scale is
the mean free path times the square root of the number of scatterings.  Since
photons travel at $c$, $N_{\rm scatter}$ scales as $c$ times the Hubble time
divided by the mean free path, so $N_{\rm scatter}^{1/2}\lambda_{\rm mfp}$
is the geometric mean of the Hubble scale and $\lambda_{\rm mfp}$.}
of the Hubble scale and the mean free path, photon diffusion during the
epoch of recombination erases the fluctuations.
A similar behaviour is seen in the polarization spectra, without the
low-$\ell$ plateau and with sharper peaks that are sourced primarily by
the quadrupole anisotropy generated during last scattering.
Not visible by eye, but included in the calculation, are slight changes
to the primordial signal due to gravitational lensing by large-scale
structure along the line of sight.

Figure~\ref{fig:cl} nicely illustrates the three conditions that make the CMB
such a powerful cosmological probe:
(i) exquisite measurements with well controlled and understood systematic
errors;
(ii) a reliable and computationally tractable framework for statistical
inference and well understood statistical errors; and
(iii) a rich phenomenology predicted by a precise theoretical model, allowing
simultaneous and tight constraints on key parameters.

The best determined parameter is $\theta_\ast$ (for which $\theta_{\rm MC}$ is a
proxy), which is constrained to better than $0.03\,\%$ by the peak and trough
positions.
Since $\theta_\ast$ is determined by the positions of the extrema, not their
amplitudes, the measurement is extremely stable and only weakly dependent
upon the model details.
One of the impressive consistency checks of the paradigm is that $\theta_\ast$
determined from the temperature power spectrum matches to high precision those 
determined from the polarization power spectrum and from the cross-spectrum
between temperature and polarization.  This limits the fraction of the
perturbations that were not adiabatic in nature.
The angular scale of the acoustic oscillations measures the ratio of the
(comoving) angular diameter distance to last scattering and the sound horizon,
$\rstar=\int c_{\rm s}\,dt=\int d\eta/\sqrt{3(1+3\rho_{\rm b}/4\rho_\gamma)}$,
with $\eta$ the conformal time.
Within the $\Lambda$CDM model, $\rstar$ depends on the sound speed and the
Hubble scale at last scattering, which is primarily determined by the baryon
and matter densities.
The angular-diameter distance depends primarily upon the late-time evolution
and geometry, and within $\Lambda$CDM this translates into a dependence on $h$
and $\omega_{\rm m}$.  Since $\omega_{\rm b}$
(which changes the mass loading of the photon-baryon fluid and hence the ratio
of gravity to pressure) is well constrained ($<1\,\%$) by the relative
amplitudes of the acoustic peaks, the $\theta_\ast$ measurement provides a
very tight constraint in the 2-dimensional $\Omega_{\rm m}$--$h$ subspace:
\begin{equation}
  \Omega_{\rm m} h^3 = 0.09633 \pm 0.00029 \quad (68\,\%\ {\rm CL}).
\label{eqn:omegah3}
\end{equation}

The direction orthogonal to the $\Omega_{\rm m} h^3$ line is less
well constrained.
Changes in $\Omega_{\rm m} h^2$ affect the damping scale and the amount by
which the gravitational potentials are determined by the cold dark matter
(which does not take part directly in the acoustic oscillations),
as opposed to the amount determined by the baryon-photon fluid.
This alters the relative heights of the peaks, 
allowing a sub-percent-level measurement of both $\Omega_{\rm m} h^2$ and $h$,
and hence constraints on $\Omega_{\rm m}$ and $\Omega_\Lambda$.

Changes in the primordial spectral index, $n_{\rm s}$, yield a corresponding
change in the observed CMB power spectrum.  Increasing $n_{\rm s}$, with the
amplitude fixed at the pivot point $k=k_0=0.05\,{\rm Mpc}^{-1}$, increases
power at $\ell>500$ while decreasing power at $\ell<500$, since modes with
$k=k_0$ project onto angular scales close to $\ell=500$.  Given the
large lever arm of \Planck, measuring three decades in wavenumber, we can
isolate this tilt precisely and have shown that it departs from scale
invariance at more than $8\,\sigma$.

Finally, reionization in the late Universe ($z\la10$) recouples the CMB
photons to the matter field (but not as tightly as before recombination,
since the matter density has dropped by six orders of magnitude
in the intervening period).  Scattering of photons off electrons in the
ionized intergalactic medium suppresses the power in the primary anisotropies
on scales smaller than the Hubble scale at reionization ($\ell>10$) by
$e^{-2\tau}$, only weakly generating new anisotropies.
More importantly for our ability to measure $\tau$, the scattering of
photons during this period generates additional polarization on large scales
(set by the angle subtended by the Hubble scale at reionization),
whose amplitude scales as $C_\ell^{EE}\propto\tau^2$.
The combination of high sensitivity with all-sky coverage allows \Planck\ to
measure this large-angle signal in order to constrain $\tau$ and limits
the redshift of reionization to $<9$ at the 95\,\% confidence level.

\begin{figure}[htbp]
\begin{center}
\resizebox{\columnwidth}{!}{\includegraphics{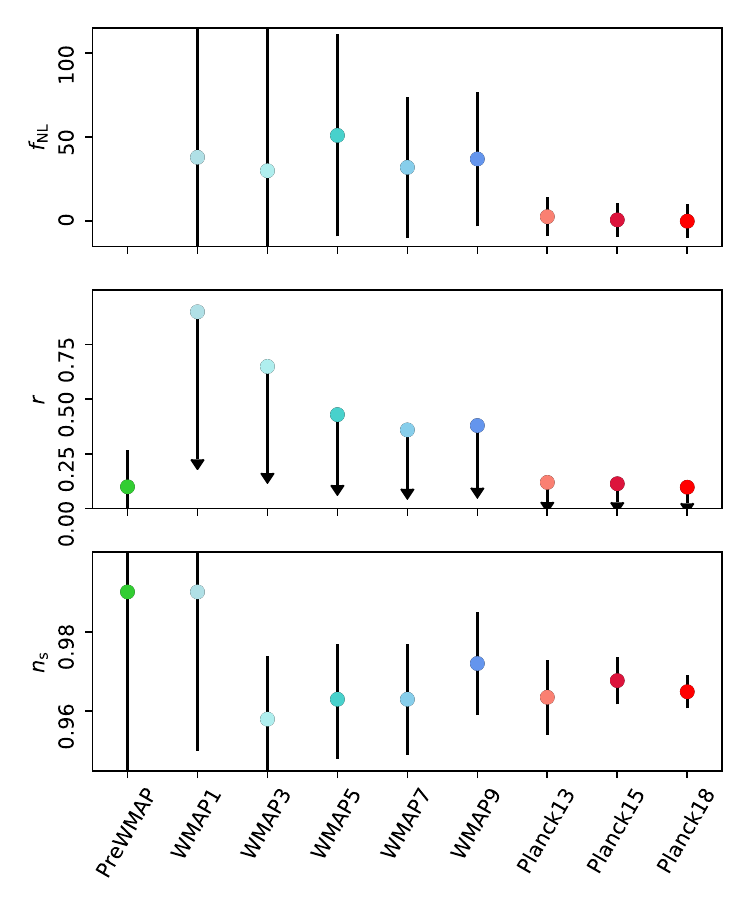}}
\end{center}
\caption{Evolution of CMB constraints on assumptions and parameters describing
``early Universe physics,'' specifically the amount of primordial, local
non-Gaussianity ($f_{\rm NL}$), the tensor-to-scalar ratio ($r$), and the
slope of the primordial power spectrum ($n_{\rm s}$).}
\label{fig:params_earlyU}
\end{figure}

\begin{figure}[htbp]
\begin{center}
\resizebox{\columnwidth}{!}{\includegraphics{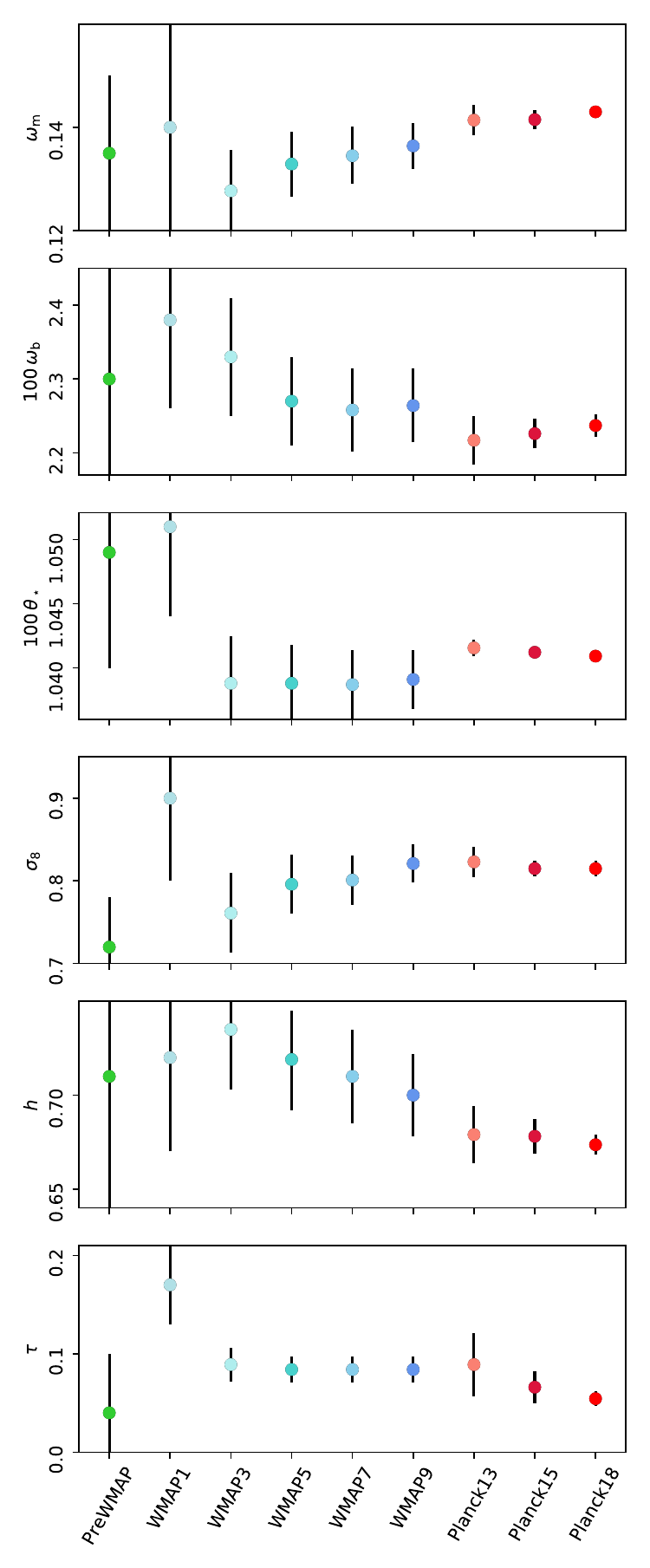}}
\end{center}
\caption{Evolution of CMB constraints on parameters describing
``late time physics,'' specifically the matter density
($\omega_{\rm m}\equiv\Omega_{\rm m}h^2$),
the baryon density ($\omega_{\rm b}\equiv\Omega_{\rm b}h^2$),
the acoustic scale ($\theta_\ast$),
the normalization of the (linear theory) matter power spectrum ($\sigma_8$),
the dimensionless Hubble constant ($h$), and the Thomson optical depth
($\tau$).}
\label{fig:params_lateU}
\end{figure}

\begin{figure}[htbp]
\begin{center}
\resizebox{\columnwidth}{!}{\includegraphics{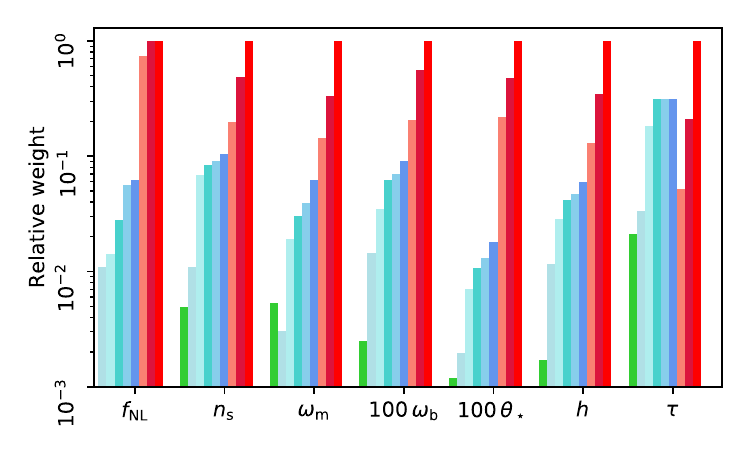}}
\end{center}
\caption{Increase in the ``statistical weight'' (i.e., $1/\sigma^2$, where
$\sigma$ for each parameter comes from marginalising over the rest of the set)
for a selection of $\Lambda$CDM parameters as a function of time.  The
bars represent the same divisions as in Figs.~\ref{fig:params_earlyU}
and \ref{fig:params_lateU}: pre-\WMAP\ (green); WMAP1, WMAP3, WMAP5, WMAP7,
and WMAP9 (blue shades); and Planck13, Planck15, and Planck18 (red
shades).}
\label{fig:params_weight}
\end{figure}

\begin{figure*}[htbp]
\begin{center}
\resizebox{\textwidth}{0.91\textheight}{\includegraphics{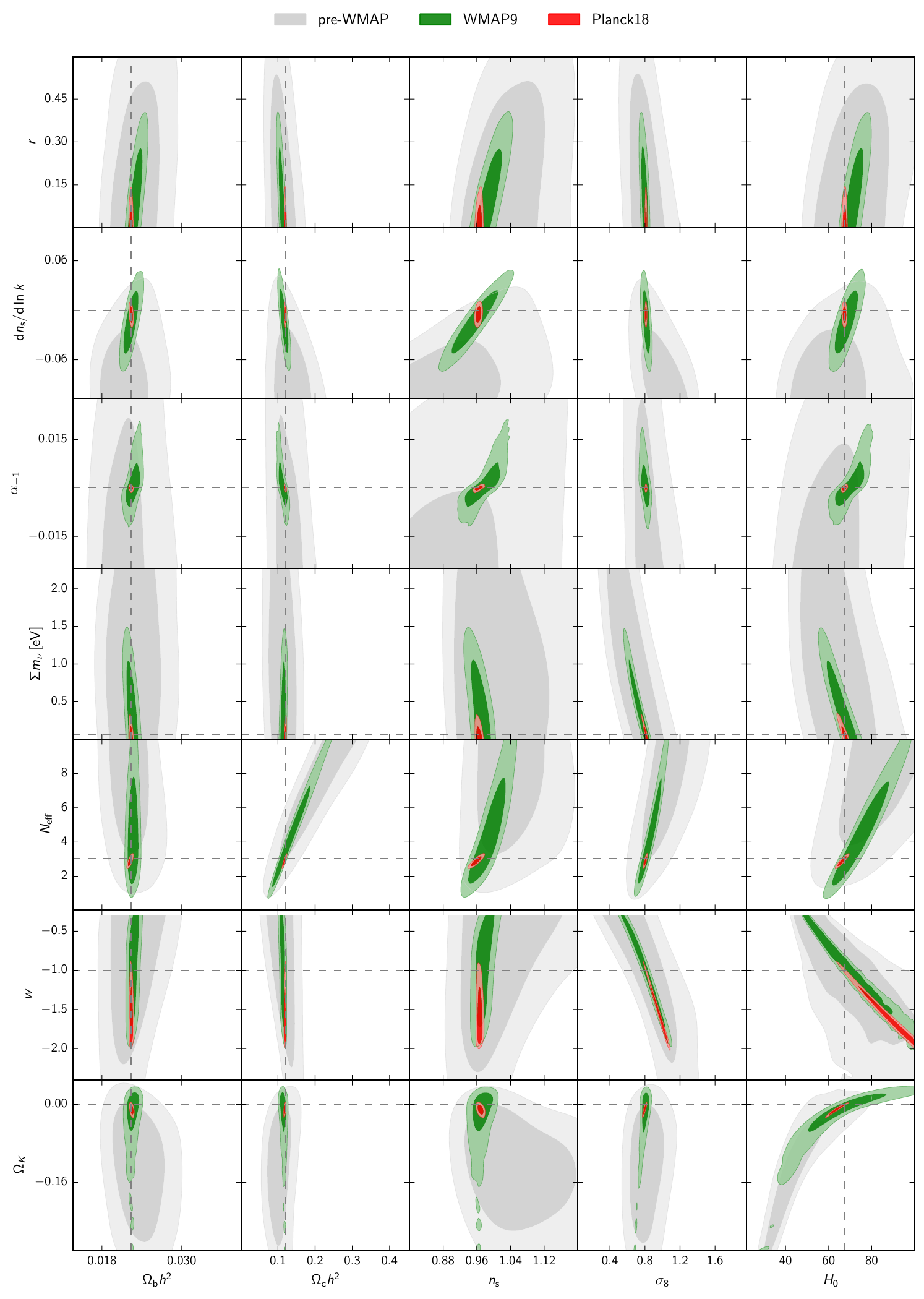}}
\end{center}
\caption{Successive reductions in the allowed parameter space for various one-parameter extensions to \LCDM, from pre-\WMAP\ (the MAXIMA, DASI, BOOMERANG, VSA, and DASI experiments) to \Planck. Each row corresponds to a specific extended model. The contours display the 68\,\% and 95\,\% confidence limits for the extra parameter versus five other base-\LCDM\ parameters. The dotted lines indicate the \LCDM\ best-fit parameters or fixed default values of the extended parameters.
\label{fig:zooming-in}}
\end{figure*}

\begin{figure}[htbp]
\begin{center}
\resizebox{\columnwidth}{!}{\includegraphics{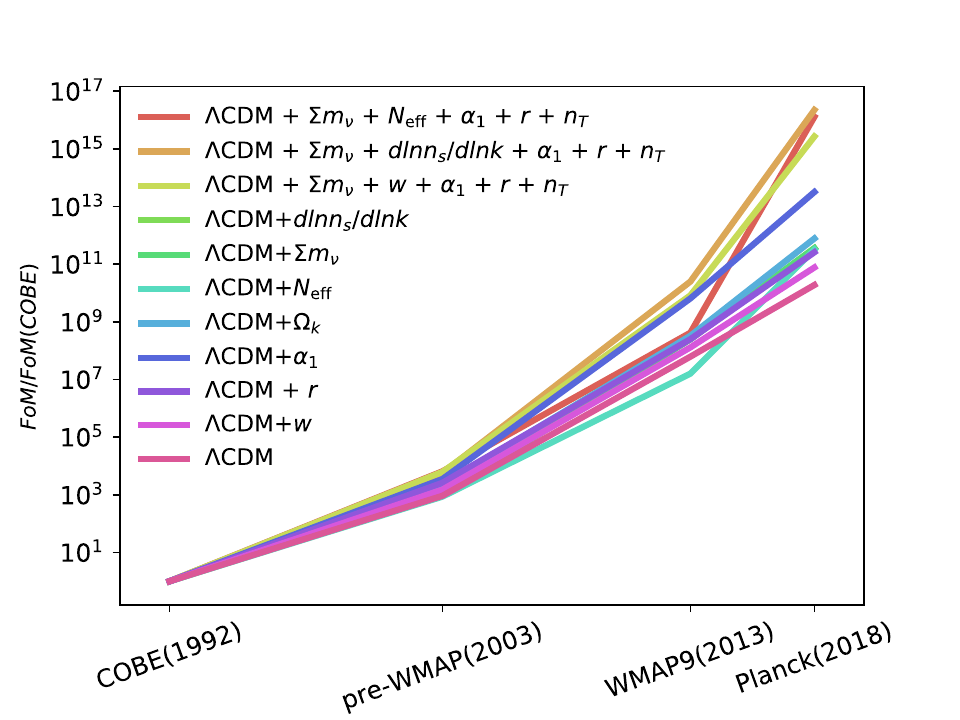}}
\end{center}
\caption{Successive improvements of the figure of merit FoM\,$= \left\{ \det \left[ {\rm Cov} \big( \Omega_{\rm b} h^2; \Omega_{\rm c} h^2; \tau, A_{\rm s}; n_{\rm s}; \ldots\big) \right] \right\}^{-1/2}$ for various models and data sets, relative to \COBE. This shows the relative improvement with respect to the anisotropy discovery experiment, \COBE\ (with first results in 1992). For \COBE, we have additionally (anachronistically) assumed a \Planck\ prior on the optical depth $ 0.055\pm0.009$. The relative reduction of the allowed phase space volume is impressive for all models, with even greater shrinkage in volume for higher-dimensional model extensions. For \LCDM, the improvement is more than $10^{10}$. For the largest model spaces, having four or five additional dimensions compared to \LCDM, this improvement is more than $10^{16}$ in 26 years, corresponding to a 6-month doubling time, three times faster than Moore's Law!  This is one reason why the study of the CMB has allowed us to address more and more ambitious questions with time, a feature that is expected to continue with future experiments.
\label{fig:FOMs} }
\end{figure}

To demonstrate the impressive advances in the field, we show in
Figs.~\ref{fig:params_earlyU} and \ref{fig:params_lateU} the evolution
of constraints on some of the parameters of the base $\Lambda$CDM model and
its most common extensions, in
Fig.~\ref{fig:params_weight} the improvement in statistical weight, and in
Figs.~\ref{fig:zooming-in} and \ref{fig:FOMs} the improvements in a number
of extensions.  Figure~\ref{fig:params_earlyU} focusses on parameters
describing ``the early Universe,'' while Fig.~\ref{fig:params_lateU}
presents late-time and derived parameters.
In order to avoid too many arbitrary choices, we have opted to plot only
CMB constraints and have started the historical development
with the pre-\WMAP\ compilations of \citet{Wang03} and \citet{Bond03}.
The values for \WMAP\ and \Planck\ are taken from
the LAMBDA archive\footnote{\url{http://lambda.gsfc.nasa.gov}} and the PLA,
respectively.

The top two panels of Fig.~\ref{fig:params_earlyU} indicate non-detections
of non-Gaussianity and primordial tensor models, respectively, with
dramatically improved precision.  The last panel shows how the primordial
power spectrum is now convincingly known to depart from scale invariance
($n_{\rm s}=1$), with more power at large scales\footnote{This is the direction
predicted by the simplest models of inflation, which invoke a scalar field
slowly rolling down an almost flat potential, with longer wavelength modes exiting the Hubble scale earlier.} than a scale-invariant spectrum.
The \Planck\ data demonstrate this departure from scale invariance in a way
that is robust to single-parameter extensions of the basic $\Lambda$CDM model.

Figure~\ref{fig:params_lateU} shows a dramatic shrinking of the error bars on
the late-time parameters, a reduction that becomes even more impressive
considering that they are all being constrained simultaneously.  Except for
the optical depth, $\tau$, the parameters are simultaneously known with
percent-level precision.

Another view of the dramatic increase in precision on these key parameters
describing our Universe is shown in Fig.~\ref{fig:params_weight}.  Here we
present, for a selection of parameters, how the ``statistical weight'' has
improved over time.  We use the inverse variance on each parameter
(marginalised over all of the others and normalized to unity for the last
\Planck\ point) as a proxy for statistical weight.
While other choices could be defended, this provides one way of seeing how
the continuing high-quality fits of the model and improvements in the data
have refined our knowledge of these key parameters.

Particularly impressive in Fig.~\ref{fig:params_weight} are the improvements
in measurement of the densities, $\omega_{\rm m}$ and $\omega_{\rm b}$,
and the present-day expansion rate, $h$, each now measured at over
$100\,\sigma$.
These parameters are key to defining both the evolution of the background
cosmology and the shape of the matter power spectrum describing large-scale
structure.  The dramatic improvements visible in Fig.~\ref{fig:params_weight}
translate directly into improvements in our ability to convert redshift into
times or distances, to measure volumes and number densities, and to characterize
the cosmic web within which all astrophysical objects (e.g., galaxies) form
and evolve.

Finally we emphasize the large step forward taken with the \Planck\ data
by showing in Fig.~\ref{fig:zooming-in} how the constraints on 1-parameter
extensions to $\Lambda$CDM have improved in going from pre-\WMAP\ to \Planck.
For pre-\WMAP, we have included the joint constraints from the BOOMERANG,
MAXIMA, DASI, VSA, and CBI experiments
\citep{BOOMERANG_data, MAXIMA_data, DASI_data, VSA_data,CBI_data}. 
Prior to \WMAP, there were few meaningful constraints on extended models,
even those with only one additional parameter.  The situation improved with
the \WMAP\ measurements, but many extensions remained highly unconstrained.
With the advent of \Planck, most of these 1-parameter extensions are now
highly constrained, and become even more so if additional data are added.

Figure~\ref{fig:FOMs} provides a different view of this same improvement,
extending farther back to \COBE\
(a data set of three bands from $\ell=2$--26;
\citealt{COBE_data}) supplemented by a \Planck\ prior on the optical depth
$\tau = 0.055\pm 0.009$. 
It shows the impressive increase of the figure of merit, defined by
${\rm FoM}^{-2}=\det\left[\mathrm{Cov}\big(\Omega_{\rm b}h^2;
 \Omega_{\rm c}h^2;\tau,A_{\rm s};n_{\rm s};\ldots\big)\right]$,
for various models and data sets, relative to \COBE.
The relative reduction of the allowed parameter space volume is impressive 
for all models. For \LCDM, the 6-dimensional space has decreased in volume
by about $10^{10}$ in the 26 years since the initial discovery.
For the 11-dimensional models that we also consider here,
the reduction is a million times larger.  

\subsection{\Planck's tests of \texorpdfstring{$\Lambda$CDM}{LCDM} assumptions}

\begin{figure}[htbp]
\begin{center}
\resizebox{\columnwidth}{!}{\includegraphics{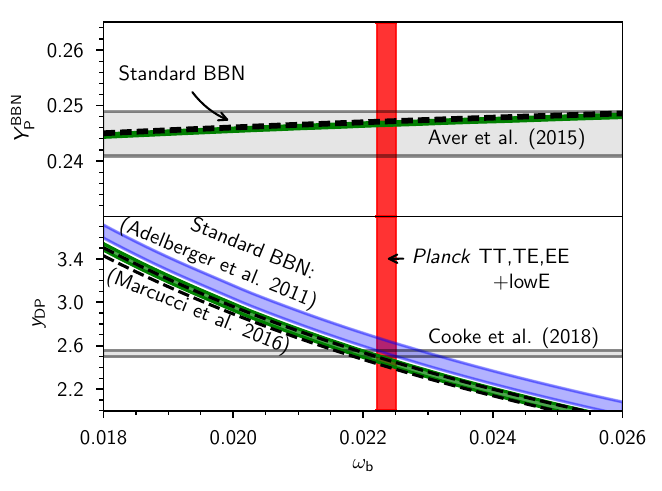}}
\end{center}
\caption{Primordial helium ($Y_{\rm P}$) and deuterium ($y_{\rm DP}$)
abundances as a function of baryon density ($\omega_{\rm b}$), from
\protect\citet{planck2016-l06}, assuming $N_{\rm eff}=3.046$.
The shaded, horizontal bands show the 68\,\% CL on the
measured values from \protect\citet{Aver15} and \protect\citet{Cooke18}.
The red vertical band is the 68\,\% CL on $\omega_{\rm b}$ from
the $\Lambda$CDM model fit to the \Planck\ temperature and polarization data,
while the other bands show the predictions from the theory of nucleosynthesis
(including uncertainties in nuclear reaction rates and neutron lifetime; see
\protect\citealt{planck2016-l06} for more details).
The excellent agreement on the inferred value of $\omega_{\rm b}$ from
processes in the first $3\,$minutes of the Universe's history with that from
the CMB at $380\,000\,$years after the big bang is one of our best
demonstrations of the universality of the laws of physics.}
\label{fig:bbn}
\end{figure}

One of the strongest pieces of evidence for the universality of physics
(point~{\sf A1} in Sect.~\ref{sec:assumptions}) comes from the agreement
between the baryon density, $\omega_{\rm b}$, as measured by the CMB and
through consideration of BBN.  The inference from
the CMB relies on the acoustic physics of the primordial plasma before
400\,000\,yr.  The inference from BBN depends upon modelling nuclear
physics in the first $3\,$minutes after the big bang, calibrated by laboratory
measurements here on Earth.  The comparison invokes all of the known forces
of nature: strong and weak nuclear, electromagnetic, and gravity.
The level of agreement is remarkable, as shown in Fig.~\ref{fig:bbn}.

The connection between cosmology and GR (point {\sf A2}) goes back
to the founding of both subjects in the early part of the 20th century.
GR has been extensively tested on the scale of the
Solar System \citep[e.g.,][]{Will06}.
The recent direct detection of gravitational waves \citep{LIGO} provides
a further confirmation of the theory in the strong gravitational fields
of merging black holes.  The detection of an optical counterpart
\citep{MultiMessenger}
provides stringent limits on the speed of propagation of gravitational
waves and thus on modified gravity theories
\citep{Lombriser15,CreVer17,MarZum17,Sakstein17,Baker17,Crisostomi17,Amendola17}.
By contrast, constraints on modifications of gravity on large scales are
weaker, although complementary because they apply to an entirely different
regime.

The structure of the peaks in the anisotropy power spectra depends upon
the gravity-driven oscillations of a relativistic fluid, and as such is
sensitive to departures from the predictions of GR.
Indeed, most modifications of gravity take as a starting point that GR be
restored in the early Universe, precisely in order to avoid modifying the
predictions of CMB anisotropies \citep{JaiKho10,Joyce15,Joyce16,Amendola18}.

In the presence of inhomogeneity, the metric of space-time is perturbed from
its Friedmann form.  It is common to parametrize the deviations to the
time-time and space-space components by two potentials
(often denoted $\Psi$ and $\Phi$, where $\Psi$ is the Newtonian potential
while $\Phi$ represents the General Relativistic effect of the bending of
space by gravity).
General Relativity predicts that, in the absence of anisotropic stresses,
$\Psi$ and $\Phi$ should be equal in magnitude
\citep{Peacock99,Dodelson03,LL09}.
The \Planck\ data alone can place a constraint on the deviation of the
two metric potentials from the GR prediction at the last-scattering surface.
Assuming that the modification to the potentials is scale independent, the
ratio of the potentials (in units of the GR prediction\footnote{Specifically
we define $\eta_{\rm slip}$ through
$k^2\left[\Phi-\eta_{\rm slip}\Psi\right]=12\pi Ga^2(\rho+p)\,\sigma$,
where $\sigma$ is the anisotropic stress.}) is
\begin{equation}
  \eta_{\rm slip}  = 1.004 \pm 0.007 \quad (68\,\%\ {\rm CL}).
\end{equation}
This shows that gravity is behaving in the early Universe exactly as
predicted by General Relativity, and is one of the tightest constraints on
the behaviour of the potentials at such early epochs.

At late times, direct constraints on modifications to GR on cosmological
scales are weaker.  \Planck\ has also served an important role in these
constraints by providing an all-sky map of lensing, which can be compared to
dynamical measurements at relatively recent epochs.
Gravitational lensing measures the combination of $\Phi$ and $\Psi$, while
the motions of non-relativistic objects such as galaxies probe only the
time-time component ($\Psi$).
It is this fact that accounts for the famous ``factor of 2'' in Einstein's
prediction for the bending of light by the Sun, as tested in the eclipse
expedition of 1919.  For this reason, a comparison of the two measures
provides a useful check of GR on cosmological scales.
The fact that the $\Lambda$CDM model provides a good fit to a wide range
of auto- and cross-correlations (Sect.~\ref{sec:concordance}) suggests that
GR passes this test.
\citet{Pullen16} and \citet{Singh18} quantified this expectation by
cross-correlating low-$z$ galaxies with the \Planck\ lensing maps,
finding consistency with the predictions of GR on tens of Mpc scales,
although with intriguing tension on very large scales.
Finally, the large-scale gravitational potentials are predicted to decay
once the expansion of the Universe begins to accelerate, leading to an
additional source of anisotropy: the integrated Sachs-Wolfe (ISW) effect
\citep{SacWol67}.  Here the blueshift of photons falling into a potential
is not precisely cancelled by the redshift upon climbing out due to the
evolution of the potential during traversal.  \Planck's measurements of
the ISW effect are consistent with expectations
\citep{planck2013-p14,planck2014-a26}.

The low level of the relative fluctuations in the CMB provides some of our
strongest evidence for the statistical homogeneity and isotropy of the Universe
(point {\sf A3}), and the impressive fit of the $\Lambda$CDM model predictions
to the observations relies on all of the above assumptions.

The blackbody nature of the CMB is the best evidence
that the Universe was once hot and dense (point {\sf A4}).
Further evidence comes from the wiggles we see in the angular power
spectrm, which arise due to acoustic oscillations in the baryon-photon
plasma.  This implies that the Universe was hot enough to ionize hydrogen and
dense enough to support acoustic oscillations without excessive dissipation.

The constituents of the Universe (point {\sf A5}) have been described
previously.  Within the $\Lambda$CDM paradigm the CMB allows measurements to
be made of the total density of these components to high precision.
We shall return to points {\sf A6} and {\sf A7} when we discuss the impact of
\Planck\ on studies of inflation and fundamental physics.
Here we simply highlight how the CMB spectra provide strong evidence that
the fluctuations from which all structure grows were laid down at very early
times.  The combination of the regular, oscillatory structure of the CMB
peaks and the relative phases of the temperature and polarization spectra
implies that the perturbations responsible for CMB anisotropies were
``primordial'' and ``apparently acausal''
\citep{Cou94,CriTur95,TestInf,HuWhiSpe97,SpeZal97}.  By $z\,{\simeq}\,1100$
the fluctuations were all in their growing mode, and there
is no evidence for ``active sources'' during this period \citep[e.g., cosmic
strings or textures, whose motion generates anisotropies,
see][]{planck2013-p20}.
Importantly, there were fluctuations in spatial curvature on scales larger
than the Hubble length at last scattering.

Two of the most significant properties of dark energy, for cosmology, are
that it be spatially nearly constant and only recently relevant.
Sections~\ref{sec:darkenergy}, as well as \citet{planck2014-a16} and
\citet{planck2016-l06} discuss \Planck's constraints on these properties.
To give just one example, the combination of \Planck\ data with other,
lower-redshift data sets demands that the dark-energy contribution must
rise from less than $10\,\%$ of the total to nearly $70\,\%$ of the total
within just the last $e$-fold of expansion and the contribution from any
``early'' dark energy must be highly sub-dominant.
The constraints on DM decays and neutrino masses are dealt with in
Sect.~\ref{sec:neutrinos} and Sect.~\ref{sec:darkmatter}.

In addition, the \Planck\ maps provide the highest quality, full-sky view of
the surface of last scattering that we have, and as such allow us to place
extremely tight constraints on departures from a globally isotropic cosmology
with trivial topology \citep[point {\sf A8};][]{planck2014-a20}.
Searches for cubic toroidal or slab topologies yield no detection, with a
scale below the diameter of the last-scattering surface.

\section{Cosmic concordance} \label{sec:concordance}

The \Planck\ data constrain the parameters of the base, 6-parameter,
$\Lambda$CDM model with high precision, without the need for any external
data sets.  With the model tested and constrained, it can then be used to
make predictions for a host of other astrophysical measurements.
Despite its apparent simplicity, the model -- with the \Planck-constrained
parameters -- has proven to be extremely successful in describing a wide range
of cosmological data across four orders of magnitude in scale and $13.8\,$Gyr
of cosmic history.

In this section we describe the extent to which the predictions of this model
are in accord with other data sets and point out where there are tensions.

\subsection{The CMB sky} \label{sec:thecmbsky}

\Planck\ is not the first experiment to measure CMB anisotropies, nor will it
be the last.  So we begin our discussion of concordance by assessing the degree
to which different measurements of the CMB sky by different experiments agree.
Our focus will be on the most recent and powerful experiments, since these
provide the most stringent tests.

Internal consistency checks and jackknife tests, including splits by spatial
and electromagnetic frequency, are discussed extensively in
\citet{planck2016-LI}, \citet{planck2016-l05}, and \citet{planck2016-l06},
and we refer the reader to those papers and to Appendix \ref{sec:changes}
for details.  A discussion of HFI-LFI consistency is given in
Appendix \ref{sec:inst-consistency}.
While some mild tensions exist, overall the data are highly consistent.
One of the newest consistency checks that is made available by the latest
\Planck\ data is a comparison of the temperature and polarization power
spectra.

\begin{figure*}[!htpb]
\begin{center}
\resizebox{\textwidth}{!}{\includegraphics{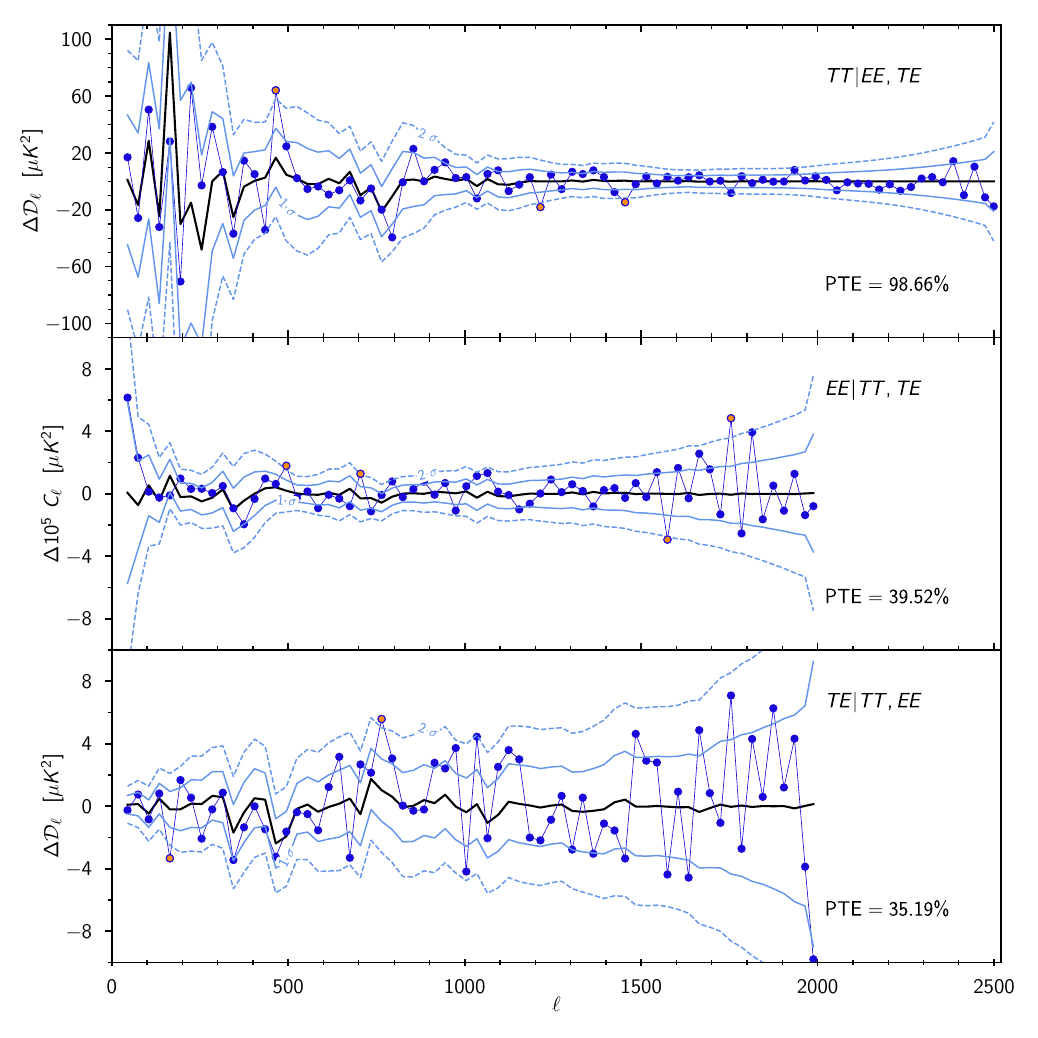}}
\end{center}
\caption{Conditional residuals for the co-added $TT$ (top panel), $EE$ (middle), and $TE$ (bottom) power spectra. The blue points show the difference between the co-added spectra and the 2018 base-\LCDM\ spectra, with the points at more than 2$\,\sigma$ coloured pink. The black lines show the difference between the conditional prediction of the spectrum and the base model. The prediction for a given spectrum is performed (within the framework of base \LCDM) conditional on the two others, e.g., in the top panel, the $TT$ prediction is conditioned on both the $TE$ and $EE$ data. The solid and dashed blue lines show the $\pm1\,\sigma$ and $\pm2\,\sigma$ contours of the prediction (around the black line), corresponding to the diagonal of the block of the conditional covariance computed from the 2018 covariance matrix and data. Probabilities to exceed (PTEs) are computed for the difference between the data and its conditional prediction using the conditional covariance for each panel. We see that any pair of spectra predicts the third one well (assuming that \LCDM\ is a good model), bringing support to the consistency of the temperature and polarization measurements within \LCDM. This is particularly true at low and intermediate multipoles (where \Planck\ is cosmic-variance limited), where the conditionals successfully predict the deviations of the co-added spectra from the theoretical base-\LCDM\ spectra.
\label{fig:cond_vs_coadd} }
\end{figure*}

That the CMB sky be linearly polarized is a direct consequence of the existence
of the anisotropies and the polarization dependence of Thomson scattering,
which itself traces back to electromagnetic gauge invariance (for a pedagogical
review see \citealt{HuWhi97}).
Since the origin of the temperature and polarization spectra are so closely
intertwined, we can use them as a test of internal consistency.
In fact we find that the \Planck-measured $TT$, $TE$, and $EE$ CMB power
spectra are completely consistent with each other under the assumptions of
$\Lambda$CDM \citep{planck2016-l05,planck2016-l06}.
The same $\Lambda$CDM models that fit the temperature provide good fits to
the polarization data and vice versa.
Figure~\ref{fig:cond_vs_coadd} shows the difference between the $TT$, $TE$, or
$EE$ spectra we measure and the spectra predicted by the $\Lambda$CDM model
that best fits the other two.
The differences are completely consistent with expectations given our noise
and sky coverage.
Not surprisingly, the $\Lambda$CDM model parameters that best fit each subset
of the spectra are consistent.
We see small shifts in the parameters as more data are added, with the size
of the shifts consistent with our expectations.
The comparison of the temperature, polarization, and lensing spectra may
provide some indication that the temperature-only results have fluctuated
``high'' in some parameters (e.g., $\sigma_8$) and that adding more data has
brought us closer to the mean.  We will see similar behaviour when we
consider the distance scale as probed by BAO (Sect.~\ref{sec:bao}), and
discuss this further in Sect.~\ref{sec:discord}.

\begin{figure*}[htbp]
\begin{center}
\resizebox{1.0\textwidth}{!}{\includegraphics{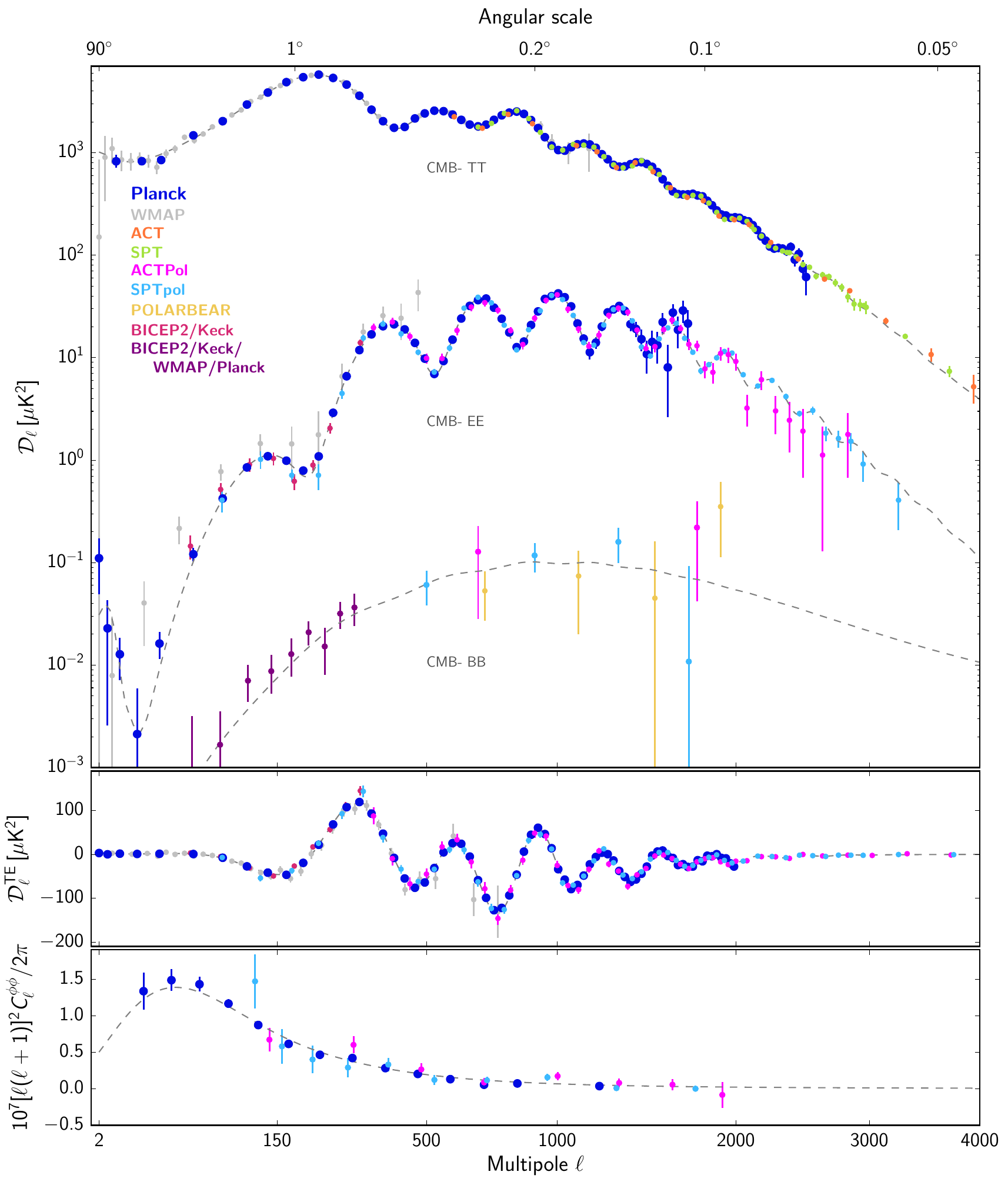}}
\vspace{-0.5cm}
\end{center}
\caption{Compilation of recent CMB angular power spectrum measurements from
which most cosmological inferences are drawn.  The upper panel shows the
power spectra of the temperature and $E$-mode and $B$-mode polarization
signals, the next panel the cross-correlation spectrum between $T$ and $E$,
while the lower panel shows the lensing deflection power spectrum.
Different colours correspond to different experiments, each retaining its
original binning.  For \Planck, ACTPol, and SPTpol, the $EE$ points
with large error bars are not plotted (to avoid clutter).
The dashed line shows the best-fit \LCDM\ model to the \Planck\ temperature,
polarization, and lensing data.  See text for details and references. 
\label{fig:cl-world} }
\end{figure*}

External consistency checks come from comparing the \Planck\ angular power
spectra to those measured by other experiments.  No single experiment can
match \Planck's sky coverage and angular resolution, but we can compare to
multiple experiments in order to test our data.
A comparison of the \Planck\ power spectrum measurements with those
of other, contemporary, experiments is given in Fig.~\ref{fig:cl-world}:
the \WMAP\ data are taken from \citet{Bennett13}; the ACT and ACTpol
data are from \citet{Das14}, \citet{Louis17}, and \citet{Sherwin17};
the SPT and SPTpol data are from \citet{George15}, \citet{Keisler15},
\citet{Story15}, and \citet{Henning18};
the PolarBear data are from \citet{PolarBear17};
and BICEP2/Keck data are from \citet{BICEPKeck15} and \citet{BICEPKeck16}.
While \Planck\ dominates the primary temperature anisotropy measurements 
and the $E$-mode polarization measurements up to $\ell\simeq 10^3$, the
other experiments' higher angular resolution and sensitivity provide better
measurements\footnote{Though \Planck\ still contributes here as a source of
calibration.} of secondary anisotropies (at high $\ell$), as well as $B$-mode
polarization.
A next generation of experiments, soon to be fielded, will also improve upon
\Planck's lensing measurement.  Visually, the impression in
Fig.~\ref{fig:cl-world} is one of concordance in all of the spectra.

The comparison of the lower-resolution and noisier 
\WMAP\ data to \Planck\ has been discussed in some detail in
\citet{planck2013-p11} and \citet{planck2014-a15}.
The agreement is excellent, multipole by multipole, for the frequencies in
common, up until the \WMAP\ data become significantly affected by noise.
The agreement between the best-fitting $\Lambda$CDM models is also quite
good, once external data (such as BAO) are introduced to break the
degeneracies that the \WMAP\ data do not have sufficient dynamic range to
break internally.  For example the constraints
$\omega_{\rm m}=0.1398\pm0.0023$, $H_0=68.14\pm0.73$, and
$\sigma_8=0.82\pm0.18$ are obtained from \WMAP\ plus BAO, whereas
$\omega_{\rm m}=0.14240\pm0.00087$, $H_0=67.66\pm0.42$, and
$\sigma_8=0.8102\pm0.0060$ come from \Planck\ plus BAO.

\citet{planck2016-LI} further investigated the discrepancy in $\Lambda$CDM
parameters between \Planck\ and \WMAP\ alone.
They found that when one carefully compares low-$\ell$ data to full-$\ell$
data, the differences are not as large as they might naively appear to be
(with probabilities to exceed of order 10\,\%).
When the lever arm of the data is reduced by only using the larger angular
scales ($\ell\,{<}\,800$), cosmological parameters are more strongly affected
by the low-$\ell$ deficit (Sect.~\ref{sec:anomalies}), that is, the apparent
lack of power at $\ell\,{\la}\,30$ compared to $\Lambda$CDM expectations.
To decrease power at $\ell\,{\la}\,30$, the best-fit $n_{\rm s}$ increases,
$A_{\rm s}e^{-2\tau}$ is then lowered to reduce power at $\ell\,{\ga}\,500$,
and $\omega_{\rm m}$ decreases to compensate the induced change of power below
$\ell\,{\simeq}\,500$, while $\omega_{\rm b}$ increases to reduce the amplitude
of the second peak (which was raised by the decrease in $\omega_{\rm m}$).
The Hubble constant is in turn pulled higher to keep the angular size of
the horizon unchanged.
In the \Planck\ data, the impact of the low-$\ell$ deficit was much reduced
by the presence of the high-$\ell$ data.  As we saw above, if BAO data
are combined with the \WMAP\ data (to reduce the geometric distance degeneracy,
wherein a change in physical scale can be traded against a change in distance
to last scattering in order to hold the angular scale fixed) the parameters
shift towards the \Planck-preferred values
\citep[see][for further comparison]{planck2016-LI}.

At the other end of the spectrum we can compare the \Planck\ data to data with higher
angular resolution and higher S/N from ACTPol \citep{Louis17} and
SPTpol \citep{Henning18}, but only over a limited area.
The ACTPol results of \citet{Louis17}, from $100\,{\rm deg}^2$ of sky, are
consistent with the $\Lambda$CDM model fit to \Planck.
Similarly, \citet{Hou18} find no evidence for systematic errors in either SPT
or \Planck\ when comparing temperature power spectra computed from the same
area of sky.
\citet{Aylor17} and \citet{planck2016-l06} compare the parameters for the
base-$\Lambda$CDM model between SPT and \Planck.
Again, restricting to the same patch of sky the agreement between the
experiments is quite good.
The \Planck\ $TE$ and $EE$ spectra are compatible with the SPT $TE$
and $EE$ spectra over the multipole range well-constrained by \Planck,
though there are hints of some differences at higher multipoles (with
limited statistical power, in a regime where foregrounds are large).
However, the $\Lambda$CDM model that best fits the \Planck\ data is
formally inconsistent with the SPT $TE+EE$ data.
These issues are discussed in more detail in \citet{planck2016-l06}.
It will be interesting to see how this discrepancy develops, and
whether it provides evidence for physics beyond $\Lambda$CDM or is due to
systematic or statistical errors in the modelling of the data.

In summary, once foreground models and calibrations are taken into account,
and allowing for mild inaccuracies in the covariance matrices, the level of
agreement between different CMB experiments is excellent.

\subsection{Large-scale structure} \label{sec:lss1}

\begin{figure*}[htbp]
\begin{center}
\resizebox{\textwidth}{!}{\includegraphics{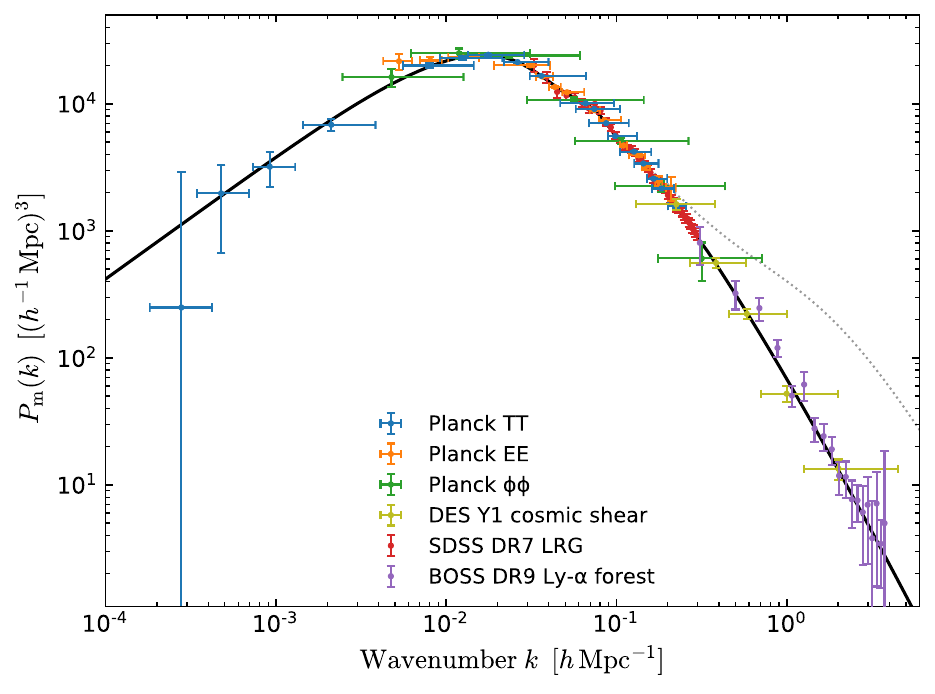}}
\end{center}
\caption{Linear-theory matter power spectrum (at $z=0$) inferred from
different cosmological probes (the dotted line shows the impact of non-linear
clustering at $z=0$).  The broad agreement of the model (black line)
with such a disparate compilation of data, spanning $14\,$Gyr in time and
three decades in scale, is an impressive testament to the explanatory power
of $\Lambda$CDM.
Earlier versions of similar plots can be found in, for
example, \citet{WSS94}, \citet{ScottSW95}, \citet{TegZal02}, and \citet{Teg04}.
A comparison with those papers shows that the evolution of the field in the
last two decades has been dramatic, with $\Lambda$CDM continuing to provide
a good fit on these scales.
\label{fig:allpk} }
\end{figure*}

Within the gravitational instability paradigm, the anisotropies that we see
in the CMB form the seeds for the large-scale structure that we observe more
locally.  It is thus interesting to ask whether these low-$z$ measurements of
inhomogeneity are consistent with what would be expected to arise from the
anisotropies seen by \Planck.

Figure~\ref{fig:allpk} shows inferences of the matter fluctuation spectrum
from a wide range of different cosmological probes, covering three orders of
magnitude in scale and much of cosmic history.
The level of agreement, assuming the $\Lambda$CDM model, is quite remarkable.
That structure grows through gravitational instability in a
dark-matter-dominated Universe seems well established, and the power of the
model to explain a wide range of different phenomena is impressive.
However, the tremendous statistical power of the \Planck\ data, and modern
probes of large-scale structure, is such that we can perform much more
detailed comparisons than this.

One consistency check, which we can make internal to the \Planck\ data set, is
to check whether the large-scale structure that lenses the CMB anisotropies at
$z\simeq 0.5$--10 has the right amplitude given the size of the anisotropies
and the constituents inferred from the acoustic oscillations.
Between the epoch of last scattering at $z\simeq1100$ and and the epoch
corresponding to the peak of the lensing kernel ($z\simeq2$--3), the
fluctuations in the matter density are predicted to grow in amplitude
by nearly three orders of magnitude.
Since for much of this time the Universe is matter dominated and the
fluctuations are in the linear regime, GR predicts the amount of growth
at the percent level, allowing a precision test of the theory.
In fact, the comparison can be done to such high accuracy that it is best
phrased as a scaling, $\Aphiphi$, of the theoretical prediction --
taking into account the distributed effects of lensing, etc.
We find $\Aphiphi=0.997\pm 0.031$, which provides a stunning confirmation
of the gravitational instability paradigm, and also allows us to constrain
constituents of the Universe that do not cluster on small scales
(such as massive neutrinos; see Sect.~\ref{sec:neutrinos}) and so reduce the
small-scale power spectrum.
Future, more precise, measurements of CMB lensing will provide strong
constraints on neutrino masses, extra relativistic degrees of freedom, and
early dark energy.

Also shown in Fig.~\ref{fig:allpk} are measurements of the matter power
spectrum inferred from galaxy clustering and the Ly$\,\alpha$ forest.
The former represents a measurement at $z\simeq 0$, although it has an
uncertain amplitude because of galaxy bias.
In plotting the SDSS galaxy clustering points, we have accounted for galaxy bias
assuming the phenomenological bias model of \citet{2010MNRAS.404...60R}.  
Specifically, we have fit this model to the \Planck\ best-fit cosmology,
yielding $\{b_0, a_1, a_2\} = \{1.23, 0.56, -0.35\}$ at a pivot wave-number
of $k_\ast = 0.2\,h\,{\rm Mpc}^{-1}$.
The agreement on the shape of the power spectrum at
$k\simeq 0.1\,h\,{\rm Mpc}^{-1}$, between the galaxy surveys at
$z\,{\simeq}\,0$ and the predictions of $\Lambda$CDM constrained by \Planck\
at $z\,{\simeq}\,10^3$, is a validation of the paradigm of gravitational
instability in a Universe with predominantly cold dark matter.
The measurements inferred from the Ly$\,\alpha$ forest are presented at $z=0$
using a scale- and redshift-dependent relation between the 1D and 3D
Ly$\,\alpha$ power spectra, coupled with the measured 3D flux
power spectrum of Ly$\,\alpha$ absorption.
The former was computed by means of hydrodynamical simulations,
for a fiducial model corresponding to the best fit values of
\citet{Palanque15};
the latter was obtained by differentiating the corresponding 1D power
spectrum using the method of \citet{Chartrand11}.
The measurements of Ly$\,\alpha$ are at higher redshift ($2\,{<}\,z\,{<}\,3$)
than galaxy clustering and probe smaller scales, but are more model-dependent.

Intermediate in redshift between the galaxy clustering and Ly$\,\alpha$
forest data are cosmic shear measurements and redshift-space distortions
\citep{Hamilton98,Weinberg13}.
Here we plot the results from the The Dark Energy Survey Y1 measurements
\citep{Troxel18a}, 
which are currently the most constraining cosmic shear measurements.
They show good agreement with the matter power spectrum inferred from
$\Lambda$CDM constrained to \Planck.
Had we used earlier data from, for example, KiDS \citep{Hildebrandt17} it would
have looked quite similar.
These points depend upon the non-linear matter power spectrum, and we have
used the method of \citet{TegZal02} based on the fitting function of
\citet{1996MNRAS.280L..19P}
to deconvolve the non-linear effects, which yields constraints sensitive to
larger scales than would would otherwise appear.  The nuisance parameters
have been fixed for the purposes of this plot.
(More detail of the calculations involved in producing Fig.~\ref{fig:allpk}
can be found in \citealt{chabanier2019}).
Bearing in mind all of these caveats, the good agreement across more than
three decades in wavenumber in Fig.~\ref{fig:allpk} is quite remarkable.

\begin{figure}[htbp]
\begin{center}
\resizebox{\columnwidth}{!}{\includegraphics{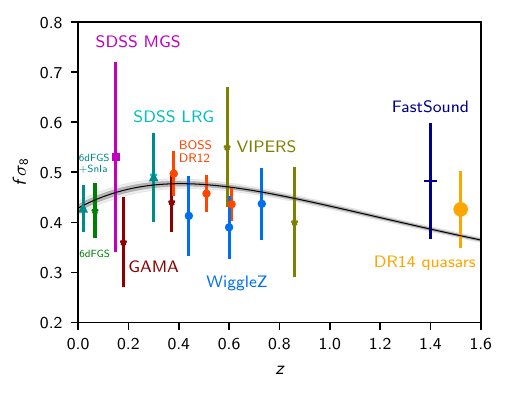}}
\end{center}
\caption{Constraints on the growth rate of fluctuations, $f\sigma_8$, as a
function of redshift, compared to the predictions of the $\Lambda$CDM model
constrained by \Planck\ (from \citealt{planck2016-l06}).
The $f\sigma_8$ measurements are:
dark cyan, 6dFGS and velocities from SNe Ia \citep{Huterer:2016uyq};
green, 6dFGRS \citep{Beutler12};
purple square, SDSS MGS \citep{Howlett:2014opa};
cyan cross, SDSS LRG \citep{Oka:2013cba};
dark red, GAMA \citep{Blake:2013nif};
red, BOSS DR12 \citep{BOSS-DR12};
blue, WiggleZ \citep{Blake:2012pj};
olive, VIPERS \citep{Pezzotta:2016gbo};
dark blue, FastSound \citep{Okumura:2015lvp};
and orange, BOSS DR14 quasars \citep{Zarrouk18}.
The agreement between the low-$z$ measures and the $\Lambda$CDM prediction is
very good, indicating that the model (constrained by observations in the
high-$z$ Universe) correctly predicts the rate of growth of large-scale
structure observed in the nearby Universe.}
\label{fig:fsig8}
\end{figure}

Figure~\ref{fig:fsig8} shows the rate\footnote{Conventionally one defines
$f$ as the logarithmic growth rate of the density perturbation $\delta$,
that is, $f=d\ln\delta/d\ln a$.  Multiplying this by the normalization,
$\sigma_8$, converts it to a growth rate per $\ln a$.} of growth, $f\sigma_8$,
determined from redshift-space distortions over the range $0\,{<}\,z\,{<}\,1.6$,
compared to the predictions of $\Lambda$CDM fit to \Planck.
Though the current constraints from redshift surveys have limited statistical
power, the agreement is quite good over the entire redshift range.  In
particular, there is little evidence that the amplitude of fluctuations in
the late Universe determined from these measurements is systematically lower
than predicted.

We shall discuss in Sect.~\ref{sec:lss} cross-correlations of CMB lensing with
other tracers and the distance scale inferred from baryon acoustic oscillations
(BAO).  In general there is very good agreement between the predictions of the
$\Lambda$CDM model and the measurements.
If there is new physics beyond base $\Lambda$CDM, then its signatures are
very weak on large scales and at early times, where the calculations are
best understood.

\subsection{Discord}
\label{sec:discord}

While there are many measurements that are consistent with the predictions
of the $\Lambda$CDM model fitted to \Planck, there are also some areas of
discordance.

Within the \Planck\ data themselves we find a preference for a larger
smoothing of the power spectrum at small scales than the $\Lambda$CDM
model predicts \citep{planck2013-p11,planck2014-a15,planck2016-l06}.
While at face value it might seem like this smoothing is the sign of an
excess amplitude of gravitational lensing, it is also possible to fit
these features through non-lensing related effects
\citep[see][for discussion]{planck2016-LI}.
The preference for these features is driven almost entirely by the CMB
spectra and not by the lensing reconstruction, which is consistent with
theoretical expectations.
The peak smoothing features are not statistically highly significant
(2--$3\,\sigma$), and could just be statistical fluctuations in the data.
Further, the level of significance depends upon choices made about the
calibration of the polarization channels, the sky fraction, and other analysis
choices, as discussed further in \citet{planck2016-l06}.  This discrepancy may
indicate that the best-fit parameters from the primary CMB have fluctuated from
their true values by a few $\sigma$, in which case the combination afforded by
multiple probes may be a more faithful measure.

We will discuss distance measurements using BAO in Sect.~\ref{sec:bao}.
There we will see (Fig.~\ref{fig:DV-BAO}) that the inferred angular diameter
distance to $z\,{\simeq}\,2$ from the auto- and cross-correlation of
Ly$\,\alpha$ measurements by the Baryon Oscillation Spectroscopic Survey
(BOSS) is discrepant with the $\Lambda$CDM predictions fit to
\Planck\ at about $2.3\,\sigma$ \citep{Bautista17,duMas17}.
Within the $\Lambda$CDM family, parameter changes that would improve
agreement with the Ly$\,\alpha$ distances are highly disfavoured by
\Planck\ and the more accurate, lower-redshift BAO measurements.  Even
within an extended class of models, it is very difficult to fit the
combination of comoving angular diameter distance, $D_{\rm M}$, and
Hubble distance, $D_{\rm H}$, inferred from the Ly$\,\alpha$ data
\citep{Aubourg15}.
This mild tension could be the result of either a statistical fluctuation
or as yet unrealized systematics in the Ly$\,\alpha$ measurements.  However
the size of the discrepancy highlights the importance of future measurements
at these redshifts.

At lower redshift, some measures of the amplitude of clustering prefer lower
values than $\Lambda$CDM normalized to \Planck.
In particular the \citet{Kohlinger17} analysis of the KiDS cosmic-shear-only
results constrains $S_8\equiv\sigma_8(\Omega_{\rm m}/0.3)^{0.5}$
to be $0.651\pm 0.058$ (which was shifted upwards to $0.772\pm 0.034$ in
an alternative analysis by \citealt{Troxel18b}).
When combined with galaxy data the results are
$0.742\pm 0.035$ or $0.800\pm0.028$ \citep{Joudaki18,vanUitert18}.
The preferred value from \Planck\ plus BAO is $0.825\pm 0.011$, which is
$2.9\,\sigma$ higher, $1.5\,\sigma$ higher, $2.3\,\sigma$ higher, or basically
consistent with these results.
The recent DES results \citep{DES-Y1} are consistent with both \Planck\ and
the earlier lensing results, $S_8=0.782\pm 0.024$, when analysed with the
same fixed neutrino mass assumption as \Planck\ \citep{planck2016-l06}.
While these data are in only modest tension with the \Planck\ best-fit
cosmology, they are consistent with each other and both pull to lower $S_8$,
which would lead to an increased significance in a joint analysis.

Some estimates of the amplitude inferred from the abundance of rich clusters
of galaxies also imply a lower $\sigma_8$.
The most difficult issue with these inferences is the dependence on the
calibration of the mass-observable relation, which \Planck\ itself can shed
little light on.
We discuss these observations further in Sect.~\ref{sec:clusters_sz}.

\begin{figure}[htbp]
\begin{center}
\resizebox{\columnwidth}{!}{\includegraphics{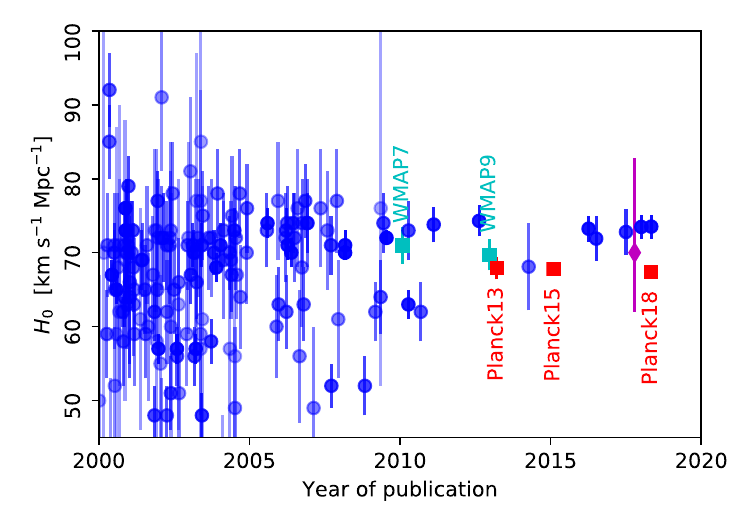}}
\end{center}
\caption{Compilation of measurements of $H_0$ since 2000, based on the
historical data assembled by J.~Huchra for the NASA-HST Key Project on the
Extragalactic Distance Scale.  The additional points since 2010 are from
\citet{Riess11}, \citet{Freedman12}, \citet{Rathna15}, \citet{Riess16},
\citet{Bonvin17}, \citet{Dhawan18}, and \citet{Riess18,Riess_Gaia}.
Blue circles show ``traditional'' measures of $H_0$, while
cyan and red squares show $H_0$ inferred from fits to CMB data from
\WMAP\ (\citealt{bennett2010,hinshaw2012}) and \Planck.
The magenta diamond shows the standard siren measurement from
\citet{StandardSiren}.  Inferences from the inverse distance ladder are
discussed in the text and Fig.~\ref{fig:H0_ladder}.  We note the tremendous
increase in precision with time, driven by improvements in methods and in
data, and the narrowing of the difference between ``high'' and `'low'' values
of $H_0$.
\label{fig:H0_history}}
\end{figure}

\begin{figure}[htbp]
\begin{center}
\resizebox{\columnwidth}{!}{\includegraphics{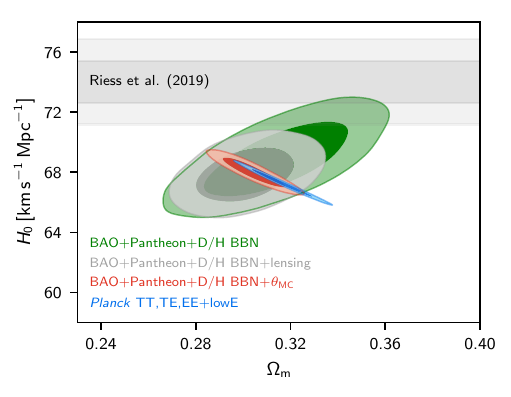}}
\end{center}
\caption{Inverse-distance-ladder constraints on $H_0$ and $\Omega_{\rm m}$ in
$\Lambda$CDM \citep[for more details see][]{planck2016-l06}.
The green region uses BAO
constraints on the distance-redshift relation relative to the sound horizon
at the drag epoch.  The sound horizon has been constrained here using a
conservative prior of $\omega_{\rm b}=0.0222\pm0.005$, based on \citet{Cooke18}
and standard BBN.  Adding \Planck\ CMB lensing gives the grey contours,
which pull to lower $H_0$ and $\Omega_{\rm m}$.
The red contours show instead the addition of a highly conservative
prior of $100\,\theta_{\rm MC}=1.0409\pm 0.0006$, which provides similar
$H_0$ constraints to the full CMB data set, but is potentially more robust.
The blue band shows the $\Lambda$CDM fits to the full CMB data set.
The horizontal grey bands show the measurement of \citet{Riess19} for
comparison.}
\label{fig:H0_ladder}
\end{figure}

There is also tension at the very lowest redshifts.
One of the notable impacts of the \Planck\ data has been a downward shift
in the value of $H_0$ compared with some earlier results
(Fig.~\ref{fig:H0_history}; plotted using the compilation of
J.\ Huchra\footnote{Available at
\url{https://www.cfa.harvard.edu/~dfabricant/huchra/hubble.plot.dat}.},
though the shift is small compared to earlier versions of the
``$H_0$ discrepancy'').
The same downward shift in $H_0$ is seen if \WMAP\ data are combined with
BAO, which serves to break the geometric degeneracy that limits the \WMAP\ data
alone (leading to $H_0=68.14\pm0.73$).  A similar shift is also seen if BAO
data are combined with inferences about $\omega_{\rm b}$ from BBN and just the
acoustic scale measured by \WMAP\ or \Planck\ \citep{planck2016-l06}, or from
\Planck\ lensing plus BAO inverse-distance-ladder results
\citep[][table 3]{planck2016-l08}.
Another view of this tension is shown in Fig.~\ref{fig:H0_ladder}
\citep[as discussed in more detail in][]{planck2016-l06},
which demonstrates how robustly the inverse-distance-ladder constraints prefer
lower $H_0$ than the measurements of \citet{Riess18,Riess_Gaia,Riess19}.

It is worth revisiting how the inference of a ``low'' $H_0$ comes about from
CMB data.  Recall that the well measured $\theta_\ast$ relates the sound
horizon and the distance to last scattering.  For $\Lambda$CDM this becomes a
tight constraint on $\Omega_{\rm m} h^3$ (Eq.~\ref{eqn:omegah3}).
An increase of $\omega_{\rm m}$ decreases the sound horizon approximately as
$\omega_{\rm m}^{-0.25}$, requiring the distance to last scattering to decrease
by the same amount.
This distance is an integral of $1/H(z)$ out to $z\simeq 1100$,
with\footnote{We have neglected the impact of massive neutrinos in this
expression for simplicity, though they are properly included in our analyses.}
$H^2(z)\propto\left\{\omega_{\rm m}\big[(1+z)^3-1\big]+h^2\right\}$ for the
dominant contribution from $z\ll z_{\rm eq}$.
Thus $h$ must decrease in order for the distance to last scattering not
to decrease too much.

The combination of absolute BAO distances calibrated to \Planck, with
relative SNe distances at overlapping redshifts, allows us to extrapolate
the distance scale from moderate redshifts to $z=0$.  This significantly
reduces the sensitivity of inferences on $H_0$ to uncertainties in the dark
energy model, but still results in a consistent $H_0$ value
\citep{planck2016-l06}.

The decrease in the inferred value of $H_0$ has resulted in tension with
some locally derived values \citep[e.g.,][]{Riess19}, as described in detail
in \citet{planck2016-l06} and our earlier papers.
In making this comparison it is important to realize that while the CMB results
are very stable to different analysis choices and data sets, $H_0$ is not
directly measured from the high-$z$ data, but rather inferred via a model.
One possibility is thus that the discrepancy between the results indicates
a failure of the $\Lambda$CDM model.  Unfortunately no simple, 1-parameter
extension of the model alleviates the tension between the measurements.
{}From the agreement between the \Planck-inferred distance scale and that
measured by BAO and SNe, it seems that any discrepancy should either be
localized to quite low $z$ or that $\Lambda$CDM is not correctly predicting
the sound horizon at the last-scattering and decoupling epochs.
It is quite difficult to change these quantities without changing other,
well measured, features of the CMB \citep[see, e.g.,][]{EisWhi04}, so 
if the discrepancy is due to new physics, it must act in a complex manner.
The more prosaic explanation is that there are under-appreciated systematics
in one or all of the data sets.  Alternatively, this could represent a
statistical fluctuation: there are six dimensions in the $\Lambda$CDM parameter
space, and many other derived-parameter directions, and hence large
fluctuations in some direction occur relatively often.
It is presently unclear what combination of statistical fluctuations,
a posteriori statistics, systematic uncertainties, and genuinely new
physics is responsible for any of the tensions seen in today's data
combinations.  Until this discrepancy is better understood we expect that
this will continue to be a fruitful area of research.

Finally, there are possible tensions on galactic and sub-galactic scales,
where the inner profiles of dark-matter halos and the abundance, orbital
properties, and structure of satellites in the Milky Way and Andromeda provide
a new avenue for testing $\Lambda$CDM.
The comparison of theory and observation in this regime is quite complex and
definitive statements are hard to make at this juncture; however, the field
is evolving rapidly both observationally and theoretically.
The challenges and possible resolutions are reviewed in \citet{Bullock17}.

\section{\Planck\ and fundamental physics} \label{sec:physics}

\subsection{Large scales and the dipole}
\label{sec:dipole}

We have already discussed the best estimate of the dipole pattern on the sky,
with the usual interpretation that it derives from Doppler boosting of the
CMB monopole, with an amplitude of $\beta T_0$.  In the standard picture
there is an ``intrinsic'' dipole of order $10^{-5}$ expected, although this
is unobservable (as well as being a small fraction of the extrinsic,
velocity-induced dipole).  However, as has been discussed previously in the
literature \citep[e.g.,][]{Turner91,Zibin08}, there is also the possibility
of an intrinsic isocurvature contribution to the observed dipole.
In addition to the usual temperature dipole (i.e., the $\ell=1$ anisotropy
pattern) on the sky, four separate effects appear at second order in
$\beta$, namely: an inferred frequency-dependent quadrupole;
an inferred frequency-dependent dipolar modulation of the CMB sky,
altering the power on all scales according to a dipole pattern; a shift in
the monopole temperature; and aberration of the CMB sky.
The first two effects are independent of the source of the CMB dipole and
therefore cannot be used to distinguish an intrinsic dipole from a boost.
The third effect is unobservable.  The last effect normally only
appears in the presence of a boost.
However, aberration is completely degenerate with an $L\,{=}\,1$ lensing mode;
in other words, a very large-scale gravitational potential fluctuation
can shift the photon directions in a dipole pattern on the sky.
Therefore, while the detection of aberration is consistent with interpreting
the CMB dipole as arising from a boost, the case against an intrinsic dipole
is not definitive (though quite compelling, since it would otherwise require an
isocurvature mode on the largest scales, despite the fact that the
fluctuations are consistent with being entirely adiabatic on all other scales).

In \citet{planck2013-pipaberration}, we performed the first experimental
verifications of the modulation and aberration effects, finding the former
to be consistent with the prediction from the CMB dipole and the latter to be
consistent with the interpretation of the dipole coming from a boost
(barring any large sources of an $L\,{=}\,1$ lensing mode).
This required treating the signal as being a frequency-dependent coupling
between adjacent $\ell$ modes.
Given that aberration and modulation effectively shift the power spectra
in the angular scale and amplitude directions, respectively,
one also needs to consider whether these boosting effects, combined with
masking part of the sky, can give any significant differences between the
\Planck-derived cosmological results and those that would come from an
unboosted sky.  Here the largest potential effect comes from aberration;
for a full-sky CMB map it would average out, but for the \Planck\ data the
need to mask the Galaxy (in an asymmetric way) biases $\theta_\ast$ at a
level estimated conservatively to be less than $0.1\,\sigma$
(\citealt{planck2016-l06}; agreeing with more detailed calculations
by \citealt{Jeong14}).  The bias can hence safely be ignored for \Planck.

The second-order quadrupole signal
(sometimes called the ``kinematic quadrupole'')
also has a frequency-dependent spectrum, as discussed by \citet{KamKnox03}.
This signal was already apparent in differences between the 2013 and 2015
\Planck\ data releases, arising from the different treatment of the expected
dipole-related quadrupole in these two data releases
\citep[see][]{planck2014-a11,planck2014-a14};
however, no estimate has been made of the amplitude of the signature, just a
check that it is broadly consistent with expectation.

\subsection{Inflation physics and constraints}
\label{sec:inflation}

A key ingredient of the standard cosmological model is the presence of
small, seed fluctuations in the very early Universe, which are amplified by
gravitational instability to form all of the structure we
see in the Universe today.
Some of the first observations of CMB anisotropies gave strong support to
an early Universe origin for the fluctuations, through the coherence of the
acoustic peaks in the power spectrum and the phasing of the temperature and
polarization anisotropies \citep{Cou94,CriTur95,TestInf,HuWhiSpe97,SpeZal97}.
In the most popular models, a period of quasi-exponential expansion in the very
early Universe pushes quantum fluctuations outside the Hubble volume, where
they become classical perturbations in the gravitational potentials and density
of the Universe \citep{LL09}.
This highly parsimonious explanation, using the inevitable quantum ``noise''
as the source of all of the observed structure, is one of the key pieces
of the ``cosmo-micro'' connection.
\Planck\ has dramatically improved upon this early legacy by firmly
establishing essentially all of the major predictions of inflation
(see Table~\ref{tab:inf_scorecard}),
while tightly constraining many specific popular models of inflation.
Whatever the true origin of the primordial fluctuations turns out to be,
it must share these features with models of inflation.

\begin{table}[htbp]
\caption{Inflationary ``scorecard,'' comparing the predictions of the
simplest inflationary models with observations.  In all cases, the tightest
observational limits come from \Planck, sometimes in combination with other
data sets (as described in the text).  Here we quote symmetric, 68\,\% CL
uncertainties or 95\,\% upper limits on each quantity, taken from
\citet{planck2014-a13}, \citet{planck2016-l06}, \citet{planck2016-l09},
and \citet{planck2016-l10}.
All quantities have their usual meanings, with $\alpha_{-1}$ the amplitude of
an isocurvature component to the fluctuations and the topological defect limit
referring specifically to Nambu-Goto cosmic strings
\citep[see table~8 of][for other cases]{planck2014-a13}.
\label{tab:inf_scorecard} }
\begingroup
\openup 4pt
\newdimen\tblskip \tblskip=5pt
\nointerlineskip
\vskip -5mm
\footnotesize
\setbox\tablebox=\vbox{
 \newdimen\digitwidth
 \setbox0=\hbox{\rm 0}
 \digitwidth=\wd0
 \catcode`*=\active
 \def*{\kern\digitwidth}
  \newdimen\dpwidth
  \setbox0=\hbox{.}
  \dpwidth=\wd0
  \catcode`!=\active
  \def!{\kern\dpwidth}
%
\halign{\tabskip 0pt#\hfil\tabskip 2.0em&#\hfil\tabskip=0pt\cr
\noalign{\doubleline}
\omit\hfil Prediction \hfil& \omit\hfil Measurement \hfil\cr
\noalign{\vskip 5pt\hrule\vskip 5pt}
A spatially flat universe& $\Omega_K=0.0007\pm0.0019$\cr
with a {\it nearly} scale-invariant (red)\hfil& \cr
spectrum of density perturbations,& $n_{\rm s}=0.967\pm 0.004$\cr
which is almost a power law,& $dn/d\ln k=-0.0042\pm 0.0067$\cr
dominated by scalar perturbations,& $r_{0.002}<0.065$\cr
which are Gaussian& $f_{\rm NL}=-0.9\pm 5.1$\cr
and adiabatic,& $\alpha_{-1}=0.00013\pm 0.00037$\cr
with negligible topological defects& $f<0.01$\cr
\noalign{\vskip 5pt\hrule\vskip 3pt}
} 
} 
\endPlancktable
\endgroup
\end{table}

\begin{figure*}[htbp]
\resizebox{\textwidth}{!}{\includegraphics{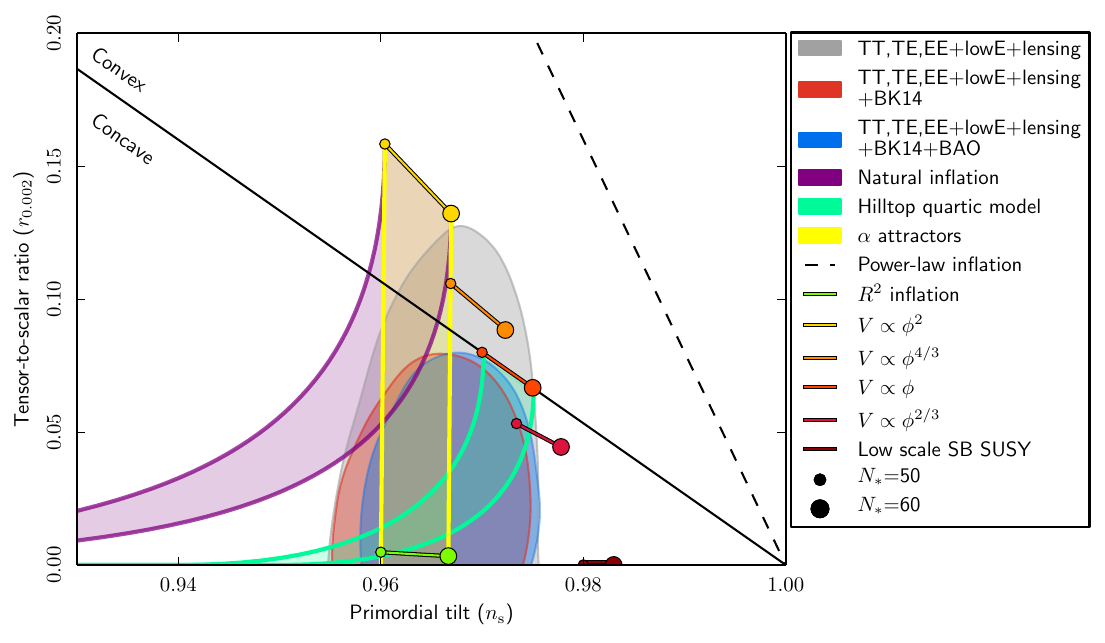}}
\caption{Limits on the tensor-to-scalar ratio, $r_{0.002}$, as a function of $n_S$ in the $\Lambda$CDM model at 95\% CL, from \Planck\ alone (grey area), or including BICEP2/Keck data 2014 (red) and BAO (blue).
Constraints assume negligible running of the inflationary consistency
relation, and the lines show the predictions of a number of models as a function of the number of $e$-folds, $N_\star$, until the end of inflation. This can be compared with the middle panel in the top row of Fig.~\ref{fig:zooming-in}, which gives a temporal perspective.
\label{fig:ns-r} }
\end{figure*}

The comparison of the \Planck\ measurements with models of inflation is
discussed in detail in \citet{planck2013-p17}, \citet{planck2014-a24},
and \citet{planck2016-l10}.
As summarized in Table~\ref{tab:inf_scorecard},  \Planck\ provides
very strong support for the inflationary paradigm, and at the same time
tightly constrains the space of allowed inflationary models
(Fig.~\ref{fig:ns-r}).  There are several points to note in the table.
First, the combination of \Planck\ data with lower-redshift data on acoustic
oscillations (measured in the distribution of galaxies) tightly constrains the
spatial hypersurfaces to be flat ($\Omega_K=0.0007\pm 0.0019$, 68\,\% CL).
In the standard interpretation, this suggests that the duration of the
slow-roll phase was not fine tuned.

\begin{figure*}[htbp]
\includegraphics[width=\textwidth]{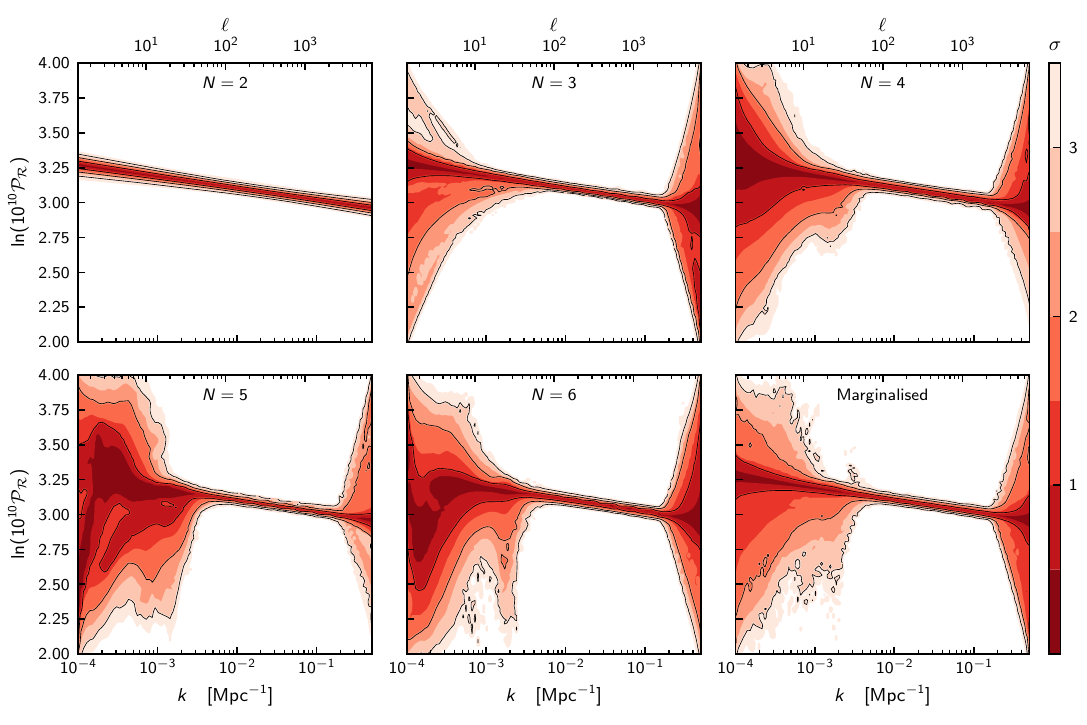}
\caption{Primordial (scalar curvature) power spectrum, reconstructed by using
the \Planck\ 2018 TTTEEE+lowE+lensing likelihood. This was done by
sampling the parameters of an extended \LCDM\ model, where the initial power
spectrum was described with a varying number of movable spline nodes
(from one to nine), rather than assumed to be a power law. The final
reconstruction (bottom right plot) is obtained by marginalising,
i.e., weighting each of the nine reconstructions by its own evidence.
With two nodes
(top left), the departure from scale invariance with $n_{\rm s}-1\simeq -0.035$
is nicely recovered. With three nodes the uncertainties at low $\ell$ (due to
the small number of modes) and high-$\ell$ (due to noise) becomes visible.
With a larger numbers of nodes, anomalies may be captured, and the most visible
departure from a pure power law reflects the well--known power deficit at
$\ell\simeq 20$--30. However, the evidence-weighted plot (bottom right panel)
shows that the evidence for such a spectral feature is actually not very
significant.  In \citet{planck2014-a24}, this spectrum was reconstructed using
three additional methods, with similar conclusions.
\label{fig:Pkrec} }
\end{figure*}

\begin{figure}[htbp]
\includegraphics[width=\columnwidth]{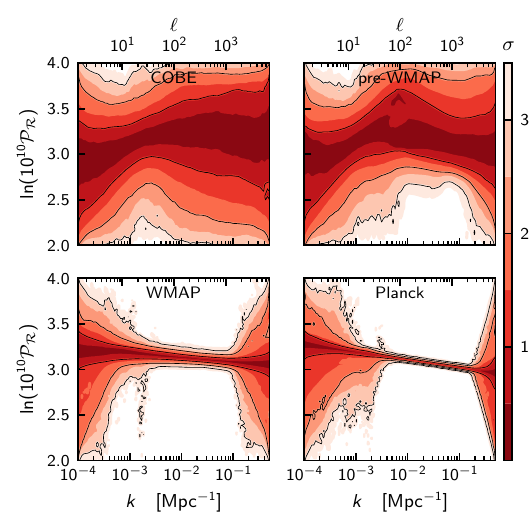}
\caption{Temporal evolution of constraints on the reconstructed primordial power spectrum. Using the same methodology as in Fig.~\ref{fig:Pkrec}, we compare the (marginalised) \planck\ 2018 reconstruction with versions based on earlier likelihoods (see text).  
\label{fig:Pkrecevol} }
\end{figure}

The primordial power spectrum shows no significant deviations from a power law
(e.g., Fig.~\ref{fig:Pkrec}).  That the simple, power-law form for the
primordial power spectrum continues to provide a good fit to the data
is quite impressive when one considers the degree to which our constraints
have improved.  Figure~\ref{fig:Pkrecevol} shows the reconstructed primordial
power spectrum, starting from the \COBE\ likelihood described in
\citet{COBE_data}, through ``pre-\WMAP'' (from the product of the previous
likelihood with those from MAXIMA, DASI, BOOMERANG, VSA, and DASI;
\citealt{MAXIMA_data,DASI_data,BOOMERANG_data,VSA_data,CBI_data}), to
``\WMAP'' (from the 9-year, final release), and to \Planck.
Even within the results from \Planck, the weak significance of
any possible features in our earlier releases has decreased even further.

Within the context of inflationary models, this implies that the inflaton
potential was featureless and relatively flat.
The power law is ``tilted'' away from scale invariance ($n_{\rm s}\,{=}\,1$),
as expected for an inflaton rolling ``down'' a potential.  \Planck\ was
the first experiment able to show that $n_{\rm s}\ne 1$ in a way that was
robust to changes in the underlying theoretical model.
In fact the CMB constraints on the scalar spectral index have improved by
about two orders of magnitude since the initial \COBE\ measurement.
Additionally,
we see no evidence for isocurvature modes, suggesting at most one (relevant)
dynamical degree of freedom, and no ``curvaton'' behaviour once the modes were
shifted outside the horizon.

\Planck\ has dramatically reduced the upper limits on non-Gaussianity
\citep{planck2013-p09a,planck2014-a19,planck2016-l09},
again suggesting a featureless inflaton potential, tightly limiting the
possibility of higher-order couplings of the inflaton field, and ruling out
a number of string-inspired models \citep[e.g.,][]{Burgess13}.
The constraints on non-Gaussianity from the CMB have improved by two orders
of magnitude from the early limits during the first decade of the millennium
\citep{spergel2007,komatsu2009}.
While consistent with the simplest models based upon slow roll of a single
degree of freedom, these limits have improved so dramatically that wide classes
of previously allowed models are now excluded.  The models that best fit
the \Planck\ data are those in which any multi-field dynamics 
present does not do much during the crucial epochs of horizon exit, and for
which the motion in field space is sub-Planckian.

With \Planck\ we have shown that scalar modes dominate the anisotropies in
the CMB by an order of magnitude (compared to tensor modes;
Fig.~\ref{fig:ns-r}).
With current CMB experiments, we are probing the class of inflationary models
for which $r\sim 1-n_{\rm s}$, excluding the popular monomial potentials
$m^2\phi^2$ and $\lambda\phi^4$ that arise in chaotic inflation at more than
the 99\,\% CL.
The combination of $n_{\rm s}<1$ and $r\ll 1$ suggests that the fluctuations
were produced near a ``special point'' in the inflaton potential
(i.e., $V'\simeq 0$ while $V''\ne 0$),
and that the space of models with ``convex'' potentials is severely limited.
Models with concave potentials, often predicting $r\sim (1-n_{\rm s})^2$,
are consistent with the \Planck\ data and include a variety of supersymmetry-
or string-inspired models with exponential potentials.
It will require dramatic increases in sensitivity, systematics control, and
foreground mitigation to probe this class of models.
Detection of tensor modes from the wide class of models with
$r\ll (1-n_{\rm s})^2$, or sub-Planckian field evolution, remains out
of reach with current or near-future technologies.

\subsection{Neutrino physics and constraints}
\label{sec:neutrinos}

As a dramatic illustration of the ``cosmo-micro'' connection, \Planck\ is able
to provide strong constraints on the properties of relic neutrinos
and additional light particles.  To discuss this further, we begin by
presenting the constraints on the masses of ordinary (``active'') neutrinos,
and then turn to discussing other light particles.
As we will see, the lower limits on neutrino masses from oscillation
experiments, combined with the upper limits from \Planck, 
leave only a narrow window at a value ($m_\nu\simeq 0.1\,$eV) that cries out
for explanation in fundamental physics.

The detection of Solar and atmospheric neutrino oscillations proves that
neutrinos are massive, with at least two species being non-relativistic by
the present day.
In the normal hierarchy ($m_1\la m_2<m_3$) the sum of the neutrino
masses must be larger than $0.06\,$eV
($\omega_\nu=\Omega_\nu h^2\simeq\sum m_\nu/93.04\,{\rm eV}\simeq 0.0006$),
while in the inverted scenario
($m_2\ga m_1\gg m_3$) the lower limit is $0.1\,$eV.
\Planck\ data provide strong upper limits on the sum of the neutrino masses,
of the same order, thus requiring $\sum m_\nu\simeq 0.1\,$eV.

The cosmological effects of neutrinos are covered in several reviews,
for example, \citet{Les13}, \citet{Pat15}, \citet{Arc17}, and \citet{LatGer17},
to which the reader is referred for more details.
For masses $\mathcal{O}(0.1\,{\rm eV})$, the neutrinos are still relativistic
at recombination and the effects on the anisotropy spectrum are
small (and primarily near the first acoustic peak, due to the evolution
of the potentials near recombination, known as the ``early ISW effect'').
The largest impact of massive neutrinos is in altering the late-time
expansion history and the shape of the matter power spectrum.
In the observationally relevant range, increasing neutrino masses increases the
expansion rate at $z>1$, changing the distance-redshift relation at low $z$. 
Since neutrinos free stream, while contributing to the background expansion,
the matter power spectrum is suppressed on small scales.
To hold $\theta_\ast$ fixed, an increase in $\sum m_\nu$ needs to
be accompanied by changes in other parameters that suppress large-scale
power.  The overall effect is thus a broad suppression of the matter power
spectrum at fixed CMB amplitude.
\Planck\ has moved us into a new regime, where the neutrino mass constraints
come not from their small effect on the primary anisotropies, but from the
measurement of the late-time potentials through gravitational lensing.
Current upper limits on $\sum m_\nu$ correspond to an $\mathcal{O}(1\,\%)$
suppression of power on sub-degree scales (unfortunately, \Planck\ is not
sensitive to mass splittings between the neutrinos).

Even tighter constraints can be obtained when combining \Planck\ data
with lower-redshift probes, in particular those that measure $H(z)$.
Increasing $\sum m_\nu$ while holding the angular scale of the acoustic
peaks fixed reduces the expansion rate at low $z$ (and increases it at
high $z$).  For fixed $\theta_\ast$ this lowers the Hubble constant and
increases the distance to $z\simeq 0.5$--1, which is tightly constrained by
BAO.  It is a testament to the incredible precision of modern cosmological
observations that neutrino masses can be constrained through such tiny
effects on the late-time expansion history.

With the improvement in the low-$\ell$ data of this final \Planck\ release,
which helps break degeneracies with $A_{\rm s}$ and $\tau$, the neutrino mass
limits have improved.  Unlike in earlier years, all three effects of massive
neutrinos -- changes in the distance to $z_{\rm dec}$, in the smoothing
of the temperature and polarization spectra, and in the shape of the
lensing spectrum -- contribute to the constraint in mutually reinforcing ways.
Thus the combination of acoustic oscillations in the early and late Universe
with the gravitational deflections of light across cosmic time provide a
tight constraint on the sum of the neutrino masses:
\begin{equation}
  \sum m_\nu < 0.12\,{\rm eV} \quad (95\,\%\ {\rm CL}).
\end{equation}
This implies that the inverted mass hierarchy is beginning to be
disfavoured by robust, cosmological data.

For this (very restricted) range of neutrino masses the impact on other
cosmological parameters is small, but not completely negligible given the
precision of the existing constraints.
As discussed in detail in \citet{planck2014-a15}, including $m_\nu$ as an
additional parameter can change the allowed values of $\Omega_{\rm m}$, $h$,
and $\sigma_8$.  However, all of the changes are correlated, so large areas
of parameter space are still excluded.  In particular, one needs to include
massive neutrinos and one other parameter (e.g., $N_{\rm eff}$) in order to
simultaneously have low values of $\sigma_8$ and high values of $h$.  Low
values of $\sigma_8$ also go with higher values of $\Omega_{\rm m}$ and lower
values of $h$, so neutrinos do not offer a solution to the discrepancy with
some (but not all) of the weak lensing or cluster count data
\citep[see, e.g., the discussion in][]{planck2016-l06}.

As well as neutrino mass, the CMB also gives sensitivity to the number of
types of neutrino.
The density of non-photon radiation in the Universe is usually parametrized
by an effective neutrino number
\begin{equation}
  \rho_{\rm rad} = \frac{7}{8}\left(\frac{4}{11}\right)^{4/3}
  \ N_{\rm eff} \, \rho_\gamma,
\end{equation}
specifying the energy density when the species are relativistic, in terms
of the neutrino temperature, assuming that three flavours of neutrinos
instantaneously decoupled.  In the standard model
$N_{\rm eff}\,{\simeq}\,3.045$--3.046 \citep{Man02,SalPas16}.
As with $m_\nu$, at \Planck\ sensitivity the best constraints on $N_{\rm eff}$
come from the distance scale.
Increasing $N_{\rm eff}$ at fixed acoustic scale ($\theta_\ast$)
and fixed $z_{\rm eq}$ increases the expansion rate before recombination.
This changes the sound horizon (approximately linearly with the age at
recombination) and the scale of photon diffusion (approximately as the
square root of the age).
The combination allows us to constrain additional relativistic species
\citep[e.g.,][]{HuWhi96}.
A tighter constraint is obtained if we include BAO data.  The increase in
$N_{\rm eff}$ (at fixed $\theta_\ast$ and $z_{\rm eq}$) increases the
expansion rate at low $z$ as well.
Although the sound horizon at the end of the baryon drag epoch, $\rdrag$,
also decreases, the combination of \Planck+BAO data still provides a strong
constraint: $N_{\rm eff}\,{=}\,3.01\pm 0.35$\ (95\,\% CL).
Imposing the constraint $N_{\rm eff}\ge 3.046$, the 95\,\% CL upper limit
on $\Delta N_{\rm eff}=N_{\rm eff}-3.046$ is $0.3$.
This mildly disfavours any light, thermal relics that froze out after the
quantum chromodynamics phase transition (which predicts
$\Delta N_{\rm eff}\,{=}\,0.3$ per degree of freedom).

The combination of robust cosmological probes has grown sufficiently
constraining that we are also able to provide limits on additional
massless relics, on top of the three active neutrinos.  Even allowing for
non-minimal neutrino masses,
$N_{\rm eff}\,{<}\,3.29$ (95\,\% CL; \citealt{planck2016-l06})
thus excluding one thermalized sterile neutrino at the $3\,\sigma$ level.

The above summary shows that \Planck\ provides evidence for a cosmic neutrino
background at very high significance.  Since the neutrinos make up a
non-negligible fraction of the total energy density near recombination
($\rho_\nu\,{\simeq}\,0.1\,\rho_{\rm tot}$),
the CMB is highly sensitive to their properties, and in particular to their
anisotropies \citep{HSSW95,Hu98,TroMel05}.
The \Planck\ data provide compelling constraints on the neutrino anisotropy
for the first time, showing that both the speed of sound in the neutrino
reference frame and the neutrino anisotropic stress are consistent with
standard predictions, $c_{\rm eff}^2=c_{\rm vis}^2=1/3$ 
\citep[to within 2\,\% and 10\,\% respectively,][]{planck2014-a15}; this limits non-standard neutrino interactions.

\subsection{Dark matter}
\label{sec:darkmatter}

Since \COBE\ first measured the amplitude of the anisotropies at the surface
of last scattering \citep{Smoot92}, the explanation of the observed
large-scale structure in the Universe through gravitational instability has
required the presence of dark matter \citep{EBW92}.  Indeed, the evolution of
the gravitational potentials and the stabilizing influence of dark matter
allow us to measure the cold dark matter density to around $1\,\%$ from the
shape of the peaks in the power spectrum.
\Planck\ has gone further and allowed us to map, in projection,
all of the dark matter back to the surface of last scattering, through
its effects on the propagation of CMB photons (i.e., gravitational lensing).
Inferences from the detailed shape of the power spectrum imply that
the dark matter must be stable, cold, and dark; moreover, if they are
thermally produced then the dark matter particles must also be massive.

If dark matter annihilates in the early Universe, and there is significant
energy in the post-decay shower at keV scales, then secondary particles can
ionize or heat the primordial gas and change the recombination history
\cite[see section 6.6 of][]{planck2014-a15}.
This can dramatically alter the CMB anisotropies
\citep{CheKam04,PadFin05}.
\Planck\ is sensitive to energy injection over the range $600\la z\la 10^3$
\citep{Fin12},
and the effects of DM annihilation can be relatively well modelled
by a single parameter that encodes the dependence on DM particle properties.
Since the main effect of DM annihilation is to increase the duration of
last scattering and enhance the ionization fraction at low $z$, a precise
measurement of polarization is particularly important.  For this reason
the \Planck\ data provide some of our tightest constraints on the
energy release per unit volume and thus DM annihilation.  For example,
they exclude a low mass ($m_\chi<44\,$GeV) thermal relic annihilating
into $e^{+}e^{-}$ pairs.

\subsection{Dark energy and modified gravity} \label{sec:darkenergy}

Of the many unexplained ingredients in our phenomenological $\Lambda$CDM
model, the cosmological constant ($\Lambda$) may be the most mysterious.
We currently lack any compelling explanation for its value, or a natural
mechanism to produce it.
In addition, the models that fit the data all predict that the present
epoch represents a ``special time'' in the history of the Universe.
Two alternatives to the introduction of a cosmological constant are to
promote $\Lambda$ to a dynamical field (or set of fields) that have an
effectively negative pressure to drive accelerated expansion (dark energy),
or to modify GR so that accelerated expansion can be
achieved with a ``standard'' stress-energy tensor (i.e., modified gravity).

\Planck, in combination with other probes, enables tests of dark energy
and modified gravity on the scales where linear theory is most applicable,
which tend to be the most theoretically robust.
In fact, many constraints on dark energy and modified gravity in cosmology
depend upon the CMB anisotropies in crucial ways.  \citet{planck2014-a16}
and \citet{planck2016-l06} discuss the \Planck\ constraints
on dark energy and modified gravity in detail.
The CMB is sensitive to these ingredients through their
effects on the expansion history, the evolution of the metric perturbations,
lensing, and the growth-rate of structure.  Since in most models dark energy
or modifications to gravity are late-time phenomena, the strongest constraints
come from combining the \Planck\ data with other data sets; however, the CMB
lensing measured by \Planck\ also provides some sensitivity.  In fact
\Planck\ lensing provided the first CMB-only evidence for dark energy
\citep{planck2016-l08}.

The background evolution can be constrained by \Planck+BAO+SNe
\citep[see][which contains details on the particular data used]{planck2016-l06}.
This provides a long enough lever arm in redshift that the geometric
degeneracy is largely broken.
Gravitational dynamics can be probed through ``growth of structure''
probes, such as redshift-space distortions.
The Weyl potential can also be probed through weak lensing of the CMB or
galaxy weak lensing.

The combination of \Planck+BAO+SNe data is compatible with $\Lambda$CDM,
and for simple models tightly constrains the dark energy equation of state,
$w\equiv p/\rho$ (e.g., $w=-1.028\pm0.032$ if it is constant).
For more flexible parameterizations, a range of equations of state remains
allowed.  Such a range in equation of state, however, does not translate into
a large uncertainty in other parameters such as $\Omega_{\rm m}$ or $\sigma_8$.
In fact, the posterior volume in the $w_0w_a$CDM model (where the equation
of state of the dark-energy component is $w=w_0+[1-a]w_a$) is not much larger
than for $\Lambda$CDM.
Interestingly, the region that is opened up by introducing new degrees of
freedom for the dark-energy evolution is not the region of reduced
$\sigma_8$ preferred by the low-$z$ probes appearing to exhibit tension with
\Planck\ (some cosmic shear measurements and some analyses of the counts of
rich clusters; Sect.~\ref{sec:discord}).
Thus evolving dark energy does not significantly impact the tension
between those measurements and \Planck.

The combination of the relative distance scale measured by SNe with the
absolute distance scale determined from CMB+BAO requires that the dark
energy density be subdominant at redshifts beyond $1$.  In most models,
the dark-energy density becomes irrelevant above $z\,{\simeq}\,2$,
and early dark energy and coupled DE models are now strongly constrained.
For example, the dark-energy density at early times must be below
$0.02\,\rho_{\rm crit}$ (95\,\% CL), even if it only plays a role below
$z\,{=}\,50$ \citep{planck2014-a14}.

The observed late-time acceleration of the cosmic expansion could be due to
modifications of GR instead of an additional component of the
energy density (e.g., recent reviews by \citealt{JaiKho10} and \citealt{Joyce15,Joyce16}).
However, at present there are no compelling models of modified gravity that
explain cosmic acceleration while being compatible with the observational
constraints, thus most explorations have tended to focus on generic
parameterizations of possible deviations from GR.
For example, within the subclass of scalar-tensor theories, the large-scale
behaviour can be effectively captured by two free functions of scale and time.

On very large scales and at late times, cosmological observations probe the
two metric potentials $\Psi$ and $\Phi$ (Sect.~\ref{sec:lcdm}),
or some combinations of them.
In \citet{planck2014-a16} and \citet{planck2016-l06}, those
potentials were allowed
to vary away from their GR values in time, holding the spatial dependence
fixed at the GR expectation.  No evidence was found for modifications to
GR, although once the relationship between the matter components and the
metric potentials is freed, lower values of $\Omega_{\rm m}$
and $\sigma_8$ are allowed by the data.

Overall, the \Planck\ data support the basic model with a spatially and
temporally constant dark-energy density (i.e., a cosmological constant) that 
is just now coming to dominate the energy density of the Universe.  The
constraints, however, are relatively weak compared to similar tests of
General Relativity on Solar System scales.  Future observations will be
required to provide stringent constraints on the plethora of models that
are currently consistent with the data.

\subsection{Isotropy and statistics; anomalies} \label{sec:anomalies}

For almost all the most important \Planck\ results, statistical isotropy
and Gaussianity of the CMB anisotropies are implicitly assumed.
This is reasonable, since when these assumptions are tested
on our CMB sky they seem to hold up well
(see \citealt{planck2016-l07}, as well as \citealt{planck2016-l09} and
\citealt{planck2016-l10}, and the earlier
papers \citealt{planck2013-p09} and \citealt{planck2014-a18}).
That is, no significant signals of statistical anisotropy or non-Gaussianity
appear, apart from those predicted by $\Lambda$CDM itself
(such as lensing and the integrated Sachs-Wolfe effect; \citealt{SacWol67})
or arising from foregrounds such as the SZ effect or the CIB.
Nevertheless, when such tests are restricted to the largest angular scales
($\ell<70$, say), some apparently 2--$3\,\sigma$ signals begin to appear,
and these have been called ``CMB anomalies.''
Specifically, it has been found that the temperature
anisotropies at the largest scales exhibit a dipolar asymmetry of power, show a
preference for odd parity modes, and contain a large cold spot in the
southern hemisphere.
The existence of these signals is not in dispute.  They appear in both \WMAP\
and \Planck, which have quite different systematics, and moreover
all of the \Planck\ results are robust with respect to the choice of
component-separated CMB map.  Thus these ``anomalies'' must be regarded
as features of the CMB temperature sky.
The main question then is whether such signals are unusual enough for physical
explanations to be sought, beyond merely being excursions in Gaussian random
skies.  This issue of ``a posteriori'' statistics is complicated by the fact
that for these scales the measurements are essentially cosmic-variance-limited, 
thus new measurements of the relevant modes will not change the
significance of the anomalies.

This final release of \Planck\ data then represents a major new opportunity,
since it contains our first comprehensive attempt at assessing the isotropy
of the Universe via an analysis of the full-mission \Planck\
polarization data.
This was not possible in earlier releases, due to residual large-scale 
systematics that required high-pass filtering of the CMB polarization maps.
Probing independent information on the sky, the large-angular-scale
polarization gives us a rare opportunity to study some of these anomalies;
however, inferences are hindered by the fact that the signal-to-noise ratio
in the \Planck\ polarization data is lower than in temperature, at large
scales the signal is very small (see Fig.~\ref{fig:cl}), and the $E$ modes
are only partially correlated with temperature.  The degree to which we expect
a signature of various claimed anomalies to appear in the polarization is
therefore somewhat model dependent.

\citet{planck2016-l07} attempts a comprehensive analysis of the statistics
of the polarization signal from large to small angular scales, using either
maps of the Stokes parameters ($Q$ and $U$) or the $E$-mode signal.  While
these studies are limited by residual systematics, a series of null tests
applied to the maps indicate that these issues do not dominate the analysis
on intermediate and large angular scales (i.e., $\ell \lesssim 400$).
In this regime, there is no unambiguous detection of cosmological
non-Gaussianity, or of anomalies corresponding to those seen in temperature.
Notably, the stacking of CMB polarization signals centred on the positions
of temperature hot and cold spots exhibits excellent agreement with the
expectations of the $\Lambda$CDM cosmological model.         
However it will require future, more sensitive, polarization observations
to fully test the models that have been advanced to explain the anomalies.

It is worth stressing that none of these so-called anomalies are
strongly inconsistent with the assumption of statistical isotropy and
Gaussianity, once one marginalises over a set of similar tests.
It would nevertheless be premature to completely dismiss all the CMB anomalies
as simple fluctuations of a pure $\Lambda$CDM cosmology, since
if any of the anomalies have a primordial origin, then their large-scale
nature would suggest an explanation rooted in fundamental physics.
Thus it is worth exploring any models that might explain an anomaly
(or even better, multiple anomalies) naturally, or with very few free
parameters.  Given a theoretical prediction, new probes of independent
modes on similar scales (obtained through more sensitive
polarization measurements, lensing,
Ly$\,\alpha$, or 21-cm studies for example) would increase the significance
of existing anomalies and allow us to develop novel probes of early Universe
physics.
So far the simplest models explaining a single anomaly are not favoured
over $\Lambda$CDM \citep[see][and references therein]{planck2016-l10}.
Further investigation of these anomalies will need to proceed on a
case-by-case basis, and will be the subject of future work.

\section{\Planck\ and structure formation} \label{sec:lss}

By cementing the gravitational instability paradigm and accurately
measuring the initial conditions and parameters determining the
subsequent growth of structure, \Planck\ provides the framework within
which to discuss the formation and evolution of large-scale
structure and galaxies, black holes, and other astrophysical objects..

With \Planck\ we have tightly constrained the densities of radiation, matter,
and baryons, as well as the amplitude and shape of the fluctuations in the
linear phase over three decades in length scale.
Our knowledge of the physical conditions and large-scale structure at
$z\,{=}\,10^3$ is better than our knowledge of such quantities at $z\,{=}\,0$.
It is for this reason that ``CMB priors'' have become an integral part of
current and future cosmological inference; indeed almost no cosmological
experiment interprets their data without adding the existing constraints
from \Planck.

\subsection{The normalization and shape of \texorpdfstring{$P(k)$}{P(k)} }
\label{sec:pk}

In cosmology we frequently refer to standard candles (objects of known
luminosity) or standard rulers (objects of known size).  However,
the CMB has provided us with a ``calibrated, standard fluctuation
spectrum,'' from which we can accurately compute how big a sample
has to be in order to be ``fair,'' how many objects constitute a
``dense'' sample, how strong clustering will be for objects of various sizes,
and the abundance of dark-matter halos as a function of mass and epoch.
By constraining the fluctuations in regions of a given volume or for
halos of a given mass, it provides quantitative answers to questions about
how well a set of objects in a sampled region embodies the average properties,
and the relative importance of sampling variance and shot noise.
Here we discuss tests enabled by this calibrated spectrum.
In the next subsections we will explore lensing cross-correlations
(Sect.~\ref{sec:lsslensing}) and discuss the acoustic features in the matter
power spectrum (BAO) that can be used as a standard ruler
(Sect.~\ref{sec:bao}).
Since the growth of structure depends sensitively on the properties of
the objects that cluster strongly (e.g., dark matter) and on those
that do not (e.g., neutrinos and dark energy), as well as on our theory of
gravity (i.e., GR), studies of clustering address many of
the most fundamental questions in cosmology.

\begin{figure}[htbp]
\begin{center}
\resizebox{\columnwidth}{!}{\includegraphics{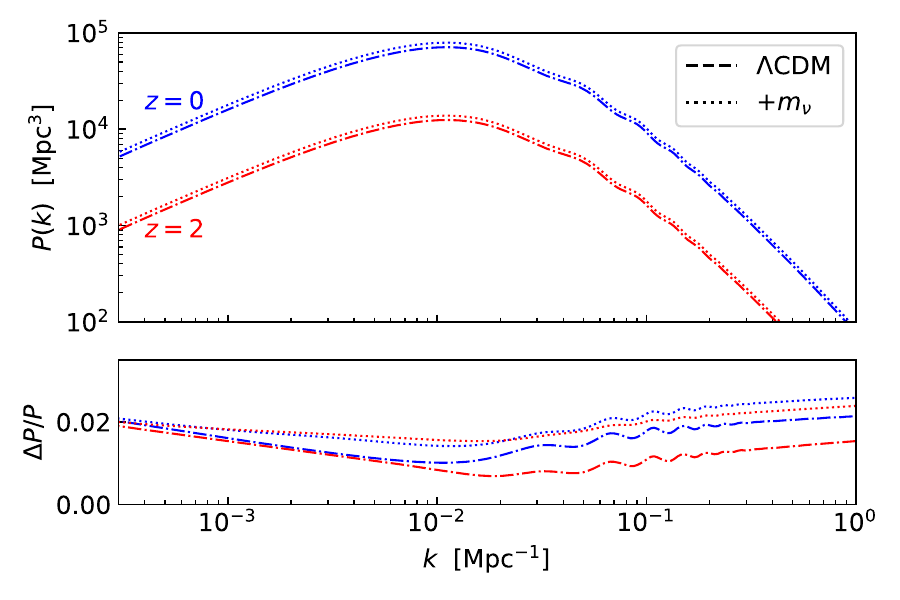}}
\end{center}
\caption{{\it Top\/}: Matter power spectrum \citep[including non-linear
corrections using the fitting form of][]{Tak12} at redshifts 0 and 2,
predicted by the $\Lambda$CDM model with a single massive neutrino of
$0.06\,$eV (dashed curve) or allowing the neutrino mass to float
(dotted curve).
{\it Bottom\/}: Fractional error in each power spectrum, 
compared to the average plotted above, due to the remaining uncertainties
in the cosmological parameters.
Uncertainty in the evolution of the scale factor at late times
(due to dark energy) leads to an additional uncertainty in the overall
amplitude, coherent across scales, which is not shown here.
With current BAO and SNe data the uncertainty in the growth from
$z\,{=}\,2$ to $z\,{=}\,0$ is 8\,\% (or 16\,\% in power).
\label{fig:pk} }
\end{figure}

In Sect.~\ref{sec:lss1} we showed that the shape of the matter power spectrum
predicted by $\Lambda$CDM fit to the \Planck\ data is in excellent agreement
with measurements at lower $z$ (Fig.~\ref{fig:allpk}).
Figure~\ref{fig:pk} shows another aspect of this, highlighting the evolution
of $P(k)$.  Within the $\Lambda$CDM
paradigm the late-time matter power spectrum is very well predicted once the
initial fluctuation spectrum and matter contents are known.  In fact $P(k)$
is sensitive to combinations of parameters that are generally well measured
by the CMB, so the final uncertainty is small.  Figure~\ref{fig:pk} shows the
non-linear matter power spectrum, over three decades in wavenumber, as predicted
by $\Lambda$CDM fit to \Planck.  We display the results with and without
variations in $m_\nu$, since this is one of the best-motivated extensions that
impacts the matter power spectrum.  We show results at $z=2$, before dark
energy becomes an appreciable fraction of the total energy density, and at
the present epoch ($z=0$).
In physical units (i.e., Mpc${}^{-1}$ rather than $h\,{\rm Mpc}^{-1}$)
the power spectrum is predicted at the few percent level up to
$k\simeq 1\,{\rm Mpc}^{-1}$ (beyond which effects from astrophysical processes
such as stellar and AGN feedback become important,
for example \citealt{White04,Zhan04,Chisari18}).
If we include dark energy with a time-varying equation of state in the model,
then the power spectrum at $z\ge 2$ is only mildly affected, but we introduce
an extra uncertainty in the amplitude of $P(k,z=0)$ at around the 16\,\% level.
The dominant uncertainty is in the amplitude, leaving the shape almost
invariant.
The fact that the constrained model predicts the spectrum to such exquisite
accuracy provides a stable platform for inferences about the lower-redshift
evolution and a target for tests of GR, the expansion history,
and the contents of the Universe.

The main feature visible in Figs.~\ref{fig:allpk} and \ref{fig:pk} is the
peak at $k\simeq 10^{-2}\,{\rm Mpc}^{-1}$.  The location of this peak
is set\footnote{Modes that are smaller than the Hubble scale during radiation
domination, $k>k_{\rm eq}$, have their growth slowed because fluctuations in
the dominant radiation component (which contribute the most to the potentials)
are stabilized by pressure and oscillate, rather than growing in amplitude.}
by the Hubble scale at matter-radiation equality, which is now extremely well
determined by the \Planck\ data: $z_{\rm eq}=3387\pm 21$.
Along with the amplitude of $P(k)$, this scale,
$k_{\rm eq}=(0.01034\pm 0.00006)\,\mathrm{Mpc}^{-1}$,
sets the characteristic volume of the Universe that needs to be surveyed in
order for a sample to be considered a ``fair'' representation of the
Universal average.

The amplitude of the spectrum and its evolution sets the level of
clustering in the Universe and, indirectly, the halo mass function.
A population of objects whose number density times large-scale bias
squared is less than the inverse peak power
(i.e., $b^2\bar{n}\ll P_{\rm peak}^{-1}$)
will always be in the shot-noise limited regime, that is, it will be a
``sparse'' tracer of large-scale structure.
This means that such a population cannot measure the large-scale structure
on a mode-by-mode basis (although it can be used to determine the statistics of
large-scale structure by averaging over many independent modes) on any scale.
For example, such a population can be used to measure $P(k)$, or in
cross-correlations, but it will not be a good choice for density-field
reconstruction or mapping the cosmic web.
When a sample becomes ``dense'' is less clear, but roughly speaking it
occurs when the number density (times $b^2$) becomes larger than the matter
power spectrum at the non-linear scale.  Surveys of such objects are dominated
by sample variance on all linear scales \citep{FKP94}.

Future galaxy, quasar, and CMB surveys will constrain $P(k)$ ever more tightly.
One immediate goal of such surveys is to look for the suppression of
small-scale power imparted by massive neutrinos (Sect.~\ref{sec:neutrinos})
or warm dark matter.  More ambitious surveys may be able to detect any running
of the spectral index, or extra relativistic degrees of freedom.  If we can
improve our knowledge of star and galaxy formation, the well-determined power
spectrum at $z\gg 1$ may enable forward modelling to the reionization epoch
(Sect.~\ref{sec:reionization}),
which can be probed by 21-cm surveys and next-generation CMB experiments.

\subsection{Lensing cross-correlations}
\label{sec:lsslensing}

Lensing provides us with both a map of all of the matter in the Universe
and a persuasive cross-check on our cosmological model.
There are three main ways in which lensing contributes:

\begin{itemize}
\item it provides better constraints on the basic parameters;
\item it tests the gravitational instability paradigm and constrains
    modifications to GR on very large scales; and
\item it allows for cross-correlations, to provide more information.
\end{itemize}

We have already discussed the first two points.
The \Planck\ lensing maps have also been used in a wide variety of
cross-correlation studies, for a number of purposes.
Since the lensing signal comes from an already well-probed redshift range and comes
from largely linear modes, it allows us to determine the bias of cosmological
objects and place constraints on their redshift distribution.

Starting with the 2013 data, the \Planck\ team has cross-correlated the
lensing maps with large-scale structure traced by radio, optical, and IR
surveys \citep{planck2013-p12}.
Other studies have correlated the \Planck\ maps with:
mid-IR selected quasars at $z\,{\simeq}\,1$
\citep{Geach13,DiPompeo15,DiPompeo16};
optical galaxies from SDSS-III
\citep{Pullen16,Giusarma18,Singh18,Doux18}, CFHT \citep{OmoriHolder15},
and DES \citep{Giannantonio16};
galaxies from 2MASS, WISE, and SuperCOSMOS \citep{BiaRei18,Rag18,PeaBil18};
the Ly$\,\alpha$ forest \citep{Doux16}; and
high-$z$ submillimetre galaxies from {\it Herschel}-ATLAS \citep{Bianchini15}.
Cross-correlations with unresolved sources include dusty star-forming
galaxies \citep{planck2013-p13}
and the $\gamma$-ray sky from {\it Fermi}-LAT \citep{Fornengo15,Feng17}.
In fact, \Planck\ even has sufficient sensitivity to detect the lensing
signal on the scale of individual dark-matter halos \citep{planck2014-a30}.

Not only have the \Planck\ lensing maps been cross-correlated with tracers
of the density field, but also with other measures of lensing, in particular
cosmic shear surveys \citep{LiuHill15,Kirk16,Har16,Har17,Miyatake17,Singh17}.
These two independent measures of the gravitational potentials from large-scale
structure promise significant complementarity, and the comparison
may aid systematic error mitigation in future surveys.  Such studies with
the \Planck\ lensing maps provide a strong proof of principle.

As large-scale structure surveys push to high redshift over large fractions
of the sky, we expect the synergies described above to become ever more
compelling.  While current and next generation experiments are expected to
significantly improve lensing maps on small scales, the \Planck\ lensing maps
are likely to remain our best tracers of the low-$\ell$ lensing modes for
some time.  In addition, the higher-frequency channels of \Planck\ will not
be surpassed for many years, and they contain valuable information on 
foregrounds that will impact temperature-based lensing reconstruction for at
least another decade.  While contaminating signals such as our galaxy, the
Sunyaev-Zeldovich effect from groups and clusters, or the cosmic infrared
background from dusty, star-forming galaxies remain a cause for concern, one
may also view them as valuable signals to be extracted.  To this end,
cross-correlations will enhance the legacy value of the \Planck\ data.

\subsection{Baryon acoustic oscillations (BAO)} \label{sec:bao}

\Planck\ has now mapped 18 acoustic peaks and an almost equal number of
troughs (in $TT$, $TE$, and $EE$ together), which form an almost harmonic
series of features in the temperature and polarization power spectra
(Table~\ref{tab:peaks}).
The peaks arise from gravity-driven acoustic oscillations in the
baryon-photon fluid prior to recombination.  The non-trivial contribution
of the baryons to the total matter content implies that an analogous series
of peaks is also visible in the matter power spectrum, leading to a special
scale that is fixed in comoving coordinates as the Universe evolves
\citep{PeeYu70,SunZel70,Dor78}.
Measurement of this scale at low redshifts, for example in large galaxy
redshift surveys or in the absorption lines imprinted by intergalactic gas
in the spectra of high-$z$ quasars, provides a ``standard ruler'' for
constraining the expansion history of the Universe.\footnote{Discussions of
the physics of BAO can be found in \citet{EisHu98}
and \citet{Meiksin99}.  For configuration space, see \citet{ESW07}.
A more recent review is \citet{Weinberg13}}.
Since the scale (approximately 150\,Mpc) is so large, it is nearly immune
to astrophysical processing and non-linear evolution.  The major obstacle to
measuring the feature in the low-$z$ Universe is that very large volumes need
to be surveyed in order to obtain a robust detection.
It is convenient that the same acoustic phenomena that give rise to the key
features in the angular power spectra also give a signature that can break
one of the few remaining (near-)degeneracies between CMB-determined
parameters, namely the angular distance degeneracy.

\begin{figure}[htbp]
\begin{center}
\resizebox{\columnwidth}{!}{\includegraphics{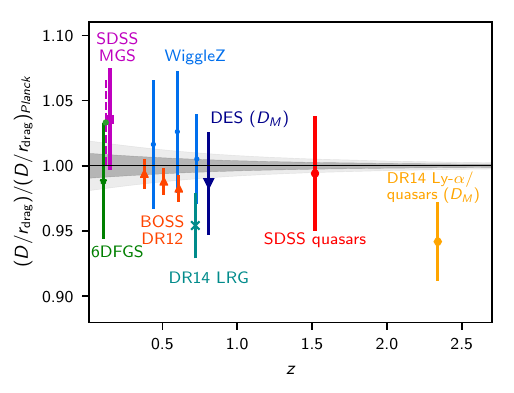}}
\end{center}
\caption{Redshift-distance relation measured by BAO surveys, compared to the
predictions of the $\Lambda$CDM model constrained by \Planck.
The grey band centred at unity shows the $\pm1$ and $\pm2\,\sigma$ confidence
regions for the \Planck\ prediction, given the remaining uncertainties in the
parameters.  This is a percent-level prediction of the distance
scale.  The BAO points are: 6dFGS, green star \citep{Beutler11};
SDSS MGS, purple square \citep{Ross15};
BOSS DR12, red triangles \citep{BOSS-DR12};
WiggleZ, blue circles \citep{Kazin14};
SDSS quasars, red circle \citep{Ata18};
and BOSS Ly$\,\alpha$, yellow cross \citep{Bautista17}.}
\label{fig:DV-BAO}
\end{figure}

Measurements of the BAO feature currently span the redshift range
$0\,{<}\,z\,{<}\,2.5$,
using either galaxies or the Ly$\,\alpha$ forest as tracers.
A comparison of the (angle-averaged) distance-redshift relation inferred from
a number of BAO measurements, to the distance scale predicted by $\Lambda$CDM
constrained by \Planck, is shown in Fig.~\ref{fig:DV-BAO}.
The agreement is excellent.
The uncertainty in the prediction, from the remaining spread
in the model parameters, is at the percent level for all redshifts.
The BAO data are approaching comparable precision, especially the BOSS DR12
data \citep{BOSS-DR12}.
Acoustic oscillations in the high- and low-$z$ Universe give a consistent,
percent-level determination of the distance scale within the $\Lambda$CDM
paradigm.
While we do not show it, the distances inferred from high-redshift Type Ia SNe
also provide a consistent distance-redshift relation.  In fact the combination
of CMB, BAO, and SNe distances allows us to establish an ``inverse distance
ladder,'' in which distances in the range $0.2\,{<}\,z\,{<}\,2$
are calibrated to the physical scale provided by the CMB at
$z\,{\simeq}\,1100$, rather than being bootstrapped
up from $z\simeq 0$ to larger redshifts.

\begin{figure*}[!btp]
\begin{center}
\resizebox{\textwidth}{!}{\includegraphics{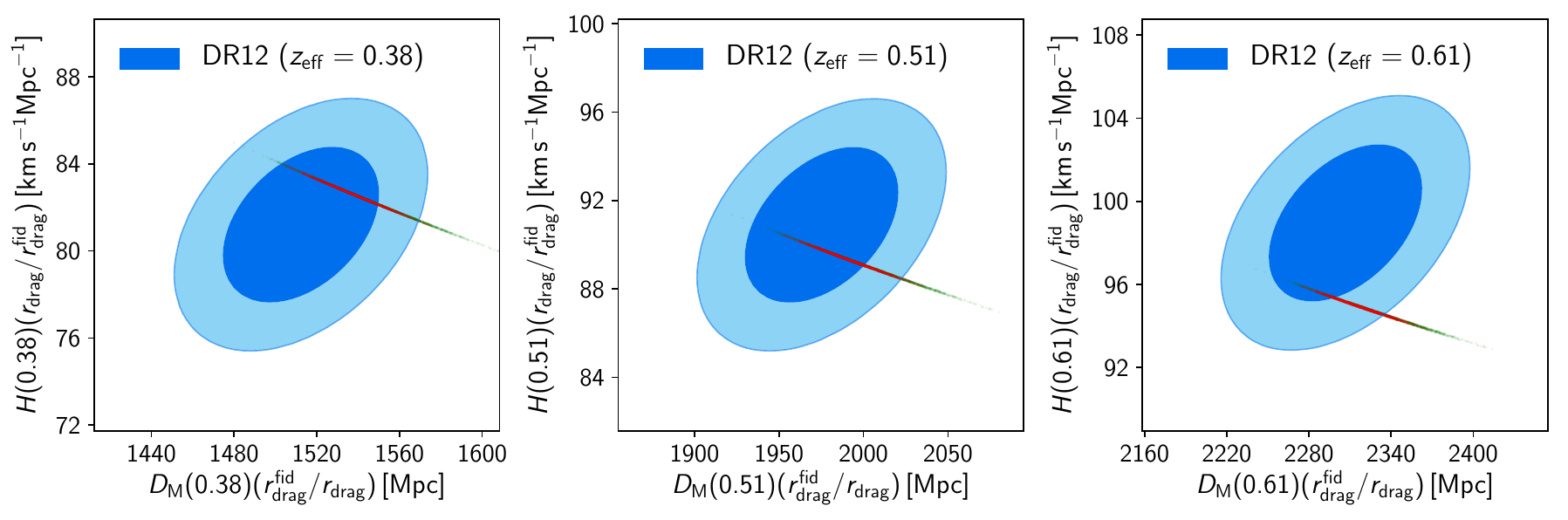}}
\end{center}
\caption{Constraints (at 68\,\% and 95\,\% CL) on the comoving angular
diameter distance ($D_{\rm M}$) and Hubble parameter ($H$) at the three
central redshifts of the BOSS DR12 analysis \protect\citep{BOSS-DR12}.
The green points show samples from the \Planck\ TT+lowE chains, while the
red points include the high-$\ell$ polarization and lensing data.
As more data are added there is a shift towards slightly lower
$D_{\rm M}$ and higher $H$, in better agreement with the BAO results.
This is also true for adding polarization and lensing separately (not shown).
}
\label{fig:DM-H-BAO}
\end{figure*}

The BAO method also provides measures of distances along the line of sight,
that is, of the Hubble parameter.  The current best measurements of the BAO
feature comes from BOSS \citep{Dawson13}, which has surveyed
$18.7\,{\rm Gpc}^3$ of the low-$z$ Universe and $150\,{\rm Gpc}^3$ of the
$z\,{\simeq}\,2.5$ Universe to provide highly significant detections of
the acoustic feature.
Figure~\ref{fig:DM-H-BAO} shows the comparison in the $D_{\rm M}$--$H$ space,
and we see that the agreement is excellent.
The thin contours show the \Planck\ $\Lambda$CDM predictions, where the
geometric degeneracy is evident.  Moving along this line, $\omega_{\rm m}$ and
$h$ are changing in concert to hold $\theta_\ast$ (almost) constant.
In Fig.~\ref{fig:DM-H-BAO} the green points show samples from the
\Planck\ TT+lowE chains, while the red points include the high-$\ell$
polarization and lensing data.  As more data are added there is a
shift towards slightly lower $D_{\rm M}$ and higher $H$, in better agreement
with the BAO results.  This is also true for adding polarization
and lensing separately (not shown).

\begin{figure}[htbp]
\begin{center}
\resizebox{\columnwidth}{!}{\includegraphics{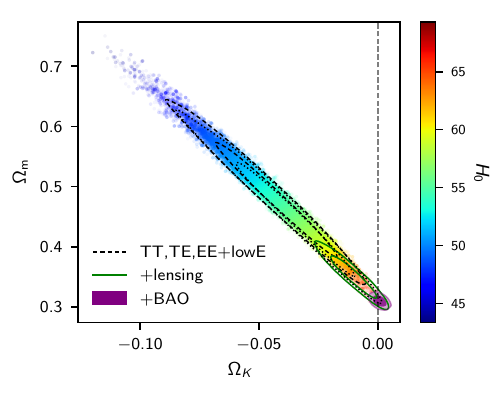}}
\end{center}
\caption{Constraints on spatial curvature from \Planck\ alone or in
combination with BAO data.  The vertical dashed line indicates a spatially
flat Universe, which is quite consistent with the combination of all of the
observations (solid, purple filled contour).
Points are coloured by the value of the Hubble constant (colour bar), dashed
lines show the 68 and 95\,\% confidence contours from the fiducial likelihood,
while dotted lines show those from the alternative ({\tt CamSpec}) likelihood
as an indication of the systematic uncertainty.
\label{fig:omega_k-omega_m}}
\end{figure}

The real power of the BAO data becomes apparent, however, when we open up
the parameter space beyond $\Lambda$CDM.  One of the key degeneracies that
enters in these extended parameter spaces is the angular scaling (often called
the ``geometric distance degeneracy''), which means that changes in the
parameters that hold the angular diameter distance to the surface of last
scattering fixed\footnote{Or more generally combinations which change
$\rstar$ and the distance so as to hold $\theta_\ast$ fixed.} are only
weakly constrained.
By providing a low-redshift distance determination, the BAO measurements
largely break this degeneracy.
One example is presented in Fig.~\ref{fig:omega_k-omega_m}, which shows the
constraints in the $\Omega_{\rm m}$--$\Omega_K$ plane.  With only the primary
CMB information, the geometric degeneracy allows a wide range of solutions.  
Including CMB lensing tightens this somewhat, but the highly precise BAO
distances break the degeneracy almost entirely (a similar effect happens
with massive neutrinos, as discussed in Sect.~\ref{sec:neutrinos}).
It is worthy of note that the constraint on $\Omega_K$ has improved by two
orders of magnitude in under two decades.

Looking at this from the point of view of BAO surveys, \Planck\ fixes
$\rdrag$ to $0.2\,\%$ (for base $\Lambda$CDM), allowing line-of-sight
BAO measurements to be translated into measures of $H(z)$ on an absolute
scale, which is limited only by our uncertainty about the high-$z$ Universe:
\begin{equation}
  \rdrag\,h \left(\frac{\Omega_{\rm m}}{0.3}\right)^{0.4} =
  \left( 101.056 \pm 0.036 \right)\,{\rm Mpc} \qquad (68\,\%\ {\rm CL}).
\end{equation}
This allows BAO experiments to provide a direct measure of the expansion
rate in physical units.

\subsection{Clusters and SZ effects}
\label{sec:clusters_sz}

\begin{figure}[htbp]
\resizebox{\columnwidth}{!}{\includegraphics{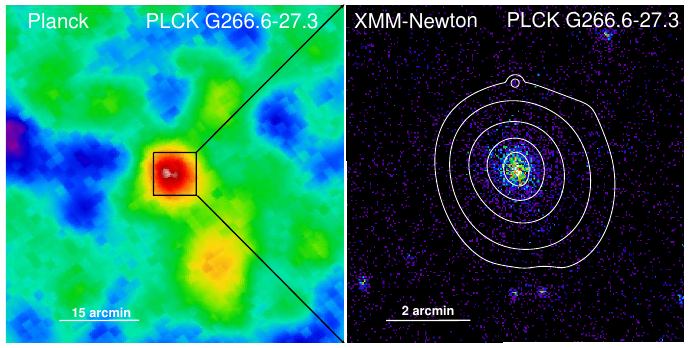}}
\caption{{\it Left\/}: \Planck\ SZ map of the candidate cluster
PLCK G266.6$-$27.3.
{\it Right\/}: XMM-{\it Newton\/} image of the central region.}
\label{fig:g266}
\end{figure}

\Planck\ has had a significant impact on the study of galaxy clusters using the Sunyaev-Zeldovich effect
(SZ; \citealt{SunZel72,SunZel80}; see \citealt{Carlstrom02} for a review).
This has contributed to \Planck's cosmological legacy, through the statistical properties of the \Planck\ SZ catalogues and maps,
as well as observations of individual objects.  Examples of the former include studies of cluster scaling relations and profiles
\citep{planck2011-5.2a,planck2011-5.2b,planck2011-5.2c,planck2012-III,planck2012-V,planck2012-XI}, while an early example of the latter was a study of the physics of gas in the Coma cluster \citep{planck2012-X}.  Another example was the discovery of an exceptionally luminous and massive cluster at
$z\,{\simeq}\,1$ via its SZ effect, an object which was verified in follow-up XMM-{\it Newton\/} observations
\citep[see][]{planck2011-5.1b,planck2012-I,planck2012-IV}.
Figure~\ref{fig:g266} shows the \Planck\ SZ map and its XMM-{\it Newton\/}
confirmation, with both images suggesting a surprisingly relaxed
cluster for this epoch \citep{planck2011-5.1c}.
More generally, the XMM-{\it Newton\/} follow up of clusters in \Planck's
first SZ catalogue \citep{planck2013-p05a} was very successful, with 51 new
clusters confirmed, spanning the redshift range 0.09 to 0.97 
\citep{planck2012-IV}.

\begin{figure}[htbp]
\resizebox{\columnwidth}{!}{\includegraphics{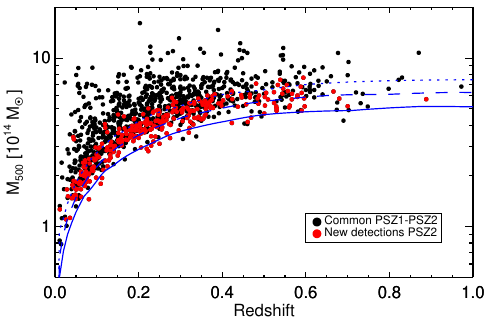}}
\caption{\Planck\ clusters in the redshift-mass $(z,M_{500})$ plane, where
$M_{500}$ indicates the mass interior to a radius where the mean enclosed
density is 500 times the critical density.
The red dots show 87 newly identified clusters from the \Planck\ 2015 SZ
catalogue paper \citep{planck2014-a36}.
Blue lines show completeness contours, at 80, 50 and 20\,\% (from the top).
}
\label{fig:new-clusts}
\end{figure}

The SZ legacy catalogue in \citet{planck2014-a36} built on the earlier versions
(\citealt{planck2011-5.1a}; \citealt{planck2013-p05a}; \citealt{planck2013-p05a-addendum}).
It contains 1\,653 detections, of which 1\,203 are confirmed
clusters with identified counterparts in external data sets.
It was the first SZ catalogue with more than 1000 confirmed clusters.
New detections, relative to the 2013 catalogue, are shown in the
redshift-mass plane in Fig.~\ref{fig:new-clusts}; these can be seen to
fit well with the completeness contours of the new survey.

\begin{figure}[htbp]
\resizebox{\columnwidth}{!}{\includegraphics{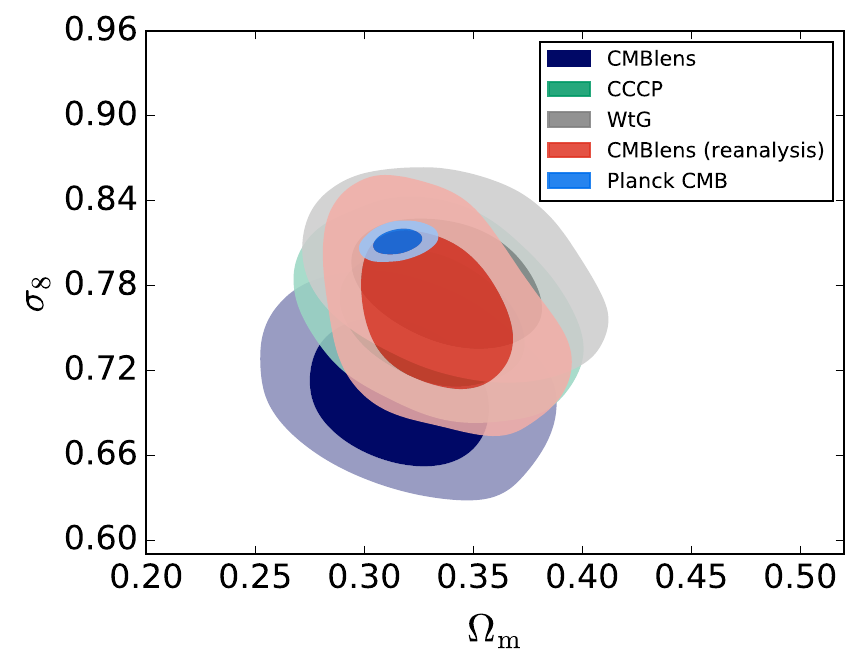}}
\caption{Probabilities in the $(\Omega_{\rm m},\sigma_8)$ plane for
different versions of the scaling relations between Compton distortion
parameter and cluster mass.  Here ``WTG'' is Weighing the Giants,
``CCCP'' is the Canadian Cluster Comparison Project, ``CMBlens''
refers to the CMB lensing method as analysed by \citet{MelBar15}
and re-analysed by \citet{ZubCha19}.
Blue contours are constraints from CMB anisotropies.}
\label{fig:sz-cosmo}
\end{figure}

The legacy catalogue enabled the subset of clusters that were used as a
sample for cosmology constraints to be substantially increased compared with
the number used in 2013, with $439$ clusters included in 2015 versus $189$ in
2013.  A key constraint that emerges from the 2015 cosmology sample
\citep{planck2014-a30} is the result for $\sigma_8$ versus $\Omega_{\rm m}$,
shown in Fig.~\ref{fig:sz-cosmo}.
The coloured contours in that figure refer to different ways of treating
the crucial scaling between the measured cluster Compton distortion parameter,
$Y_{500}$, and the cluster mass, $M_{500}$ (both defined within a radius
where the mean enclosed density is 500 times the critical density).
This is a complex procedure, in which numerous possible systematic and
statistical errors have to be taken into account, and is at the heart of
any attempt to use cluster abundance data for cosmology.
The \Planck\ data themselves provide only weak constraints on
this scaling, so external data are typically required.
An additional uncertainty comes from the choice of ``mass function,'' that is,
the function that predicts the abundance of clusters of different
mass for varying cosmological parameters.  This is generally derived from
fits of dark-matter halo abundances in numerical simulations,
ideally accounting for the effects of the baryonic component.

In the 2013 results paper \citep{planck2013-p15}, the scaling was carried out by
using an X-ray-defined version of the Compton parameter $Y_{500}$
(called $Y_{\rm X}$) as an intermediary, and using the
$Y_{\rm X}$--$M_{500}$ relation,
assumed known up to some so-called mass bias factor $(1-b)$, in order to
calibrate the $Y_{500}$-$M_{500}$ relation in the cosmological sample.
Leaving aside other factors, the relation found was of the form
\begin{equation}
  Y_{500} \propto \left[ (1-b) M_{500} \right]^\alpha,
\end{equation}
with $\alpha\,{\simeq}\,1.8$.
The factor $(1-b)$ arises from an expected miscalibration of the local sample used to calibrate the X-ray relation,
due to deviation of the clusters from the assumption of
hydrostatic equilibrium, but also encompasses other systematic errors.
Various values of $(1-b)$ were considered in the 2013 results paper, some
motivated from simulations, and the analysis was carried out assuming
a baseline of $(1-b)$ varying over the range $[0.7,1.0]$ with a flat prior.
This yielded an equivalent to Fig.~\ref{fig:sz-cosmo}, which showed quite
strong discrepancy between the confidence contours for $\sigma_8$ and
$\Omega_{\rm m}$ coming from the \Planck\ primordial CMB results, and
those from the SZ cluster analysis.
It was possible to reconcile the two, but only by moving $(1-b)$ to
lower values than suggested by numerical simulations.
Specifically, asking for agreement between the \Planck\ primordial CMB
results and SZ cluster counts, yielded a ``measurement'' of $(1-b)$ of
$0.59\pm0.05$, definitely lower that the 10 to 20\,\% bias away from
hydrostatic equilibrium expected previously from simulations.

In the \Planck\ 2018 results shown in Fig.~\ref{fig:sz-cosmo}, we see that
the situation has alleviated somewhat, in that for some versions of the
procedure to establish the scaling relation between observed SZ value
and mass, there is effectively no discrepancy with the primordial CMB values
of $\sigma_8$ and $\Omega_{\rm m}$.
In particular, the mass scaling implied by the ``Weighing the Giants''
programme \citep{vdL14}, based on the availability of high-quality
gravitational shear information for 22 clusters from the \Planck\ 2013
cosmology sample, gives $(1-b)=0.688\pm 0.072$, and
therefore little evidence of any tension with primary CMB results.
On the other hand, we see that some alternative methods do still give some
tension.  The violet contours of Fig.~\ref{fig:sz-cosmo} refer to a mass
calibration carried out using lensing of the CMB itself by the clusters
\citep{MelBar15}, which finds $1/(1-b) = 0.99\pm 0.19$
(the CMB lensing method constrains the reciprocal of the quantity found from
the shear measurements). Since this implies a small hydrostatic-equilibrium
bias, then it follows that there is a fairly large discrepancy between the
results in the $(\Omega_{\rm m},\sigma_8)$ plane using this method,
and the CMB anisotropy values.  In contrast a recent reanalysis of the CMB
lensing data by \citet{ZubCha19}, shown as the red contours
in Fig.~\ref{fig:sz-cosmo}, implies no such discrepancy.
Other recent determinations show a similar diversity.  For example,
\citet{Applegate16} find consistency between hydrostatic and weak-lensing
mass measurements of massive, dynamically relaxed clusters,
\citet{Okabe16} obtain different mass measurements for some clusters
than the ``Weighing the Giants'' programme,
\citet{Medezinski18} use a Hyper Suprime-Cam weak-lensing sample of five
\Planck\ clusters to infer $1-b=0.80\pm 0.14$,
\citet{Penna-Lima17} use weak-lensing masses from the Cluster Lensing And
Supernova survey with Hubble to infer $1-b=0.75\pm0.10$, while
\citet{Sereno17} use CFHTLenS and RCSLenS to infer
$1-b=0.60\pm0.16$ for the cosmology sub-sample (and provide a summary of
other determinations, their table~5).

The situation overall, therefore, is not yet wholly clear.
In particular, are the residual discrepancies caused by uncorrected
systematics or remaining biases in the astrophysical assumptions and
simulations, or are they perhaps a hint of something more important, such as
the first signs of new physics?

\begin{figure}[htbp]
\resizebox{\columnwidth}{!}{\includegraphics{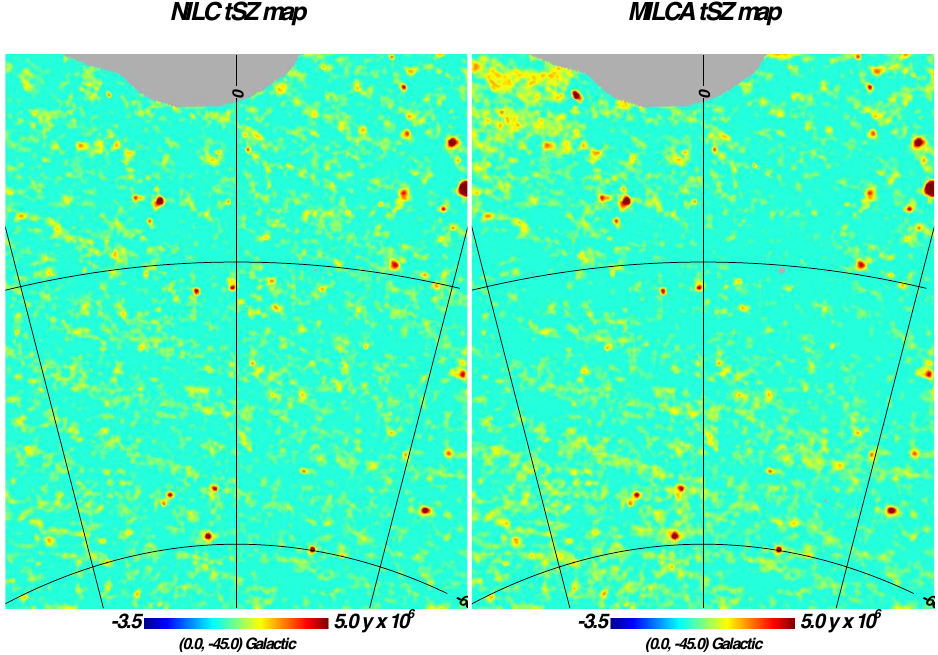}}
\caption{Region of the southern sky reconstructed in the thermal SZ
Compton $y$ parameter.
The results from two different reconstruction methods are shown. }
\label{fig:sz-map-detail}
\end{figure}

As well as building up a catalogue of individual SZ clusters, it is possible
to make a map of the Compton $y$-parameter over the whole sky.
This was presented in \cite{planck2013-p05b}, updated in
\citet{planck2014-a28}, and is available as a \Planck\ product via the PLA.
A sub-region of this map is shown in Fig.~\ref{fig:sz-map-detail}, and
illustrates the combination of individual clusters (plus possible diffuse
SZ effect regions) that is visible.
An important question is whether the power spectrum of this map agrees
with the conclusions from the catalogue-based analysis discussed above.

\begin{figure}[htbp]
\resizebox{\columnwidth}{!}{\includegraphics{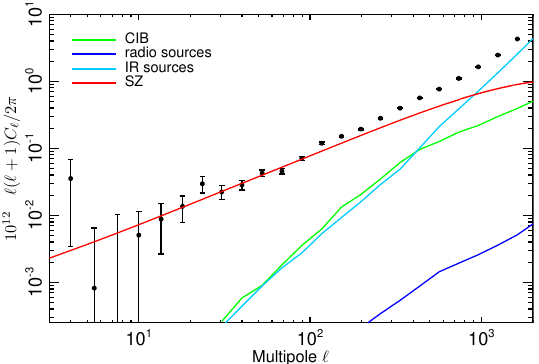}}
\caption{Thermal SZ power spectrum from \Planck\ data.
Black points with errors show the $\mathcal{C}_{\ell}$ spectrum
formed from the all-sky $y$ map of \citet{planck2014-a28}.
Blue, green, and cyan lines represent a set of physically motivated
foregrounds, with red being the best-fit SZ model from that paper.}
\label{fig:sz-map-spectra}
\end{figure}

Figure~\ref{fig:sz-map-spectra} shows power spectra derived from
the all-sky $y$ map for a division into components consisting of clustered
CIB, infrared sources, radio sources, and a thermal SZ model.
We can see that overall a quite reasonable fit is obtained.
One can then use the SZ spectrum to set constraints on $\Omega_{\rm m}$ and
$\sigma_8$, and compare these with the main CMB values, as above.
Again this will depend on assumptions about mass bias, and the result is
best expressed in terms of the combination
$\sigma_8 \left(\Omega_{\rm m}/0.3\right)^{3/8}$.
With a mass bias of $(1-b)=0.8$, a result of
$\sigma_8 \left(\Omega_{\rm m}/0.3\right)^{3/8}=0.78^{+0.01}_{-0.03}$
is obtained, while with $(1-b)=0.6$, the result is
$\sigma_8 \left(\Omega_{\rm m}/0.3\right)^{3/8}=0.86^{+0.01}_{-0.03}$.
For the \Planck\ CMB anisotropy value of $\Omega_{\rm m}\,{=}\,0.3156$
(using TTTEEE+lowP, as in \citealt{planck2014-a36}), the former result
gives $\sigma_8\,{=}\,0.76$, while the the latter gives $\sigma_8\,{=}\,0.86$.
Recently \citet{HorSel17} have re-analysed the thermal SZ power spectrum,
including the effects of feedback and the tri-spectrum contribution to the
uncertainties, finding
$\sigma_8=0.81^{+0.021}_{-0.009}\left(\Omega_{\rm m}/0.3\right)^{0.4}$
when fixing other parameters to their central values.
This is in excellent agreement with the results of the anisotropy analysis.

Another important contribution to cluster physics from \Planck\ has been
work on the average density profile in clusters.
\citet{planck2012-V} showed that by stacking individual clusters, the
resolution and sensitivity of \Planck\ allowed the construction of an
average profile out to a radius of $3\times R_{500}$, giving for the first
time a quantitative description of the thermal pressure distribution in
the outer regions of clusters.
Using fits to a generalized profile, this study showed that the average
pressure profile is slightly flatter than most predictions from numerical
simulations, indicating the need for more detailed modelling of baryonic
physics in cluster outskirts.
The gas fraction values found appeared to converge well to the expected
cosmological value of
$f_{\rm gas}=\Omega_{\rm b} h^2/\Omega_{\rm m} h^2=0.156$
(using values from \citealt{planck2014-a15} for the same TTTEEE+lowP
combination as above).

A further area of SZ studies where \Planck\ has contributed
significantly is the kinetic SZ effect, where peculiar
velocities of material encountered by CMB photons on their way to us result
in a frequency-independent shift of the CMB spectrum to a slightly different
temperature.
In \citet{planck2013-XIII}, searches were carried out for evidence of
large-scale bulk flows in the Universe, including those on very large
scales for which there had previously been some suggestions in the literature,
most prominently a dipole signal in $z\,{\simeq}\,0.1$ clusters claimed by
\citet{Kashlinsky08} and \citet{Atrio13}.
\cite{planck2013-XIII} found tight constraints on such dipoles
by carefully fitting to filtered multi-frequency \Planck\ maps at the
positions of known clusters, constraining bulk flows to be below
$250\,{\rm km}\,{\rm s}^{-1}$ (at 95\,\% confidence).
\cite{planck2013-XIII} also describes how results similar to those seen in
the \citet{Kashlinsky08} and \citet{Atrio13} studies can be
found when performing the same analysis steps, but suggests that this approach
is non-optimal and that the apparent signal should be attributed to
residuals (mostly of CMB origin) in the filtered map rather than to the
clusters' peculiar motion.  \Planck's lack of a kSZ dipole shows that
the Universe is not more inhomogeneous on gigaparsec scales
than expected in the $\Lambda$CDM model.
In addition, the lack of a strong kSZ monopole signal (from outward flows),
provides extremely stringent constraints on those inhomogeneous cosmologies
that attempt to explain the apparent cosmic acceleration by placing us at the
centre of a giant void \citep[e.g., figure~13 in][]{planck2013-XIII}.

In \citet{planck2015-XXXVII}, searches were carried out for the kinetic SZ
effects around the
positions of galaxies from the ``Central Galaxy Catalogue,'' which are
expected to be the central galaxies of their dark-matter halos.
This provided evidence for unbound diffuse gas at twice the
mean virial radius of halos, supporting the idea that the majority of
baryons lie outside this radius; however, the specific correlations found
between SZ and velocity fields suggest that the gas both inside and outside
the central galaxy host halos is comoving with the overall matter flows.

A further statistical use of the \Planck\ data for investigating the kSZ
effect is to determine the excess kSZ variance at the positions of
clusters (\citealt{planck2017-LIII}, see also \citealt{hill2016})
compared to random positions.
Interpreted as a velocity dispersion the result is
$\big\langle v^2\big\rangle=(120\pm70)\times10^3({\rm km}\,{\rm s}^{-1})^2$,
which is consistent with results from other large-scale structure studies
\citep[e.g.,][]{Scrimgeour2016}.

\Planck\ maps do not have the sensitivity or resolution to measure individual
kSZ cluster signals, but nevertheless, the all-sky nature of the \Planck\
observations, coupled with the redshift-independence of the SZ signal, have
enabled statistical insights to be gained on the thermal contents of cluster
gas and the homogeneity of the Universe on large scales.  Future CMB
observations with higher resolution and lower noise will be able to mine even
more information from studies like these.

\subsection{Cosmic infrared background anisotropies} \label{sec:cib}

The high-frequency channels of \Planck\ have enabled very precise
measurements of anisotropy in the cosmic infrared background (CIB).
Discovered in 1996 \citep{Puget_1996}, the CIB is the cumulative far-IR
emission from all galaxies throughout cosmic history, containing an
equal amount of energy as from direct starlight \citep{dwek1998,dole_2006,hill2018} and
implying a considerable amount of star formation in dust-enshrouded galaxies
\citep[e.g.,][]{gispert_2000}.

Since dusty star-forming galaxies trace large-scale structure, one expects
anisotropy in the CIB \citep{Knox_2001}, and these theoretical expectations
were confirmed by early measurements
\citep{Lagache_2000, Matsuhara_2000,Lagache_2007,Grossan_2007,Viero2009}.
Compared to these early detections, \Planck\ (and {\it Herschel\/}) provide
greater sky area, lower systematics, and longer wavelengths (and thus a more
favourable ratio of CIB signal over Galactic dust contamination).
The anisotropy measurements have been presented in
\citet{Amblard_2011}, \citet{planck2011-6.6}, \citet{Viero_2013},
and \citet{planck2013-p05a},
and modelled by \citet{shang_2012}, \citet{bethermin_2013},
\citet{thacker_2013}, and \citet{Maniyar_2018}.  The models imply that the mass
of the ``typical'' dark matter halo contributing to the CIB at $z\,{=}\,2$
is $\log(M_{\rm h}/{\rm M}_{\odot})=12.77_{-0.125}^{+0.128}$
\citep{Maniyar_2018}.  Such modelling predicts that dusty star-forming galaxies
at high redshift are highly biased.
The cosmic abundance of dust is $\Omega_{\rm dust}=(1{-}8)\times 10^{-6}$
for $z\,{\simeq}\,0$--3 \citep{thacker_2013,schmidt_2015}.
This implies that the dust-to-stellar-mass ratio increases from about
0.2\,\% at $z\,{=}\,0$ to 1\,\% at $z\,{\simeq}\,2$.
The modelling of \citet{Maniyar_2018} implies that obscured star formation
dominates unobscured up to at least $z\,{=}\,4$, with obscured and
unobscured contributions becoming comparable at $z\,{=}\,5$.

As described in Sect.~\ref{sec:lensingspectra}, the large-scale structure
traced by dusty galaxies lenses the primary CMB anisotropies.  Since the CIB
probes the structure at intermediate redshift, the two are highly correlated
\citep{Song_2003,Holder_2013,planck2013-p13,planck2016-l08}. 
In \citet{planck2016-l08} we present a joint analysis of lensing reconstruction
and the CIB, with the latter providing our best current picture of the lensing
modes on small scales.

\subsection{Reionization} \label{sec:reionization}

The CMB spectra of Fig.~\ref{fig:cl-world} provide the critical context for our
understanding of reionization.  The presence of a series of acoustic peaks in
the angular power spectra of the CMB indicates that the Universe was dense
and ionized at early times and then  underwent a rapid transition to being
(largely) neutral at $z\,{\simeq}\,1100$.  This neutral period lasted for
a significant time.  Had this transition not occurred, or lasted only a short
time, multiple scatterings would have erased the anisotropies on scales
smaller than the Hubble scale \citep[e.g.,][]{Efs88}.
The presence of an enhancement in the $E$-mode power spectrum at low $\ell$
indicates that the Universe became ionized again at $z\,{\simeq}\,10$.
This second transition is known as ``reionization,'' and is often referred
to as the end of the dark ages.

The picture described above is consistent with numerous observations
(see, e.g., \citealt{Furlanetto06}, \citealt{Becker15}, \citealt{Bouwens15}, \citealt{McQuinn16}, and \citealt{Mes16} for reviews)
which can be used to constrain the sources of reionization and the
manner in which the process occurred.
By providing a measurement of the (integrated) optical depth to Thomson
scattering, $\tau$, and constraints on the kinetic SZ effect, the CMB can
provide limits on the epoch and duration of the reionization process that
are highly complementary to those obtained from other probes
\citep{planck2014-a25}.

In currently popular models, ultraviolet photons from massive stars in
relatively low mass early galaxies reionize hydrogen progressively
throughout the Universe between $z\,{\simeq}\,12$ and $z\,{\simeq}\,6$,
while quasars take over to reionize helium from $z\,{\simeq}\,6$ to
$z\,{\simeq}\,2$.  The combination of measurements indirectly constrains the
nature of the sources driving reionization, and hence the formation of
early stars and galaxies.
The current observations point towards a ``late and fast'' reionization
period, though with considerable uncertainty.

\begin{figure}[htbp]
\begin{center}
\resizebox{\columnwidth}{!}{\includegraphics{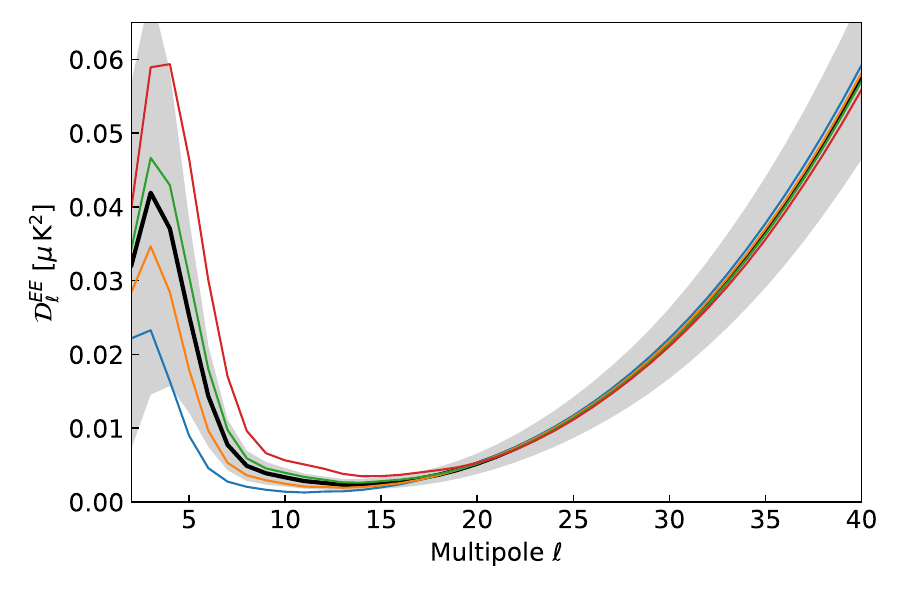}}
\end{center}
\caption{The polarization angular power spectrum, $\mathcal{D}_\ell^{EE}$,
for different optical depths, $\tau$, running from $\tau=0.04$ to $0.07$
in steps of 0.01.  The thick black line shows the fiducial value
$\tau=0.056$, while the grey shading shows the $\pm1\,\sigma$ sample variance
band for $f_{\rm sky}=0.67$.
\label{fig:cl_tau} }
\end{figure}

The amplitude of the large-scale anisotropies in polarization is particularly
sensitive to the value of $\tau$
($C_\ell^{EE}\propto\tau^2$; Fig.~\ref{fig:cl_tau}),
with the shape of the low-$\ell$ bump encoding information about how the
Universe reionized.
This measurement is very demanding, since the expected level of the $E$-mode
polarization power spectrum at low multipoles ($\ell\,{\la}\,10$) is
only a few times $10^{-2}\muK^2$, lower by more than two orders
of magnitude than the level of the temperature anisotropy power spectrum.
For such weak signals, the difficulty is not only to have enough detector
sensitivity, but also to reduce and control both instrumental systematic
effects and foreground residuals to very low levels.
Our best estimate \citep{planck2016-l06} is
\begin{eqnarray}
  \tau &=& n_{\rm H}(0)\,\sigma_{\rm T}\,c
  \int_0^{z_{\rm max}}\frac{(1+z)^2\,dz}{H(z)}\ x_{\rm e}(z) \\
       &=& 0.056\pm 0.007 ,
\label{eqn:tau_value}
\end{eqnarray}
where $\sigma_{\rm T}$ is the Thomson scattering cross-section,
$n_{\rm H}(0)$ is the number density of hydrogen nuclei today and
$x_{\rm e}$ is the ionized fraction.\footnote{To be more specific,
this neglects the residual $x_{\rm e}$ from recombination, and includes singly
ionized helium.  In principle, $\tau$ is a mass-weighted quantity, whereas the
porosity often used in reionization studies is a volume-weighted quantity.
For a homogeneous Universe the distinction is irrelevant, but it could be
important at $z\,{\simeq}\,6$--10 when structure is well developed.
Nevertheless, the distinction is not relevant for inferences based on
low-$\ell$ CMB anisotropy.} At low $\tau$ the measurement becomes very
difficult.  Indeed for the low values of $\tau$ currently favoured, the
CMB cannot give tight constraints on details of the reionization process,
although early reionization models are disfavoured.

On smaller scales, reionization generates CMB temperature anisotropies
through the kinetic Sunyaev-Zeldovich (kSZ) effect, that is, the Doppler shift
of photons scattering off electrons moving with bulk velocities. 
Currently we have only upper limits on the kSZ effect arising from the Universe
during reionization, which suggest that reionization happened relatively
quickly.

Given the extreme difficulty of the measurement, and the trend of measured
$\tau$ values to drop with time (Fig.~\ref{fig:params_lateU}) as measurement uncertainties have decreased, it is
encouraging that \Planck\ provides another channel for constraining $\tau$.
Though more model dependent, the lensing of the CMB provides an independent,
consistent measurement of $\tau$.  Within $\Lambda$CDM the peak of the
$\tau$ posterior from lensing peaks at slightly higher values than
Eq.~(\ref{eqn:tau_value}), but is consistent at the $1.4\,\sigma$ level.

\begin{figure}[htbp]
\begin{center}
\resizebox{\columnwidth}{!}{\includegraphics{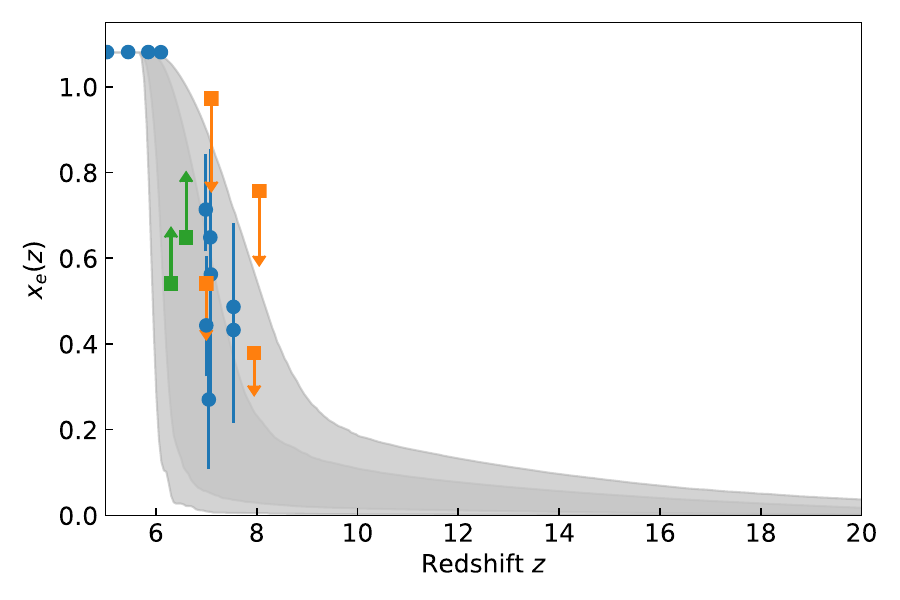}}
\end{center}
\caption{Free electron fraction, $x_{\rm e}(z)$, constrained from \Planck
\ using the ``FlexKnots'' method of \citet{planck2016-l06} and
\citet{2018arXiv180408476M}, along with astrophysical constraints, as tabulated
in \citet{Bouwens15}, updated to include \citet{Greig17}, \citet{Banados18},
\citet{Mason18}, and \citet{Davies18}.
\label{fig:reion_history} }
\end{figure}

The latest results confirm that reionization occurred rather late, leaving
little room for any significant ionization at high redshift
(the optical depth from $z\,{>}\,15$ is less than 1\,\%;
\citealt{planck2016-l06}).
This is consistent with suggestions from other probes
\citep{Becker15,Bouwens15,McQuinn16,Mes16},
as shown in Fig.~\ref{fig:reion_history}. 
The \Planck\ results strongly reduce the need for a significant contribution
of Lyman continuum emission at early times.  Non-standard early galaxies or
significantly evolving escape and clumping factors are no longer required,
nor do the \Planck\ results require any emission from high-redshift
($z=10$--15) galaxies.

The lower optical depth measured by \Planck, in concert with the rapidly
declining abundance of bright galaxies measured in UV luminosity functions
at high redshift \citep[e.g.,][]{Oesch18} is consistent with this simple,
galaxy-driven scenario.  Indeed an extrapolation of the measured UV LFs
to galaxies in halos above the atomic cooling threshold 
($T_{\rm vir}\,{>}\,10^4\,$K) provides enough photons to reionize the Universe
by $z\,{\simeq}\,6$ if the escape fraction of ionizing photons is
$f_{\rm esc}\,{\simeq}\,0.1$ \citep{Bouwens15}.

Measurements of redshifted 21-cm radiation promise to provide a
complementary view to the one provided by the intergalactic medium, galaxy,
and Thomson-scattering constraints.  The recent, claimed detection
of a larger-than-expected feature in the sky-averaged (that is, global) 21-cm
signal by EDGES \citep{Bowman18} would require a colder IGM than most models
predict, or some other change in the conditions at high $z$.  Since many
mechanisms for explaining the signal would also generate some ionization at
high $z$, the low optical depth measured by \Planck\ serves to limit
candidates.  Future observations of this global signal, and the fluctuations
in the background, will be able to shed further light on this apparent
discrepancy, and provide constraints that are complementary to those
coming from the CMB.

\section{Post-\Planck\ landscape}

\Planck\ was designed to measure the CMB temperature anisotropies to
fundamental limits over the range of scales defined by the Hubble radius and
the diffusion damping scale.  It achieved this through a combination of
instrument design, experimental optimization, sophisticated and iterative
instrument modelling, and analysis.  
\Planck\ validated the standard cosmological model ($\Lambda$CDM) and
provided simultaneous, precise measurements of its key parameters, as well as
tight constraints on possible extensions.
However, multiple questions remain.  For example:
\begin{itemize}
\item What is the mechanism for the generation of fluctuations
      in the early Universe?
\item If it is inflation, as we suspect, what is the inflaton,
      what determines the initial state, and how does inflation end?
\item How did baryogenesis occur and why is $\omega_{\rm b}\sim\omega_{\rm c}$?
\item What is the dark matter?
\item Are there additional, light, relic particles?
\item What is causing the accelerated expansion of the Universe today?
\item How did the Universe reionize?
\item How do astrophysical objects form and evolve in the cosmic web?
\end{itemize}

Absent a breakthrough in our theoretical understanding, the route forward on
all of these questions is improved measurements, in which further observations
of CMB anisotropies will play a key role.  Advances in detector technology
and in processing power will enable much higher-sensitivity observations of
the CMB than \Planck\ was able to provide.  Given that \Planck\ has effectively
mined the information in the primary temperature anisotropies, the focus of
CMB research is now shifting to studies of polarization and secondary effects
such as CMB lensing.

Many of the lessons learned\footnote{A report may be found at
{\tt https://www.cosmos.esa.int/} {\tt web/planck/lessons-learned}.} from
\Planck\ remain relevant in this post-\Planck\ landscape.
Wide frequency coverage and excellent control of systematic effects are
prerequisites, but these must be coupled with a thorough modelling of the
instrument, detailed simulations, and a sophisticated and efficient analysis
pipeline.
The \Planck\ experience was that detailed simulations were invaluable, but
required an enormous effort and needed attention from an early stage.
Redundant methods for critical steps, including reconciling areas of
disagreement, was important for verification of the final results.
The understanding of the data and the extraction of the core science were
closely intertwined, requiring large, integrated data-processing pipelines.
Calibration, mapmaking, component separation, and analysis needed to be
treated as a single, tightly-coupled problem.

The \Planck\ 2018 papers, including this one, represent the final word from
the \Planck\ collaboration, but do not mark the end of developments of
\Planck\ products.  The activity leading to this release was circumscribed
by time and funding constraints, not by perfection of the data products.
We expect that contributions from individuals, both within and external to the
\Planck\ collaboration, will continue to build upon the \Planck\ legacy
and that the \Planck\ data will prove invaluable for a wide range of future
cosmological studies.

\section{Conclusions} \label{sec:conclusions}

\Planck\ was the third-generation space mission dedicated to measurement of
CMB anisotropies.  It delivered on its promise to provide a measurement of the
primary CMB temperature anisotropies between the Hubble scale and the damping
scale to fundamental limits, and provides some of our most important
constraints on models of cosmology and fundamental physics.
The \Planck\ temperature auto-spectrum is cosmic variance limited to
$\ell\simeq 1600$.  
In this respect, \Planck\ has ended a phase in primary temperature
anisotropy studies that was opened by \COBE\ in 1992.  

The study of the CMB has been central to the story of cosmology since
its discovery provided some of the earliest evidence for the hot-big-bang
model of an expanding Universe.
Building upon a legacy of earlier experiments and decades of theoretical
development, \Planck\ has now measured the properties of the Universe to
percent-level fidelity and tested our cosmological model to high precision.
\Planck\ data provide the strongest evidence we have that dark matter cannot
be entirely baryonic (luminous or dark) and that the observed fluctuations
were laid down at very early times, proportionally in all of the constituents
of the Universe.

One of the major scientific legacies of \Planck\ has been to test and highly
constrain the $\Lambda$CDM model.
The inflationary $\Lambda$CDM model was first developed in the 1980s
\citep{Peebles84,VittorioSilk85} and rose to prominence during the 1990s
\citep{Efstathiou90,OstSte95,KraTur95,LLVW96}.
The discovery of the accelerated expansion of the Universe at the end
of that decade \citep{Riess98,Perlmutter99}, and the measurement of the
acoustic peaks soon after \citep{deBernardis00,Balbi00,Miller02},
cemented it as the ``standard model'' of cosmology.
In the three decades since it was first developed, the quality and quantity
of cosmological data have exploded.  The model has weathered all of these
challenges, and remains our best (phenomenological) description of the Universe.

The \Planck\ data have been particularly important in this regard.  The
$\Lambda$CDM model, with parameters fixed by pre-\Planck\ experiments, made
solid predictions for the range of anisotropy spectra that \Planck\ would
measure.  The dramatic improvement in angular resolution, sensitivity, foreground
cleaning, and systematics control from the \Planck\ data provide the most
stringent test of the model yet, and have allowed us to measure its parameters
to high precision (Table~\ref{tab:LCDMparams}).

A network of tests established the consistency of the measurements and
enabled many different tests of the model.
The small amplitude fluctuations traced by \Planck\ at $z\,{\simeq}\,1100$, the
gravitational potentials traced by \Planck\ lensing at $z\,{\simeq}\,0.5$--10,
and the matter fluctuations probed by large-scale structure surveys at $z<1$
are part of a consistent picture.

The flatness of the spatial hypersurfaces has been established at the
$5\times 10^{-3}$ level.  Neutrino masses have been constrained to be
$\mathcal{O}(0.1\,{\rm eV})$.  The number of relativistic species is
consistent with three light neutrinos and disfavours any light, thermal
relics that froze out after the QCD phase transition.
The baryon density inferred from the acoustic oscillations up to
$t=400\,000\,$yr is consistent with that inferred from BBN at $t=3\,$min.
Dark-matter annihilations are tightly constrained.
Dark energy is consistent with being a cosmological constant that dominates
only recently.

The pattern of acoustic oscillations in temperature and polarization power
spectra implies an early-Universe origin for the fluctuations, as in the
inflationary framework.  The primordial fluctuations are Gaussian to an
exceptional degree.
There are no gravitational waves at the 5\,\% level, suggesting the
energy scale of an inflationary epoch was below the Planck scale.

The ability of the $\Lambda$CDM model to explain the \Planck\ data,
and a wealth of other astrophysical observations, indicates that our
understanding of physics is good enough to model $14\,$Gyr of cosmic
history and explain observations out to the edges of the observable Universe.
However, the surprising ingredients required by this model suggest that our
understanding is highly incomplete in several areas.

Despite these successes, some puzzling tensions and open question remain.
While many measures of the matter perturbations at low redshift are in
excellent agreement with the predictions of \LCDM\ fit to the \Planck\ data,
this is not true of all of them.  In particular, measurements of the
fluctuation amplitude from cosmic shear tend to lie low compared to the
\Planck\ predictions.  Measures of the distance scale from nearby Type Ia SNe
remain discordant with the inferences from the inverse distance ladder.
We expect these areas will see continued attention from the community,
which will determine whether these tensions point to statistical fluctuations,
misestimated systematic uncertainties, or physics beyond \LCDM.

\paragraph{Acknowledgements}
\noindent The development of \Planck\ has been supported by: ESA; CNES and
CNRS/INSU-IN2P3-INP (France); ASI, CNR, and INAF (Italy); NASA and DoE (USA);
STFC and UKSA (UK); CSIC, MICINN, JA, and RES (Spain); Tekes, AoF, and CSC
(Finland); DLR and MPG (Germany); CSA (Canada); DTU Space (Denmark); SER/SSO
(Switzerland); RCN (Norway); SFI (Ireland); FCT/MCTES (Portugal); and PRACE
(EU). A description of the \Planck\ Collaboration and a list of its members,
including the technical or scientific activities in which they have been
involved, can be found at  \href{http://www.cosmos.esa.int/web/planck/planck-collaboration}{\texttt{http://www.cosmos.esa.int/web/planck/ planck-collaboration}}. In addition, we thank Sol\`{e}ne Chabanier and Nathalie Palanque-Delabrouille for computing the Ly$\,\alpha$ forest constraints we have used in Fig.~\ref{fig:allpk} and Inigo Zubeldia for preparing Fig.~\ref{fig:sz-cosmo}.


\bibliographystyle{aat}
\bibliography{Planck_bib,ms}

\def\eprinttmppp@#1arXiv:@{#1}
\providecommand{\arxivlink[1]}{\href{http://arxiv.org/abs/#1}{arXiv:#1}}
\def\eprinttmp@#1arXiv:#2 [#3]#4@{\ifthenelse{\equal{#3}{x}}{\ifthenelse{
\equal{#1}{}}{\arxivlink{\eprinttmppp@#2@}}{\arxivlink{#1}}}{\arxivlink{#2}
  [#3]}}
\providecommand{\eprintlink}[1]{\eprinttmp@#1arXiv: [x]@}
\providecommand{\eprint}[1]{\eprintlink{#1}}
\providecommand{\adsurl}[1]{\href{#1}{ADS}}
\begin{thebibliography}{370}
\expandafter\ifx\csname natexlab\endcsname\relax\def\natexlab#1{#1}\fi

\bibitem[{Abbott {et~al.}(2016)Abbott, Abbott, Abbott, Abernathy, Acernese,
  Ackley, Adams, Adams, Addesso, Adhikari, Adya, Affeldt, Agathos, Agatsuma,
  Aggarwal, Aguiar, Aiello, Ain, Ajith, Allen, Allocca, Altin, Anderson,
  Anderson, Arai, Arain, Araya, Arceneaux, Areeda, Arnaud, Arun, Ascenzi,
  Ashton, Ast, Aston, Astone, Aufmuth, Aulbert, Babak, Bacon, Bader, Baker,
  Baldaccini, Ballardin, Ballmer, Barayoga, Barclay, Barish, Barker, Barone,
  Barr, Barsotti, Barsuglia, Barta, Bartlett, Barton, Bartos, Bassiri, Basti,
  Batch, Baune, Bavigadda, Bazzan, Behnke, Bejger, Belczynski, Bell, Bell,
  Berger, Bergman, Bergmann, Berry, Bersanetti, Bertolini, Betzwieser, Bhagwat,
  Bhandare, Bilenko, Billingsley, Birch, Birney, Birnholtz, Biscans, Bisht,
  Bitossi, Biwer, Bizouard, Blackburn, Blair, Blair, Blair, Bloemen, Bock,
  Bodiya, Boer, Bogaert, Bogan, Bohe, Bojtos, Bond, Bondu, Bonnand, Boom, Bork,
  Boschi, Bose, Bouffanais, Bozzi, Bradaschia, Brady, Braginsky, Branchesi,
  Brau, Briant, Brillet, Brinkmann, Brisson, Brockill, Brooks, Brown, Brown,
  Brown, Buchanan, Buikema, Bulik, Bulten, Buonanno, Buskulic, Buy, Byer,
  Cabero, Cadonati, Cagnoli, Cahillane, Bustillo, Callister, Calloni, Camp,
  Cannon, Cao, Capano, Capocasa, Carbognani, Caride, Diaz, Casentini, Caudill,
  Cavagli\`a, Cavalier, Cavalieri, Cella, Cepeda, Baiardi, Cerretani, Cesarini,
  Chakraborty, Chalermsongsak, Chamberlin, Chan, Chao, Charlton,
  Chassande-Mottin, Chen, Chen, Cheng, Chincarini, Chiummo, Cho, Cho, Chow,
  Christensen, Chu, Chua, Chung, Ciani, Clara, Clark, Cleva, Coccia, Cohadon,
  Colla, Collette, Cominsky, Constancio, Conte, Conti, Cook, Corbitt, Cornish,
  Corsi, Cortese, Costa, Coughlin, Coughlin, Coulon, Countryman, Couvares,
  Cowan, Coward, Cowart, Coyne, Coyne, Craig, Creighton, Creighton, Cripe,
  Crowder, Cruise, Cumming, Cunningham, Cuoco, Canton, Danilishin, D'Antonio,
  Danzmann, Darman, Da~Silva~Costa, Dattilo, Dave, Daveloza, Davier, Davies,
  Daw, Day, De, DeBra, Debreczeni, Degallaix, De~Laurentis, Del\'eglise,
  Del~Pozzo, Denker, Dent, Dereli, Dergachev, DeRosa, De~Rosa, DeSalvo,
  Dhurandhar, D\'{\i}az, Di~Fiore, Di~Giovanni, Di~Lieto, Di~Pace, Di~Palma,
  Di~Virgilio, Dojcinoski, Dolique, Donovan, Dooley, Doravari, Douglas, Downes,
  Drago, Drever, Driggers, Du, Ducrot, Dwyer, Edo, Edwards, Effler, Eggenstein,
  Ehrens, Eichholz, Eikenberry, Engels, Essick, Etzel, Evans, Evans, Everett,
  Factourovich, Fafone, Fair, Fairhurst, Fan, Fang, Farinon, Farr, Farr,
  Favata, Fays, Fehrmann, Fejer, Feldbaum, Ferrante, Ferreira, Ferrini,
  Fidecaro, Finn, Fiori, Fiorucci, Fisher, Flaminio, Fletcher, Fong, Fournier,
  Franco, Frasca, Frasconi, Frede, Frei, Freise, Frey, Frey, Fricke, Fritschel,
  Frolov, Fulda, Fyffe, Gabbard, Gair, Gammaitoni, Gaonkar, Garufi, Gatto,
  Gaur, Gehrels, Gemme, Gendre, Genin, Gennai, George, Gergely, Germain, Ghosh,
  Ghosh, Ghosh, Giaime, Giardina, Giazotto, Gill, Glaefke, Gleason, Goetz,
  Goetz, Gondan, Gonz\'alez, Castro, Gopakumar, Gordon, Gorodetsky, Gossan,
  Gosselin, Gouaty, Graef, Graff, Granata, Grant, Gras, Gray, Greco, Green,
  Greenhalgh, Groot, Grote, Grunewald, Guidi, Guo, Gupta, Gupta, Gushwa,
  Gustafson, Gustafson, Hacker, Hall, Hall, Hammond, Haney, Hanke, Hanks,
  Hanna, Hannam, Hanson, Hardwick, Harms, Harry, Harry, Hart, Hartman, Haster,
  Haughian, Healy, Heefner, Heidmann, Heintze, Heinzel, Heitmann, Hello,
  Hemming, Hendry, Heng, Hennig, Heptonstall, Heurs, Hild, Hoak, Hodge, Hofman,
  Hollitt, Holt, Holz, Hopkins, Hosken, Hough, Houston, Howell, Hu, Huang,
  Huerta, Huet, Hughey, Husa, Huttner, Huynh-Dinh, Idrisy, Indik, Ingram, Inta,
  Isa, Isac, Isi, Islas, Isogai, Iyer, Izumi, Jacobson, Jacqmin, Jang, Jani,
  Jaranowski, Jawahar, Jim\'enez-Forteza, Johnson, Johnson-McDaniel, Jones,
  Jones, Jonker, Ju, Haris, Kalaghatgi, Kalogera, Kandhasamy, Kang, Kanner,
  Karki, Kasprzack, Katsavounidis, Katzman, Kaufer, Kaur, Kawabe, Kawazoe,
  K\'ef\'elian, Kehl, Keitel, Kelley, Kells, Kennedy, Keppel, Key,
  Khalaidovski, Khalili, Khan, Khan, Khan, Khazanov, Kijbunchoo, Kim, Kim, Kim,
  Kim, Kim, Kim, King, King, Kinzel, Kissel, Kleybolte, Klimenko, Koehlenbeck,
  Kokeyama, Koley, Kondrashov, Kontos, Koranda, Korobko, Korth, Kowalska,
  Kozak, Kringel, Krishnan, Kr\'olak, Krueger, Kuehn, Kumar, Kumar, Kuo,
  Kutynia, Kwee, Lackey, Landry, Lange, Lantz, Lasky, Lazzarini, Lazzaro,
  Leaci, Leavey, Lebigot, Lee, Lee, Lee, Lee, Lenon, Leonardi, Leong, Leroy,
  Letendre, Levin, Levine, Li, Libson, Littenberg, Lockerbie, Logue, Lombardi,
  London, Lord, Lorenzini, Loriette, Lormand, Losurdo, Lough, Lousto, Lovelace,
  L\"uck, Lundgren, Luo, Lynch, Ma, MacDonald, Machenschalk, MacInnis, Macleod,
  Maga\~na Sandoval, Magee, Mageswaran, Majorana, Maksimovic, Malvezzi, Man,
  Mandel, Mandic, Mangano, Mansell, Manske, Mantovani, Marchesoni, Marion,
  M\'arka, M\'arka, Markosyan, Maros, Martelli, Martellini, Martin, Martin,
  Martynov, Marx, Mason, Masserot, Massinger, Masso-Reid, Matichard, Matone,
  Mavalvala, Mazumder, Mazzolo, McCarthy, McClelland, McCormick, McGuire,
  McIntyre, McIver, McManus, McWilliams, Meacher, Meadors, Meidam, Melatos,
  Mendell, Mendoza-Gandara, Mercer, Merilh, Merzougui, Meshkov, Messenger,
  Messick, Meyers, Mezzani, Miao, Michel, Middleton, Mikhailov, Milano, Miller,
  Millhouse, Minenkov, Ming, Mirshekari, Mishra, Mitra, Mitrofanov,
  Mitselmakher, Mittleman, Moggi, Mohan, Mohapatra, Montani, Moore, Moore,
  Moraru, Moreno, Morriss, Mossavi, Mours, Mow-Lowry, Mueller, Mueller, Muir,
  Mukherjee, Mukherjee, Mukherjee, Mukund, Mullavey, Munch, Murphy, Murray,
  Mytidis, Nardecchia, Naticchioni, Nayak, Necula, Nedkova, Nelemans, Neri,
  Neunzert, Newton, Nguyen, Nielsen, Nissanke, Nitz, Nocera, Nolting,
  Normandin, Nuttall, Oberling, Ochsner, O'Dell, Oelker, Ogin, Oh, Oh, Ohme,
  Oliver, Oppermann, Oram, O'Reilly, O'Shaughnessy, Ott, Ottaway, Ottens,
  Overmier, Owen, Pai, Pai, Palamos, Palashov, Palomba, Pal-Singh, Pan, Pan,
  Pankow, Pannarale, Pant, Paoletti, Paoli, Papa, Paris, Parker, Pascucci,
  Pasqualetti, Passaquieti, Passuello, Patricelli, Patrick, Pearlstone,
  Pedraza, Pedurand, Pekowsky, Pele, Penn, Perreca, Pfeiffer, Phelps, Piccinni,
  Pichot, Pickenpack, Piergiovanni, Pierro, Pillant, Pinard, Pinto, Pitkin,
  Poeld, Poggiani, Popolizio, Post, Powell, Prasad, Predoi, Premachandra,
  Prestegard, Price, Prijatelj, Principe, Privitera, Prix, Prodi, Prokhorov,
  Puncken, Punturo, Puppo, P\"urrer, Qi, Qin, Quetschke, Quintero,
  Quitzow-James, Raab, Rabeling, Radkins, Raffai, Raja, Rakhmanov, Ramet,
  Rapagnani, Raymond, Razzano, Re, Read, Reed, Regimbau, Rei, Reid, Reitze,
  Rew, Reyes, Ricci, Riles, Robertson, Robie, Robinet, Rocchi, Rolland,
  Rollins, Roma, Romano, Romano, Romanov, Romie, Rosi\ifmmode~\acute{n}\else
  \'{n}\fi{}ska, Rowan, R\"udiger, Ruggi, Ryan, Sachdev, Sadecki, Sadeghian,
  Salconi, Saleem, Salemi, Samajdar, Sammut, Sampson, Sanchez, Sandberg,
  Sandeen, Sanders, Sanders, Sassolas, Sathyaprakash, Saulson, Sauter, Savage,
  Sawadsky, Schale, Schilling, Schmidt, Schmidt, Schnabel, Schofield,
  Sch\"onbeck, Schreiber, Schuette, Schutz, Scott, Scott, Sellers, Sengupta,
  Sentenac, Sequino, Sergeev, Serna, Setyawati, Sevigny, Shaddock, Shaffer,
  Shah, Shahriar, Shaltev, Shao, Shapiro, Shawhan, Sheperd, Shoemaker,
  Shoemaker, Siellez, Siemens, Sigg, Silva, Simakov, Singer, Singer, Singh,
  Singh, Singhal, Sintes, Slagmolen, Smith, Smith, Smith, Smith, Son, Sorazu,
  Sorrentino, Souradeep, Srivastava, Staley, Steinke, Steinlechner,
  Steinlechner, Steinmeyer, Stephens, Stevenson, Stone, Strain, Straniero,
  Stratta, Strauss, Strigin, Sturani, Stuver, Summerscales, Sun, Sutton,
  Swinkels, Szczepa\ifmmode~\acute{n}\else \'{n}\fi{}czyk, Tacca, Talukder,
  Tanner, T\'apai, Tarabrin, Taracchini, Taylor, Theeg, Thirugnanasambandam,
  Thomas, Thomas, Thomas, Thorne, Thorne, Thrane, Tiwari, Tiwari, Tokmakov,
  Tomlinson, Tonelli, Torres, Torrie, T\"oyr\"a, Travasso, Traylor, Trifir\`o,
  Tringali, Trozzo, Tse, Turconi, Tuyenbayev, Ugolini, Unnikrishnan, Urban,
  Usman, Vahlbruch, Vajente, Valdes, Vallisneri, van Bakel, van Beuzekom,
  van~den Brand, Van Den~Broeck, Vander-Hyde, van~der Schaaf, van Heijningen,
  van Veggel, Vardaro, Vass, Vas\'uth, Vaulin, Vecchio, Vedovato, Veitch,
  Veitch, Venkateswara, Verkindt, Vetrano, Vicer\'e, Vinciguerra, Vine, Vinet,
  Vitale, Vo, Vocca, Vorvick, Voss, Vousden, Vyatchanin, Wade, Wade, Wade,
  Waldman, Walker, Wallace, Walsh, Wang, Wang, Wang, Wang, Wang, Ward, Ward,
  Warner, Was, Weaver, Wei, Weinert, Weinstein, Weiss, Welborn, Wen,
  We\ss{}els, Westphal, Wette, Whelan, Whitcomb, White, Whiting, Wiesner,
  Wilkinson, Willems, Williams, Williams, Williamson, Willis, Willke, Wimmer,
  Winkelmann, Winkler, Wipf, Wiseman, Wittel, Woan, Worden, Wright, Wu, Yablon,
  Yakushin, Yam, Yamamoto, Yancey, Yap, Yu, Yvert, Zadro\ifmmode~\dot{z}\else
  \.{z}\fi{}ny, Zangrando, Zanolin, Zendri, Zevin, Zhang, Zhang, Zhang, Zhang,
  Zhao, Zhou, Zhou, Zhu, Zucker, Zuraw, \& Zweizig}]{LIGO}
Abbott, B.~P., Abbott, R., Abbott, T.~D., {et~al.}, Observation of
  Gravitational Waves from a Binary Black Hole Merger. 2016, \prl, 116, 061102

\bibitem[{{Abbott} {et~al.}(2017{\natexlab{a}}){Abbott}, {Abbott}, {Abbott},
  {Acernese}, {Ackley}, {Adams}, {Adams}, {Addesso}, {Adhikari}, {Adya}, \&
  et~al.}]{StandardSiren}
{Abbott}, B.~P., {Abbott}, R., {Abbott}, T.~D., {et~al.}, {A gravitational-wave
  standard siren measurement of the Hubble constant}. 2017{\natexlab{a}}, \nat,
  551, 85, \eprint{1710.05835}

\bibitem[{{Abbott} {et~al.}(2017{\natexlab{b}}){Abbott}, {Abbott}, {Abbott},
  {Acernese}, {Ackley}, {Adams}, {Adams}, {Addesso}, {Adhikari}, {Adya}, \&
  et~al.}]{MultiMessenger}
{Abbott}, B.~P., {Abbott}, R., {Abbott}, T.~D., {et~al.}, {Multi-messenger
  Observations of a Binary Neutron Star Merger}. 2017{\natexlab{b}}, \apjl,
  848, L12, \eprint{1710.05833}

\bibitem[{{Alam} {et~al.}(2017){Alam}, {Ata}, {Bailey}, {Beutler}, {Bizyaev},
  {Blazek}, {Bolton}, {Brownstein}, {Burden}, {Chuang}, {Comparat}, {Cuesta},
  {Dawson}, {Eisenstein}, {Escoffier}, {Gil-Mar{\'{\i}}n}, {Grieb}, {Hand},
  {Ho}, {Kinemuchi}, {Kirkby}, {Kitaura}, {Malanushenko}, {Malanushenko},
  {Maraston}, {McBride}, {Nichol}, {Olmstead}, {Oravetz}, {Padmanabhan},
  {Palanque-Delabrouille}, {Pan}, {Pellejero-Ibanez}, {Percival}, {Petitjean},
  {Prada}, {Price-Whelan}, {Reid}, {Rodr{\'{\i}}guez-Torres}, {Roe}, {Ross},
  {Ross}, {Rossi}, {Rubi{\~n}o-Mart{\'{\i}}n}, {Saito}, {Salazar-Albornoz},
  {Samushia}, {S{\'a}nchez}, {Satpathy}, {Schlegel}, {Schneider},
  {Sc{\'o}ccola}, {Seo}, {Sheldon}, {Simmons}, {Slosar}, {Strauss}, {Swanson},
  {Thomas}, {Tinker}, {Tojeiro}, {Maga{\~n}a}, {Vazquez}, {Verde}, {Wake},
  {Wang}, {Weinberg}, {White}, {Wood-Vasey}, {Y{\`e}che}, {Zehavi}, {Zhai}, \&
  {Zhao}}]{BOSS-DR12}
{Alam}, S., {Ata}, M., {Bailey}, S., {et~al.}, {The clustering of galaxies in
  the completed SDSS-III Baryon Oscillation Spectroscopic Survey: cosmological
  analysis of the DR12 galaxy sample}. 2017, \mnras, 470, 2617,
  \eprint{1607.03155}

\bibitem[{{Amblard} {et~al.}(2011){Amblard}, {Cooray}, {Serra}, {Altieri},
  {Arumugam}, {Aussel}, {Blain}, {Bock}, {Boselli}, {Buat},
  {Castro-Rodr{\'{\i}}guez}, {Cava}, {Chanial}, {Chapin}, {Clements}, {Conley},
  {Conversi}, {Dowell}, {Dwek}, {Eales}, {Elbaz}, {Farrah}, {Franceschini},
  {Gear}, {Glenn}, {Griffin}, {Halpern}, {Hatziminaoglou}, {Ibar}, {Isaak},
  {Ivison}, {Khostovan}, {Lagache}, {Levenson}, {Lu}, {Madden}, {Maffei},
  {Mainetti}, {Marchetti}, {Marsden}, {Mitchell-Wynne}, {Nguyen}, {O'Halloran},
  {Oliver}, {Omont}, {Page}, {Panuzzo}, {Papageorgiou}, {Pearson},
  {P{\'e}rez-Fournon}, {Pohlen}, {Rangwala}, {Roseboom}, {Rowan-Robinson},
  {Portal}, {Schulz}, {Scott}, {Seymour}, {Shupe}, {Smith}, {Stevens},
  {Symeonidis}, {Trichas}, {Tugwell}, {Vaccari}, {Valiante}, {Valtchanov},
  {Vieira}, {Vigroux}, {Wang}, {Ward}, {Wright}, {Xu}, \&
  {Zemcov}}]{Amblard_2011}
{Amblard}, A., {Cooray}, A., {Serra}, P., {et~al.}, {Submillimetre galaxies
  reside in dark matter haloes with masses greater than $3{\times}10^{11}$
  solar masses}. 2011, \nat, 470, 510, \eprint{1101.1080}

\bibitem[{Amendola {et~al.}(2018{\natexlab{a}})Amendola, Appleby, Avgoustidis,
  Bacon, Baker, Baldi, Bartolo, Blanchard, Bonvin, Borgani, Branchini, Burrage,
  Camera, Carbone, Casarini, Cropper, de~Rham, Dietrich, Di~Porto, Durrer,
  Ealet, Ferreira, Finelli, Garc{\'i}a-Bellido, Giannantonio, Guzzo, Heavens,
  Heisenberg, Heymans, Hoekstra, Hollenstein, Holmes, Hwang, Jahnke, Kitching,
  Koivisto, Kunz, La~Vacca, Linder, March, Marra, Martins, Majerotto, Markovic,
  Marsh, Marulli, Massey, Mellier, Montanari, Mota, Nunes, Percival, Pettorino,
  Porciani, Quercellini, Read, Rinaldi, Sapone, Sawicki, Scaramella, Skordis,
  Simpson, Taylor, Thomas, Trotta, Verde, Vernizzi, Vollmer, Wang, Weller, \&
  Zlosnik}]{Amendola18}
Amendola, L., Appleby, S., Avgoustidis, A., {et~al.}, Cosmology and fundamental
  physics with the Euclid satellite. 2018{\natexlab{a}}, Living Reviews in
  Relativity, 21, 2

\bibitem[{Amendola {et~al.}(2018{\natexlab{b}})Amendola, Kunz, Saltas, \&
  Sawicki}]{Amendola17}
Amendola, L., Kunz, M., Saltas, I.~D., \& Sawicki, I., {Fate of Large-Scale
  Structure in Modified Gravity After GW170817 and GRB170817A}.
  2018{\natexlab{b}}, \prl, 120, 131101, \eprint{1711.04825}

\bibitem[{{Applegate} {et~al.}(2016){Applegate}, {Mantz}, {Allen}, {von der
  Linden}, {Morris}, {Hilbert}, {Kelly}, {Burke}, {Ebeling}, {Rapetti}, \&
  {Schmidt}}]{Applegate16}
{Applegate}, D.~E., {Mantz}, A., {Allen}, S.~W., {et~al.}, {Cosmology and
  astrophysics from relaxed galaxy clusters - IV. Robustly calibrating
  hydrostatic masses with weak lensing}. 2016, \mnras, 457, 1522,
  \eprint{1509.02162}

\bibitem[{{Archidiacono} {et~al.}(2017){Archidiacono}, {Brinckmann},
  {Lesgourgues}, \& {Poulin}}]{Arc17}
{Archidiacono}, M., {Brinckmann}, T., {Lesgourgues}, J., \& {Poulin}, V.,
  {Physical effects involved in the measurements of neutrino masses with future
  cosmological data}. 2017, \jcap, 2, 052, \eprint{1610.09852}

\bibitem[{{Ata} {et~al.}(2018){Ata}, {Baumgarten}, {Bautista}, {Beutler},
  {Bizyaev}, {Blanton}, {Blazek}, {Bolton}, {Brinkmann}, {Brownstein},
  {Burtin}, {Chuang}, {Comparat}, {Dawson}, {de la Macorra}, {Du}, {du Mas des
  Bourboux}, {Eisenstein}, {Gil-Mar{\'{\i}}n}, {Grabowski}, {Guy}, {Hand},
  {Ho}, {Hutchinson}, {Ivanov}, {Kitaura}, {Kneib}, {Laurent}, {Le Goff},
  {McEwen}, {Mueller}, {Myers}, {Newman}, {Palanque-Delabrouille}, {Pan},
  {P{\^a}ris}, {Pellejero-Ibanez}, {Percival}, {Petitjean}, {Prada}, {Prakash},
  {Rodr{\'{\i}}guez-Torres}, {Ross}, {Rossi}, {Ruggeri}, {S{\'a}nchez},
  {Satpathy}, {Schlegel}, {Schneider}, {Seo}, {Slosar}, {Streblyanska},
  {Tinker}, {Tojeiro}, {Vargas Maga{\~n}a}, {Vivek}, {Wang}, {Y{\`e}che}, {Yu},
  {Zarrouk}, {Zhao}, {Zhao}, \& {Zhu}}]{Ata18}
{Ata}, M., {Baumgarten}, F., {Bautista}, J., {et~al.}, {The clustering of the
  SDSS-IV extended Baryon Oscillation Spectroscopic Survey DR14 quasar sample:
  first measurement of baryon acoustic oscillations between redshift 0.8 and
  2.2}. 2018, \mnras, 473, 4773, \eprint{1705.06373}

\bibitem[{{Atrio-Barandela}(2013)}]{Atrio13}
{Atrio-Barandela}, F., {On the statistical significance of the bulk flow
  measured by the Planck satellite}. 2013, \aap, 557, A116, \eprint{1303.6614}

\bibitem[{{Aubourg} {et~al.}(2015){Aubourg}, {Bailey}, {Bautista}, {Beutler},
  {Bhardwaj}, {Bizyaev}, {Blanton}, {Blomqvist}, {Bolton}, {Bovy},
  {Brewington}, {Brinkmann}, {Brownstein}, {Burden}, {Busca}, {Carithers},
  {Chuang}, {Comparat}, {Croft}, {Cuesta}, {Dawson}, {Delubac}, {Eisenstein},
  {Font-Ribera}, {Ge}, {Le Goff}, {Gontcho}, {Gott}, {Gunn}, {Guo}, {Guy},
  {Hamilton}, {Ho}, {Honscheid}, {Howlett}, {Kirkby}, {Kitaura}, {Kneib},
  {Lee}, {Long}, {Lupton}, {Maga{\~n}a}, {Malanushenko}, {Malanushenko},
  {Manera}, {Maraston}, {Margala}, {McBride}, {Miralda-Escud{\'e}}, {Myers},
  {Nichol}, {Noterdaeme}, {Nuza}, {Olmstead}, {Oravetz}, {P{\^a}ris},
  {Padmanabhan}, {Palanque-Delabrouille}, {Pan}, {Pellejero-Ibanez},
  {Percival}, {Petitjean}, {Pieri}, {Prada}, {Reid}, {Rich}, {Roe}, {Ross},
  {Ross}, {Rossi}, {Rubi{\~n}o-Mart{\'{\i}}n}, {S{\'a}nchez}, {Samushia},
  {G{\'e}nova-Santos}, {Sc{\'o}ccola}, {Schlegel}, {Schneider}, {Seo},
  {Sheldon}, {Simmons}, {Skibba}, {Slosar}, {Strauss}, {Thomas}, {Tinker},
  {Tojeiro}, {Vazquez}, {Viel}, {Wake}, {Weaver}, {Weinberg}, {Wood-Vasey},
  {Y{\`e}che}, {Zehavi}, {Zhao}, \& {BOSS Collaboration}}]{Aubourg15}
{Aubourg}, {\'E}., {Bailey}, S., {Bautista}, J.~E., {et~al.}, {Cosmological
  implications of baryon acoustic oscillation measurements}. 2015, \prd, 92,
  123516, \eprint{1411.1074}

\bibitem[{{Aver} {et~al.}(2015){Aver}, {Olive}, \& {Skillman}}]{Aver15}
{Aver}, E., {Olive}, K.~A., \& {Skillman}, E.~D., {The effects of He I
  {$\lambda$}10830 on helium abundance determinations}. 2015, \jcap, 7, 011,
  \eprint{1503.08146}

\bibitem[{{Aylor} {et~al.}(2017){Aylor}, {Hou}, {Knox}, {Story}, {Benson},
  {Bleem}, {Carlstrom}, {Chang}, {Cho}, {Chown}, {Crawford}, {Crites}, {de
  Haan}, {Dobbs}, {Everett}, {George}, {Halverson}, {Harrington}, {Holder},
  {Holzapfel}, {Hrubes}, {Keisler}, {Lee}, {Leitch}, {Luong-Van}, {Marrone},
  {McMahon}, {Meyer}, {Millea}, {Mocanu}, {Mohr}, {Natoli}, {Omori}, {Padin},
  {Pryke}, {Reichardt}, {Ruhl}, {Sayre}, {Schaffer}, {Shirokoff},
  {Staniszewski}, {Stark}, {Vanderlinde}, {Vieira}, \& {Williamson}}]{Aylor17}
{Aylor}, K., {Hou}, Z., {Knox}, L., {et~al.}, {A Comparison of Cosmological
  Parameters Determined from CMB Temperature Power Spectra from the South Pole
  Telescope and the Planck Satellite}. 2017, \apj, 850, 101,
  \eprint{1706.10286}

\bibitem[{{Ba{\~n}ados} {et~al.}(2018){Ba{\~n}ados}, {Venemans},
  {Mazzucchelli}, {Farina}, {Walter}, {Wang}, {Decarli}, {Stern}, {Fan},
  {Davies}, {Hennawi}, {Simcoe}, {Turner}, {Rix}, {Yang}, {Kelson}, {Rudie}, \&
  {Winters}}]{Banados18}
{Ba{\~n}ados}, E., {Venemans}, B.~P., {Mazzucchelli}, C., {et~al.}, {An
  800-million-solar-mass black hole in a significantly neutral Universe at a
  redshift of 7.5}. 2018, \nat, 553, 473, \eprint{1712.01860}

\bibitem[{{Baker} {et~al.}(2017){Baker}, {Bellini}, {Ferreira}, {Lagos},
  {Noller}, \& {Sawicki}}]{Baker17}
{Baker}, T., {Bellini}, E., {Ferreira}, P.~G., {et~al.}, {Strong Constraints on
  Cosmological Gravity from GW170817 and GRB 170817A}. 2017, Physical Review
  Letters, 119, 251301, \eprint{1710.06394}

\bibitem[{{Balbi} {et~al.}(2000){Balbi}, {Ade}, {Bock}, {Borrill}, {Boscaleri},
  {De Bernardis}, {Ferreira}, {Hanany}, {Hristov}, {Jaffe}, {Lee}, {Oh},
  {Pascale}, {Rabii}, {Richards}, {Smoot}, {Stompor}, {Winant}, \&
  {Wu}}]{Balbi00}
{Balbi}, A., {Ade}, P., {Bock}, J., {et~al.}, {Constraints on Cosmological
  Parameters from MAXIMA-1}. 2000, \apjl, 545, L1

\bibitem[{{Basak} \& {Delabrouille}(2013)}]{BasDel13}
{Basak}, S. \& {Delabrouille}, J., {A needlet ILC analysis of WMAP 9-year
  polarization data: CMB polarization power spectra}. 2013, \mnras, 435, 18,
  \eprint{1204.0292}

\bibitem[{{Bautista} {et~al.}(2017){Bautista}, {Busca}, {Guy}, {Rich},
  {Blomqvist}, {du Mas des Bourboux}, {Pieri}, {Font-Ribera}, {Bailey},
  {Delubac}, {Kirkby}, {Le Goff}, {Margala}, {Slosar}, {Vazquez}, {Brownstein},
  {Dawson}, {Eisenstein}, {Miralda-Escud{\'e}}, {Noterdaeme},
  {Palanque-Delabrouille}, {P{\^a}ris}, {Petitjean}, {Ross}, {Schneider},
  {Weinberg}, \& {Y{\`e}che}}]{Bautista17}
{Bautista}, J.~E., {Busca}, N.~G., {Guy}, J., {et~al.}, {Measurement of baryon
  acoustic oscillation correlations at z = 2.3 with SDSS DR12
  Ly{$\alpha$}-Forests}. 2017, \aap, 603, A12, \eprint{1702.00176}

\bibitem[{{Becker} {et~al.}(2015){Becker}, {Bolton}, \& {Lidz}}]{Becker15}
{Becker}, G.~D., {Bolton}, J.~S., \& {Lidz}, A., {Reionisation and
  High-Redshift Galaxies: The View from Quasar Absorption Lines}. 2015, \pasa,
  32, e045, \eprint{1510.03368}

\bibitem[{{Bennett} {et~al.}(1996){Bennett}, {Banday}, {Gorski}, {Hinshaw},
  {Jackson}, {Keegstra}, {Kogut}, {Smoot}, {Wilkinson}, \&
  {Wright}}]{COBE_data}
{Bennett}, C.~L., {Banday}, A.~J., {Gorski}, K.~M., {et~al.}, {Four-Year COBE
  DMR Cosmic Microwave Background Observations: Maps and Basic Results}. 1996,
  \apjl, 464, L1, \eprint{astro-ph/9601067}

\bibitem[{{Bennett} {et~al.}(2003){Bennett}, {Halpern}, {Hinshaw}, {Jarosik},
  {Kogut}, {Limon}, {Meyer}, {Page}, {Spergel}, {Tucker}, {Wollack}, {Wright},
  {Barnes}, {Greason}, {Hill}, {Komatsu}, {Nolta}, {Odegard}, {Peiris},
  {Verde}, \& {Weiland}}]{bennett2003a}
{Bennett}, C.~L., {Halpern}, M., {Hinshaw}, G., {et~al.}, {First-Year Wilkinson
  Microwave Anisotropy Probe (WMAP) Observations: Preliminary Maps and Basic
  Results}. 2003, \apjs, 148, 1, \eprint{astro-ph/0302207}

\bibitem[{{Bennett} {et~al.}(2011){Bennett}, {Hill}, {Hinshaw}, {Larson},
  {Smith}, {Dunkley}, {Gold}, {Halpern}, {Jarosik}, {Kogut}, {Komatsu},
  {Limon}, {Meyer}, {Nolta}, {Odegard}, {Page}, {Spergel}, {Tucker}, {Weiland},
  {Wollack}, \& {Wright}}]{bennett2010}
{Bennett}, C.~L., {Hill}, R.~S., {Hinshaw}, G., {et~al.}, {Seven-year Wilkinson
  Microwave Anisotropy Probe (WMAP) Observations: Are There Cosmic Microwave
  Background Anomalies?} 2011, \apjs, 192, 17, \eprint{1001.4758}

\bibitem[{{Bennett} {et~al.}(2013{\natexlab{a}}){Bennett}, {Larson}, {Weiland},
  {Jarosik}, {Hinshaw}, {Odegard}, {Smith}, {Hill}, {Gold}, {Halpern},
  {Komatsu}, {Nolta}, {Page}, {Spergel}, {Wollack}, {Dunkley}, {Kogut},
  {Limon}, {Meyer}, {Tucker}, \& {Wright}}]{Bennett13}
{Bennett}, C.~L., {Larson}, D., {Weiland}, J.~L., {et~al.}, {Nine-year
  Wilkinson Microwave Anisotropy Probe (WMAP) Observations: Final Maps and
  Results}. 2013{\natexlab{a}}, \apjs, 208, 20, \eprint{1212.5225}

\bibitem[{{Bennett} {et~al.}(2013{\natexlab{b}}){Bennett}, {Larson}, {Weiland},
  {Jarosik}, {Hinshaw}, {Odegard}, {Smith}, {Hill}, {Gold}, {Halpern},
  {Komatsu}, {Nolta}, {Page}, {Spergel}, {Wollack}, {Dunkley}, {Kogut},
  {Limon}, {Meyer}, {Tucker}, \& {Wright}}]{bennett2012}
{Bennett}, C.~L., {Larson}, D., {Weiland}, J.~L., {et~al.}, {Nine-year
  Wilkinson Microwave Anisotropy Probe (WMAP) Observations: Final Maps and
  Results}. 2013{\natexlab{b}}, \apjs, 208, 20, \eprint{1212.5225}

\bibitem[{{B{\'e}thermin} {et~al.}(2012){B{\'e}thermin}, {Daddi}, {Magdis},
  {Sargent}, {Hezaveh}, {Elbaz}, {Le Borgne}, {Mullaney}, {Pannella}, {Buat},
  {Charmandaris}, {Lagache}, \& {Scott}}]{Bethermin_2012}
{B{\'e}thermin}, M., {Daddi}, E., {Magdis}, G., {et~al.}, {A Unified Empirical
  Model for Infrared Galaxy Counts Based on the Observed Physical Evolution of
  Distant Galaxies}. 2012, \apjl, 757, L23, \eprint{1208.6512}

\bibitem[{{B{\'e}thermin} {et~al.}(2013){B{\'e}thermin}, {Wang}, {Dor{\'e}},
  {Lagache}, {Sargent}, {Daddi}, {Cousin}, \& {Aussel}}]{bethermin_2013}
{B{\'e}thermin}, M., {Wang}, L., {Dor{\'e}}, O., {et~al.}, {The redshift
  evolution of the distribution of star formation among dark matter halos as
  seen in the infrared}. 2013, \aap, 557, A66, \eprint{1304.3936}

\bibitem[{{Beutler} {et~al.}(2011){Beutler}, {Blake}, {Colless}, {Jones},
  {Staveley-Smith}, {Campbell}, {Parker}, {Saunders}, \& {Watson}}]{Beutler11}
{Beutler}, F., {Blake}, C., {Colless}, M., {et~al.}, {The 6dF Galaxy Survey:
  baryon acoustic oscillations and the local Hubble constant}. 2011, \mnras,
  416, 3017, \eprint{1106.3366}

\bibitem[{Beutler {et~al.}(2012)Beutler, Blake, Colless, Jones, Staveley-Smith,
  {et~al.}}]{Beutler12}
Beutler, F., Blake, C., Colless, M., {et~al.}, {The 6dF Galaxy Survey: $z
  \approx 0$ measurement of the growth rate and $\sigma_8$}. 2012, \mnras, 423,
  3430, \eprint{1204.4725}

\bibitem[{{Bianchini} {et~al.}(2015){Bianchini}, {Bielewicz}, {Lapi},
  {Gonzalez-Nuevo}, {Baccigalupi}, {de Zotti}, {Danese}, {Bourne}, {Cooray},
  {Dunne}, {Dye}, {Eales}, {Ivison}, {Maddox}, {Negrello}, {Scott}, {Smith}, \&
  {Valiante}}]{Bianchini15}
{Bianchini}, F., {Bielewicz}, P., {Lapi}, A., {et~al.}, {Cross-correlation
  between the CMB Lensing Potential Measured by Planck and High-z Submillimeter
  Galaxies Detected by the Herschel-Atlas Survey}. 2015, \apj, 802, 64,
  \eprint{1410.4502}

\bibitem[{{Bianchini} \& {Reichardt}(2018)}]{BiaRei18}
{Bianchini}, F. \& {Reichardt}, C.~L., {Constraining gravity at large scales
  with the 2MASS Photometric Redshift catalogue and Planck lensing}. 2018,
  ArXiv e-prints, \eprint{1801.03736}

\bibitem[{{BICEP2 and Keck Array Collaborations} {et~al.}(2015){BICEP2 and Keck
  Array Collaborations}, {Ade}, {Ahmed}, {Aikin}, {Alexander}, {Barkats},
  {Benton}, {Bischoff}, {Bock}, {Brevik}, {Buder}, {Bullock}, {Buza},
  {Connors}, {Crill}, {Dowell}, {Dvorkin}, {Duband}, {Filippini}, {Fliescher},
  {Golwala}, {Halpern}, {Harrison}, {Hasselfield}, {Hildebrandt}, {Hilton},
  {Hristov}, {Hui}, {Irwin}, {Karkare}, {Kaufman}, {Keating}, {Kefeli},
  {Kernasovskiy}, {Kovac}, {Kuo}, {Leitch}, {Lueker}, {Mason}, {Megerian},
  {Netterfield}, {Nguyen}, {O'Brient}, {Ogburn}, {Orlando}, {Pryke},
  {Reintsema}, {Richter}, {Schwarz}, {Sheehy}, {Staniszewski}, {Sudiwala},
  {Teply}, {Thompson}, {Tolan}, {Turner}, {Vieregg}, {Weber}, {Willmert},
  {Wong}, \& {Yoon}}]{BICEPKeck15}
{BICEP2 and Keck Array Collaborations}, {Ade}, P.~A.~R., {Ahmed}, Z., {et~al.},
  {BICEP2/Keck Array V: Measurements of B-mode Polarization at Degree Angular
  Scales and 150 GHz by the Keck Array}. 2015, \apj, 811, 126,
  \eprint{1502.00643}

\bibitem[{{BICEP2 Collaboration} {et~al.}(2016){BICEP2 Collaboration}, {Keck
  Array Collaboration}, {Ade}, {Ahmed}, {Aikin}, {Alexander}, {Barkats},
  {Benton}, {Bischoff}, {Bock}, {Bowens-Rubin}, {Brevik}, {Buder}, {Bullock},
  {Buza}, {Connors}, {Crill}, {Duband}, {Dvorkin}, {Filippini}, {Fliescher},
  {Grayson}, {Halpern}, {Harrison}, {Hilton}, {Hui}, {Irwin}, {Karkare},
  {Karpel}, {Kaufman}, {Keating}, {Kefeli}, {Kernasovskiy}, {Kovac}, {Kuo},
  {Leitch}, {Lueker}, {Megerian}, {Netterfield}, {Nguyen}, {O'Brient},
  {Ogburn}, {Orlando}, {Pryke}, {Richter}, {Schwarz}, {Sheehy}, {Staniszewski},
  {Steinbach}, {Sudiwala}, {Teply}, {Thompson}, {Tolan}, {Tucker}, {Turner},
  {Vieregg}, {Weber}, {Wiebe}, {Willmert}, {Wong}, {Wu}, \&
  {Yoon}}]{BICEPKeck16}
{BICEP2 Collaboration}, {Keck Array Collaboration}, {Ade}, P.~A.~R., {et~al.},
  {Improved Constraints on Cosmology and Foregrounds from BICEP2 and Keck Array
  Cosmic Microwave Background Data with Inclusion of 95 GHz Band}. 2016,
  Physical Review Letters, 116, 031302, \eprint{1510.09217}

\bibitem[{Blake {et~al.}(2012)}]{Blake:2012pj}
Blake, C. {et~al.}, {The WiggleZ Dark Energy Survey: Joint measurements of the
  expansion and growth history at $z < 1$}. 2012, \mnras, 425, 405,
  \eprint{1204.3674}

\bibitem[{Blake {et~al.}(2013)}]{Blake:2013nif}
Blake, C. {et~al.}, {Galaxy And Mass Assembly (GAMA): improved cosmic growth
  measurements using multiple tracers of large-scale structure}. 2013, \mnras,
  436, 3089, \eprint{1309.5556}

\bibitem[{{Blanchard} \& {Schneider}(1987)}]{1987A&A...184....1B}
{Blanchard}, A. \& {Schneider}, J., {Gravitational lensing effect on the
  fluctuations of the cosmic background radiation}. 1987, \aap, 184, 1

\bibitem[{{Bond} {et~al.}(2003){Bond}, {Contaldi}, \& {Pogosyan}}]{Bond03}
{Bond}, J.~R., {Contaldi}, C., \& {Pogosyan}, D., {Cosmic microwave background
  snapshots: pre-WMAP and post-WMAP}. 2003, Philosophical Transactions of the
  Royal Society of London Series A, 361, 2435, \eprint{astro-ph/0310735}

\bibitem[{{Bonvin} {et~al.}(2017){Bonvin}, {Courbin}, {Suyu}, {Marshall},
  {Rusu}, {Sluse}, {Tewes}, {Wong}, {Collett}, {Fassnacht}, {Treu}, {Auger},
  {Hilbert}, {Koopmans}, {Meylan}, {Rumbaugh}, {Sonnenfeld}, \&
  {Spiniello}}]{Bonvin17}
{Bonvin}, V., {Courbin}, F., {Suyu}, S.~H., {et~al.}, {H0LiCOW - V. New
  COSMOGRAIL time delays of HE 0435-1223: H$_{0}$ to 3.8 per cent precision
  from strong lensing in a flat {$\Lambda$}CDM model}. 2017, \mnras, 465, 4914,
  \eprint{1607.01790}

\bibitem[{{Bouwens} {et~al.}(2015){Bouwens}, {Illingworth}, {Oesch}, {Caruana},
  {Holwerda}, {Smit}, \& {Wilkins}}]{Bouwens15}
{Bouwens}, R.~J., {Illingworth}, G.~D., {Oesch}, P.~A., {et~al.}, {Reionization
  After Planck: The Derived Growth of the Cosmic Ionizing Emissivity Now
  Matches the Growth of the Galaxy UV Luminosity Density}. 2015, \apj, 811,
  140, \eprint{1503.08228}

\bibitem[{{Bowman} {et~al.}(2018){Bowman}, {Rogers}, {Monsalve}, {Mozdzen}, \&
  {Mahesh}}]{Bowman18}
{Bowman}, J.~D., {Rogers}, A.~E.~E., {Monsalve}, R.~A., {Mozdzen}, T.~J., \&
  {Mahesh}, N., {An absorption profile centred at 78 megahertz in the
  sky-averaged spectrum}. 2018, \nat, 555, 67

\bibitem[{{Bullock} \& {Boylan-Kolchin}(2017)}]{Bullock17}
{Bullock}, J.~S. \& {Boylan-Kolchin}, M., {Small-Scale Challenges to the
  {$\Lambda$}CDM Paradigm}. 2017, \araa, 55, 343, \eprint{1707.04256}

\bibitem[{{Burgess} {et~al.}(2013){Burgess}, {Cicoli}, \&
  {Quevedo}}]{Burgess13}
{Burgess}, C.~P., {Cicoli}, M., \& {Quevedo}, F., {String inflation after
  Planck 2013}. 2013, \jcap, 11, 003, \eprint{1306.3512}

\bibitem[{{Ca{\~n}ameras} {et~al.}(2015){Ca{\~n}ameras}, {Nesvadba}, {Guery},
  {McKenzie}, {K{\"o}nig}, {Petitpas}, {Dole}, {Frye}, {Flores-Cacho},
  {Montier}, {Negrello}, {Beelen}, {Boone}, {Dicken}, {Lagache}, {Le Floc'h},
  {Altieri}, {B{\'e}thermin}, {Chary}, {de Zotti}, {Giard}, {Kneissl}, {Krips},
  {Malhotra}, {Martinache}, {Omont}, {Pointecouteau}, {Puget}, {Scott},
  {Soucail}, {Valtchanov}, {Welikala}, \& {Yan}}]{Canameras2015}
{Ca{\~n}ameras}, R., {Nesvadba}, N.~P.~H., {Guery}, D., {et~al.}, {Planck's
  dusty GEMS: The brightest gravitationally lensed galaxies discovered with the
  Planck all-sky survey}. 2015, \aap, 581, A105, \eprint{1506.01962}

\bibitem[{{Carlstrom} {et~al.}(2002){Carlstrom}, {Holder}, \&
  {Reese}}]{Carlstrom02}
{Carlstrom}, J.~E., {Holder}, G.~P., \& {Reese}, E.~D., {Cosmology with the
  Sunyaev-Zel'dovich Effect}. 2002, \araa, 40, 643, \eprint{astro-ph/0208192}

\bibitem[{{Chabanier} {et~al.}(2019){Chabanier}, {Millea}, \&
  {Palanque-Delabrouille}}]{chabanier2019}
{Chabanier}, S., {Millea}, M., \& {Palanque-Delabrouille}, N., {Matter power
  spectrum: from Ly {\ensuremath{\alpha}} forest to CMB scales}. 2019, \mnras,
  489, 2247, \eprint{1905.08103}

\bibitem[{{Chartrand}(2011)}]{Chartrand11}
{Chartrand}, R., {Numerical differentiation of noisy, non-smooth data}. 2011,
  ISRN Applied Mathematics, 2011, 164564

\bibitem[{{Chen} \& {Kamionkowski}(2004)}]{CheKam04}
{Chen}, X. \& {Kamionkowski}, M., {Particle decays during the cosmic dark
  ages}. 2004, \prd, 70, 043502, \eprint{astro-ph/0310473}

\bibitem[{{Chisari} {et~al.}(2018){Chisari}, {Richardson}, {Devriendt},
  {Dubois}, {Schneider}, {Le Brun}, {Beckmann}, {Peirani}, {Slyz}, \&
  {Pichon}}]{Chisari18}
{Chisari}, N.~E., {Richardson}, M.~L.~A., {Devriendt}, J., {et~al.}, {The
  impact of baryons on the matter power spectrum from the Horizon-AGN
  cosmological hydrodynamical simulation}. 2018, ArXiv e-prints,
  \eprint{1801.08559}

\bibitem[{{Chluba} \& {Sunyaev}(2012)}]{ChlubaSunyaev12}
{Chluba}, J. \& {Sunyaev}, R.~A., {The evolution of CMB spectral distortions in
  the early Universe}. 2012, \mnras, 419, 1294, \eprint{1109.6552}

\bibitem[{{Clark} {et~al.}(2015){Clark}, {Hill}, {Peek}, {Putman}, \&
  {Babler}}]{Clark2015}
{Clark}, S.~E., {Hill}, J.~C., {Peek}, J.~E.~G., {Putman}, M.~E., \& {Babler},
  B.~L., {Neutral Hydrogen Structures Trace Dust Polarization Angle:
  Implications for Cosmic Microwave Background Foregrounds}. 2015, Physical
  Review Letters, 115, 241302, \eprint{1508.07005}

\bibitem[{{Cooke} {et~al.}(2018){Cooke}, {Pettini}, \& {Steidel}}]{Cooke18}
{Cooke}, R.~J., {Pettini}, M., \& {Steidel}, C.~C., {One Percent Determination
  of the Primordial Deuterium Abundance}. 2018, \apj, 855, 102,
  \eprint{1710.11129}

\bibitem[{{Corasaniti} \& {Melchiorri}(2008)}]{Corasaniti08}
{Corasaniti}, P.~S. \& {Melchiorri}, A., {Testing cosmology with cosmic sound
  waves}. 2008, \prd, 77, 103507, \eprint{0711.4119}

\bibitem[{{Coulson} {et~al.}(1994){Coulson}, {Ferreira}, {Graham}, \&
  {Turok}}]{Cou94}
{Coulson}, D., {Ferreira}, P., {Graham}, P., \& {Turok}, N., {Microwave
  anisotropies from cosmic defects}. 1994, \nat, 368, 27,
  \eprint{hep-ph/9310322}

\bibitem[{{Courteau} \& {van den Bergh}(1999)}]{courteau1999}
{Courteau}, S. \& {van den Bergh}, S., {The Solar Motion Relative to the Local
  Group}. 1999, \aj, 118, 337, \eprint{astro-ph/9903298}

\bibitem[{{Creminelli} \& {Vernizzi}(2017)}]{CreVer17}
{Creminelli}, P. \& {Vernizzi}, F., {Dark Energy after GW170817 and
  GRB170817A}. 2017, Physical Review Letters, 119, 251302, \eprint{1710.05877}

\bibitem[{Crisostomi \& Koyama(2018)}]{Crisostomi17}
Crisostomi, M. \& Koyama, K., {Self-accelerating universe in scalar-tensor
  theories after GW170817}. 2018, Phys. Rev., D97, 084004, \eprint{1712.06556}

\bibitem[{{Crittenden} \& {Turok}(1995)}]{CriTur95}
{Crittenden}, R.~G. \& {Turok}, N., {Doppler Peaks from Cosmic Texture}. 1995,
  Physical Review Letters, 75, 2642, \eprint{astro-ph/9505120}

\bibitem[{{Das} {et~al.}(2014){Das}, {Louis}, {Nolta}, {Addison},
  {Battistelli}, {Bond}, {Calabrese}, {Crichton}, {Devlin}, {Dicker},
  {Dunkley}, {D{\"u}nner}, {Fowler}, {Gralla}, {Hajian}, {Halpern},
  {Hasselfield}, {Hilton}, {Hincks}, {Hlozek}, {Huffenberger}, {Hughes},
  {Irwin}, {Kosowsky}, {Lupton}, {Marriage}, {Marsden}, {Menanteau}, {Moodley},
  {Niemack}, {Page}, {Partridge}, {Reese}, {Schmitt}, {Sehgal}, {Sherwin},
  {Sievers}, {Spergel}, {Staggs}, {Swetz}, {Switzer}, {Thornton}, {Trac}, \&
  {Wollack}}]{Das14}
{Das}, S., {Louis}, T., {Nolta}, M.~R., {et~al.}, {The Atacama Cosmology
  Telescope: temperature and gravitational lensing power spectrum measurements
  from three seasons of data}. 2014, \jcap, 4, 014, \eprint{1301.1037}

\bibitem[{{Davies} {et~al.}(2018){Davies}, {Hennawi}, {Ba{\~n}ados},
  {Luki{\'c}}, {Decarli}, {Fan}, {Farina}, {Mazzucchelli}, {Rix}, {Venemans},
  {Walter}, {Wang}, \& {Yang}}]{Davies18}
{Davies}, F.~B., {Hennawi}, J.~F., {Ba{\~n}ados}, E., {et~al.}, {Quantitative
  Constraints on the Reionization History from the IGM Damping Wing Signature
  in Two Quasars at $z>7$}. 2018, ArXiv e-prints, \eprint{1802.06066}

\bibitem[{{Dawson} {et~al.}(2013){Dawson}, {Schlegel}, {Ahn}, {Anderson},
  {Aubourg}, {Bailey}, {Barkhouser}, {Bautista}, {Beifiori}, {Berlind},
  {Bhardwaj}, {Bizyaev}, {Blake}, {Blanton}, {Blomqvist}, {Bolton}, {Borde},
  {Bovy}, {Brandt}, {Brewington}, {Brinkmann}, {Brown}, {Brownstein}, {Bundy},
  {Busca}, {Carithers}, {Carnero}, {Carr}, {Chen}, {Comparat}, {Connolly},
  {Cope}, {Croft}, {Cuesta}, {da Costa}, {Davenport}, {Delubac}, {de Putter},
  {Dhital}, {Ealet}, {Ebelke}, {Eisenstein}, {Escoffier}, {Fan}, {Filiz Ak},
  {Finley}, {Font-Ribera}, {G{\'e}nova-Santos}, {Gunn}, {Guo}, {Haggard},
  {Hall}, {Hamilton}, {Harris}, {Harris}, {Ho}, {Hogg}, {Holder}, {Honscheid},
  {Huehnerhoff}, {Jordan}, {Jordan}, {Kauffmann}, {Kazin}, {Kirkby}, {Klaene},
  {Kneib}, {Le Goff}, {Lee}, {Long}, {Loomis}, {Lundgren}, {Lupton}, {Maia},
  {Makler}, {Malanushenko}, {Malanushenko}, {Mandelbaum}, {Manera}, {Maraston},
  {Margala}, {Masters}, {McBride}, {McDonald}, {McGreer}, {McMahon}, {Mena},
  {Miralda-Escud{\'e}}, {Montero-Dorta}, {Montesano}, {Muna}, {Myers},
  {Naugle}, {Nichol}, {Noterdaeme}, {Nuza}, {Olmstead}, {Oravetz}, {Oravetz},
  {Owen}, {Padmanabhan}, {Palanque-Delabrouille}, {Pan}, {Parejko},
  {P{\^a}ris}, {Percival}, {P{\'e}rez-Fournon}, {P{\'e}rez-R{\`a}fols},
  {Petitjean}, {Pfaffenberger}, {Pforr}, {Pieri}, {Prada}, {Price-Whelan},
  {Raddick}, {Rebolo}, {Rich}, {Richards}, {Rockosi}, {Roe}, {Ross}, {Ross},
  {Rossi}, {Rubi{\~n}o-Martin}, {Samushia}, {S{\'a}nchez}, {Sayres}, {Schmidt},
  {Schneider}, {Sc{\'o}ccola}, {Seo}, {Shelden}, {Sheldon}, {Shen}, {Shu},
  {Slosar}, {Smee}, {Snedden}, {Stauffer}, {Steele}, {Strauss}, {Streblyanska},
  {Suzuki}, {Swanson}, {Tal}, {Tanaka}, {Thomas}, {Tinker}, {Tojeiro},
  {Tremonti}, {Vargas Maga{\~n}a}, {Verde}, {Viel}, {Wake}, {Watson}, {Weaver},
  {Weinberg}, {Weiner}, {West}, {White}, {Wood-Vasey}, {Yeche}, {Zehavi},
  {Zhao}, \& {Zheng}}]{Dawson13}
{Dawson}, K.~S., {Schlegel}, D.~J., {Ahn}, C.~P., {et~al.}, {The Baryon
  Oscillation Spectroscopic Survey of SDSS-III}. 2013, \aj, 145, 10,
  \eprint{1208.0022}

\bibitem[{{de Bernardis} {et~al.}(2002){de Bernardis}, {Ade}, {Bock}, {Bond},
  {Borrill}, {Boscaleri}, {Coble}, {Contaldi}, {Crill}, {De Troia}, {Farese},
  {Ganga}, {Giacometti}, {Hivon}, {Hristov}, {Iacoangeli}, {Jaffe}, {Jones},
  {Lange}, {Martinis}, {Masi}, {Mason}, {Mauskopf}, {Melchiorri}, {Montroy},
  {Netterfield}, {Pascale}, {Piacentini}, {Pogosyan}, {Polenta}, {Pongetti},
  {Prunet}, {Romeo}, {Ruhl}, \& {Scaramuzzi}}]{deBernardis02}
{de Bernardis}, P., {Ade}, P.~A.~R., {Bock}, J.~J., {et~al.}, {Multiple Peaks
  in the Angular Power Spectrum of the Cosmic Microwave Background:
  Significance and Consequences for Cosmology}. 2002, \apj, 564, 559,
  \eprint{astro-ph/0105296}

\bibitem[{{de Bernardis} {et~al.}(2000){de Bernardis}, {Ade}, {Bock}, {Bond},
  {Borrill}, {Boscaleri}, {Coble}, {Crill}, {De Gasperis}, {Farese},
  {Ferreira}, {Ganga}, {Giacometti}, {Hivon}, {Hristov}, {Iacoangeli}, {Jaffe},
  {Lange}, {Martinis}, {Masi}, {Mason}, {Mauskopf}, {Melchiorri}, {Miglio},
  {Montroy}, {Netterfield}, {Pascale}, {Piacentini}, {Pogosyan}, {Prunet},
  {Rao}, {Romeo}, {Ruhl}, {Scaramuzzi}, {Sforna}, \&
  {Vittorio}}]{deBernardis00}
{de Bernardis}, P., {Ade}, P.~A.~R., {Bock}, J.~J., {et~al.}, {A flat Universe
  from high-resolution maps of the cosmic microwave background radiation}.
  2000, \nat, 404, 955, \eprint{astro-ph/0004404}

\bibitem[{{de Salas} \& {Pastor}(2016)}]{SalPas16}
{de Salas}, P.~F. \& {Pastor}, S., {Relic neutrino decoupling with flavour
  oscillations revisited}. 2016, \jcap, 7, 051, \eprint{1606.06986}

\bibitem[{{DES Collaboration} {et~al.}(2017){DES Collaboration}, {Abbott},
  {Abdalla}, {Alarcon}, {Aleksi{\'c}}, {Allam}, {Allen}, {Amara}, {Annis},
  {Asorey}, {Avila}, {Bacon}, {Balbinot}, {Banerji}, {Banik}, {Barkhouse},
  {Baumer}, {Baxter}, {Bechtol}, {Becker}, {Benoit-L{\'e}vy}, {Benson},
  {Bernstein}, {Bertin}, {Blazek}, {Bridle}, {Brooks}, {Brout}, {Buckley-Geer},
  {Burke}, {Busha}, {Capozzi}, {Carnero Rosell}, {Carrasco Kind}, {Carretero},
  {Castander}, {Cawthon}, {Chang}, {Chen}, {Childress}, {Choi}, {Conselice},
  {Crittenden}, {Crocce}, {Cunha}, {D'Andrea}, {da Costa}, {Das}, {Davis},
  {Davis}, {De Vicente}, {DePoy}, {DeRose}, {Desai}, {Diehl}, {Dietrich},
  {Dodelson}, {Doel}, {Drlica-Wagner}, {Eifler}, {Elliott}, {Elsner},
  {Elvin-Poole}, {Estrada}, {Evrard}, {Fang}, {Fernandez}, {Fert{\'e}},
  {Finley}, {Flaugher}, {Fosalba}, {Friedrich}, {Frieman},
  {Garc{\'{\i}}a-Bellido}, {Garcia-Fernandez}, {Gatti}, {Gaztanaga}, {Gerdes},
  {Giannantonio}, {Gill}, {Glazebrook}, {Goldstein}, {Gruen}, {Gruendl},
  {Gschwend}, {Gutierrez}, {Hamilton}, {Hartley}, {Hinton}, {Honscheid},
  {Hoyle}, {Huterer}, {Jain}, {James}, {Jarvis}, {Jeltema}, {Johnson},
  {Johnson}, {Kacprzak}, {Kent}, {Kim}, {King}, {Kirk}, {Kokron}, {Kovacs},
  {Krause}, {Krawiec}, {Kremin}, {Kuehn}, {Kuhlmann}, {Kuropatkin}, {Lacasa},
  {Lahav}, {Li}, {Liddle}, {Lidman}, {Lima}, {Lin}, {MacCrann}, {Maia},
  {Makler}, {Manera}, {March}, {Marshall}, {Martini}, {McMahon}, {Melchior},
  {Menanteau}, {Miquel}, {Miranda}, {Mudd}, {Muir}, {M{\"o}ller}, {Neilsen},
  {Nichol}, {Nord}, {Nugent}, {Ogando}, {Palmese}, {Peacock}, {Peiris},
  {Peoples}, {Percival}, {Petravick}, {Plazas}, {Porredon}, {Prat}, {Pujol},
  {Rau}, {Refregier}, {Ricker}, {Roe}, {Rollins}, {Romer}, {Roodman},
  {Rosenfeld}, {Ross}, {Rozo}, {Rykoff}, {Sako}, {Salvador}, {Samuroff},
  {S{\'a}nchez}, {Sanchez}, {Santiago}, {Scarpine}, {Schindler}, {Scolnic},
  {Secco}, {Serrano}, {Sevilla-Noarbe}, {Sheldon}, {Smith}, {Smith}, {Smith},
  {Soares-Santos}, {Sobreira}, {Suchyta}, {Tarle}, {Thomas}, {Troxel},
  {Tucker}, {Tucker}, {Uddin}, {Varga}, {Vielzeuf}, {Vikram}, {Vivas},
  {Walker}, {Wang}, {Wechsler}, {Weller}, {Wester}, {Wolf}, {Yanny}, {Yuan},
  {Zenteno}, {Zhang}, {Zhang}, \& {Zuntz}}]{DES-Y1}
{DES Collaboration}, {Abbott}, T.~M.~C., {Abdalla}, F.~B., {et~al.}, {Dark
  Energy Survey Year 1 Results: Cosmological Constraints from Galaxy Clustering
  and Weak Lensing}. 2017, ArXiv e-prints, \eprint{1708.01530}

\bibitem[{{Dhawan} {et~al.}(2018){Dhawan}, {Jha}, \& {Leibundgut}}]{Dhawan18}
{Dhawan}, S., {Jha}, S.~W., \& {Leibundgut}, B., {Measuring the Hubble constant
  with Type Ia supernovae as near-infrared standard candles}. 2018, \aap, 609,
  A72, \eprint{1707.00715}

\bibitem[{{Diaz} {et~al.}(2014){Diaz}, {Koposov}, {Irwin}, {Belokurov}, \&
  {Evans}}]{diaz2014}
{Diaz}, J.~D., {Koposov}, S.~E., {Irwin}, M., {Belokurov}, V., \& {Evans},
  N.~W., {Balancing mass and momentum in the Local Group}. 2014, \mnras, 443,
  1688, \eprint{1405.3662}

\bibitem[{{Dicke} {et~al.}(1965){Dicke}, {Peebles}, {Roll}, \&
  {Wilkinson}}]{DPRW65}
{Dicke}, R.~H., {Peebles}, P.~J.~E., {Roll}, P.~G., \& {Wilkinson}, D.~T.,
  {Cosmic Black-Body Radiation.} 1965, \apj, 142, 414

\bibitem[{{DiPompeo} {et~al.}(2016){DiPompeo}, {Hickox}, \&
  {Myers}}]{DiPompeo16}
{DiPompeo}, M.~A., {Hickox}, R.~C., \& {Myers}, A.~D., {Updated measurements of
  the dark matter halo masses of obscured quasars with improved WISE and Planck
  data}. 2016, \mnras, 456, 924, \eprint{1511.04469}

\bibitem[{{DiPompeo} {et~al.}(2015){DiPompeo}, {Myers}, {Hickox}, {Geach},
  {Holder}, {Hainline}, \& {Hall}}]{DiPompeo15}
{DiPompeo}, M.~A., {Myers}, A.~D., {Hickox}, R.~C., {et~al.}, {Weighing
  obscured and unobscured quasar hosts with the cosmic microwave background}.
  2015, \mnras, 446, 3492, \eprint{1411.0527}

\bibitem[{{Dodelson}(2003)}]{Dodelson03}
{Dodelson}, S. 2003, {Modern cosmology} (Academic Press)

\bibitem[{{Dole} {et~al.}(2006){Dole}, {Lagache}, {Puget}, {Caputi},
  {Fern{\'a}ndez-Conde}, {Le Floc'h}, {Papovich}, {P{\'e}rez-Gonz{\'a}lez},
  {Rieke}, \& {Blaylock}}]{dole_2006}
{Dole}, H., {Lagache}, G., {Puget}, J.-L., {et~al.}, {The cosmic infrared
  background resolved by Spitzer. Contributions of mid-infrared galaxies to the
  far-infrared background}. 2006, \aap, 451, 417, \eprint{astro-ph/0603208}

\bibitem[{{Doroshkevich} {et~al.}(1978){Doroshkevich}, {Zel'dovich}, \&
  {Syunyaev}}]{Dor78}
{Doroshkevich}, A.~G., {Zel'dovich}, Y.~B., \& {Syunyaev}, R.~A., {Fluctuations
  of the microwave background radiation in the adiabatic and entropic theories
  of galaxy formation}. 1978, \sovast, 22, 523

\bibitem[{{Doux} {et~al.}(2017){Doux}, {Penna-Lima}, {Vitenti}, {Tr{\'e}guer},
  {Aubourg}, \& {Ganga}}]{Doux18}
{Doux}, C., {Penna-Lima}, M., {Vitenti}, S.~D.~P., {et~al.}, {Cosmological
  constraints from a joint analysis of cosmic microwave background and
  large-scale structure}. 2017, ArXiv e-prints, \eprint{1706.04583}

\bibitem[{{Doux} {et~al.}(2016){Doux}, {Schaan}, {Aubourg}, {Ganga}, {Lee},
  {Spergel}, \& {Tr{\'e}guer}}]{Doux16}
{Doux}, C., {Schaan}, E., {Aubourg}, E., {et~al.}, {First detection of cosmic
  microwave background lensing and Lyman-{$\alpha$} forest bispectrum}. 2016,
  \prd, 94, 103506, \eprint{1607.03625}

\bibitem[{{du Mas des Bourboux} {et~al.}(2017){du Mas des Bourboux}, {Le Goff},
  {Blomqvist}, {Busca}, {Guy}, {Rich}, {Y{\`e}che}, {Bautista}, {Burtin},
  {Dawson}, {Eisenstein}, {Font-Ribera}, {Kirkby}, {Miralda-Escud{\'e}},
  {Noterdaeme}, {Palanque-Delabrouille}, {P{\^a}ris}, {Petitjean},
  {P{\'e}rez-R{\`a}fols}, {Pieri}, {Ross}, {Schlegel}, {Schneider}, {Slosar},
  {Weinberg}, \& {Zarrouk}}]{duMas17}
{du Mas des Bourboux}, H., {Le Goff}, J.-M., {Blomqvist}, M., {et~al.}, {Baryon
  acoustic oscillations from the complete SDSS-III Ly{$\alpha$}-quasar
  cross-correlation function at z = 2.4}. 2017, \aap, 608, A130,
  \eprint{1708.02225}

\bibitem[{{Durrer} {et~al.}(2003){Durrer}, {Novosyadlyj}, \&
  {Apunevych}}]{Durrer03}
{Durrer}, R., {Novosyadlyj}, B., \& {Apunevych}, S., {Acoustic Peaks and Dips
  in the Cosmic Microwave Background Power Spectrum: Observational Data and
  Cosmological Constraints}. 2003, \apj, 583, 33, \eprint{astro-ph/0111594}

\bibitem[{{Dwek} {et~al.}(1998){Dwek}, {Arendt}, {Hauser}, {Fixsen}, {Kelsall},
  {Leisawitz}, {Pei}, {Wright}, {Mather}, {Moseley}, {Odegard}, {Shafer},
  {Silverberg}, \& {Weiland}}]{dwek1998}
{Dwek}, E., {Arendt}, R.~G., {Hauser}, M.~G., {et~al.}, {The COBE Diffuse
  Infrared Background Experiment Search for the Cosmic Infrared Background. IV.
  Cosmological Implications}. 1998, \apj, 508, 106, \eprint{astro-ph/9806129}

\bibitem[{{Efstathiou}(1988)}]{Efs88}
{Efstathiou}, G. 1988, in Large-Scale Motions in the Universe: A Vatican study
  Week, ed. V.~C. {Rubin} \& G.~V. {Coyne} (Princeton University Press),
  299--319

\bibitem[{{Efstathiou}(2006)}]{2006MNRAS.370..343E}
{Efstathiou}, G., {Hybrid estimation of cosmic microwave background
  polarization power spectra}. 2006, \mnras, 370, 343,
  \eprint{astro-ph/0601107}

\bibitem[{{Efstathiou} {et~al.}(1992){Efstathiou}, {Bond}, \& {White}}]{EBW92}
{Efstathiou}, G., {Bond}, J.~R., \& {White}, S.~D.~M., {COBE background
  radiation anisotropies and large-scale structure in the universe}. 1992,
  \mnras, 258, 1P

\bibitem[{{Efstathiou} {et~al.}(1990){Efstathiou}, {Sutherland}, \&
  {Maddox}}]{Efstathiou90}
{Efstathiou}, G., {Sutherland}, W.~J., \& {Maddox}, S.~J., {The cosmological
  constant and cold dark matter}. 1990, \nat, 348, 705

\bibitem[{{Eisenstein} \& {White}(2004)}]{EisWhi04}
{Eisenstein}, D. \& {White}, M., {Theoretical uncertainty in baryon
  oscillations}. 2004, \prd, 70, 103523

\bibitem[{{Eisenstein} \& {Hu}(1998)}]{EisHu98}
{Eisenstein}, D.~J. \& {Hu}, W., {Baryonic Features in the Matter Transfer
  Function}. 1998, \apj, 496, 605, \eprint{astro-ph/9709112}

\bibitem[{{Eisenstein} {et~al.}(2007){Eisenstein}, {Seo}, \& {White}}]{ESW07}
{Eisenstein}, D.~J., {Seo}, H.-J., \& {White}, M., {On the Robustness of the
  Acoustic Scale in the Low-Redshift Clustering of Matter}. 2007, \apj, 664,
  660, \eprint{astro-ph/0604361}

\bibitem[{{Eriksen} {et~al.}(2008){Eriksen}, {Jewell}, {Dickinson}, {Banday},
  {G{\'o}rski}, \& {Lawrence}}]{eriksen2008}
{Eriksen}, H.~K., {Jewell}, J.~B., {Dickinson}, C., {et~al.}, {Joint Bayesian
  Component Separation and CMB Power Spectrum Estimation}. 2008, \apj, 676, 10,
  \eprint{0709.1058}

\bibitem[{Ezquiaga \& Zumalac\'arregui(2017)}]{MarZum17}
Ezquiaga, J.~M. \& Zumalac\'arregui, M., Dark Energy After GW170817: Dead Ends
  and the Road Ahead. 2017, \prl, 119, 251304

\bibitem[{{Feldman} {et~al.}(1994){Feldman}, {Kaiser}, \& {Peacock}}]{FKP94}
{Feldman}, H.~A., {Kaiser}, N., \& {Peacock}, J.~A., {Power-spectrum analysis
  of three-dimensional redshift surveys}. 1994, \apj, 426, 23,
  \eprint{astro-ph/9304022}

\bibitem[{{Feng} {et~al.}(2017){Feng}, {Cooray}, \& {Keating}}]{Feng17}
{Feng}, C., {Cooray}, A., \& {Keating}, B., {Planck Lensing and Cosmic Infrared
  Background Cross-correlation with Fermi-LAT: Tracing Dark Matter Signals in
  the Gamma-ray Background}. 2017, \apj, 836, 127, \eprint{1608.04351}

\bibitem[{{Fern{\'a}ndez-Cobos} {et~al.}(2012){Fern{\'a}ndez-Cobos}, {Vielva},
  {Barreiro}, \& {Mart{\'{\i}}nez-Gonz{\'a}lez}}]{fernandez2012}
{Fern{\'a}ndez-Cobos}, R., {Vielva}, P., {Barreiro}, R.~B., \&
  {Mart{\'{\i}}nez-Gonz{\'a}lez}, E., {Multiresolution internal template
  cleaning: an application to the Wilkinson Microwave Anisotropy Probe 7-yr
  polarization data}. 2012, \mnras, 420, 2162, \eprint{1106.2016}

\bibitem[{{Finkbeiner} {et~al.}(2012){Finkbeiner}, {Galli}, {Lin}, \&
  {Slatyer}}]{Fin12}
{Finkbeiner}, D.~P., {Galli}, S., {Lin}, T., \& {Slatyer}, T.~R., {Searching
  for dark matter in the CMB: A compact parametrization of energy injection
  from new physics}. 2012, \prd, 85, 043522, \eprint{1109.6322}

\bibitem[{{Fixsen}(2009)}]{fixsen2009}
{Fixsen}, D.~J., {The Temperature of the Cosmic Microwave Background}. 2009,
  \apj, 707, 916, \eprint{0911.1955}

\bibitem[{{Fixsen} {et~al.}(1996){Fixsen}, {Cheng}, {Gales}, {Mather},
  {Shafer}, \& {Wright}}]{Fixsen96}
{Fixsen}, D.~J., {Cheng}, E.~S., {Gales}, J.~M., {et~al.}, {The Cosmic
  Microwave Background Spectrum from the Full COBE FIRAS Data Set}. 1996, \apj,
  473, 576, \eprint{astro-ph/9605054}

\bibitem[{{Fornengo} {et~al.}(2015){Fornengo}, {Perotto}, {Regis}, \&
  {Camera}}]{Fornengo15}
{Fornengo}, N., {Perotto}, L., {Regis}, M., \& {Camera}, S., {Evidence of
  Cross-correlation between the CMB Lensing and the {$\gamma$}-Ray Sky}. 2015,
  \apjl, 802, L1, \eprint{1410.4997}

\bibitem[{{Freedman} {et~al.}(2012){Freedman}, {Madore}, {Scowcroft}, {Burns},
  {Monson}, {Persson}, {Seibert}, \& {Rigby}}]{Freedman12}
{Freedman}, W.~L., {Madore}, B.~F., {Scowcroft}, V., {et~al.}, {Carnegie Hubble
  Program: A Mid-infrared Calibration of the Hubble Constant}. 2012, \apj, 758,
  24, \eprint{1208.3281}

\bibitem[{{Furlanetto} {et~al.}(2006){Furlanetto}, {Oh}, \&
  {Briggs}}]{Furlanetto06}
{Furlanetto}, S.~R., {Oh}, S.~P., \& {Briggs}, F.~H., {Cosmology at low
  frequencies: The 21 cm transition and the high-redshift Universe}. 2006,
  \physrep, 433, 181, \eprint{astro-ph/0608032}

\bibitem[{{Geach} {et~al.}(2013){Geach}, {Hickox}, {Bleem}, {Brodwin},
  {Holder}, {Aird}, {Benson}, {Bhattacharya}, {Carlstrom}, {Chang}, {Cho},
  {Crawford}, {Crites}, {de Haan}, {Dobbs}, {Dudley}, {George}, {Hainline},
  {Halverson}, {Holzapfel}, {Hoover}, {Hou}, {Hrubes}, {Keisler}, {Knox},
  {Lee}, {Leitch}, {Lueker}, {Luong-Van}, {Marrone}, {McMahon}, {Mehl},
  {Meyer}, {Millea}, {Mohr}, {Montroy}, {Myers}, {Padin}, {Plagge}, {Pryke},
  {Reichardt}, {Ruhl}, {Sayre}, {Schaffer}, {Shaw}, {Shirokoff}, {Spieler},
  {Staniszewski}, {Stark}, {Story}, {van Engelen}, {Vanderlinde}, {Vieira},
  {Williamson}, \& {Zahn}}]{Geach13}
{Geach}, J.~E., {Hickox}, R.~C., {Bleem}, L.~E., {et~al.}, {A Direct
  Measurement of the Linear Bias of Mid-infrared-selected Quasars at z of 1
  Using Cosmic Microwave Background Lensing}. 2013, \apjl, 776, L41,
  \eprint{1307.1706}

\bibitem[{{George} {et~al.}(2015){George}, {Reichardt}, {Aird}, {Benson},
  {Bleem}, {Carlstrom}, {Chang}, {Cho}, {Crawford}, {Crites}, {de Haan},
  {Dobbs}, {Dudley}, {Halverson}, {Harrington}, {Holder}, {Holzapfel}, {Hou},
  {Hrubes}, {Keisler}, {Knox}, {Lee}, {Leitch}, {Lueker}, {Luong-Van},
  {McMahon}, {Mehl}, {Meyer}, {Millea}, {Mocanu}, {Mohr}, {Montroy}, {Padin},
  {Plagge}, {Pryke}, {Ruhl}, {Schaffer}, {Shaw}, {Shirokoff}, {Spieler},
  {Staniszewski}, {Stark}, {Story}, {van Engelen}, {Vanderlinde}, {Vieira},
  {Williamson}, \& {Zahn}}]{George15}
{George}, E.~M., {Reichardt}, C.~L., {Aird}, K.~A., {et~al.}, {A Measurement of
  Secondary Cosmic Microwave Background Anisotropies from the 2500
  Square-degree SPT-SZ Survey}. 2015, \apj, 799, 177, \eprint{1408.3161}

\bibitem[{{Ghosh} {et~al.}(2017){Ghosh}, {Boulanger}, {Martin}, {Bracco},
  {Vansyngel}, {Aumont}, {Bock}, {Dor{\'e}}, {Haud}, {Kalberla}, \&
  {Serra}}]{Ghosh2017}
{Ghosh}, T., {Boulanger}, F., {Martin}, P.~G., {et~al.}, {Modelling and
  simulation of large-scale polarized dust emission over the southern Galactic
  cap using the GASS Hi data}. 2017, \aap, 601, A71, \eprint{1611.02418}

\bibitem[{{Giannantonio} {et~al.}(2016){Giannantonio}, {Fosalba}, {Cawthon},
  {Omori}, {Crocce}, {Elsner}, {Leistedt}, {Dodelson}, {Benoit-L{\'e}vy},
  {Gazta{\~n}aga}, {Holder}, {Peiris}, {Percival}, {Kirk}, {Bauer}, {Benson},
  {Bernstein}, {Carretero}, {Crawford}, {Crittenden}, {Huterer}, {Jain},
  {Krause}, {Reichardt}, {Ross}, {Simard}, {Soergel}, {Stark}, {Story},
  {Vieira}, {Weller}, {Abbott}, {Abdalla}, {Allam}, {Armstrong}, {Banerji},
  {Bernstein}, {Bertin}, {Brooks}, {Buckley-Geer}, {Burke}, {Capozzi},
  {Carlstrom}, {Carnero Rosell}, {Carrasco Kind}, {Castander}, {Chang},
  {Cunha}, {da Costa}, {D'Andrea}, {DePoy}, {Desai}, {Diehl}, {Dietrich},
  {Doel}, {Eifler}, {Evrard}, {Neto}, {Fernandez}, {Finley}, {Flaugher},
  {Frieman}, {Gerdes}, {Gruen}, {Gruendl}, {Gutierrez}, {Holzapfel},
  {Honscheid}, {James}, {Kuehn}, {Kuropatkin}, {Lahav}, {Li}, {Lima}, {March},
  {Marshall}, {Martini}, {Melchior}, {Miquel}, {Mohr}, {Nichol}, {Nord},
  {Ogando}, {Plazas}, {Romer}, {Roodman}, {Rykoff}, {Sako}, {Saliwanchik},
  {Sanchez}, {Schubnell}, {Sevilla-Noarbe}, {Smith}, {Soares-Santos},
  {Sobreira}, {Suchyta}, {Swanson}, {Tarle}, {Thaler}, {Thomas}, {Vikram},
  {Walker}, {Wechsler}, \& {Zuntz}}]{Giannantonio16}
{Giannantonio}, T., {Fosalba}, P., {Cawthon}, R., {et~al.}, {CMB lensing
  tomography with the DES Science Verification galaxies}. 2016, \mnras, 456,
  3213, \eprint{1507.05551}

\bibitem[{{Gispert} {et~al.}(2000){Gispert}, {Lagache}, \&
  {Puget}}]{gispert_2000}
{Gispert}, R., {Lagache}, G., \& {Puget}, J.~L., {Implications of the cosmic
  infrared background for light production and the star formation history in
  the Universe}. 2000, \aap, 360, 1, \eprint{astro-ph/0005554}

\bibitem[{{Giusarma} {et~al.}(2018){Giusarma}, {Vagnozzi}, {Ho}, {Ferraro},
  {Freese}, {Kamen-Rubio}, \& {Luk}}]{Giusarma18}
{Giusarma}, E., {Vagnozzi}, S., {Ho}, S., {et~al.}, {Scale-dependent galaxy
  bias, CMB lensing-galaxy cross-correlation, and neutrino masses}. 2018, ArXiv
  e-prints, \eprint{1802.08694}

\bibitem[{{G{\'o}rski} {et~al.}(2005){G{\'o}rski}, {Hivon}, {Banday},
  {Wandelt}, {Hansen}, {Reinecke}, \& {Bartelmann}}]{gorski2005}
{G{\'o}rski}, K.~M., {Hivon}, E., {Banday}, A.~J., {et~al.}, {HEALPix: A
  Framework for High-Resolution Discretization and Fast Analysis of Data
  Distributed on the Sphere}. 2005, \apj, 622, 759, \eprint{astro-ph/0409513}

\bibitem[{{Greig} {et~al.}(2017){Greig}, {Mesinger}, {Haiman}, \&
  {Simcoe}}]{Greig17}
{Greig}, B., {Mesinger}, A., {Haiman}, Z., \& {Simcoe}, R.~A., {Are we
  witnessing the epoch of reionisation at z = 7.1 from the spectrum of
  J1120+0641?} 2017, \mnras, 466, 4239, \eprint{1606.00441}

\bibitem[{{Grossan} \& {Smoot}(2007)}]{Grossan_2007}
{Grossan}, B. \& {Smoot}, G.~F., {Power spectrum analysis of far-IR background
  fluctuations in 160 {$\mu$}m maps from the multiband imaging photometer for
  Spitzer}. 2007, \aap, 474, 731, \eprint{astro-ph/0604512}

\bibitem[{{Halverson} {et~al.}(2002){Halverson}, {Leitch}, {Pryke}, {Kovac},
  {Carlstrom}, {Holzapfel}, {Dragovan}, {Cartwright}, {Mason}, {Padin},
  {Pearson}, {Readhead}, \& {Shepherd}}]{DASI_data}
{Halverson}, N.~W., {Leitch}, E.~M., {Pryke}, C., {et~al.}, {Degree Angular
  Scale Interferometer First Results: A Measurement of the Cosmic Microwave
  Background Angular Power Spectrum}. 2002, \apj, 568, 38,
  \eprint{astro-ph/0104489}

\bibitem[{{Hamilton}(1998)}]{Hamilton98}
{Hamilton}, A.~J.~S. 1998, in Astrophysics and Space Science Library, Vol. 231,
  The Evolving Universe, ed. D.~{Hamilton}, 185

\bibitem[{{Hanany} {et~al.}(2000){Hanany}, {Ade}, {Balbi}, {Bock}, {Borrill},
  {Boscaleri}, {de Bernardis}, {Ferreira}, {Hristov}, {Jaffe}, {Lange}, {Lee},
  {Mauskopf}, {Netterfield}, {Oh}, {Pascale}, {Rabii}, {Richards}, {Smoot},
  {Stompor}, {Winant}, \& {Wu}}]{MAXIMA_data}
{Hanany}, S., {Ade}, P., {Balbi}, A., {et~al.}, {MAXIMA-1: A Measurement of the
  Cosmic Microwave Background Anisotropy on Angular Scales of 10'-5{\deg}}.
  2000, \apjl, 545, L5, \eprint{astro-ph/0005123}

\bibitem[{{Hancock} \& {Rocha}(1997)}]{Hancock97}
{Hancock}, S. \& {Rocha}, G. 1997, in Microwave Background Anisotropies, ed.
  F.~R. {Bouchet}, R.~{Gispert}, B.~{Guiderdoni}, \& J.~{Tr{\^a}n Thanh
  V{\^a}n}, 179--188

\bibitem[{{Hanson} {et~al.}(2010){Hanson}, {Challinor}, \& {Lewis}}]{Hanson10}
{Hanson}, D., {Challinor}, A., \& {Lewis}, A., {Weak lensing of the CMB}. 2010,
  General Relativity and Gravitation, 42, 2197, \eprint{0911.0612}

\bibitem[{{Harnois-D{\'e}raps} {et~al.}(2017){Harnois-D{\'e}raps},
  {Tr{\"o}ster}, {Chisari}, {Heymans}, {van Waerbeke}, {Asgari}, {Bilicki},
  {Choi}, {Erben}, {Hildebrandt}, {Hoekstra}, {Joudaki}, {Kuijken}, {Merten},
  {Miller}, {Robertson}, {Schneider}, \& {Viola}}]{Har17}
{Harnois-D{\'e}raps}, J., {Tr{\"o}ster}, T., {Chisari}, N.~E., {et~al.},
  {KiDS-450: tomographic cross-correlation of galaxy shear with Planck
  lensing}. 2017, \mnras, 471, 1619, \eprint{1703.03383}

\bibitem[{{Harnois-D{\'e}raps} {et~al.}(2016){Harnois-D{\'e}raps},
  {Tr{\"o}ster}, {Hojjati}, {van Waerbeke}, {Asgari}, {Choi}, {Erben},
  {Heymans}, {Hildebrandt}, {Kitching}, {Miller}, {Nakajima}, {Viola},
  {Arnouts}, {Coupon}, \& {Moutard}}]{Har16}
{Harnois-D{\'e}raps}, J., {Tr{\"o}ster}, T., {Hojjati}, A., {et~al.}, {CFHTLenS
  and RCSLenS cross-correlation with Planck lensing detected in fourier and
  configuration space}. 2016, \mnras, 460, 434, \eprint{1603.07723}

\bibitem[{{Haslam} {et~al.}(1982){Haslam}, {Salter}, {Stoffel}, \&
  {Wilson}}]{haslam1982}
{Haslam}, C.~G.~T., {Salter}, C.~J., {Stoffel}, H., \& {Wilson}, W.~E., {A 408
  MHz all-sky continuum survey. II - The atlas of contour maps}. 1982, \aaps,
  47, 1

\bibitem[{{Henning} {et~al.}(2018){Henning}, {Sayre}, {Reichardt}, {Ade},
  {Anderson}, {Austermann}, {Beall}, {Bender}, {Benson}, {Bleem}, {Carlstrom},
  {Chang}, {Chiang}, {Cho}, {Citron}, {Corbett Moran}, {Crawford}, {Crites},
  {de Haan}, {Dobbs}, {Everett}, {Gallicchio}, {George}, {Gilbert},
  {Halverson}, {Harrington}, {Hilton}, {Holder}, {Holzapfel}, {Hoover}, {Hou},
  {Hrubes}, {Huang}, {Hubmayr}, {Irwin}, {Keisler}, {Knox}, {Lee}, {Leitch},
  {Li}, {Lowitz}, {Manzotti}, {McMahon}, {Meyer}, {Mocanu}, {Montgomery},
  {Nadolski}, {Natoli}, {Nibarger}, {Novosad}, {Padin}, {Pryke}, {Ruhl},
  {Saliwanchik}, {Schaffer}, {Sievers}, {Smecher}, {Stark}, {Story}, {Tucker},
  {Vanderlinde}, {Veach}, {Vieira}, {Wang}, {Whitehorn}, {Wu}, \&
  {Yefremenko}}]{Henning18}
{Henning}, J.~W., {Sayre}, J.~T., {Reichardt}, C.~L., {et~al.}, {Measurements
  of the Temperature and E-mode Polarization of the CMB from 500 Square Degrees
  of SPTpol Data}. 2018, \apj, 852, 97, \eprint{1707.09353}

\bibitem[{{Hildebrandt} {et~al.}(2017){Hildebrandt}, {Viola}, {Heymans},
  {Joudaki}, {Kuijken}, {Blake}, {Erben}, {Joachimi}, {Klaes}, {Miller},
  {Morrison}, {Nakajima}, {Verdoes Kleijn}, {Amon}, {Choi}, {Covone}, {de
  Jong}, {Dvornik}, {Fenech Conti}, {Grado}, {Harnois-D{\'e}raps}, {Herbonnet},
  {Hoekstra}, {K{\"o}hlinger}, {McFarland}, {Mead}, {Merten}, {Napolitano},
  {Peacock}, {Radovich}, {Schneider}, {Simon}, {Valentijn}, {van den Busch},
  {van Uitert}, \& {Van Waerbeke}}]{Hildebrandt17}
{Hildebrandt}, H., {Viola}, M., {Heymans}, C., {et~al.}, {KiDS-450:
  cosmological parameter constraints from tomographic weak gravitational
  lensing}. 2017, \mnras, 465, 1454, \eprint{1606.05338}

\bibitem[{{Hill} {et~al.}(2016){Hill}, {Ferraro}, {Battaglia}, {Liu}, \&
  {Spergel}}]{hill2016}
{Hill}, J.~C., {Ferraro}, S., {Battaglia}, N., {Liu}, J., \& {Spergel}, D.~N.,
  {Kinematic Sunyaev-Zel'dovich Effect with Projected Fields: A Novel Probe of
  the Baryon Distribution with Planck, WMAP, and WISE Data}. 2016, Physical
  Review Letters, 117, 051301, \eprint{1603.01608}

\bibitem[{{Hill} {et~al.}(2018){Hill}, {Masui}, \& {Scott}}]{hill2018}
{Hill}, R., {Masui}, K.~W., \& {Scott}, D., {The Spectrum of the Universe}.
  2018, Applied Spectroscopy, 72, 663, \eprint{1802.03694}

\bibitem[{{Hinshaw} {et~al.}(2013){Hinshaw}, {Larson}, {Komatsu}, {Spergel},
  {Bennett}, {Dunkley}, {Nolta}, {Halpern}, {Hill}, {Odegard}, {Page}, {Smith},
  {Weiland}, {Gold}, {Jarosik}, {Kogut}, {Limon}, {Meyer}, {Tucker}, {Wollack},
  \& {Wright}}]{hinshaw2012}
{Hinshaw}, G., {Larson}, D., {Komatsu}, E., {et~al.}, {Nine-year Wilkinson
  Microwave Anisotropy Probe (WMAP) Observations: Cosmological Parameter
  Results}. 2013, \apjs, 208, 19, \eprint{1212.5226}

\bibitem[{{Hinshaw} {et~al.}(2007){Hinshaw}, {Nolta}, {Bennett}, {Bean},
  {Dor{\'e}}, {Greason}, {Halpern}, {Hill}, {Jarosik}, {Kogut}, {Komatsu},
  {Limon}, {Odegard}, {Meyer}, {Page}, {Peiris}, {Spergel}, {Tucker}, {Verde},
  {Weiland}, {Wollack}, \& {Wright}}]{hinshaw2007}
{Hinshaw}, G., {Nolta}, M.~R., {Bennett}, C.~L., {et~al.}, {Three-Year
  Wilkinson Microwave Anisotropy Probe (WMAP) Observations: Temperature
  Analysis}. 2007, \apjs, 170, 288, \eprint{astro-ph/0603451}

\bibitem[{{Hinshaw} {et~al.}(2009){Hinshaw}, {Weiland}, {Hill}, {Odegard},
  {Larson}, {Bennett}, {Dunkley}, {Gold}, {Greason}, {Jarosik}, {Komatsu},
  {Nolta}, {Page}, {Spergel}, {Wollack}, {Halpern}, {Kogut}, {Limon}, {Meyer},
  {Tucker}, \& {Wright}}]{hinshaw2009}
{Hinshaw}, G., {Weiland}, J.~L., {Hill}, R.~S., {et~al.}, {Five-Year Wilkinson
  Microwave Anisotropy Probe (WMAP) Observations: Data Processing, Sky Maps,
  and Basic Results}. 2009, \apjs, 180, 225, \eprint{0803.0732}

\bibitem[{Hirata \& Seljak(2003)}]{Hirata:2002jy}
Hirata, C.~M. \& Seljak, U., {Analyzing weak lensing of the cosmic microwave
  background using the likelihood function}. 2003, Phys. Rev., D67, 043001,
  \eprint{astro-ph/0209489}

\bibitem[{{Hivon} {et~al.}(2017){Hivon}, {Mottet}, \&
  {Ponthieu}}]{2017A&A...598A..25H}
{Hivon}, E., {Mottet}, S., \& {Ponthieu}, N., {QuickPol: Fast calculation of
  effective beam matrices for CMB polarization}. 2017, \aap, 598, A25

\bibitem[{{Holder} {et~al.}(2013){Holder}, {Viero}, {Zahn}, {Aird}, {Benson},
  {Bhattacharya}, {Bleem}, {Bock}, {Brodwin}, {Carlstrom}, {Chang}, {Cho},
  {Conley}, {Crawford}, {Crites}, {de Haan}, {Dobbs}, {Dudley}, {George},
  {Halverson}, {Holzapfel}, {Hoover}, {Hou}, {Hrubes}, {Keisler}, {Knox},
  {Lee}, {Leitch}, {Lueker}, {Luong-Van}, {Marsden}, {Marrone}, {McMahon},
  {Mehl}, {Meyer}, {Millea}, {Mohr}, {Montroy}, {Padin}, {Plagge}, {Pryke},
  {Reichardt}, {Ruhl}, {Sayre}, {Schaffer}, {Schulz}, {Shaw}, {Shirokoff},
  {Spieler}, {Staniszewski}, {Stark}, {Story}, {van Engelen}, {Vanderlinde},
  {Vieira}, {Williamson}, \& {Zemcov}}]{Holder_2013}
{Holder}, G.~P., {Viero}, M.~P., {Zahn}, O., {et~al.}, {A Cosmic Microwave
  Background Lensing Mass Map and Its Correlation with the Cosmic Infrared
  Background}. 2013, \apjl, 771, L16, \eprint{1303.5048}

\bibitem[{{Horowitz} \& {Seljak}(2017)}]{HorSel17}
{Horowitz}, B. \& {Seljak}, U., {Cosmological constraints from thermal
  Sunyaev-Zeldovich power spectrum revisited}. 2017, \mnras, 469, 394,
  \eprint{1609.01850}

\bibitem[{{Hou} {et~al.}(2018){Hou}, {Aylor}, {Benson}, {Bleem}, {Carlstrom},
  {Chang}, {Cho}, {Chown}, {Crawford}, {Crites}, {de Haan}, {Dobbs}, {Everett},
  {Follin}, {George}, {Halverson}, {Harrington}, {Holder}, {Holzapfel},
  {Hrubes}, {Keisler}, {Knox}, {Lee}, {Leitch}, {Luong-Van}, {Marrone},
  {McMahon}, {Meyer}, {Millea}, {Mocanu}, {Mohr}, {Natoli}, {Omori}, {Padin},
  {Pryke}, {Reichardt}, {Ruhl}, {Sayre}, {Schaffer}, {Shirokoff},
  {Staniszewski}, {Stark}, {Story}, {Vanderlinde}, {Vieira}, \&
  {Williamson}}]{Hou18}
{Hou}, Z., {Aylor}, K., {Benson}, B.~A., {et~al.}, {A Comparison of Maps and
  Power Spectra Determined from South Pole Telescope and Planck Data}. 2018,
  \apj, 853, 3, \eprint{1704.00884}

\bibitem[{Howlett {et~al.}(2015)Howlett, Ross, Samushia, Percival, \&
  Manera}]{Howlett:2014opa}
Howlett, C., Ross, A., Samushia, L., Percival, W., \& Manera, M., {The
  clustering of the SDSS main galaxy sample – II. Mock galaxy catalogues and
  a measurement of the growth of structure from redshift space distortions at
  $z = 0.15$}. 2015, \mnras, 449, 848, \eprint{1409.3238}

\bibitem[{{Hu}(1998)}]{Hu98}
{Hu}, W., {Structure Formation with Generalized Dark Matter}. 1998, \apj, 506,
  485, \eprint{astro-ph/9801234}

\bibitem[{Hu \& Okamoto(2002)}]{Hu:2001kj}
Hu, W. \& Okamoto, T., {Mass Reconstruction with CMB Polarization}. 2002, \apj,
  574, 566, \eprint{astro-ph/0111606}

\bibitem[{{Hu} {et~al.}(1995){Hu}, {Scott}, {Sugiyama}, \& {White}}]{HSSW95}
{Hu}, W., {Scott}, D., {Sugiyama}, N., \& {White}, M., {Effect of physical
  assumptions on the calculation of microwave background anisotropies}. 1995,
  \prd, 52, 5498, \eprint{astro-ph/9505043}

\bibitem[{{Hu} {et~al.}(1997){Hu}, {Spergel}, \& {White}}]{HuWhiSpe97}
{Hu}, W., {Spergel}, D.~N., \& {White}, M., {Distinguishing causal seeds from
  inflation}. 1997, \prd, 55, 3288, \eprint{astro-ph/9605193}

\bibitem[{{Hu} \& {White}(1996{\natexlab{a}})}]{TestInf}
{Hu}, W. \& {White}, M., {A New Test of Inflation}. 1996{\natexlab{a}},
  Physical Review Letters, 77, 1687, \eprint{astro-ph/9602020}

\bibitem[{{Hu} \& {White}(1996{\natexlab{b}})}]{HuWhi96}
{Hu}, W. \& {White}, M., {Acoustic Signatures in the Cosmic Microwave
  Background}. 1996{\natexlab{b}}, \apj, 471, 30, \eprint{astro-ph/9602019}

\bibitem[{{Hu} \& {White}(1997)}]{HuWhi97}
{Hu}, W. \& {White}, M., {A CMB polarization primer}. 1997, \na, 2, 323,
  \eprint{astro-ph/9706147}

\bibitem[{Huterer {et~al.}(2017)Huterer, Shafer, Scolnic, \&
  Schmidt}]{Huterer:2016uyq}
Huterer, D., Shafer, D., Scolnic, D., \& Schmidt, F., {Testing $\Lambda$CDM at
  the lowest redshifts with SN Ia and galaxy velocities}. 2017, \jcap, 1705,
  015, \eprint{1611.09862}

\bibitem[{{Jain} \& {Khoury}(2010)}]{JaiKho10}
{Jain}, B. \& {Khoury}, J., {Cosmological tests of gravity}. 2010, Annals of
  Physics, 325, 1479, \eprint{1004.3294}

\bibitem[{{Jeong} {et~al.}(2014){Jeong}, {Chluba}, {Dai}, {Kamionkowski}, \&
  {Wang}}]{Jeong14}
{Jeong}, D., {Chluba}, J., {Dai}, L., {Kamionkowski}, M., \& {Wang}, X.,
  {Effect of aberration on partial-sky measurements of the cosmic microwave
  background temperature power spectrum}. 2014, \prd, 89, 023003,
  \eprint{1309.2285}

\bibitem[{{Jones} {et~al.}(2006){Jones}, {Ade}, {Bock}, {Bond}, {Borrill},
  {Boscaleri}, {Cabella}, {Contaldi}, {Crill}, {de Bernardis}, {De Gasperis},
  {de Oliveira-Costa}, {De Troia}, {di Stefano}, {Hivon}, {Jaffe}, {Kisner},
  {Lange}, {MacTavish}, {Masi}, {Mauskopf}, {Melchiorri}, {Montroy}, {Natoli},
  {Netterfield}, {Pascale}, {Piacentini}, {Pogosyan}, {Polenta}, {Prunet},
  {Ricciardi}, {Romeo}, {Ruhl}, {Santini}, {Tegmark}, {Veneziani}, \&
  {Vittorio}}]{Jones06}
{Jones}, W.~C., {Ade}, P.~A.~R., {Bock}, J.~J., {et~al.}, {A Measurement of the
  Angular Power Spectrum of the CMB Temperature Anisotropy from the 2003 Flight
  of BOOMERANG}. 2006, \apj, 647, 823, \eprint{astro-ph/0507494}

\bibitem[{{Joudaki} {et~al.}(2018){Joudaki}, {Blake}, {Johnson}, {Amon},
  {Asgari}, {Choi}, {Erben}, {Glazebrook}, {Harnois-D{\'e}raps}, {Heymans},
  {Hildebrandt}, {Hoekstra}, {Klaes}, {Kuijken}, {Lidman}, {Mead}, {Miller},
  {Parkinson}, {Poole}, {Schneider}, {Viola}, \& {Wolf}}]{Joudaki18}
{Joudaki}, S., {Blake}, C., {Johnson}, A., {et~al.}, {KiDS-450 + 2dFLenS:
  Cosmological parameter constraints from weak gravitational lensing tomography
  and overlapping redshift-space galaxy clustering}. 2018, \mnras, 474, 4894,
  \eprint{1707.06627}

\bibitem[{{Joyce} {et~al.}(2015){Joyce}, {Jain}, {Khoury}, \&
  {Trodden}}]{Joyce15}
{Joyce}, A., {Jain}, B., {Khoury}, J., \& {Trodden}, M., {Beyond the
  cosmological standard model}. 2015, \physrep, 568, 1, \eprint{1407.0059}

\bibitem[{{Joyce} {et~al.}(2016){Joyce}, {Lombriser}, \& {Schmidt}}]{Joyce16}
{Joyce}, A., {Lombriser}, L., \& {Schmidt}, F., {Dark Energy Versus Modified
  Gravity}. 2016, Annual Review of Nuclear and Particle Science, 66, 95,
  \eprint{1601.06133}

\bibitem[{{Kamionkowski} \& {Knox}(2003)}]{KamKnox03}
{Kamionkowski}, M. \& {Knox}, L., {Aspects of the cosmic microwave background
  dipole}. 2003, \prd, 67, 063001, \eprint{astro-ph/0210165}

\bibitem[{{Kamionkowski} {et~al.}(1997){Kamionkowski}, {Kosowsky}, \&
  {Stebbins}}]{Kam97}
{Kamionkowski}, M., {Kosowsky}, A., \& {Stebbins}, A., {Statistics of cosmic
  microwave background polarization}. 1997, \prd, 55, 7368,
  \eprint{astro-ph/9611125}

\bibitem[{{Kashlinsky} {et~al.}(2008){Kashlinsky}, {Atrio-Barandela},
  {Kocevski}, \& {Ebeling}}]{Kashlinsky08}
{Kashlinsky}, A., {Atrio-Barandela}, F., {Kocevski}, D., \& {Ebeling}, H., {A
  Measurement of Large-Scale Peculiar Velocities of Clusters of Galaxies:
  Results and Cosmological Implications}. 2008, \apjl, 686, L49,
  \eprint{0809.3734}

\bibitem[{{Kazin} {et~al.}(2014){Kazin}, {Koda}, {Blake}, {Padmanabhan},
  {Brough}, {Colless}, {Contreras}, {Couch}, {Croom}, {Croton}, {Davis},
  {Drinkwater}, {Forster}, {Gilbank}, {Gladders}, {Glazebrook}, {Jelliffe},
  {Jurek}, {Li}, {Madore}, {Martin}, {Pimbblet}, {Poole}, {Pracy}, {Sharp},
  {Wisnioski}, {Woods}, {Wyder}, \& {Yee}}]{Kazin14}
{Kazin}, E.~A., {Koda}, J., {Blake}, C., {et~al.}, {The WiggleZ Dark Energy
  Survey: improved distance measurements to z = 1 with reconstruction of the
  baryonic acoustic feature}. 2014, \mnras, 441, 3524, \eprint{1401.0358}

\bibitem[{{Keisler} {et~al.}(2015){Keisler}, {Hoover}, {Harrington}, {Henning},
  {Ade}, {Aird}, {Austermann}, {Beall}, {Bender}, {Benson}, {Bleem},
  {Carlstrom}, {Chang}, {Chiang}, {Cho}, {Citron}, {Crawford}, {Crites}, {de
  Haan}, {Dobbs}, {Everett}, {Gallicchio}, {Gao}, {George}, {Gilbert},
  {Halverson}, {Hanson}, {Hilton}, {Holder}, {Holzapfel}, {Hou}, {Hrubes},
  {Huang}, {Hubmayr}, {Irwin}, {Knox}, {Lee}, {Leitch}, {Li}, {Luong-Van},
  {Marrone}, {McMahon}, {Mehl}, {Meyer}, {Mocanu}, {Natoli}, {Nibarger},
  {Novosad}, {Padin}, {Pryke}, {Reichardt}, {Ruhl}, {Saliwanchik}, {Sayre},
  {Schaffer}, {Shirokoff}, {Smecher}, {Stark}, {Story}, {Tucker},
  {Vanderlinde}, {Vieira}, {Wang}, {Whitehorn}, {Yefremenko}, \&
  {Zahn}}]{Keisler15}
{Keisler}, R., {Hoover}, S., {Harrington}, N., {et~al.}, {Measurements of
  Sub-degree B-mode Polarization in the Cosmic Microwave Background from 100
  Square Degrees of SPTpol Data}. 2015, \apj, 807, 151, \eprint{1503.02315}

\bibitem[{{Kirk} {et~al.}(2016){Kirk}, {Omori}, {Benoit-L{\'e}vy}, {Cawthon},
  {Chang}, {Larsen}, {Amara}, {Bacon}, {Crawford}, {Dodelson}, {Fosalba},
  {Giannantonio}, {Holder}, {Jain}, {Kacprzak}, {Lahav}, {MacCrann}, {Nicola},
  {Refregier}, {Sheldon}, {Story}, {Troxel}, {Vieira}, {Vikram}, {Zuntz},
  {Abbott}, {Abdalla}, {Becker}, {Benson}, {Bernstein}, {Bernstein}, {Bleem},
  {Bonnett}, {Bridle}, {Brooks}, {Buckley-Geer}, {Burke}, {Capozzi},
  {Carlstrom}, {Rosell}, {Kind}, {Carretero}, {Crocce}, {Cunha}, {D'Andrea},
  {da Costa}, {Desai}, {Diehl}, {Dietrich}, {Doel}, {Eifler}, {Evrard},
  {Flaugher}, {Frieman}, {Gerdes}, {Goldstein}, {Gruen}, {Gruendl},
  {Honscheid}, {James}, {Jarvis}, {Kent}, {Kuehn}, {Kuropatkin}, {Lima},
  {March}, {Martini}, {Melchior}, {Miller}, {Miquel}, {Nichol}, {Ogando},
  {Plazas}, {Reichardt}, {Roodman}, {Rozo}, {Rykoff}, {Sako}, {Sanchez},
  {Scarpine}, {Schubnell}, {Sevilla-Noarbe}, {Simard}, {Smith},
  {Soares-Santos}, {Sobreira}, {Suchyta}, {Swanson}, {Tarle}, {Thomas},
  {Wechsler}, \& {Weller}}]{Kirk16}
{Kirk}, D., {Omori}, Y., {Benoit-L{\'e}vy}, A., {et~al.}, {Cross-correlation of
  gravitational lensing from DES Science Verification data with SPT and Planck
  lensing}. 2016, \mnras, 459, 21, \eprint{1512.04535}

\bibitem[{{Klein} \& {Roodman}(2005)}]{Klein2005}
{Klein}, J.~R. \& {Roodman}, A., {Blind Analysis in Nuclear and Particle
  Physics}. 2005, Annual Review of Nuclear and Particle Science, 55, 141

\bibitem[{{Knox} {et~al.}(2001){Knox}, {Cooray}, {Eisenstein}, \&
  {Haiman}}]{Knox_2001}
{Knox}, L., {Cooray}, A., {Eisenstein}, D., \& {Haiman}, Z., {Probing Early
  Structure Formation with Far-Infrared Background Correlations}. 2001, \apj,
  550, 7, \eprint{astro-ph/0009151}

\bibitem[{{Knox} \& {Page}(2000{\natexlab{a}})}]{KnoxPage00}
{Knox}, L. \& {Page}, L., {Characterizing the Peak in the Cosmic Microwave
  Background Angular Power Spectrum}. 2000{\natexlab{a}}, Physical Review
  Letters, 85, 1366, \eprint{astro-ph/0002162}

\bibitem[{{Knox} \& {Page}(2000{\natexlab{b}})}]{Knox00}
{Knox}, L. \& {Page}, L., {Characterizing the Peak in the Cosmic Microwave
  Background Angular Power Spectrum}. 2000{\natexlab{b}}, Physical Review
  Letters, 85, 1366, \eprint{astro-ph/0002162}

\bibitem[{{Kogut} {et~al.}(1993){Kogut}, {Lineweaver}, {Smoot}, {Bennett},
  {Banday}, {Boggess}, {Cheng}, {de Amici}, {Fixsen}, {Hinshaw}, {Jackson},
  {Janssen}, {Keegstra}, {Loewenstein}, {Lubin}, {Mather}, {Tenorio}, {Weiss},
  {Wilkinson}, \& {Wright}}]{kogut1993}
{Kogut}, A., {Lineweaver}, C., {Smoot}, G.~F., {et~al.}, {Dipole Anisotropy in
  the COBE Differential Microwave Radiometers First-Year Sky Maps}. 1993, \apj,
  419, 1, \eprint{astro-ph/9312056}

\bibitem[{{Kogut} {et~al.}(2003){Kogut}, {Spergel}, {Barnes}, {Bennett},
  {Halpern}, {Hinshaw}, {Jarosik}, {Limon}, {Meyer}, {Page}, {Tucker},
  {Wollack}, \& {Wright}}]{kogut2003}
{Kogut}, A., {Spergel}, D.~N., {Barnes}, C., {et~al.}, {First-Year Wilkinson
  Microwave Anisotropy Probe (WMAP) Observations: Temperature-Polarization
  Correlation}. 2003, \apjs, 148, 161, \eprint{astro-ph/0302213}

\bibitem[{{K{\"o}hlinger} {et~al.}(2017){K{\"o}hlinger}, {Viola}, {Joachimi},
  {Hoekstra}, {van Uitert}, {Hildebrandt}, {Choi}, {Erben}, {Heymans},
  {Joudaki}, {Klaes}, {Kuijken}, {Merten}, {Miller}, {Schneider}, \&
  {Valentijn}}]{Kohlinger17}
{K{\"o}hlinger}, F., {Viola}, M., {Joachimi}, B., {et~al.}, {KiDS-450: the
  tomographic weak lensing power spectrum and constraints on cosmological
  parameters}. 2017, \mnras, 471, 4412, \eprint{1706.02892}

\bibitem[{{Komatsu} {et~al.}(2009){Komatsu}, {Dunkley}, {Nolta}, {Bennett},
  {Gold}, {Hinshaw}, {Jarosik}, {Larson}, {Limon}, {Page}, {Spergel},
  {Halpern}, {Hill}, {Kogut}, {Meyer}, {Tucker}, {Weiland}, {Wollack}, \&
  {Wright}}]{komatsu2009}
{Komatsu}, E., {Dunkley}, J., {Nolta}, M.~R., {et~al.}, {Five-Year Wilkinson
  Microwave Anisotropy Probe (WMAP) Observations: Cosmological Interpretation}.
  2009, \apjs, 180, 330, \eprint{0803.0547}

\bibitem[{{Krauss} \& {Turner}(1995)}]{KraTur95}
{Krauss}, L.~M. \& {Turner}, M.~S., {The cosmological constant is back}. 1995,
  General Relativity and Gravitation, 27, 1137, \eprint{astro-ph/9504003}

\bibitem[{{Lagache} {et~al.}(2007){Lagache}, {Bavouzet}, {Fernandez-Conde},
  {Ponthieu}, {Rodet}, {Dole}, {Miville-Desch{\^e}nes}, \&
  {Puget}}]{Lagache_2007}
{Lagache}, G., {Bavouzet}, N., {Fernandez-Conde}, N., {et~al.}, {Correlated
  Anisotropies in the Cosmic Far-Infrared Background Detected by the Multiband
  Imaging Photometer for Spitzer: Constraint on the Bias}. 2007, \apjl, 665,
  L89, \eprint{0707.2443}

\bibitem[{{Lagache} \& {Puget}(2000)}]{Lagache_2000}
{Lagache}, G. \& {Puget}, J.~L., {Detection of the extra-Galactic background
  fluctuations at 170 mu m}. 2000, \aap, 355, 17, \eprint{astro-ph/9910255}

\bibitem[{{Lattanzi} \& {Gerbino}(2017)}]{LatGer17}
{Lattanzi}, M. \& {Gerbino}, M., {Status of neutrino properties and future
  prospects - Cosmological and astrophysical constraints}. 2017, ArXiv
  e-prints, \eprint{1712.07109}

\bibitem[{{Leach} {et~al.}(2008){Leach}, {Cardoso}, {Baccigalupi}, {Barreiro},
  {Betoule}, {Bobin}, {Bonaldi}, {Delabrouille}, {de Zotti}, {Dickinson},
  {Eriksen}, {Gonz{\'a}lez-Nuevo}, {Hansen}, {Herranz}, {Le Jeune},
  {L{\'o}pez-Caniego}, {Mart{\'{\i}}nez-Gonz{\'a}lez}, {Massardi}, {Melin},
  {Miville-Desch{\^e}nes}, {Patanchon}, {Prunet}, {Ricciardi}, {Salerno},
  {Sanz}, {Starck}, {Stivoli}, {Stolyarov}, {Stompor}, \& {Vielva}}]{leach2008}
{Leach}, S.~M., {Cardoso}, J., {Baccigalupi}, C., {et~al.}, {Component
  separation methods for the PLANCK mission}. 2008, \aap, 491, 597,
  \eprint{0805.0269}

\bibitem[{{Lesgourgues} {et~al.}(2013){Lesgourgues}, {Mangano}, {Miele}, \&
  {Pastor}}]{Les13}
{Lesgourgues}, J., {Mangano}, G., {Miele}, G., \& {Pastor}, S. 2013, {Neutrino
  Cosmology} (Cambridge University Press)

\bibitem[{{Lewis} \& {Challinor}(2006)}]{LewisChallinor06}
{Lewis}, A. \& {Challinor}, A., {Weak gravitational lensing of the CMB}. 2006,
  \physrep, 429, 1, \eprint{astro-ph/0601594}

\bibitem[{{Liddle} {et~al.}(1996){Liddle}, {Lyth}, {Viana}, \&
  {White}}]{LLVW96}
{Liddle}, A.~R., {Lyth}, D.~H., {Viana}, P.~T.~P., \& {White}, M., {Cold dark
  matter models with a cosmological constant}. 1996, \mnras, 282, 281,
  \eprint{astro-ph/9512102}

\bibitem[{{Lineweaver} {et~al.}(1996){Lineweaver}, {Tenorio}, {Smoot},
  {Keegstra}, {Banday}, \& {Lubin}}]{lineweaver1996}
{Lineweaver}, C.~H., {Tenorio}, L., {Smoot}, G.~F., {et~al.}, {The Dipole
  Observed in the COBE DMR 4 Year Data}. 1996, \apj, 470, 38,
  \eprint{astro-ph/9601151}

\bibitem[{{Liu} \& {Hill}(2015)}]{LiuHill15}
{Liu}, J. \& {Hill}, J.~C., {Cross-correlation of Planck CMB lensing and
  CFHTLenS galaxy weak lensing maps}. 2015, \prd, 92, 063517,
  \eprint{1504.05598}

\bibitem[{Lombriser \& Taylor(2016)}]{Lombriser15}
Lombriser, L. \& Taylor, A., {Breaking a Dark Degeneracy with Gravitational
  Waves}. 2016, JCAP, 1603, 031, \eprint{1509.08458}

\bibitem[{{Louis} {et~al.}(2017){Louis}, {Grace}, {Hasselfield}, {Lungu},
  {Maurin}, {Addison}, {Ade}, {Aiola}, {Allison}, {Amiri}, {Angile},
  {Battaglia}, {Beall}, {de Bernardis}, {Bond}, {Britton}, {Calabrese}, {Cho},
  {Choi}, {Coughlin}, {Crichton}, {Crowley}, {Datta}, {Devlin}, {Dicker},
  {Dunkley}, {D{\"u}nner}, {Ferraro}, {Fox}, {Gallardo}, {Gralla}, {Halpern},
  {Henderson}, {Hill}, {Hilton}, {Hilton}, {Hincks}, {Hlozek}, {Ho}, {Huang},
  {Hubmayr}, {Huffenberger}, {Hughes}, {Infante}, {Irwin}, {Muya Kasanda},
  {Klein}, {Koopman}, {Kosowsky}, {Li}, {Madhavacheril}, {Marriage}, {McMahon},
  {Menanteau}, {Moodley}, {Munson}, {Naess}, {Nati}, {Newburgh}, {Nibarger},
  {Niemack}, {Nolta}, {Nu{\~n}ez}, {Page}, {Pappas}, {Partridge}, {Rojas},
  {Schaan}, {Schmitt}, {Sehgal}, {Sherwin}, {Sievers}, {Simon}, {Spergel},
  {Staggs}, {Switzer}, {Thornton}, {Trac}, {Treu}, {Tucker}, {Van Engelen},
  {Ward}, \& {Wollack}}]{Louis17}
{Louis}, T., {Grace}, E., {Hasselfield}, M., {et~al.}, {The Atacama Cosmology
  Telescope: two-season ACTPol spectra and parameters}. 2017, \jcap, 6, 031,
  \eprint{1610.02360}

\bibitem[{{Lynden-Bell} {et~al.}(1988){Lynden-Bell}, {Faber}, {Burstein},
  {Davies}, {Dressler}, {Terlevich}, \& {Wegner}}]{lyndenbell1988}
{Lynden-Bell}, D., {Faber}, S.~M., {Burstein}, D., {et~al.}, {Spectroscopy and
  photometry of elliptical galaxies. V - Galaxy streaming toward the new
  supergalactic center}. 1988, \apj, 326, 19

\bibitem[{{Lyth} \& {Liddle}(2009)}]{LL09}
{Lyth}, D.~H. \& {Liddle}, A.~R. 2009, {The Primordial Density Perturbation}
  (Cambridge University Press)

\bibitem[{{Maccoun} \& {Perlmutter}(2015)}]{Maccoun2015}
{Maccoun}, R. \& {Perlmutter}, S., {Blind analysis: Hide results to seek the
  truth}. 2015, \nat, 526, 187

\bibitem[{{Mac{\'{\i}}as-P{\'e}rez} {et~al.}(2007){Mac{\'{\i}}as-P{\'e}rez},
  {Lagache}, {Maffei}, {Ganga}, {Bourrachot}, {Ade}, {Amblard}, {Ansari},
  {Aubourg}, {Aumont}, {Bargot}, {Bartlett}, {Beno{\^i}t}, {Bernard}, {Bhatia},
  {Blanchard}, {Bock}, {Boscaleri}, {Bouchet}, {Camus}, {Cardoso}, {Couchot},
  {de Bernardis}, {Delabrouille}, {D{\'e}sert}, {Dor{\'e}}, {Douspis},
  {Dumoulin}, {Dupac}, {Filliatre}, {Fosalba}, {Gannaway}, {Gautier}, {Giard},
  {Giraud-H{\'e}raud}, {Gispert}, {Guglielmi}, {Hamilton}, {Hanany},
  {Henrot-Versill{\'e}}, {Hristov}, {Kaplan}, {Lamarre}, {Lange}, {Madet},
  {Magneville}, {Marrone}, {Masi}, {Mayet}, {Murphy}, {Naraghi}, {Nati},
  {Patanchon}, {Perdereau}, {Perrin}, {Plaszczynski}, {Piat}, {Ponthieu},
  {Prunet}, {Puget}, {Renault}, {Rosset}, {Santos}, {Starobinsky}, {Strukov},
  {Sudiwala}, {Teyssier}, {Tristram}, {Tucker}, {Vanel}, {Vibert}, {Wakui}, \&
  {Yvon}}]{macias2007}
{Mac{\'{\i}}as-P{\'e}rez}, J.~F., {Lagache}, G., {Maffei}, B., {et~al.},
  {Archeops in-flight performance, data processing, and map making}. 2007,
  \aap, 467, 1313, \eprint{astro-ph/0603665}

\bibitem[{{Mangano} {et~al.}(2002){Mangano}, {Miele}, {Pastor}, \&
  {Peloso}}]{Man02}
{Mangano}, G., {Miele}, G., {Pastor}, S., \& {Peloso}, M., {A precision
  calculation of the effective number of cosmological neutrinos}. 2002, Phys.
  Lett. B, 534, 8, \eprint{astro-ph/0111408}

\bibitem[{{Maniyar} {et~al.}(2018){Maniyar}, {B{\'e}thermin}, \&
  {Lagache}}]{Maniyar_2018}
{Maniyar}, A., {B{\'e}thermin}, M., \& {Lagache}, G., {The history of star
  formation from the cosmic infrared background anisotropies}. 2018, ArXiv
  e-prints, \eprint{1801.10146}

\bibitem[{{Mason} {et~al.}(2018){Mason}, {Treu}, {Dijkstra}, {Mesinger},
  {Trenti}, {Pentericci}, {de Barros}, \& {Vanzella}}]{Mason18}
{Mason}, C.~A., {Treu}, T., {Dijkstra}, M., {et~al.}, {The Universe Is
  Reionizing at $z\sim 7$: Bayesian Inference of the IGM Neutral Fraction Using
  Ly$\alpha$ Emission from Galaxies}. 2018, \apj, 856, 2, \eprint{1709.05356}

\bibitem[{{Matsuhara} {et~al.}(2000){Matsuhara}, {Kawara}, {Sato}, {Taniguchi},
  {Okuda}, {Matsumoto}, {Sofue}, {Wakamatsu}, {Cowie}, {Joseph}, \&
  {Sanders}}]{Matsuhara_2000}
{Matsuhara}, H., {Kawara}, K., {Sato}, Y., {et~al.}, {ISO deep far-infrared
  survey in the ``Lockman Hole``. II. Power spectrum analysis: evidence of a
  strong evolution in number counts}. 2000, \aap, 361, 407,
  \eprint{astro-ph/0006444}

\bibitem[{{McMillan}(2011)}]{mcmillan2011}
{McMillan}, P.~J., {Mass models of the Milky Way}. 2011, \mnras, 414, 2446,
  \eprint{1102.4340}

\bibitem[{{McQuinn}(2016)}]{McQuinn16}
{McQuinn}, M., {The Evolution of the Intergalactic Medium}. 2016, \araa, 54,
  313, \eprint{1512.00086}

\bibitem[{{Medezinski} {et~al.}(2018){Medezinski}, {Battaglia}, {Umetsu},
  {Oguri}, {Miyatake}, {Nishizawa}, {Sif{\'o}n}, {Spergel}, {Chiu}, {Lin},
  {Bahcall}, \& {Komiyama}}]{Medezinski18}
{Medezinski}, E., {Battaglia}, N., {Umetsu}, K., {et~al.}, {Planck
  Sunyaev-Zel'dovich cluster mass calibration using Hyper Suprime-Cam weak
  lensing}. 2018, \pasj, 70, S28, \eprint{1706.00434}

\bibitem[{{Meiksin} {et~al.}(1999){Meiksin}, {White}, \& {Peacock}}]{Meiksin99}
{Meiksin}, A., {White}, M., \& {Peacock}, J.~A., {Baryonic signatures in
  large-scale structure}. 1999, \mnras, 304, 851, \eprint{astro-ph/9812214}

\bibitem[{{Melin} \& {Bartlett}(2015)}]{MelBar15}
{Melin}, J.-B. \& {Bartlett}, J.~G., {Measuring cluster masses with CMB
  lensing: a statistical approach}. 2015, \aap, 578, A21, \eprint{1408.5633}

\bibitem[{{Mesinger}(2016)}]{Mes16}
{Mesinger}, A., {Understanding the Epoch of Cosmic Reionization}. 2016,
  Understanding the Epoch of Cosmic Reionization: Challenges and Progress, 423

\bibitem[{{Mikulizky}(2015)}]{mikulizky2015}
{Mikulizky}, Z. 2015, PhD thesis, Technion -- Israel Institute of Technology

\bibitem[{{Millea} \& {Bouchet}(2018)}]{2018arXiv180408476M}
{Millea}, M. \& {Bouchet}, F., {Cosmic Microwave Background Constraints in
  Light of Priors Over Reionization Histories}. 2018, ArXiv e-prints,
  \eprint{1804.08476}

\bibitem[{{Miller} {et~al.}(2002){Miller}, {Beach}, {Bradley}, {Caldwell},
  {Chapman}, {Devlin}, {Dorwart}, {Herbig}, {Jones}, {Monnelly}, {Netterfield},
  {Nolta}, {Page}, {Puchalla}, {Robertson}, {Torbet}, {Tran}, \&
  {Vinje}}]{Miller02}
{Miller}, A., {Beach}, J., {Bradley}, S., {et~al.}, {The QMAP and MAT/TOCO
  Experiments for Measuring Anisotropy in the Cosmic Microwave Background}.
  2002, \apjs, 140, 115, \eprint{astro-ph/0108030}

\bibitem[{{Miyatake} {et~al.}(2017){Miyatake}, {Madhavacheril}, {Sehgal},
  {Slosar}, {Spergel}, {Sherwin}, \& {van Engelen}}]{Miyatake17}
{Miyatake}, H., {Madhavacheril}, M.~S., {Sehgal}, N., {et~al.}, {Measurement of
  a Cosmographic Distance Ratio with Galaxy and Cosmic Microwave Background
  Lensing}. 2017, Physical Review Letters, 118, 161301, \eprint{1605.05337}

\bibitem[{{Mukhanov}(2005)}]{2005pfc..book.....M}
{Mukhanov}, V. 2005, {Physical Foundations of Cosmology} (Cambridge University
  Press)

\bibitem[{{Netterfield} {et~al.}(2002){Netterfield}, {Ade}, {Bock}, {Bond},
  {Borrill}, {Boscaleri}, {Coble}, {Contaldi}, {Crill}, {de Bernardis},
  {Farese}, {Ganga}, {Giacometti}, {Hivon}, {Hristov}, {Iacoangeli}, {Jaffe},
  {Jones}, {Lange}, {Martinis}, {Masi}, {Mason}, {Mauskopf}, {Melchiorri},
  {Montroy}, {Pascale}, {Piacentini}, {Pogosyan}, {Pongetti}, {Prunet},
  {Romeo}, {Ruhl}, \& {Scaramuzzi}}]{BOOMERANG_data}
{Netterfield}, C.~B., {Ade}, P.~A.~R., {Bock}, J.~J., {et~al.}, {A Measurement
  by BOOMERANG of Multiple Peaks in the Angular Power Spectrum of the Cosmic
  Microwave Background}. 2002, \apj, 571, 604, \eprint{astro-ph/0104460}

\bibitem[{{Oesch} {et~al.}(2018){Oesch}, {Bouwens}, {Illingworth}, {Labb{\'e}},
  \& {Stefanon}}]{Oesch18}
{Oesch}, P.~A., {Bouwens}, R.~J., {Illingworth}, G.~D., {Labb{\'e}}, I., \&
  {Stefanon}, M., {The Dearth of $z{\sim}10$ Galaxies in All HST Legacy Fields
  - The Rapid Evolution of the Galaxy Population in the First 500 Myr}. 2018,
  \apj, 855, 105, \eprint{1710.11131}

\bibitem[{Oka {et~al.}(2014)Oka, Saito, Nishimichi, Taruya, \&
  Yamamoto}]{Oka:2013cba}
Oka, A., Saito, S., Nishimichi, T., Taruya, A., \& Yamamoto, K., {Simultaneous
  constraints on the growth of structure and cosmic expansion from the
  multipole power spectra of the SDSS DR7 LRG sample}. 2014, \mnras, 439, 2515,
  \eprint{1310.2820}

\bibitem[{{Okabe} \& {Smith}(2016)}]{Okabe16}
{Okabe}, N. \& {Smith}, G.~P., {LoCuSS: weak-lensing mass calibration of galaxy
  clusters}. 2016, \mnras, 461, 3794, \eprint{1507.04493}

\bibitem[{Okumura {et~al.}(2016)}]{Okumura:2015lvp}
Okumura, T. {et~al.}, {The Subaru FMOS galaxy redshift survey (FastSound). IV.
  New constraint on gravity theory from redshift space distortions at $z\sim
  1.4$}. 2016, PASJ, 68, 24, \eprint{1511.08083}

\bibitem[{{Omori} \& {Holder}(2015)}]{OmoriHolder15}
{Omori}, Y. \& {Holder}, G., {Cross-Correlation of CFHTLenS Galaxy Number
  Density and Planck CMB Lensing}. 2015, ArXiv e-prints, \eprint{1502.03405}

\bibitem[{{Ostriker} \& {Steinhardt}(1995)}]{OstSte95}
{Ostriker}, J.~P. \& {Steinhardt}, P.~J., {The observational case for a
  low-density Universe with a non-zero cosmological constant}. 1995, \nat, 377,
  600

\bibitem[{{Padmanabhan} \& {Finkbeiner}(2005)}]{PadFin05}
{Padmanabhan}, N. \& {Finkbeiner}, D.~P., {Detecting dark matter annihilation
  with CMB polarization: Signatures and experimental prospects}. 2005, \prd,
  72, 023508, \eprint{astro-ph/0503486}

\bibitem[{{Page} {et~al.}(2003){Page}, {Nolta}, {Barnes}, {Bennett}, {Halpern},
  {Hinshaw}, {Jarosik}, {Kogut}, {Limon}, {Meyer}, {Peiris}, {Spergel},
  {Tucker}, {Wollack}, \& {Wright}}]{page2003b}
{Page}, L., {Nolta}, M.~R., {Barnes}, C., {et~al.}, {First-Year Wilkinson
  Microwave Anisotropy Probe (WMAP) Observations: Interpretation of the TT and
  TE Angular Power Spectrum Peaks}. 2003, \apjs, 148, 233,
  \eprint{astro-ph/0302220}

\bibitem[{{Palanque-Delabrouille} {et~al.}(2015){Palanque-Delabrouille},
  {Y{\`e}che}, {Baur}, {Magneville}, {Rossi}, {Lesgourgues}, {Borde}, {Burtin},
  {LeGoff}, {Rich}, {Viel}, \& {Weinberg}}]{Palanque15}
{Palanque-Delabrouille}, N., {Y{\`e}che}, C., {Baur}, J., {et~al.}, {Neutrino
  masses and cosmology with Lyman-alpha forest power spectrum}. 2015, \jcap,
  11, 011, \eprint{1506.05976}

\bibitem[{{Partridge}(1995)}]{Partridge95}
{Partridge}, R.~B. 1995, {3K: The Cosmic Microwave Background Radiation}
  (Cambridge University Press), 393

\bibitem[{{Patterson}(2015)}]{Pat15}
{Patterson}, R.~B., {Prospects for Measurement of the Neutrino Mass Hierarchy}.
  2015, Annual Review of Nuclear and Particle Science, 65, 177,
  \eprint{1506.07917}

\bibitem[{{Peacock}(1999)}]{Peacock99}
{Peacock}, J.~A. 1999, {Cosmological Physics} (Cambridge University Press), 704

\bibitem[{{Peacock} \& {Bilicki}(2018)}]{PeaBil18}
{Peacock}, J.~A. \& {Bilicki}, M., {Wide-area tomography of CMB lensing and the
  growth of cosmological density fluctuations}. 2018, ArXiv e-prints,
  \eprint{1805.11525}

\bibitem[{{Peacock} \& {Dodds}(1996)}]{1996MNRAS.280L..19P}
{Peacock}, J.~A. \& {Dodds}, S.~J., {Non-linear evolution of cosmological power
  spectra}. 1996, \mnras, 280, L19, \eprint{astro-ph/9603031}

\bibitem[{{Pearson} {et~al.}(2003){Pearson}, {Mason}, {Readhead}, {Shepherd},
  {Sievers}, {Udomprasert}, {Cartwright}, {Farmer}, {Padin}, {Myers}, {Bond},
  {Contaldi}, {Pen}, {Prunet}, {Pogosyan}, {Carlstrom}, {Kovac}, {Leitch},
  {Pryke}, {Halverson}, {Holzapfel}, {Altamirano}, {Bronfman}, {Casassus},
  {May}, \& {Joy}}]{CBI_data}
{Pearson}, T.~J., {Mason}, B.~S., {Readhead}, A.~C.~S., {et~al.}, {The
  Anisotropy of the Microwave Background to l = 3500: Mosaic Observations with
  the Cosmic Background Imager}. 2003, \apj, 591, 556,
  \eprint{astro-ph/0205388}

\bibitem[{{Peebles}(1984)}]{Peebles84}
{Peebles}, P.~J.~E., {Tests of cosmological models constrained by inflation}.
  1984, \apj, 284, 439

\bibitem[{{Peebles} {et~al.}(2009){Peebles}, {Page}, \&
  {Partridge}}]{Peebles2009}
{Peebles}, P.~J.~E., {Page}, Jr., L.~A., \& {Partridge}, R.~B. 2009, {Finding
  the Big Bang} (Cambridge University Press)

\bibitem[{{Peebles} \& {Yu}(1970)}]{PeeYu70}
{Peebles}, P.~J.~E. \& {Yu}, J.~T., {Primeval Adiabatic Perturbation in an
  Expanding Universe}. 1970, \apj, 162, 815

\bibitem[{{Penna-Lima} {et~al.}(2017){Penna-Lima}, {Bartlett}, {Rozo}, {Melin},
  {Merten}, {Evrard}, {Postman}, \& {Rykoff}}]{Penna-Lima17}
{Penna-Lima}, M., {Bartlett}, J.~G., {Rozo}, E., {et~al.}, {Calibrating the
  Planck cluster mass scale with CLASH}. 2017, \aap, 604, A89,
  \eprint{1608.05356}

\bibitem[{{Penzias} \& {Wilson}(1965)}]{PenWil65}
{Penzias}, A.~A. \& {Wilson}, R.~W., {A Measurement of Excess Antenna
  Temperature at 4080 Mc/s.} 1965, \apj, 142, 419

\bibitem[{{Perlmutter} {et~al.}(1999){Perlmutter}, {Aldering}, {Goldhaber},
  {Knop}, {Nugent}, {Castro}, {Deustua}, {Fabbro}, {Goobar}, {Groom}, {Hook},
  {Kim}, {Kim}, {Lee}, {Nunes}, {Pain}, {Pennypacker}, {Quimby}, {Lidman},
  {Ellis}, {Irwin}, {McMahon}, {Ruiz-Lapuente}, {Walton}, {Schaefer}, {Boyle},
  {Filippenko}, {Matheson}, {Fruchter}, {Panagia}, {Newberg}, {Couch}, \&
  {Project}}]{Perlmutter99}
{Perlmutter}, S., {Aldering}, G., {Goldhaber}, G., {et~al.}, {Measurements of
  {$\Omega$} and {$\Lambda$} from 42 High-Redshift Supernovae}. 1999, \apj,
  517, 565, \eprint{astro-ph/9812133}

\bibitem[{Peter \& Uzan(2009)}]{Peter:1208401}
Peter, P. \& Uzan, J.-P. 2009, {Primordial cosmology}, Oxford Graduate Texts
  (Oxford: Oxford Univ. Press)

\bibitem[{Pezzotta {et~al.}(2017)}]{Pezzotta:2016gbo}
Pezzotta, A. {et~al.}, {The VIMOS Public Extragalactic Redshift Survey
  (VIPERS): The growth of structure at $0.5 < z < 1.2$ from redshift-space
  distortions in the clustering of the PDR-2 final sample}. 2017, \aap, 604,
  A33, \eprint{1612.05645}

\bibitem[{{Pierpaoli} {et~al.}(2000){Pierpaoli}, {Scott}, \&
  {White}}]{Pierpaoli00}
{Pierpaoli}, E., {Scott}, D., \& {White}, M., {How Flat is the Universe?} 2000,
  Science, 287, 2171

\bibitem[{{Planck Collaboration} {et~al.}(2017){Planck Collaboration},
  {Akrami}, {Ashdown}, {Aumont}, {Baccigalupi}, {Ballardini}, {Band ay},
  {Barreiro}, {Bartolo}, {Basak}, {Benabed}, {Bernard}, {Bersanelli},
  {Bielewicz}, {Bonavera}, {Bond}, {Borrill}, {Bouchet}, {Boulanger}, {Bucher},
  {Burigana}, {Butler}, {Calabrese}, {Cardoso}, {Carron}, {Chiang}, {Colombo},
  {Comis}, {Couchot}, {Coulais}, {Crill}, {Curto}, {Cuttaia}, {de Bernardis},
  {de Rosa}, {de Zotti}, {Delabrouille}, {Di Valentino}, {Dickinson}, {Diego},
  {Dor{\'e}}, {Ducout}, {Dupac}, {Elsner}, {En{\ss}lin}, {Eriksen},
  {Falgarone}, {Fantaye}, {Finelli}, {Frailis}, {Fraisse}, {Franceschi},
  {Frolov}, {Galeotta}, {Galli}, {Ganga}, {G{\'e}nova-Santos}, {Gerbino},
  {Gonz{\'a}lez-Nuevo}, {G{\'o}rski}, {Gruppuso}, {Gudmundsson}, {Hansen},
  {Helou}, {Henrot-Versill{\'e}}, {Herranz}, {Hivon}, {Jaffe}, {Jones},
  {Keih{\"a}nen}, {Keskitalo}, {Kiiveri}, {Kim}, {Kisner}, {Krachmalnicoff},
  {Kunz}, {Kurki-Suonio}, {Lagache}, {Lamarre}, {Lasenby}, {Lattanzi},
  {Lawrence}, {Le Jeune}, {Lellouch}, {Levrier}, {Liguori}, {Lilje},
  {Lindholm}, {L{\'o}pez-Caniego}, {Ma}, {Mac{\'\i}as-P{\'e}rez}, {Maggio},
  {Maino}, {Mand olesi}, {Maris}, {Martin}, {Mart{\'\i}nez-Gonz{\'a}lez},
  {Matarrese}, {Mauri}, {McEwen}, {Melchiorri}, {Mennella}, {Migliaccio},
  {Miville-Desch{\^e}nes}, {Molinari}, {Moneti}, {Montier}, {Moreno},
  {Morgante}, {Natoli}, {Oxborrow}, {Paoletti}, {Partridge}, {Patanchon},
  {Patrizii}, {Perdereau}, {Piacentini}, {Plaszczynski}, {Polenta}, {Rachen},
  {Racine}, {Reinecke}, {Remazeilles}, {Renzi}, {Rocha}, {Romelli}, {Rosset},
  {Roudier}, {Rubi{\~n}o-Mart{\'\i}n}, {Ruiz-Granados}, {Salvati}, {Sandri},
  {Savelainen}, {Scott}, {Sirri}, {Spencer}, {Suur-Uski}, {Tauber},
  {Tavagnacco}, {Tenti}, {Toffolatti}, {Tomasi}, {Tristram}, {Trombetti},
  {Valiviita}, {Van Tent}, {Vielva}, {Villa}, {Wehus}, \&
  {Zacchei}}]{2017A&A...607A.122P}
{Planck Collaboration}, {Akrami}, Y., {Ashdown}, M., {et~al.}, {Planck
  intermediate results. LII. Planet flux densities}. 2017, \aap, 607, A122,
  \eprint{1612.07151}

\bibitem[{{Planck Collaboration ES}(2015)}]{planck2014-ES}
{Planck Collaboration ES}. 2015, {The Explanatory Supplement to the \Planck\
  2015 results, \url{http://wiki.cosmos.esa.int/planckpla2015}} ({ESA})

\bibitem[{{Planck Collaboration ES}(2018)}]{planck2016-ES}
{Planck Collaboration ES}. 2018, {The Legacy Explanatory Supplement,
  \href{http://wiki.cosmos.esa.int/planck-legacy-archive}{\texttt{http://wiki.cosmos.esa.int/planck-legacy-archive}}}
  ({ESA})

\bibitem[{{\sorthelp{Planck Collaboration 2011A}}{Planck Collaboration
  I}(2011)}]{planck2011-1.1}
{\sorthelp{Planck Collaboration 2011A}}{Planck Collaboration I},
  {\textit{Planck} early results. I. The Planck mission}. 2011, \aap, 536, A1,
  \eprint{1101.2022}

\bibitem[{{\sorthelp{Planck Collaboration 2011G}}{Planck Collaboration
  VII}(2011)}]{planck2011-1.10}
{\sorthelp{Planck Collaboration 2011G}}{Planck Collaboration VII},
  {\textit{Planck} early results. VII. The Early Release Compact Source
  Catalogue}. 2011, \aap, 536, A7, \eprint{1101.2041}

\bibitem[{{\sorthelp{Planck Collaboration 2011H}}{Planck Collaboration
  VIII}(2011)}]{planck2011-5.1a}
{\sorthelp{Planck Collaboration 2011H}}{Planck Collaboration VIII},
  {\textit{Planck} early results. VIII. The all-sky early Sunyaev-Zeldovich
  cluster sample}. 2011, \aap, 536, A8, \eprint{1101.2024}

\bibitem[{{\sorthelp{Planck Collaboration 2011I}}{Planck Collaboration
  IX}(2011)}]{planck2011-5.1b}
{\sorthelp{Planck Collaboration 2011I}}{Planck Collaboration IX},
  {\textit{Planck} early results. IX. XMM-\textit{Newton} follow-up validation
  programme of \textit{Planck} cluster candidates}. 2011, \aap, 536, A9,
  \eprint{1101.2025}

\bibitem[{{\sorthelp{Planck Collaboration 2011J}}{Planck Collaboration
  X}(2011)}]{planck2011-5.2a}
{\sorthelp{Planck Collaboration 2011J}}{Planck Collaboration X},
  {\textit{Planck} early results. X. Statistical analysis of Sunyaev-Zeldovich
  scaling relations for X-ray galaxy clusters}. 2011, \aap, 536, A10,
  \eprint{1101.2043}

\bibitem[{{\sorthelp{Planck Collaboration 2011K}}{Planck Collaboration
  XI}(2011)}]{planck2011-5.2b}
{\sorthelp{Planck Collaboration 2011K}}{Planck Collaboration XI},
  {\textit{Planck} early results. XI. Calibration of the local galaxy cluster
  Sunyaev-Zeldovich scaling relations}. 2011, \aap, 536, A11,
  \eprint{1101.2026}

\bibitem[{{\sorthelp{Planck Collaboration 2011L}}{Planck Collaboration
  XII}(2011)}]{planck2011-5.2c}
{\sorthelp{Planck Collaboration 2011L}}{Planck Collaboration XII},
  {\textit{Planck} early results. XII. Cluster Sunyaev-Zeldovich optical
  scaling relations}. 2011, \aap, 536, A12, \eprint{1101.2027}

\bibitem[{{\sorthelp{Planck Collaboration 2011M}}{Planck Collaboration
  XIII}(2011)}]{planck2011-6.1}
{\sorthelp{Planck Collaboration 2011M}}{Planck Collaboration XIII},
  {\textit{Planck} early results. XIII. Statistical properties of extragalactic
  radio sources in the Planck Early Release Compact Source Catalogue}. 2011,
  \aap, 536, A13, \eprint{1101.2044}

\bibitem[{{\sorthelp{Planck Collaboration 2011N}}{Planck Collaboration
  XIV}(2011)}]{planck2011-6.2}
{\sorthelp{Planck Collaboration 2011N}}{Planck Collaboration XIV},
  {\textit{Planck} early results. XIV. ERCSC validation and extreme radio
  sources}. 2011, \aap, 536, A14, \eprint{1101.1721}

\bibitem[{{\sorthelp{Planck Collaboration 2011O}}{Planck Collaboration
  XV}(2011)}]{planck2011-6.3a}
{\sorthelp{Planck Collaboration 2011O}}{Planck Collaboration XV},
  {\textit{Planck} early results. XV. Spectral energy distributions and radio
  continuum spectra of northern extragalactic radio sources}. 2011, \aap, 536,
  A15, \eprint{1101.2047}

\bibitem[{{\sorthelp{Planck Collaboration 2011P}}{Planck Collaboration
  XVI}(2011)}]{planck2011-6.4a}
{\sorthelp{Planck Collaboration 2011P}}{Planck Collaboration XVI},
  {\textit{Planck} early results. XVI. The \textit{Planck} view of nearby
  galaxies}. 2011, \aap, 536, A16, \eprint{1101.2045}

\bibitem[{{\sorthelp{Planck Collaboration 2011R}}{Planck Collaboration
  XVIII}(2011)}]{planck2011-6.6}
{\sorthelp{Planck Collaboration 2011R}}{Planck Collaboration XVIII},
  {\textit{Planck} early results. XVIII. The power spectrum of cosmic infrared
  background anisotropies}. 2011, \aap, 536, A18, \eprint{1101.2028}

\bibitem[{{\sorthelp{Planck Collaboration 2011Z}}{Planck Collaboration
  XXVI}(2011)}]{planck2011-5.1c}
{\sorthelp{Planck Collaboration 2011Z}}{Planck Collaboration XXVI},
  {\textit{Planck} early results. XXVI. Detection with \textit{Planck} and
  confirmation by XMM-\textit{Newton} of PLCK G266.6-27.3, an exceptionally
  X-ray luminous and massive galaxy cluster at \textit{z}$\sim$1}. 2011, \aap,
  536, A26, \eprint{1106.1376}

\bibitem[{{\sorthelp{Planck Collaboration 2014A}}{Planck Collaboration
  I}(2014)}]{planck2013-p01}
{\sorthelp{Planck Collaboration 2014A}}{Planck Collaboration I},
  {\textit{Planck} 2013 results. I. Overview of products and scientific
  results}. 2014, \aap, 571, A1, \eprint{1303.5062}

\bibitem[{{\sorthelp{Planck Collaboration 2014B}}{Planck Collaboration
  II}(2014)}]{planck2013-p02}
{\sorthelp{Planck Collaboration 2014B}}{Planck Collaboration II},
  {\textit{Planck} 2013 results. II. Low Frequency Instrument data processing}.
  2014, \aap, 571, A2, \eprint{1303.5063}

\bibitem[{{\sorthelp{Planck Collaboration 2014F}}{Planck Collaboration
  VI}(2014)}]{planck2013-p03}
{\sorthelp{Planck Collaboration 2014F}}{Planck Collaboration VI},
  {\textit{Planck} 2013 results. VI. High Frequency Instrument data
  processing}. 2014, \aap, 571, A6, \eprint{1303.5067}

\bibitem[{{\sorthelp{Planck Collaboration 2014H}}{Planck Collaboration
  VIII}(2014)}]{planck2013-p03f}
{\sorthelp{Planck Collaboration 2014H}}{Planck Collaboration VIII},
  {\textit{Planck} 2013 results. VIII. HFI photometric calibration and
  mapmaking}. 2014, \aap, 571, A8, \eprint{1303.5069}

\bibitem[{{\sorthelp{Planck Collaboration 2014L}}{Planck Collaboration
  XII}(2014)}]{planck2013-p06}
{\sorthelp{Planck Collaboration 2014L}}{Planck Collaboration XII},
  {\textit{Planck} 2013 results. XII. Diffuse component separation}. 2014,
  \aap, 571, A12, \eprint{1303.5072}

\bibitem[{{\sorthelp{Planck Collaboration 2014P}}{Planck Collaboration
  XVI}(2014)}]{planck2013-p11}
{\sorthelp{Planck Collaboration 2014P}}{Planck Collaboration XVI},
  {\textit{Planck} 2013 results. XVI. Cosmological parameters}. 2014, \aap,
  571, A16, \eprint{1303.5076}

\bibitem[{{\sorthelp{Planck Collaboration 2014Q}}{Planck Collaboration
  XVII}(2014)}]{planck2013-p12}
{\sorthelp{Planck Collaboration 2014Q}}{Planck Collaboration XVII},
  {\textit{Planck} 2013 results. XVII. Gravitational lensing by large-scale
  structure}. 2014, \aap, 571, A17, \eprint{1303.5077}

\bibitem[{{\sorthelp{Planck Collaboration 2014R}}{Planck Collaboration
  XVIII}(2014)}]{planck2013-p13}
{\sorthelp{Planck Collaboration 2014R}}{Planck Collaboration XVIII},
  {\textit{Planck} 2013 results. XVIII. The gravitational lensing-infrared
  background correlation}. 2014, \aap, 571, A18, \eprint{1303.5078}

\bibitem[{{\sorthelp{Planck Collaboration 2014S}}{Planck Collaboration
  XIX}(2014)}]{planck2013-p14}
{\sorthelp{Planck Collaboration 2014S}}{Planck Collaboration XIX},
  {\textit{Planck} 2013 results. XIX. The integrated Sachs-Wolfe effect}. 2014,
  \aap, 571, A19, \eprint{1303.5079}

\bibitem[{{\sorthelp{Planck Collaboration 2014T}}{Planck Collaboration
  XX}(2014)}]{planck2013-p15}
{\sorthelp{Planck Collaboration 2014T}}{Planck Collaboration XX},
  {\textit{Planck} 2013 results. XX. Cosmology from Sunyaev-Zeldovich cluster
  counts}. 2014, \aap, 571, A20, \eprint{1303.5080}

\bibitem[{{\sorthelp{Planck Collaboration 2014U}}{Planck Collaboration
  XXI}(2014)}]{planck2013-p05b}
{\sorthelp{Planck Collaboration 2014U}}{Planck Collaboration XXI},
  {\textit{Planck} 2013 results. XXI. Power spectrum and high-order statistics
  of the \textit{Planck} all-sky Compton parameter map}. 2014, \aap, 571, A21,
  \eprint{1303.5081}

\bibitem[{{\sorthelp{Planck Collaboration 2014V}}{Planck Collaboration
  XXII}(2014)}]{planck2013-p17}
{\sorthelp{Planck Collaboration 2014V}}{Planck Collaboration XXII},
  {\textit{Planck} 2013 results. XXII. Constraints on inflation}. 2014, \aap,
  571, A22, \eprint{1303.5082}

\bibitem[{{\sorthelp{Planck Collaboration 2014W}}{Planck Collaboration
  XXIII}(2014)}]{planck2013-p09}
{\sorthelp{Planck Collaboration 2014W}}{Planck Collaboration XXIII},
  {\textit{Planck} 2013 results. XXIII. Isotropy and statistics of the CMB}.
  2014, \aap, 571, A23, \eprint{1303.5083}

\bibitem[{{\sorthelp{Planck Collaboration 2014X}}{Planck Collaboration
  XXIV}(2014)}]{planck2013-p09a}
{\sorthelp{Planck Collaboration 2014X}}{Planck Collaboration XXIV},
  {\textit{Planck} 2013 results. XXIV. Constraints on primordial
  non-Gaussianity}. 2014, \aap, 571, A24, \eprint{1303.5084}

\bibitem[{{\sorthelp{Planck Collaboration 2014Y}}{Planck Collaboration
  XXV}(2014)}]{planck2013-p20}
{\sorthelp{Planck Collaboration 2014Y}}{Planck Collaboration XXV},
  {\textit{Planck} 2013 results. XXV. Searches for cosmic strings and other
  topological defects}. 2014, \aap, 571, A25, \eprint{1303.5085}

\bibitem[{{\sorthelp{Planck Collaboration 2014ZB}}{Planck Collaboration
  XXVII}(2014)}]{planck2013-pipaberration}
{\sorthelp{Planck Collaboration 2014ZB}}{Planck Collaboration XXVII},
  {\textit{Planck} 2013 results. XXVII. Doppler boosting of the CMB: Eppur si
  muove}. 2014, \aap, 571, A27, \eprint{1303.5087}

\bibitem[{{\sorthelp{Planck Collaboration 2014ZD}}{Planck Collaboration
  XXIX}(2014)}]{planck2013-p05a}
{\sorthelp{Planck Collaboration 2014ZD}}{Planck Collaboration XXIX},
  {\textit{Planck} 2013 results. XXIX. The Planck catalogue of
  Sunyaev-Zeldovich sources}. 2014, \aap, 571, A29, \eprint{1303.5089}

\bibitem[{{\sorthelp{Planck Collaboration 2014ZE}}{Planck Collaboration
  XXX}(2014)}]{planck2013-pip56}
{\sorthelp{Planck Collaboration 2014ZE}}{Planck Collaboration XXX},
  {\textit{Planck} 2013 results. XXX. Cosmic infrared background measurements
  and implications for star formation}. 2014, \aap, 571, A30,
  \eprint{1309.0382}

\bibitem[{{\sorthelp{Planck Collaboration 2014ZF}}{Planck Collaboration
  XXXI}(2014)}]{planck2013-p01a}
{\sorthelp{Planck Collaboration 2014ZF}}{Planck Collaboration XXXI},
  {\textit{Planck} 2013 results. XXXI. Consistency of the \textit{Planck}
  data}. 2014, \aap, 571, A31, \eprint{1508.03375}

\bibitem[{{\sorthelp{Planck Collaboration 2014ZG}}{Planck Collaboration
  XXXII}(2015)}]{planck2013-p05a-addendum}
{\sorthelp{Planck Collaboration 2014ZG}}{Planck Collaboration XXXII},
  {\textit{Planck} 2013 results. XXXII. The updated Planck catalogue of
  Sunyaev-Zeldovich sources}. 2015, \aap, 581, A14, \eprint{1502.00543}

\bibitem[{{\sorthelp{Planck Collaboration 2015A}}{Planck Collaboration
  I}(2016)}]{planck2014-a01}
{\sorthelp{Planck Collaboration 2015A}}{Planck Collaboration I},
  {\textit{Planck} 2015 results. I. Overview of products and results}. 2016,
  \aap, 594, A1, \eprint{1502.01582}

\bibitem[{{\sorthelp{Planck Collaboration 2015B}}{Planck Collaboration
  II}(2016)}]{planck2014-a03}
{\sorthelp{Planck Collaboration 2015B}}{Planck Collaboration II},
  {\textit{Planck} 2015 results. II. Low Frequency Instrument data processing}.
  2016, \aap, 594, A2, \eprint{1502.01583}

\bibitem[{{\sorthelp{Planck Collaboration 2015D}}{Planck Collaboration
  IV}(2016)}]{planck2014-a05}
{\sorthelp{Planck Collaboration 2015D}}{Planck Collaboration IV},
  {\textit{Planck} 2015 results. IV. LFI beams and window functions}. 2016,
  \aap, 594, A4, \eprint{1502.01584}

\bibitem[{{\sorthelp{Planck Collaboration 2015E}}{Planck Collaboration
  V}(2016)}]{planck2014-a06}
{\sorthelp{Planck Collaboration 2015E}}{Planck Collaboration V},
  {\textit{Planck} 2015 results. V. LFI calibration}. 2016, \aap, 594, A5,
  \eprint{1505.08022}

\bibitem[{{\sorthelp{Planck Collaboration 2015G}}{Planck Collaboration
  VII}(2016)}]{planck2014-a08}
{\sorthelp{Planck Collaboration 2015G}}{Planck Collaboration VII},
  {\textit{Planck} 2015 results. VII. High Frequency Instrument data
  processing: Time-ordered information and beam processing}. 2016, \aap, 594,
  A7, \eprint{1502.01586}

\bibitem[{{\sorthelp{Planck Collaboration 2015H}}{Planck Collaboration
  VIII}(2016)}]{planck2014-a09}
{\sorthelp{Planck Collaboration 2015H}}{Planck Collaboration VIII},
  {\textit{Planck} 2015 results. VIII. High Frequency Instrument data
  processing: Calibration and maps}. 2016, \aap, 594, A8, \eprint{1502.01587}

\bibitem[{{\sorthelp{Planck Collaboration 2015I}}{Planck Collaboration
  IX}(2016)}]{planck2014-a11}
{\sorthelp{Planck Collaboration 2015I}}{Planck Collaboration IX},
  {\textit{Planck} 2015 results. IX. Diffuse component separation: CMB maps}.
  2016, \aap, 594, A9, \eprint{1502.05956}

\bibitem[{{\sorthelp{Planck Collaboration 2015J}}{Planck Collaboration
  X}(2016)}]{planck2014-a12}
{\sorthelp{Planck Collaboration 2015J}}{Planck Collaboration X},
  {\textit{Planck} 2015 results. X. Diffuse component separation: Foreground
  maps}. 2016, \aap, 594, A10, \eprint{1502.01588}

\bibitem[{{\sorthelp{Planck Collaboration 2015K}}{Planck Collaboration
  XI}(2016)}]{planck2014-a13}
{\sorthelp{Planck Collaboration 2015K}}{Planck Collaboration XI},
  {\textit{Planck} 2015 results. XI. CMB power spectra, likelihoods, and
  robustness of parameters}. 2016, \aap, 594, A11, \eprint{1507.02704}

\bibitem[{{\sorthelp{Planck Collaboration 2015L}}{Planck Collaboration
  XII}(2016)}]{planck2014-a14}
{\sorthelp{Planck Collaboration 2015L}}{Planck Collaboration XII},
  {\textit{Planck} 2015 results. XII. Full Focal Plane simulations}. 2016,
  \aap, 594, A12, \eprint{1509.06348}

\bibitem[{{\sorthelp{Planck Collaboration 2015M}}{Planck Collaboration
  XIII}(2016)}]{planck2014-a15}
{\sorthelp{Planck Collaboration 2015M}}{Planck Collaboration XIII},
  {\textit{Planck} 2015 results. XIII. Cosmological parameters}. 2016, \aap,
  594, A13, \eprint{1502.01589}

\bibitem[{{\sorthelp{Planck Collaboration 2015N}}{Planck Collaboration
  XIV}(2016)}]{planck2014-a16}
{\sorthelp{Planck Collaboration 2015N}}{Planck Collaboration XIV},
  {\textit{Planck} 2015 results. XIV. Dark energy and modified gravity}. 2016,
  \aap, 594, A14, \eprint{1502.01590}

\bibitem[{{\sorthelp{Planck Collaboration 2015O}}{Planck Collaboration
  XV}(2016)}]{planck2014-a17}
{\sorthelp{Planck Collaboration 2015O}}{Planck Collaboration XV},
  {\textit{Planck} 2015 results. XV. Gravitational lensing}. 2016, \aap, 594,
  A15, \eprint{1502.01591}

\bibitem[{{\sorthelp{Planck Collaboration 2015P}}{Planck Collaboration
  XVI}(2016)}]{planck2014-a18}
{\sorthelp{Planck Collaboration 2015P}}{Planck Collaboration XVI},
  {\textit{Planck} 2015 results. XVI. Isotropy and statistics of the CMB}.
  2016, \aap, 594, A16, \eprint{1506.07135}

\bibitem[{{\sorthelp{Planck Collaboration 2015Q}}{Planck Collaboration
  XVII}(2016)}]{planck2014-a19}
{\sorthelp{Planck Collaboration 2015Q}}{Planck Collaboration XVII},
  {\textit{Planck} 2015 results. XVII. Constraints on primordial
  non-Gaussianity}. 2016, \aap, 594, A17, \eprint{1502.01592}

\bibitem[{{\sorthelp{Planck Collaboration 2015R}}{Planck Collaboration
  XVIII}(2016)}]{planck2014-a20}
{\sorthelp{Planck Collaboration 2015R}}{Planck Collaboration XVIII},
  {\textit{Planck} 2015 results. XVIII. Background geometry and topology of the
  Universe}. 2016, \aap, 594, A18, \eprint{1502.01593}

\bibitem[{{\sorthelp{Planck Collaboration 2015T}}{Planck Collaboration
  XX}(2016)}]{planck2014-a24}
{\sorthelp{Planck Collaboration 2015T}}{Planck Collaboration XX},
  {\textit{Planck} 2015 results. XX. Constraints on inflation}. 2016, \aap,
  594, A20, \eprint{1502.02114}

\bibitem[{{\sorthelp{Planck Collaboration 2015U}}{Planck Collaboration
  XXI}(2016)}]{planck2014-a26}
{\sorthelp{Planck Collaboration 2015U}}{Planck Collaboration XXI},
  {\textit{Planck} 2015 results. XXI. The integrated Sachs-Wolfe effect}. 2016,
  \aap, 594, A21, \eprint{1502.01595}

\bibitem[{{\sorthelp{Planck Collaboration 2015V}}{Planck Collaboration
  XXII}(2016)}]{planck2014-a28}
{\sorthelp{Planck Collaboration 2015V}}{Planck Collaboration XXII},
  {\textit{Planck} 2015 results. XXII. A map of the thermal Sunyaev-Zeldovich
  effect}. 2016, \aap, 594, A22, \eprint{1502.01596}

\bibitem[{{\sorthelp{Planck Collaboration 2015X}}{Planck Collaboration
  XXIV}(2016)}]{planck2014-a30}
{\sorthelp{Planck Collaboration 2015X}}{Planck Collaboration XXIV},
  {\textit{Planck} 2015 results. XXIV. Cosmology from Sunyaev-Zeldovich cluster
  counts}. 2016, \aap, 594, A24, \eprint{1502.01597}

\bibitem[{{\sorthelp{Planck Collaboration 2015Y}}{Planck Collaboration
  XXV}(2016)}]{planck2014-a31}
{\sorthelp{Planck Collaboration 2015Y}}{Planck Collaboration XXV},
  {\textit{Planck} 2015 results. XXV. Diffuse, low-frequency Galactic
  foregrounds}. 2016, \aap, 594, A25, \eprint{1506.06660}

\bibitem[{{\sorthelp{Planck Collaboration 2015ZA}}{Planck Collaboration
  XXVI}(2016)}]{planck2014-a35}
{\sorthelp{Planck Collaboration 2015ZA}}{Planck Collaboration XXVI},
  {\textit{Planck} 2015 results. XXVI. The Second Planck Catalogue of Compact
  Sources}. 2016, \aap, 594, A26, \eprint{1507.02058}

\bibitem[{{\sorthelp{Planck Collaboration 2015ZB}}{Planck Collaboration
  XXVII}(2016)}]{planck2014-a36}
{\sorthelp{Planck Collaboration 2015ZB}}{Planck Collaboration XXVII},
  {\textit{Planck} 2015 results. XXVII. The Second Planck Catalogue of
  Sunyaev-Zeldovich Sources}. 2016, \aap, 594, A27, \eprint{1502.01598}

\bibitem[{{\sorthelp{Planck Collaboration 2018B}}{Planck Collaboration
  II}(2018)}]{planck2016-l02}
{\sorthelp{Planck Collaboration 2018B}}{Planck Collaboration II},
  {\textit{Planck} 2018 results. II. Low Frequency Instrument data processing}.
  2018, \aap, submitted

\bibitem[{{\sorthelp{Planck Collaboration 2018C}}{Planck Collaboration
  III}(2018)}]{planck2016-l03}
{\sorthelp{Planck Collaboration 2018C}}{Planck Collaboration III},
  {\textit{Planck} 2018 results. III. High Frequency Instrument data
  processing}. 2018, \aap, submitted

\bibitem[{{\sorthelp{Planck Collaboration 2018D}}{Planck Collaboration
  IV}(2018)}]{planck2016-l04}
{\sorthelp{Planck Collaboration 2018D}}{Planck Collaboration IV},
  {\textit{Planck} 2018 results. IV. CMB and foreground extraction}. 2018,
  \aap, submitted

\bibitem[{{\sorthelp{Planck Collaboration 2018E}}{Planck Collaboration
  V}(2018)}]{planck2016-l05}
{\sorthelp{Planck Collaboration 2018E}}{Planck Collaboration V},
  {\textit{Planck} 2018 results. V. Power spectra and likelihoods}. 2018, \aap,
  in preparation

\bibitem[{{\sorthelp{Planck Collaboration 2018F}}{Planck Collaboration
  VI}(2018)}]{planck2016-l06}
{\sorthelp{Planck Collaboration 2018F}}{Planck Collaboration VI},
  {\textit{Planck} 2018 results. VI. Cosmological parameters}. 2018, \aap,
  submitted

\bibitem[{{\sorthelp{Planck Collaboration 2018G}}{Planck Collaboration
  VII}(2018)}]{planck2016-l07}
{\sorthelp{Planck Collaboration 2018G}}{Planck Collaboration VII},
  {\textit{Planck} 2018 results. VII. Isotropy and statistics}. 2018, \aap, in
  preparation

\bibitem[{{\sorthelp{Planck Collaboration 2018H}}{Planck Collaboration
  VIII}(2018)}]{planck2016-l08}
{\sorthelp{Planck Collaboration 2018H}}{Planck Collaboration VIII},
  {\textit{Planck} 2018 results. VIII. Gravitational lensing}. 2018, \aap,
  submitted

\bibitem[{{\sorthelp{Planck Collaboration 2018I}}{Planck Collaboration
  IX}(2018)}]{planck2016-l09}
{\sorthelp{Planck Collaboration 2018I}}{Planck Collaboration IX},
  {\textit{Planck} 2018 results. IX. Constraints on primordial
  non-Gaussianity}. 2018, \aap, in preparation

\bibitem[{{\sorthelp{Planck Collaboration 2018J}}{Planck Collaboration
  X}(2018)}]{planck2016-l10}
{\sorthelp{Planck Collaboration 2018J}}{Planck Collaboration X},
  {\textit{Planck} 2018 results. X. Constraints on inflation}. 2018, \aap,
  submitted

\bibitem[{{\sorthelp{Planck Collaboration 2018K}}{Planck Collaboration
  XI}(2018)}]{planck2016-l11A}
{\sorthelp{Planck Collaboration 2018K}}{Planck Collaboration XI},
  {\textit{Planck} 2018 results. XI. Polarized dust foregrounds}. 2018, \aap,
  submitted, \eprint{1801.04945}

\bibitem[{{\sorthelp{Planck Collaboration 2018L}}{Planck Collaboration
  XII}(2018)}]{planck2016-l11B}
{\sorthelp{Planck Collaboration 2018L}}{Planck Collaboration XII},
  {\textit{Planck} 2018 results. XII. Galactic astrophysics using polarized
  dust emission}. 2018, \aap, submitted

\bibitem[{{\sorthelp{Planck Collaboration IntA}}{Planck Collaboration Int.
  I}(2012)}]{planck2012-I}
{\sorthelp{Planck Collaboration IntA}}{Planck Collaboration Int. I},
  {\textit{Planck} intermediate results. I. Further validation of new
  \textit{Planck} clusters with XMM-\textit{Newton}}. 2012, \aap, 543, A102,
  \eprint{1112.5595}

\bibitem[{{\sorthelp{Planck Collaboration IntC}}{Planck Collaboration Int.
  III}(2013)}]{planck2012-III}
{\sorthelp{Planck Collaboration IntC}}{Planck Collaboration Int. III},
  {\textit{Planck} intermediate results. III. The relation between galaxy
  cluster mass and Sunyaev-Zeldovich signal}. 2013, \aap, 550, A129,
  \eprint{1204.2743}

\bibitem[{{\sorthelp{Planck Collaboration IntD}}{Planck Collaboration Int.
  IV}(2013)}]{planck2012-IV}
{\sorthelp{Planck Collaboration IntD}}{Planck Collaboration Int. IV},
  {\textit{Planck} intermediate results. IV. The XMM-\textit{Newton} validation
  programme for new \textit{Planck} clusters}. 2013, \aap, 550, A130,
  \eprint{1205.3376}

\bibitem[{{\sorthelp{Planck Collaboration IntEA}}{Planck Collaboration Int.
  V}(2013)}]{planck2012-V}
{\sorthelp{Planck Collaboration IntEA}}{Planck Collaboration Int. V},
  {\textit{Planck} intermediate results. V. Pressure profiles of galaxy
  clusters from the Sunyaev-Zeldovich effect}. 2013, \aap, 550, A131,
  \eprint{1207.4061}

\bibitem[{{\sorthelp{Planck Collaboration IntG}}{Planck Collaboration Int.
  VII}(2013)}]{planck2012-VII}
{\sorthelp{Planck Collaboration IntG}}{Planck Collaboration Int. VII},
  {\textit{Planck} intermediate results. VII. Statistical properties of
  infrared and radio extragalactic sources from the Planck Early Release
  Compact Source Catalogue at frequencies between 100 and 857\,GHz}. 2013,
  \aap, 550, A133, \eprint{1207.4706}

\bibitem[{{\sorthelp{Planck Collaboration IntJ}}{Planck Collaboration Int.
  X}(2013)}]{planck2012-X}
{\sorthelp{Planck Collaboration IntJ}}{Planck Collaboration Int. X},
  {\textit{Planck} intermediate results. X. Physics of the hot gas in the Coma
  cluster}. 2013, \aap, 554, A140, \eprint{1208.3611}

\bibitem[{{\sorthelp{Planck Collaboration IntK}}{Planck Collaboration Int.
  XI}(2013)}]{planck2012-XI}
{\sorthelp{Planck Collaboration IntK}}{Planck Collaboration Int. XI},
  {\textit{Planck} intermediate results. XI. The gas content of dark matter
  halos: the Sunyaev-Zeldovich-stellar mass relation for locally brightest
  galaxies}. 2013, \aap, 557, A52, \eprint{1212.4131}

\bibitem[{{\sorthelp{Planck Collaboration IntM}}{Planck Collaboration Int.
  XIII}(2014)}]{planck2013-XIII}
{\sorthelp{Planck Collaboration IntM}}{Planck Collaboration Int. XIII},
  {\textit{Planck} intermediate results. XIII. Constraints on peculiar
  velocities}. 2014, \aap, 561, A97, \eprint{1303.5090}

\bibitem[{{\sorthelp{Planck Collaboration IntS}}{Planck Collaboration Int.
  XIX}(2015)}]{planck2014-XIX}
{\sorthelp{Planck Collaboration IntS}}{Planck Collaboration Int. XIX},
  {\textit{Planck} intermediate results. XIX. An overview of the polarized
  thermal emission from Galactic dust}. 2015, \aap, 576, A104,
  \eprint{1405.0871}

\bibitem[{{\sorthelp{Planck Collaboration IntZB}}{Planck Collaboration Int.
  XXVII}(2015)}]{planck2014-XXVII}
{\sorthelp{Planck Collaboration IntZB}}{Planck Collaboration Int. XXVII},
  {\textit{Planck} intermediate results. XXVII. High-redshift infrared galaxy
  overdensity candidates and lensed sources discovered by \textit{Planck} and
  confirmed by \textit{Herschel}-SPIRE}. 2015, \aap, 582, A30,
  \eprint{1503.08773}

\bibitem[{{\sorthelp{Planck Collaboration IntZE}}{Planck Collaboration Int.
  XXX}(2016)}]{planck2014-XXX}
{\sorthelp{Planck Collaboration IntZE}}{Planck Collaboration Int. XXX},
  {\textit{Planck} intermediate results. XXX. The angular power spectrum of
  polarized dust emission at intermediate and high Galactic latitudes}. 2016,
  \aap, 586, A133, \eprint{1409.5738}

\bibitem[{{\sorthelp{Planck Collaboration IntZG}}{Planck Collaboration Int.
  XXXII}(2016)}]{planck2014-XXXII}
{\sorthelp{Planck Collaboration IntZG}}{Planck Collaboration Int. XXXII},
  {\textit{Planck} intermediate results. XXXII. The relative orientation
  between the magnetic field and structures traced by interstellar dust}. 2016,
  \aap, 586, A135, \eprint{1409.6728}

\bibitem[{{\sorthelp{Planck Collaboration IntZJ}}{Planck Collaboration Int.
  XXXV}(2016)}]{planck2015-XXXV}
{\sorthelp{Planck Collaboration IntZJ}}{Planck Collaboration Int. XXXV},
  {\textit{Planck} intermediate results. XXXV. Probing the role of the magnetic
  field in the formation of structure in molecular clouds}. 2016, \aap, 586,
  A138, \eprint{1502.04123}

\bibitem[{{\sorthelp{Planck Collaboration IntZL}}{Planck Collaboration Int.
  XXXVII}(2016)}]{planck2015-XXXVII}
{\sorthelp{Planck Collaboration IntZL}}{Planck Collaboration Int. XXXVII},
  {\textit{Planck} intermediate results. XXXVII. Evidence of unbound gas from
  the kinetic Sunyaev-Zeldovich effect}. 2016, \aap, 586, A140,
  \eprint{1504.03339}

\bibitem[{{\sorthelp{Planck Collaboration IntZM}}{Planck Collaboration Int.
  XXXVIII}(2016)}]{planck2015-XXXVIII}
{\sorthelp{Planck Collaboration IntZM}}{Planck Collaboration Int. XXXVIII},
  {\textit{Planck} intermediate results. XXXVIII. \textit{E}- and
  \textit{B}-modes of dust polarization from the magnetized filamentary
  structure of the interstellar medium}. 2016, \aap, 586, A141,
  \eprint{1505.02779}

\bibitem[{{\sorthelp{Planck Collaboration IntZN}}{Planck Collaboration Int.
  XXXIX}(2016)}]{planck2015-XXXIX}
{\sorthelp{Planck Collaboration IntZN}}{Planck Collaboration Int. XXXIX},
  {\textit{Planck} intermediate results. XXXIX. The Planck List of
  High-redshift Source Candidates}. 2016, \aap, 596, A100, \eprint{1508.04171}

\bibitem[{{\sorthelp{Planck Collaboration IntZT}}{Planck Collaboration Int.
  XLV}(2016)}]{planck2016-XLV}
{\sorthelp{Planck Collaboration IntZT}}{Planck Collaboration Int. XLV},
  {\textit{Planck} intermediate results. XLV. Radio spectra of northern
  extragalactic radio sources}. 2016, \aap, 596, A106, \eprint{1606.05120}

\bibitem[{{\sorthelp{Planck Collaboration IntZU}}{Planck Collaboration Int.
  XLVI}(2016)}]{planck2014-a10}
{\sorthelp{Planck Collaboration IntZU}}{Planck Collaboration Int. XLVI},
  {\textit{Planck} intermediate results. XLVI. Reduction of large-scale
  systematic effects in HFI polarization maps and estimation of the
  reionization optical depth}. 2016, \aap, 596, A107, \eprint{1605.02985}

\bibitem[{{\sorthelp{Planck Collaboration IntZV}}{Planck Collaboration Int.
  XLVII}(2016)}]{planck2014-a25}
{\sorthelp{Planck Collaboration IntZV}}{Planck Collaboration Int. XLVII},
  {\textit{Planck} intermediate results. XLVII. Constraints on reionization
  history}. 2016, \aap, 596, A108, \eprint{1605.03507}

\bibitem[{{\sorthelp{Planck Collaboration IntZY}}{Planck Collaboration Int.
  L}(2017)}]{planck2016-L}
{\sorthelp{Planck Collaboration IntZY}}{Planck Collaboration Int. L},
  {\textit{Planck} intermediate results. L. Evidence of spatial variation of
  the polarized thermal dust spectral energy distribution and implications for
  CMB \textit{B}-mode analysis}. 2017, \aap, 599, A51, \eprint{1606.07335}

\bibitem[{{\sorthelp{Planck Collaboration IntZZA}}{Planck Collaboration Int.
  LI}(2017)}]{planck2016-LI}
{\sorthelp{Planck Collaboration IntZZA}}{Planck Collaboration Int. LI},
  {\textit{Planck} intermediate results. LI. Features in the cosmic microwave
  background temperature power spectrum and shifts in cosmological parameters}.
  2017, \aap, 607, A95, \eprint{1608.02487}

\bibitem[{{\sorthelp{Planck Collaboration IntZZB}}{Planck Collaboration Int.
  LII}(2017)}]{planck2016-LII}
{\sorthelp{Planck Collaboration IntZZB}}{Planck Collaboration Int. LII},
  {\textit{Planck} intermediate results. LII. Planet flux densities}. 2017,
  \aap, 607, A122, \eprint{1612.07151}

\bibitem[{{\sorthelp{Planck Collaboration IntZZC}}{Planck Collaboration Int.
  LIII}(2018)}]{planck2017-LIII}
{\sorthelp{Planck Collaboration IntZZC}}{Planck Collaboration Int. LIII},
  {\textit{Planck} intermediate results. LIII. Detection of velocity dispersion
  from the kinetic Sunyaev-Zeldovich effect}. 2018, \aap, in press,
  \eprint{1707.00132}

\bibitem[{{\sorthelp{Planck Collaboration IntZZD}}{Planck Collaboration Int.
  LIV}(2018)}]{planck2018-LIV}
{\sorthelp{Planck Collaboration IntZZD}}{Planck Collaboration Int. LIV},
  {\textit{Planck} intermediate results. LIV. The Planck Multi-frequency
  Catalogue of Non-thermal Sources}. 2018, \aap, submitted, \eprint{1802.0864}

\bibitem[{{POLARBEAR Collaboration} {et~al.}(2017){POLARBEAR Collaboration},
  {Ade}, {Aguilar}, {Akiba}, {Arnold}, {Baccigalupi}, {Barron}, {Beck},
  {Bianchini}, {Boettger}, {Borrill}, {Chapman}, {Chinone}, {Crowley},
  {Cukierman}, {D{\"u}nner}, {Dobbs}, {Ducout}, {Elleflot}, {Errard},
  {Fabbian}, {Feeney}, {Feng}, {Fujino}, {Galitzki}, {Gilbert},
  {Goeckner-Wald}, {Groh}, {Hall}, {Halverson}, {Hamada}, {Hasegawa}, {Hazumi},
  {Hill}, {Howe}, {Inoue}, {Jaehnig}, {Jaffe}, {Jeong}, {Kaneko}, {Katayama},
  {Keating}, {Keskitalo}, {Kisner}, {Krachmalnicoff}, {Kusaka}, {Le Jeune},
  {Lee}, {Leitch}, {Leon}, {Linder}, {Lowry}, {Matsuda}, {Matsumura}, {Minami},
  {Montgomery}, {Navaroli}, {Nishino}, {Paar}, {Peloton}, {Pham}, {Poletti},
  {Puglisi}, {Reichardt}, {Richards}, {Ross}, {Segawa}, {Sherwin},
  {Silva-Feaver}, {Siritanasak}, {Stebor}, {Stompor}, {Suzuki}, {Tajima},
  {Takakura}, {Takatori}, {Tanabe}, {Teply}, {Tomaru}, {Tucker}, {Whitehorn},
  \& {Zahn}}]{PolarBear17}
{POLARBEAR Collaboration}, {Ade}, P.~A.~R., {Aguilar}, M., {et~al.}, {A
  Measurement of the Cosmic Microwave Background B-mode Polarization Power
  Spectrum at Subdegree Scales from Two Years of polarbear Data}. 2017, \apj,
  848, 121, \eprint{1705.02907}

\bibitem[{{Pryke} {et~al.}(2009){Pryke}, {Ade}, {Bock}, {Bowden}, {Brown},
  {Cahill}, {Castro}, {Church}, {Culverhouse}, {Friedman}, {Ganga}, {Gear},
  {Gupta}, {Hinderks}, {Kovac}, {Lange}, {Leitch}, {Melhuish}, {Memari},
  {Murphy}, {Orlando}, {Schwarz}, {O'Sullivan}, {Piccirillo}, {Rajguru},
  {Rusholme}, {Taylor}, {Thompson}, {Turner}, {Wu}, \& {Zemcov}}]{Pryke09}
{Pryke}, C., {Ade}, P., {Bock}, J., {et~al.}, {Second and Third Season QUaD
  Cosmic Microwave Background Temperature and Polarization Power Spectra}.
  2009, \apj, 692, 1247, \eprint{0805.1944}

\bibitem[{{Puget} {et~al.}(1996){Puget}, {Abergel}, {Bernard}, {Boulanger},
  {Burton}, {Desert}, \& {Hartmann}}]{Puget_1996}
{Puget}, J.-L., {Abergel}, A., {Bernard}, J.-P., {et~al.}, {Tentative detection
  of a cosmic far-infrared background with COBE.} 1996, \aap, 308, L5

\bibitem[{{Pullen} {et~al.}(2016){Pullen}, {Alam}, {He}, \& {Ho}}]{Pullen16}
{Pullen}, A.~R., {Alam}, S., {He}, S., \& {Ho}, S., {Constraining gravity at
  the largest scales through CMB lensing and galaxy velocities}. 2016, \mnras,
  460, 4098, \eprint{1511.04457}

\bibitem[{{Raghunathan} {et~al.}(2017){Raghunathan}, {Bianchini}, \&
  {Reichardt}}]{Rag18}
{Raghunathan}, S., {Bianchini}, F., \& {Reichardt}, C.~L., {Imprints of
  gravitational lensing in the Planck CMB data at the location of WISExSCOS
  galaxies}. 2017, ArXiv e-prints, \eprint{1710.09770}

\bibitem[{{Rathna Kumar} {et~al.}(2015){Rathna Kumar}, {Stalin}, \&
  {Prabhu}}]{Rathna15}
{Rathna Kumar}, S., {Stalin}, C.~S., \& {Prabhu}, T.~P., {H$_{0}$ from ten
  well-measured time delay lenses}. 2015, \aap, 580, A38, \eprint{1404.2920}

\bibitem[{{Readhead} {et~al.}(2004){Readhead}, {Myers}, {Pearson}, {Sievers},
  {Mason}, {Contaldi}, {Bond}, {Bustos}, {Altamirano}, {Achermann}, {Bronfman},
  {Carlstrom}, {Cartwright}, {Casassus}, {Dickinson}, {Holzapfel}, {Kovac},
  {Leitch}, {May}, {Padin}, {Pogosyan}, {Pospieszalski}, {Pryke}, {Reeves},
  {Shepherd}, \& {Torres}}]{Readhead04}
{Readhead}, A.~C.~S., {Myers}, S.~T., {Pearson}, T.~J., {et~al.}, {Polarization
  Observations with the Cosmic Background Imager}. 2004, Science, 306, 836,
  \eprint{astro-ph/0409569}

\bibitem[{{Reid} {et~al.}(2010){Reid}, {Percival}, {Eisenstein}, {Verde},
  {Spergel}, {Skibba}, {Bahcall}, {Budavari}, {Frieman}, {Fukugita}, {Gott},
  {Gunn}, {Ivezi{\'c}}, {Knapp}, {Kron}, {Lupton}, {McKay}, {Meiksin},
  {Nichol}, {Pope}, {Schlegel}, {Schneider}, {Stoughton}, {Strauss}, {Szalay},
  {Tegmark}, {Vogeley}, {Weinberg}, {York}, \& {Zehavi}}]{2010MNRAS.404...60R}
{Reid}, B.~A., {Percival}, W.~J., {Eisenstein}, D.~J., {et~al.}, {Cosmological
  constraints from the clustering of the Sloan Digital Sky Survey DR7 luminous
  red galaxies}. 2010, \mnras, 404, 60, \eprint{0907.1659}

\bibitem[{{Remazeilles} {et~al.}(2011){Remazeilles}, {Delabrouille}, \&
  {Cardoso}}]{GNILC}
{Remazeilles}, M., {Delabrouille}, J., \& {Cardoso}, J.-F., {Foreground
  component separation with generalized Internal Linear Combination}. 2011,
  \mnras, 418, 467, \eprint{1103.1166}

\bibitem[{{Riess} {et~al.}(2018{\natexlab{a}}){Riess}, {Casertano}, {Yuan},
  {Macri}, {Anderson}, {MacKenty}, {Bowers}, {Clubb}, {Filippenko}, {Jones}, \&
  {Tucker}}]{Riess18}
{Riess}, A.~G., {Casertano}, S., {Yuan}, W., {et~al.}, {New Parallaxes of
  Galactic Cepheids from Spatially Scanning the Hubble Space Telescope:
  Implications for the Hubble Constant}. 2018{\natexlab{a}}, \apj, 855, 136,
  \eprint{1801.01120}

\bibitem[{{Riess} {et~al.}(2018{\natexlab{b}}){Riess}, {Casertano}, {Yuan},
  {Macri}, {Bucciarelli}, {Lattanzi}, {MacKenty}, {Bowers}, {Zheng},
  {Filippenko}, {Huang}, \& {Anderson}}]{Riess_Gaia}
{Riess}, A.~G., {Casertano}, S., {Yuan}, W., {et~al.}, {Milky Way Cepheid
  Standards for Measuring Cosmic Distances and Application to Gaia DR2:
  Implications for the Hubble Constant}. 2018{\natexlab{b}}, ArXiv e-prints,
  \eprint{1804.10655}

\bibitem[{{Riess} {et~al.}(2019){Riess}, {Casertano}, {Yuan}, {Macri}, \&
  {Scolnic}}]{Riess19}
{Riess}, A.~G., {Casertano}, S., {Yuan}, W., {Macri}, L.~M., \& {Scolnic}, D.,
  {Large Magellanic Cloud Cepheid Standards Provide a 1\% Foundation for the
  Determination of the Hubble Constant and Stronger Evidence for Physics beyond
  {\ensuremath{\Lambda}}CDM}. 2019, \apj, 876, 85, \eprint{1903.07603}

\bibitem[{{Riess} {et~al.}(1998){Riess}, {Filippenko}, {Challis},
  {Clocchiatti}, {Diercks}, {Garnavich}, {Gilliland}, {Hogan}, {Jha},
  {Kirshner}, {Leibundgut}, {Phillips}, {Reiss}, {Schmidt}, {Schommer},
  {Smith}, {Spyromilio}, {Stubbs}, {Suntzeff}, \& {Tonry}}]{Riess98}
{Riess}, A.~G., {Filippenko}, A.~V., {Challis}, P., {et~al.}, {Observational
  Evidence from Supernovae for an Accelerating Universe and a Cosmological
  Constant}. 1998, \aj, 116, 1009, \eprint{astro-ph/9805201}

\bibitem[{{Riess} {et~al.}(2011){Riess}, {Macri}, {Casertano}, {Lampeitl},
  {Ferguson}, {Filippenko}, {Jha}, {Li}, \& {Chornock}}]{Riess11}
{Riess}, A.~G., {Macri}, L., {Casertano}, S., {et~al.}, {A 3\% Solution:
  Determination of the Hubble Constant with the Hubble Space Telescope and Wide
  Field Camera 3}. 2011, \apj, 730, 119, \eprint{1103.2976}

\bibitem[{{Riess} {et~al.}(2016){Riess}, {Macri}, {Hoffmann}, {Scolnic},
  {Casertano}, {Filippenko}, {Tucker}, {Reid}, {Jones}, {Silverman},
  {Chornock}, {Challis}, {Yuan}, {Brown}, \& {Foley}}]{Riess16}
{Riess}, A.~G., {Macri}, L.~M., {Hoffmann}, S.~L., {et~al.}, {A 2.4\%
  Determination of the Local Value of the Hubble Constant}. 2016, \apj, 826,
  56, \eprint{1604.01424}

\bibitem[{{Ross} {et~al.}(2015){Ross}, {Samushia}, {Howlett}, {Percival},
  {Burden}, \& {Manera}}]{Ross15}
{Ross}, A.~J., {Samushia}, L., {Howlett}, C., {et~al.}, {The clustering of the
  SDSS DR7 main Galaxy sample - I. A 4 per cent distance measure at z = 0.15}.
  2015, \mnras, 449, 835, \eprint{1409.3242}

\bibitem[{{Sachs} \& {Wolfe}(1967)}]{SacWol67}
{Sachs}, R.~K. \& {Wolfe}, A.~M., {Perturbations of a Cosmological Model and
  Angular Variations of the Microwave Background}. 1967, \apj, 147, 73

\bibitem[{{Sakstein} \& {Jain}(2017)}]{Sakstein17}
{Sakstein}, J. \& {Jain}, B., {Implications of the Neutron Star Merger GW170817
  for Cosmological Scalar-Tensor Theories}. 2017, \prl, 119, 251303,
  \eprint{1710.05893}

\bibitem[{{Schmidt} {et~al.}(2015){Schmidt}, {M{\'e}nard}, {Scranton},
  {Morrison}, {Rahman}, \& {Hopkins}}]{schmidt_2015}
{Schmidt}, S.~J., {M{\'e}nard}, B., {Scranton}, R., {et~al.}, {Inferring the
  redshift distribution of the cosmic infrared background}. 2015, \mnras, 446,
  2696, \eprint{1407.0031}

\bibitem[{{Sch{\"o}nrich} {et~al.}(2010){Sch{\"o}nrich}, {Binney}, \&
  {Dehnen}}]{schonrich2010}
{Sch{\"o}nrich}, R., {Binney}, J., \& {Dehnen}, W., {Local kinematics and the
  local standard of rest}. 2010, \mnras, 403, 1829, \eprint{0912.3693}

\bibitem[{{Scott} {et~al.}(2016){Scott}, {Contreras}, {Narimani}, \&
  {Ma}}]{ScottCNM}
{Scott}, D., {Contreras}, D., {Narimani}, A., \& {Ma}, Y.-Z., {The information
  content of cosmic microwave background anisotropies}. 2016, \jcap, 6, 046,
  \eprint{1603.03550}

\bibitem[{{Scott} {et~al.}(1995){Scott}, {Silk}, \& {White}}]{ScottSW95}
{Scott}, D., {Silk}, J., \& {White}, M., {From Microwave Anisotropies to
  Cosmology}. 1995, Science, 268, 829, \eprint{astro-ph/9505015}

\bibitem[{{Scott} \& {White}(1994)}]{Scott94}
{Scott}, D. \& {White}, M. 1994, in CMB Anisotropies Two Years after COBE:
  Observations, Theory and the Future, ed. L.~M. {Krauss}, 214

\bibitem[{{Scott} {et~al.}(2003){Scott}, {Carreira}, {Cleary}, {Davies},
  {Davis}, {Dickinson}, {Grainge}, {Guti{\'e}rrez}, {Hobson}, {Jones},
  {Kneissl}, {Lasenby}, {Maisinger}, {Pooley}, {Rebolo}, {Rubi{\~n}o-Martin},
  {Sosa Molina}, {Rusholme}, {Saunders}, {Savage}, {Slosar}, {Taylor},
  {Titterington}, {Waldram}, {Watson}, \& {Wilkinson}}]{VSA_data}
{Scott}, P.~F., {Carreira}, P., {Cleary}, K., {et~al.}, {First results from the
  Very Small Array - III. The cosmic microwave background power spectrum}.
  2003, \mnras, 341, 1076, \eprint{astro-ph/0205380}

\bibitem[{{Scrimgeour} {et~al.}(2016){Scrimgeour}, {Davis}, {Blake},
  {Staveley-Smith}, {Magoulas}, {Springob}, {Beutler}, {Colless}, {Johnson},
  {Jones}, {Koda}, {Lucey}, {Ma}, {Mould}, \& {Poole}}]{Scrimgeour2016}
{Scrimgeour}, M.~I., {Davis}, T.~M., {Blake}, C., {et~al.}, {The 6dF Galaxy
  Survey: bulk flows on 50-70 h$^{-1}$ Mpc scales}. 2016, \mnras, 455, 386,
  \eprint{1511.06930}

\bibitem[{Seljak(1996)}]{Seljak:1995ve}
Seljak, U., {Gravitational lensing effect on cosmic microwave background
  anisotropies: A Power spectrum approach}. 1996, \apj, 463, 1,
  \eprint{astro-ph/9505109}

\bibitem[{{Seljak}(1997)}]{Sel97}
{Seljak}, U., {Measuring Polarization in the Cosmic Microwave Background}.
  1997, \apj, 482, 6, \eprint{astro-ph/9608131}

\bibitem[{{Sereno} {et~al.}(2017){Sereno}, {Covone}, {Izzo}, {Ettori},
  {Coupon}, \& {Lieu}}]{Sereno17}
{Sereno}, M., {Covone}, G., {Izzo}, L., {et~al.}, {PSZ2LenS. Weak lensing
  analysis of the Planck clusters in the CFHTLenS and in the RCSLenS}. 2017,
  \mnras, 472, 1946, \eprint{1703.06886}

\bibitem[{{Shang} {et~al.}(2012){Shang}, {Haiman}, {Knox}, \&
  {Oh}}]{shang_2012}
{Shang}, C., {Haiman}, Z., {Knox}, L., \& {Oh}, S.~P., {Improved models for
  cosmic infrared background anisotropies: new constraints on the infrared
  galaxy population}. 2012, \mnras, 421, 2832, \eprint{1109.1522}

\bibitem[{{Sherwin} {et~al.}(2017){Sherwin}, {van Engelen}, {Sehgal},
  {Madhavacheril}, {Addison}, {Aiola}, {Allison}, {Battaglia}, {Becker},
  {Beall}, {Bond}, {Calabrese}, {Datta}, {Devlin}, {D{\"u}nner}, {Dunkley},
  {Fox}, {Gallardo}, {Halpern}, {Hasselfield}, {Henderson}, {Hill}, {Hilton},
  {Hubmayr}, {Hughes}, {Hincks}, {Hlozek}, {Huffenberger}, {Koopman},
  {Kosowsky}, {Louis}, {Maurin}, {McMahon}, {Moodley}, {Naess}, {Nati},
  {Newburgh}, {Niemack}, {Page}, {Sievers}, {Spergel}, {Staggs}, {Thornton},
  {Van Lanen}, {Vavagiakis}, \& {Wollack}}]{Sherwin17}
{Sherwin}, B.~D., {van Engelen}, A., {Sehgal}, N., {et~al.}, {Two-season
  Atacama Cosmology Telescope polarimeter lensing power spectrum}. 2017, \prd,
  95, 123529, \eprint{1611.09753}

\bibitem[{{Singh} {et~al.}(2018){Singh}, {Alam}, {Mandelbaum}, {Seljak},
  {Rodriguez-Torres}, \& {Ho}}]{Singh18}
{Singh}, S., {Alam}, S., {Mandelbaum}, R., {et~al.}, {Probing gravity with a
  joint analysis of galaxy and CMB lensing and SDSS spectroscopy}. 2018, ArXiv
  e-prints, \eprint{1803.08915}

\bibitem[{{Singh} {et~al.}(2017){Singh}, {Mandelbaum}, \&
  {Brownstein}}]{Singh17}
{Singh}, S., {Mandelbaum}, R., \& {Brownstein}, J.~R., {Cross-correlating
  Planck CMB lensing with SDSS: lensing-lensing and galaxy-lensing
  cross-correlations}. 2017, \mnras, 464, 2120, \eprint{1606.08841}

\bibitem[{{Smoot} {et~al.}(1992){Smoot}, {Bennett}, {Kogut}, {Wright}, {Aymon},
  {Boggess}, {Cheng}, {de Amici}, {Gulkis}, {Hauser}, {Hinshaw}, {Jackson},
  {Janssen}, {Kaita}, {Kelsall}, {Keegstra}, {Lineweaver}, {Loewenstein},
  {Lubin}, {Mather}, {Meyer}, {Moseley}, {Murdock}, {Rokke}, {Silverberg},
  {Tenorio}, {Weiss}, \& {Wilkinson}}]{Smoot92}
{Smoot}, G.~F., {Bennett}, C.~L., {Kogut}, A., {et~al.}, {Structure in the COBE
  differential microwave radiometer first-year maps}. 1992, \apjl, 396, L1

\bibitem[{{Song} {et~al.}(2003){Song}, {Cooray}, {Knox}, \&
  {Zaldarriaga}}]{Song_2003}
{Song}, Y.-S., {Cooray}, A., {Knox}, L., \& {Zaldarriaga}, M., {The
  Far-Infrared Background Correlation with Cosmic Microwave Background
  Lensing}. 2003, \apj, 590, 664, \eprint{astro-ph/0209001}

\bibitem[{{Spergel} {et~al.}(2007){Spergel}, {Bean}, {Dor{\'e}}, {Nolta},
  {Bennett}, {Dunkley}, {Hinshaw}, {Jarosik}, {Komatsu}, {Page}, {Peiris},
  {Verde}, {Halpern}, {Hill}, {Kogut}, {Limon}, {Meyer}, {Odegard}, {Tucker},
  {Weiland}, {Wollack}, \& {Wright}}]{spergel2007}
{Spergel}, D.~N., {Bean}, R., {Dor{\'e}}, O., {et~al.}, {Three-Year Wilkinson
  Microwave Anisotropy Probe (WMAP) Observations: Implications for Cosmology}.
  2007, \apjs, 170, 377, \eprint{astro-ph/0603449}

\bibitem[{{Spergel} \& {Zaldarriaga}(1997)}]{SpeZal97}
{Spergel}, D.~N. \& {Zaldarriaga}, M., {Cosmic Microwave Background
  Polarization as a Direct Test of Inflation}. 1997, Physical Review Letters,
  79, 2180, \eprint{astro-ph/9705182}

\bibitem[{{Story} {et~al.}(2015){Story}, {Hanson}, {Ade}, {Aird}, {Austermann},
  {Beall}, {Bender}, {Benson}, {Bleem}, {Carlstrom}, {Chang}, {Chiang}, {Cho},
  {Citron}, {Crawford}, {Crites}, {de Haan}, {Dobbs}, {Everett}, {Gallicchio},
  {Gao}, {George}, {Gilbert}, {Halverson}, {Harrington}, {Henning}, {Hilton},
  {Holder}, {Holzapfel}, {Hoover}, {Hou}, {Hrubes}, {Huang}, {Hubmayr},
  {Irwin}, {Keisler}, {Knox}, {Lee}, {Leitch}, {Li}, {Liang}, {Luong-Van},
  {McMahon}, {Mehl}, {Meyer}, {Mocanu}, {Montroy}, {Natoli}, {Nibarger},
  {Novosad}, {Padin}, {Pryke}, {Reichardt}, {Ruhl}, {Saliwanchik}, {Sayre},
  {Schaffer}, {Smecher}, {Stark}, {Tucker}, {Vanderlinde}, {Vieira}, {Wang},
  {Whitehorn}, {Yefremenko}, \& {Zahn}}]{Story15}
{Story}, K.~T., {Hanson}, D., {Ade}, P.~A.~R., {et~al.}, {A Measurement of the
  Cosmic Microwave Background Gravitational Lensing Potential from 100 Square
  Degrees of SPTpol Data}. 2015, \apj, 810, 50, \eprint{1412.4760}

\bibitem[{{Sunyaev} \& {Zeldovich}(1980)}]{SunZel80}
{Sunyaev}, R.~A. \& {Zeldovich}, I.~B., {The velocity of clusters of galaxies
  relative to the microwave background - The possibility of its measurement}.
  1980, \mnras, 190, 413

\bibitem[{{Sunyaev} \& {Zeldovich}(1970)}]{SunZel70}
{Sunyaev}, R.~A. \& {Zeldovich}, Y.~B., {Small-Scale Fluctuations of Relic
  Radiation}. 1970, \apss, 7, 3

\bibitem[{{Sunyaev} \& {Zeldovich}(1972)}]{SunZel72}
{Sunyaev}, R.~A. \& {Zeldovich}, Y.~B., {The Observations of Relic Radiation as
  a Test of the Nature of X-Ray Radiation from the Clusters of Galaxies}. 1972,
  Comments on Astrophysics and Space Physics, 4, 173

\bibitem[{{Takahashi} {et~al.}(2012){Takahashi}, {Sato}, {Nishimichi},
  {Taruya}, \& {Oguri}}]{Tak12}
{Takahashi}, R., {Sato}, M., {Nishimichi}, T., {Taruya}, A., \& {Oguri}, M.,
  {Revising the Halofit Model for the Nonlinear Matter Power Spectrum}. 2012,
  \apj, 761, 152, \eprint{1208.2701}

\bibitem[{{Tauber} {et~al.}(2010){Tauber}, {Mandolesi}, {Puget}, {Banos},
  {Bersanelli}, {Bouchet}, {Butler}, {Charra}, {Crone}, {Dodsworth}, \&
  et~al.}]{tauber2010a}
{Tauber}, J.~A., {Mandolesi}, N., {Puget}, J., {et~al.}, {\textit{Planck}
  pre-launch status: The Planck mission}. 2010, \aap, 520, A1

\bibitem[{{Tegmark} {et~al.}(2004){Tegmark}, {Blanton}, {Strauss}, {Hoyle},
  {Schlegel}, {Scoccimarro}, {Vogeley}, {Weinberg}, {Zehavi}, {Berlind},
  {Budavari}, {Connolly}, {Eisenstein}, {Finkbeiner}, {Frieman}, {Gunn},
  {Hamilton}, {Hui}, {Jain}, {Johnston}, {Kent}, {Lin}, {Nakajima}, {Nichol},
  {Ostriker}, {Pope}, {Scranton}, {Seljak}, {Sheth}, {Stebbins}, {Szalay},
  {Szapudi}, {Verde}, {Xu}, {Annis}, {Bahcall}, {Brinkmann}, {Burles},
  {Castander}, {Csabai}, {Loveday}, {Doi}, {Fukugita}, {Gott}, {Hennessy},
  {Hogg}, {Ivezi{\'c}}, {Knapp}, {Lamb}, {Lee}, {Lupton}, {McKay}, {Kunszt},
  {Munn}, {O'Connell}, {Peoples}, {Pier}, {Richmond}, {Rockosi}, {Schneider},
  {Stoughton}, {Tucker}, {Vanden Berk}, {Yanny}, {York}, \& {SDSS
  Collaboration}}]{Teg04}
{Tegmark}, M., {Blanton}, M.~R., {Strauss}, M.~A., {et~al.}, {The
  Three-Dimensional Power Spectrum of Galaxies from the Sloan Digital Sky
  Survey}. 2004, \apj, 606, 702, \eprint{astro-ph/0310725}

\bibitem[{{Tegmark} \& {Zaldarriaga}(2002)}]{TegZal02}
{Tegmark}, M. \& {Zaldarriaga}, M., {Separating the early universe from the
  late universe: Cosmological parameter estimation beyond the black box}. 2002,
  \prd, 66, 103508, \eprint{astro-ph/0207047}

\bibitem[{{Thacker} {et~al.}(2013){Thacker}, {Cooray}, {Smidt}, {De Bernardis},
  {Mitchell-Wynne}, {Amblard}, {Auld}, {Baes}, {Clements}, {Dariush}, {De
  Zotti}, {Dunne}, {Eales}, {Hopwood}, {Hoyos}, {Ibar}, {Jarvis}, {Maddox},
  {Micha{\l}owski}, {Pascale}, {Scott}, {Serjeant}, {Smith}, {Valiante}, \&
  {van der Werf}}]{thacker_2013}
{Thacker}, C., {Cooray}, A., {Smidt}, J., {et~al.}, {H-ATLAS: The Cosmic
  Abundance of Dust from the Far-infrared Background Power Spectrum}. 2013,
  \apj, 768, 58, \eprint{1212.2211}

\bibitem[{{Trotta} \& {Melchiorri}(2005)}]{TroMel05}
{Trotta}, R. \& {Melchiorri}, A., {Indication for Primordial Anisotropies in
  the Neutrino Background from the Wilkinson Microwave Anisotropy Probe and the
  Sloan Digital Sky Survey}. 2005, \prl, 95, 011305, \eprint{astro-ph/0412066}

\bibitem[{{Troxel} {et~al.}(2018){Troxel}, {Krause}, {Chang}, {Eifler},
  {Friedrich}, {Gruen}, {MacCrann}, {Chen}, {Davis}, {DeRose}, {Dodelson},
  {Gatti}, {Hoyle}, {Huterer}, {Jarvis}, {Lacasa}, {Peiris}, {Prat},
  {Samuroff}, {S{\'a}nchez}, {Sheldon}, {Vielzeuf}, {Wang}, {Zuntz}, {Abdalla},
  {Allam}, {Annis}, {Avila}, {Bertin}, {Brooks}, {Burke}, {Carnero Rosell},
  {Carrasco Kind}, {Carretero}, {Crocce}, {Cunha}, {D'Andrea}, {da Costa}, {De
  Vicente}, {Diehl}, {Doel}, {Evrard}, {Flaugher}, {Fosalba}, {Frieman},
  {Garc{\'{\i}}a-Bellido}, {Gaztanaga}, {Gerdes}, {Gruendl}, {Gschwend},
  {Gutierrez}, {Hartley}, {Hollowood}, {Honscheid}, {James}, {Kirk}, {Kuehn},
  {Kuropatkin}, {Li}, {Lima}, {March}, {Menanteau}, {Miquel}, {Mohr}, {Ogando},
  {Plazas}, {Roodman}, {Sanchez}, {Scarpine}, {Schindler}, {Sevilla-Noarbe},
  {Smith}, {Soares-Santos}, {Sobreira}, {Suchyta}, {Swanson}, {Thomas},
  {Walker}, \& {Wechsler}}]{Troxel18b}
{Troxel}, M.~A., {Krause}, E., {Chang}, C., {et~al.}, {Survey geometry and the
  internal consistency of recent cosmic shear measurements}. 2018, ArXiv
  e-prints, \eprint{1804.10663}

\bibitem[{{Troxel} {et~al.}(2017){Troxel}, {MacCrann}, {Zuntz}, {Eifler},
  {Krause}, {Dodelson}, {Gruen}, {Blazek}, {Friedrich}, {Samuroff}, {Prat},
  {Secco}, {Davis}, {Fert{\'e}}, {DeRose}, {Alarcon}, {Amara}, {Baxter},
  {Becker}, {Bernstein}, {Bridle}, {Cawthon}, {Chang}, {Choi}, {De Vicente},
  {Drlica-Wagner}, {Elvin-Poole}, {Frieman}, {Gatti}, {Hartley}, {Honscheid},
  {Hoyle}, {Huff}, {Huterer}, {Jain}, {Jarvis}, {Kacprzak}, {Kirk}, {Kokron},
  {Krawiec}, {Lahav}, {Liddle}, {Peacock}, {Rau}, {Refregier}, {Rollins},
  {Rozo}, {Rykoff}, {S{\'a}nchez}, {Sevilla-Noarbe}, {Sheldon}, {Stebbins},
  {Varga}, {Vielzeuf}, {Wang}, {Wechsler}, {Yanny}, {Abbott}, {Abdalla},
  {Allam}, {Annis}, {Bechtol}, {Benoit-L{\'e}vy}, {Bertin}, {Brooks},
  {Buckley-Geer}, {Burke}, {Carnero Rosell}, {Carrasco Kind}, {Carretero},
  {Castander}, {Crocce}, {Cunha}, {D'Andrea}, {da Costa}, {DePoy}, {Desai},
  {Diehl}, {Dietrich}, {Doel}, {Fernandez}, {Flaugher}, {Fosalba},
  {Garc{\'{\i}}a-Bellido}, {Gaztanaga}, {Gerdes}, {Giannantonio}, {Goldstein},
  {Gruendl}, {Gschwend}, {Gutierrez}, {James}, {Jeltema}, {Johnson}, {Johnson},
  {Kent}, {Kuehn}, {Kuhlmann}, {Kuropatkin}, {Li}, {Lima}, {Lin}, {Maia},
  {March}, {Marshall}, {Martini}, {Melchior}, {Menanteau}, {Miquel}, {Mohr},
  {Neilsen}, {Nichol}, {Nord}, {Petravick}, {Plazas}, {Romer}, {Roodman},
  {Sako}, {Sanchez}, {Scarpine}, {Schindler}, {Schubnell}, {Smith}, {Smith},
  {Soares-Santos}, {Sobreira}, {Suchyta}, {Swanson}, {Tarle}, {Thomas},
  {Tucker}, {Vikram}, {Walker}, {Weller}, \& {Zhang}}]{Troxel18a}
{Troxel}, M.~A., {MacCrann}, N., {Zuntz}, J., {et~al.}, {Dark Energy Survey
  Year 1 Results: Cosmological Constraints from Cosmic Shear}. 2017, ArXiv
  e-prints, \eprint{1708.01538}

\bibitem[{{Tully} {et~al.}(2008){Tully}, {Shaya}, {Karachentsev}, {Courtois},
  {Kocevski}, {Rizzi}, \& {Peel}}]{tully2008}
{Tully}, R.~B., {Shaya}, E.~J., {Karachentsev}, I.~D., {et~al.}, {Our Peculiar
  Motion Away from the Local Void}. 2008, \apj, 676, 184, \eprint{0705.4139}

\bibitem[{{Turner}(1991)}]{Turner91}
{Turner}, M.~S., {Tilted Universe and other remnants of the preinflationary
  Universe}. 1991, \prd, 44, 3737

\bibitem[{{van Uitert} {et~al.}(2018){van Uitert}, {Joachimi}, {Joudaki},
  {Amon}, {Heymans}, {K{\"o}hlinger}, {Asgari}, {Blake}, {Choi}, {Erben},
  {Farrow}, {Harnois-D{\'e}raps}, {Hildebrandt}, {Hoekstra}, {Kitching},
  {Klaes}, {Kuijken}, {Merten}, {Miller}, {Nakajima}, {Schneider}, {Valentijn},
  \& {Viola}}]{vanUitert18}
{van Uitert}, E., {Joachimi}, B., {Joudaki}, S., {et~al.}, {KiDS+GAMA:
  Cosmology constraints from a joint analysis of cosmic shear, galaxy-galaxy
  lensing and angular clustering}. 2018, \mnras, \eprint{1706.05004}

\bibitem[{{Viero} {et~al.}(2009){Viero}, {Ade}, {Bock}, {Chapin}, {Devlin},
  {Griffin}, {Gundersen}, {Halpern}, {Hargrave}, {Hughes}, {Klein},
  {MacTavish}, {Marsden}, {Martin}, {Mauskopf}, {Moncelsi}, {Negrello},
  {Netterfield}, {Olmi}, {Pascale}, {Patanchon}, {Rex}, {Scott}, {Semisch},
  {Thomas}, {Truch}, {Tucker}, {Tucker}, \& {Wiebe}}]{Viero2009}
{Viero}, M.~P., {Ade}, P.~A.~R., {Bock}, J.~J., {et~al.}, {BLAST: Correlations
  in the Cosmic Far-Infrared Background at 250, 350, and 500 {$\mu$}m Reveal
  Clustering of Star-forming Galaxies}. 2009, \apj, 707, 1766,
  \eprint{0904.1200}

\bibitem[{{Viero} {et~al.}(2013){Viero}, {Wang}, {Zemcov}, {Addison},
  {Amblard}, {Arumugam}, {Aussel}, {B{\'e}thermin}, {Bock}, {Boselli}, {Buat},
  {Burgarella}, {Casey}, {Clements}, {Conley}, {Conversi}, {Cooray}, {De
  Zotti}, {Dowell}, {Farrah}, {Franceschini}, {Glenn}, {Griffin},
  {Hatziminaoglou}, {Heinis}, {Ibar}, {Ivison}, {Lagache}, {Levenson},
  {Marchetti}, {Marsden}, {Nguyen}, {O'Halloran}, {Oliver}, {Omont}, {Page},
  {Papageorgiou}, {Pearson}, {P{\'e}rez-Fournon}, {Pohlen}, {Rigopoulou},
  {Roseboom}, {Rowan-Robinson}, {Schulz}, {Scott}, {Seymour}, {Shupe}, {Smith},
  {Symeonidis}, {Vaccari}, {Valtchanov}, {Vieira}, {Wardlow}, \&
  {Xu}}]{Viero_2013}
{Viero}, M.~P., {Wang}, L., {Zemcov}, M., {et~al.}, {HerMES: Cosmic Infrared
  Background Anisotropies and the Clustering of Dusty Star-forming Galaxies}.
  2013, \apj, 772, 77, \eprint{1208.5049}

\bibitem[{{Vittorio} \& {Silk}(1985)}]{VittorioSilk85}
{Vittorio}, N. \& {Silk}, J., {Can a relic cosmological constant reconcile
  inflationary predictions with the observations?} 1985, \apjl, 297, L1

\bibitem[{{von der Linden} {et~al.}(2014){von der Linden}, {Mantz}, {Allen},
  {Applegate}, {Kelly}, {Morris}, {Wright}, {Allen}, {Burchat}, {Burke},
  {Donovan}, \& {Ebeling}}]{vdL14}
{von der Linden}, A., {Mantz}, A., {Allen}, S.~W., {et~al.}, {Robust
  weak-lensing mass calibration of Planck galaxy clusters}. 2014, \mnras, 443,
  1973, \eprint{1402.2670}

\bibitem[{{Wang} {et~al.}(2003){Wang}, {Tegmark}, {Jain}, \&
  {Zaldarriaga}}]{Wang03}
{Wang}, X., {Tegmark}, M., {Jain}, B., \& {Zaldarriaga}, M., {Last stand before
  WMAP: Cosmological parameters from lensing, CMB, and galaxy clustering}.
  2003, \prd, 68, 123001, \eprint{astro-ph/0212417}

\bibitem[{{Weinberg} {et~al.}(2013){Weinberg}, {Mortonson}, {Eisenstein},
  {Hirata}, {Riess}, \& {Rozo}}]{Weinberg13}
{Weinberg}, D.~H., {Mortonson}, M.~J., {Eisenstein}, D.~J., {et~al.},
  {Observational probes of cosmic acceleration}. 2013, \physrep, 530, 87,
  \eprint{1201.2434}

\bibitem[{{White}(2004)}]{White04}
{White}, M., {Baryons and weak lensing power spectra}. 2004, Astroparticle
  Physics, 22, 211, \eprint{astro-ph/0405593}

\bibitem[{{White} {et~al.}(1994){White}, {Scott}, \& {Silk}}]{WSS94}
{White}, M., {Scott}, D., \& {Silk}, J., {Anisotropies in the Cosmic Microwave
  Background}. 1994, \araa, 32, 319

\bibitem[{{Will}(2006)}]{Will06}
{Will}, C.~M., {The Confrontation between General Relativity and Experiment}.
  2006, Living Reviews in Relativity, 9, \eprint{gr-qc/0510072}

\bibitem[{{Yahil} {et~al.}(1977){Yahil}, {Tammann}, \& {Sandage}}]{yahil1977}
{Yahil}, A., {Tammann}, G.~A., \& {Sandage}, A., {The Local Group - The solar
  motion relative to its centroid}. 1977, \apj, 217, 903

\bibitem[{{Zaldarriaga} \& {Seljak}(1997)}]{ZalSel97}
{Zaldarriaga}, M. \& {Seljak}, U., {All-sky analysis of polarization in the
  microwave background}. 1997, \prd, 55, 1830, \eprint{astro-ph/9609170}

\bibitem[{Zaldarriaga \& Seljak(1998)}]{Zaldarriaga:1998ar}
Zaldarriaga, M. \& Seljak, U., {Gravitational Lensing Effect on Cosmic
  Microwave Background Polarization}. 1998, Phys. Rev., D58, 023003,
  \eprint{astro-ph/9803150}

\bibitem[{Zarrouk {et~al.}(2018)}]{Zarrouk18}
Zarrouk, P. {et~al.}, {The clustering of the SDSS-IV extended Baryon
  Oscillation Spectroscopic Survey DR14 quasar sample: measurement of the
  growth rate of structure from the anisotropic correlation function between
  redshift 0.8 and 2.2}. 2018, \mnras, \eprint{1801.03062}

\bibitem[{{Zhan} \& {Knox}(2004)}]{Zhan04}
{Zhan}, H. \& {Knox}, L., {Effect of Hot Baryons on the Weak-Lensing Shear
  Power Spectrum}. 2004, \apjl, 616, L75, \eprint{astro-ph/0409198}

\bibitem[{{Zibin} \& {Scott}(2008)}]{Zibin08}
{Zibin}, J.~P. \& {Scott}, D., {Gauging the cosmic microwave background}. 2008,
  \prd, 78, 123529, \eprint{0808.2047}

\bibitem[{{Zubeldia} \& {Challinor}(2019)}]{ZubCha19}
{Zubeldia}, I. \& {Challinor}, A., {Cosmological constraints from Planck galaxy
  clusters with CMB lensing mass bias calibration}. 2019, arXiv e-prints,
  \eprint{1904.07887}

\end{thebibliography}

\appendix 

\section{The 2018 release} \label{sec:therelease}

\subsection{Papers in the 2018 release}

The characteristics, processing, and analysis of the \Planck\ data,
as well as a number of scientific results, are described in a series
of papers released with the data.
The titles of the papers begin with ``Planck 2018 results.'',
followed by the specific titles given in Table~\ref{tab:legPaps}.

While this is the last release of the Planck Collaboration, that does not
mean we have reached the point at which no significant improvements would
be possible.  Time was simply up. In particular we believe that the frequency
maps can be improved, further reducing systematic effect residuals, which
would in turn permit the production of improved component maps, likelihoods,
and their scientific implications.

\begin{table}[!htbp]
\caption{\label{tab:legPaps} The \Planck\ legacy release (i.e., ``2018
results'') set of papers.}
\begingroup
\newdimen\tblskip \tblskip=5pt
\nointerlineskip
\vskip -2mm
\setbox\tablebox=\vbox{
 \newdimen\digitwidth
 \setbox0=\hbox{\rm 0}
 \digitwidth=\wd0
 \catcode`*=\active
 \def*{\kern\digitwidth}
 \newdimen\dpwidth
 \setbox0=\hbox{.}
 \dpwidth=\wd0
 \catcode`!=\active
 \def!{\kern\dpwidth}
\halign{\tabskip 0pt#\hfil\tabskip 2.0em&#\hfil\tabskip=0pt\cr
\noalign{\vskip 5pt\hrule\vskip 5pt}
I.   & Overview and the cosmological legacy of \Planck\ (this paper)\cr
II.  & Low Frequency Instrument data processing\cr
III. & High Frequency Instrument data processing\cr
IV.  & Diffuse component separation\cr
V.   & Power spectra and likelihoods\cr
VI.  & Cosmological parameters\cr
VII. & Isotropy and statistics of the CMB\cr
VIII.& Gravitational lensing\cr
IX.  & Constraints on primordial non-Gaussianity\cr
X.   & Constraints on inflation\cr
XI.  & Polarized dust foregrounds\cr
XII.  & Galactic astrophysics using polarized dust emission\cr
\noalign{\vskip 5pt\hrule\vskip 3pt}
} 
} 
\endPlancktable
\endgroup
\end{table}

\subsection{Data products in the 2018 release} \label{sec:dataprods}

The 2018 distribution of released products, freely accessible via the
PLA interface, contains the following items.
\begin{itemize}
\item A reduced instrument model (RIMO), containing the effective beam window
      functions for temperature and polarization detector assemblies for both
      auto- and cross-spectra.  The RIMO also contains beam error eigenmodes
      and their covariance matrices.
\item Cleaned and calibrated data time-lines for each LFI detector.
\item Cleaned and calibrated {\tt HEALpix} data rings for each HFI detector.
\item Maps of the sky at nine frequencies in temperature and seven frequencies
      in polarization.  Additional products serve to quantify
      the characteristics of the maps to a level adequate for the science
      results being presented, including noise maps, masks, and
      instrument characteristics, as well as bandpass-leakage-correction
      maps and gain templates for LFI, and simulated CO bias maps for HFI.
\item Effective beams for LFI and HFI.
\item High-resolution maps of the CMB sky, in temperature and polarization,
      from a variety of different component-separation approaches, including
      an SZ-free CMB map from \SMICA, and CMB maps at several frequencies from
      \SEVEM.
\item A low-resolution CMB map used in the low-$\ell$ likelihood, with an
      associated set of foreground maps (in polarization) and characterization
      of products.
\item Filtered maps of polarized fluctuations in thermal dust and synchrotron emission, and
      thermal dust temperatures in temperature and polarization.
\item A map of the estimated lensing potential and  several types of lensing
      components (SZ, CIB, and $B$ modes).
\item A map of the SZ effect, i.e.,the Compton $y$ parameter.
\item A suite of simulations, including noise and the CMB only, the fiducial
      sky and processed noise, and the CMB run through the four
      component-separation pipelines.
\item A likelihood code and data package used for testing cosmological models
      against the \Planck\ data.
\item Markov chain Monte Carlo samples used in determining the cosmological
      parameters from \Planck\ data.
\end{itemize}
All of these are linked to the \Planck\ Explanatory Supplement
\citep{planck2016-ES}.
The current data release does not include single-bolometer maps,
which limits our ability to robustly perform foreground separation; thus our
temperature foreground results do not supersede the corresponding 2015
products.

\section{Changes for the 2018 release} \label{sec:changes}

The 2018 release uses the same raw, full-mission data as the 2015 release,
but with improved data processing and analysis procedures.  Here we describe
the major refinements in the processing, and discuss where further
improvements may still be made.

\subsection{2018 LFI processing improvements} \label{sec:lfi_improvements}

The most important change to the LFI pipeline for the 2018 data release
concerns the calibration approach. 
For the 2015 release, the main calibration source for LFI was the \Planck\ orbital dipole convolved with a model
of the 4$\pi$ beam response, properly weighted according to the bandpass of
each single radiometer (see \citealt{planck2014-a06} for details). 

The 2018 calibration procedure \citep[see][]{planck2016-l02}
includes Galactic emission along with the CMB dipole in the
calibration model.  Indeed a detailed analysis of the 2015 data
demonstrated that the Galactic contribution could be important,
especially near dipole minima.  The new approach is
iterative and involves all of the calibration, mapmaking, and
component-separation steps. Schematically:
\begin{enumerate}
\item take $T_\mathrm{sky}$ to be the best-fit \Planck\ 2015
  astrophysical model \citep{planck2014-a12}, which 
  includes the CMB, synchrotron, free-free, thermal and spinning dust, and CO
  emissions for temperature, as well as the CMB, synchrotron, and thermal dust
  in polarization;
\item estimate the calibration factor $G$, including in the Galactic model
  both the temperature and polarization components of the sky, as
  well as the Solar and orbital dipoles;
\item apply gains and construct frequency maps;
\item determine a new astrophysical model from the frequency maps using
  {\tt{Commander}} (including only LFI frequencies);
\item iterate steps (2) to (4).
\end{enumerate}

This approach is quite demanding computationally, and each iteration typically requires one week to complete. In practice, the iterative process was stopped after four iterations, by which point good convergence had been achieved.
This approach worked well at 30 and 44\GHz\ but failed at 70\GHz.
This is because for the foreground modelling the 30- and 44-GHz channels is
signal-dominated, while the 70-GHz channel is noise dominated, resulting in
a diverging process (with the algorithm partly calibrating on noise rather
than signal).
  
Another, more minor change in the LFI DPC pipeline is a revision of
the flagging procedure.  This resulted in more conservative criteria, which
discarded additional samples, especially in the first 200 operational days.

\subsection{2018 HFI processing improvements} \label{sec:improvements}

The raw HFI data for this 2018 release are identical to those of the \Planck\ 2015 release \citep[see][]{planck2014-a09} with one exception, namely that approximately 22\,days of data were dropped from the analyses. These data were taken in the final days of HFI observations at a time of increasing Solar activity and of some HFI end-of-life changes in the cryogenic chain operations. These 22\,days correspond to 1000 rings, for which the data were affected significantly more than in any earlier period of similar length during the mission, as revealed by the statistics of the $C_{\ell}$ at low multipoles ($\ell=3$ to 20).  This last period is the farthest outlier of the 27 blocks of 1000 rings.

The main differences in the data processing are the use of a new map-making and calibration algorithm called \sroll. A first version of this algorithm was  introduced in \citet{planck2014-a10}, which used the very low multipoles from HFI to extract the $\tau$ parameter. \sroll\ is based on a generalized polarization destriper that uses the redundancy in the data to extract a number of instrumental systematic-effect parameters directly from the sky data, for example parameters associated with intensity-to-polarization leakage. The spectral transmissions or bandpasses of the HFI band filters were measured on the ground, but with insufficient accuracy for the legacy mapmaking.  Bandpass mismatch between two detectors sensitive to orthogonal polarizations results in a ``leakage'' of temperature foreground signals (with a different spectral energy distribution from the CMB). into polarization.  This effect is taken into account in the map-making. In the 2015 release we used bandpass-mismatch coefficients computed from the ground measurements.  In \sroll\ these are obtained from the sky data, using foreground-map templates from the previous release to derive relative values of the coefficients between detectors, and taking advantage of the redundancy of polarization measurements of the same sky pixel.  This brings significant improvements, as demonstrated by the end-to-end simulations and by the reduction of the systematic effects in null tests.  The power of the \sroll\ map-making was tested a posteriori, after the maps were frozen, by using as input templates for the CO lines two maps of the Taurus molecular cloud in the $^{12}$CO and $^{13}$CO molecules. After the recovery of the relative response coefficients and the reconstruction of all-sky maps of the CO foregrounds, these maps were tested on other radio-astronomy data at high latitudes, and showed a significant improvement over the 2015 maps.

Similarly, CMB calibration errors between detectors sensitive to orthogonal polarization states will also induce spurious polarization.  Changes in detector response over time can be measured using the large-amplitude CMB dipole signals, averaged over short periods during the mission, and show larger variations of the response with time than expected. Moreover, the non-linear part of the analogue-to-digital-converter (ADC) response was not known with sufficient accuracy. To mitigate this, apparent gain variations per optimized time periods were extracted in \sroll\ using redundancy in the surveys over the mission, leading to better maps than when corrections for non-linearity of the ADCs were performed in the time-ordered data.

The improved measurement of the CMB Solar dipole discussed in Sect.~\ref{sec:thesky} allows us to perform an extremely accurate check of the calibration error through end-to-end simulations. In turn this shows that the dispersion between the full-mission-averaged photometric calibration of bolometers within a frequency band is also induced by the temporal variations themselves, and fully accounted for by the uncertainties in the ADC non-linearity correction.
The systematic effects in relative calibration revealed by the Solar dipole between the ``a'' versus ``b'' detectors within polarization-sensitive bolometers (PSBs) of the same frequency seen in 2015 \citep{planck2014-a09} are no longer detectable in the 2018 release (see \citealt{planck2016-l03} for details).

The introduction of these sky-extracted systematic-effect parameters led to a major improvement in null tests, as can be seen in \citet{planck2016-l03} for the lower frequency CMB channels (100 to 217\,GHz), especially at large scales. However, for 353\,GHz the ADC non-linearity is not the dominant systematic effect. The very long time constants (around 30\,s) of the bolometers (which primarily affect the dipoles and for which a correction was already implemented in 2015) also affect other low harmonics of the spin frequency. This systematic effect dominates the low multipoles of the power spectra at 353\,GHz. It was detected through the \sroll\ destriper at 353\,GHz; however, the correction introduced is not very accurate, and leaves artefacts that are still detectable in odd-even survey null tests.

These main improvements introduced in the 2018 HFI data release with respect to 2015 are described in detail in \citet{planck2016-l03}.

\subsection{Simulations} \label{sec:simulations}

At the level of precision reached by \Planck, the best method for
conveying our knowledge of the maps in relation to the sky emission is
through end-to-end simulations of the sky observations and data processing.
Since these simulations are the best characterization of the statistical properties of
the data that we have, we have made available to the community detailed
simulations of the full focal plane; these are called the ``FFP'' series,
the ones used in 2018 being ``FFP10.''

We have used detailed instrumental simulations to estimate the
level of residual systematic effects (see Fig.~\ref{fig:syst_sims}),
and decide which of these needed to be included in the full end-to-end
simulations.
Each FFP10 simulation comprises a single ``fiducial'' realization
(CMB, astrophysical foregrounds, and noise),
together with separate Monte Carlo (MC) realizations of the CMB and noise.
To mimic the \Planck\ data as closely as possible, the simulations use
the actual pointing, data flags, detector bandpasses, beams, and noise
properties of the mission. For the fiducial realization, maps were made of
the total observation (CMB, foregrounds, and noise) at each frequency. In
addition, maps were made of each component separately, of subsets of
detectors at each frequency, and of half-ring and single-survey
subsets of the data. The noise and CMB MC realization-sets
include all detectors, as well as subsets of detectors (so-called ``DetSets'')
at each frequency, and full and half-ring\footnote{A half-ring is the
co-added data of either the first or second half of each stable pointing
period; see \citet{planck2013-p02} and \citet{planck2013-p03}.} 
data sets for each detector combination.

\subsection{Map analysis improvements\label{sec:mapimprovements}}

The improvements summarized in Appendices~\ref{sec:lfi_improvements}
and \ref{sec:improvements} translate directly into lower instrumental
systematics in the corresponding derived sky maps
\citep{planck2016-l02,planck2016-l03}, and thereby more robust
component-separation \citep{planck2016-l04} and likelihood
\citep{planck2016-l05} results. For the purposes of CMB temperature
reconstruction, these updates have a relatively minor practical impact,
due to the very high signal-to-noise ratio of the \Planck\ temperature
observations, where already the \Planck\ 2015 temperature products had
residual errors significantly below the limit set by cosmic variance
for nearly all cosmologically relevant angular scales.

The same does not hold true for polarization reconstruction. In this
case, \Planck's sensitivity corresponds roughly to a signal-to-noise
ratio of unity or less for the (unbinned) CMB $E$-mode power spectrum, and
a fraction thereof for the $B$-mode reconstruction. At the same time
the astrophysical foreground signal from polarized thermal dust and
synchrotron emission is brighter than the $E$-mode signal by more than
an order of magnitude at frequencies below 40\,GHz and above 200\,GHz,
and comparable to it even in the foreground minimum around 70--100\,GHz
(see Fig.~\ref{fig:foregrounds}). The greatest gains deriving from the
\Planck\ 2018 processing are therefore observed in terms of more robust
polarization component-separation and likelihood products, in
particular on large angular scales.

Starting with the CMB component-separation products, this is immediately
highlighted by the fact that the cleaned \Planck\ 2018 CMB polarization maps
include information at all angular scales, from $\ell=2$ to 3000
\citep{planck2016-l04}. In comparison, the corresponding 2015 products were
high-pass filtered below $\ell=40$ in order to remove obvious
instrumental contamination \citep{planck2014-a11}.
Furthermore, for the first time the new CMB polarization maps appear
statistically consistent with detailed end-to-end CMB-plus-noise simulations
(see Sect.~\ref{sec:simulations}) on large angular scales, in terms of power
spectra and basic higher-order statistics.

However, it is critical to note that while the new maps are consistent
with end-to-end simulations, they are \emph{not} consistent with naive
white noise simulations. The \Planck\ noise properties are complicated and
spatially correlated, both because of intrinsic $1/f$ noise and transfer
function effects, and because of gain fluctuations coupled to the actual
sky signal, in particular via the bright CMB dipole.
In the current release, we therefore provide 1000 CMB
realizations processed through the full end-to-end analysis pipeline,
as well as 300 noise simulations per data split. Detailed scientific
analyses of the \Planck\ 2018 CMB products should be accompanied with
a corresponding analysis of these simulations.

Similar improvements are observed in terms of polarized foreground
products. Indeed, the \Planck\ 2018 maps represent the first
reduction of the \Planck\ data that allows even preliminary estimation
of the spectral index of thermal dust emission location-by-location on a
degree smoothing scale.  Such analyses are in general highly sensitive
to the presence of large-scale systematics, since they simultaneously
depend on all angular scales. The fact that no obvious instrumental
artefacts are seen in the polarized thermal dust spectral index map
derived from the \Planck\ 2018 observations (see figure~29 in
\citealp{planck2016-l04}) provides evidence for a high degree of
internal consistency between the 143, 217, and 353\,GHz frequency channels.

For reconstructing temperature foregrounds, the \Planck\ 2018
data release is not an improvement, due to the
lack of single-bolometer sky maps (see Sect.~3.1.2 of
\citealt{planck2016-l03} for details). First, this strongly limits our
ability to model and extract CO line emission, which in turn also
affects the robustness of other correlated components, including
thermal dust, free-free, spinning dust, and synchrotron
emission. Second, it also precludes the possibility of removing single
channels that are particularly strongly affected by specific
instrumental systematic errors, such as bandpass-mismatch or far-sidelobe
contamination.  (See \citet{planck2014-a12} for an example of
such analysis based on the 2015 measurements.)  For these reasons, we do
not provide an updated, comprehensive \commander-based foreground model
in intensity in the 2018 release, but instead suggest
continued usage of the corresponding 2015 model. We consider the 2015
thermal dust model to be more robust than the new model also for
\texttt{GNILC}.  However, for CMB temperature extraction
purposes these issues are of minor concern, since the accuracy of this
process only depends on the sum of the foregrounds, and not on each
individual component. As shown in \citet{planck2016-l04}, the CMB
temperature maps derived from the \Planck\ 2018 frequency products are
consistent with the corresponding 2015 temperature maps.

\subsection{CMB power spectra and likelihood improvements}

The likelihoods have seen many changes and improvements since 2015, as
listed and discussed in detail in \citet{planck2016-l05} for the CMB
spectra, and \citet{planck2016-l08} for lensing. 

As in 2013 and 2015, the cosmological constraints are obtained using an approximate likelihood. Different mathematical approximations and different data-selection choices are needed at different scales to correctly evaluate the likelihood. For this reason, the overall \Planck\ likelihood is formed using a hybridization of different approximations, neglecting the correlations between the different parts of the likelihood. The impact of this hybrid approach has been extensively discussed in the literature \citep[e.g.,][]{2006MNRAS.370..343E, planck2014-a13}. In the following, we only discuss the specific improvements and changes for each part of this hybrid approach.

The 2018 baseline hybridization scheme relies on a different data mix than in 2015. In 2015, residual unresolved systematics and a conservative approach led us to recommend the use of the \commander\ large-scale $TT$ map, the LFI large-scale polarization maps, and the small-scale HFI temperature maps, while the reconstructed lensing map was only used in some specific analyses and the small-scale, HFI polarization maps were used in a preliminary manner.
In 2018, the baseline data for cosmology now consist of the \commander\ large-scale $TT$ map, the HFI large-scale polarization maps (using the $EE$ and $BB$ spectra only), the HFI small-scale temperature and polarization maps, and the lensing reconstruction map.
The LFI large-scale polarization map is now used for cross-validation and some
specific analyses.  These changes provide a very significant improvement on
the constraining power of the \Planck\ data, as seen for example in
Fig.~\ref{fig:params_weight}.

The likelihoods used are labelled by the data that go into them, as described
in \citet{planck2016-l05} and \citet{planck2016-l06}.  For example
TT,TE,EE+lowE uses the combination of temperature and $E$-mode polarization
auto-spectra with the TE cross-correlation at high $\ell$ and the TT and
EE spectra at low $\ell$ computed from \commander\ and \simall\ respectively.

\subsubsection{Large-scale temperature and the \commander{} likelihood}

The framework of the \commander\ component-separation method, described in \citet{planck2016-l04}, allows for a joint Bayesian sampling of an explicit parametric model that includes 
both the cosmological CMB signal and non-cosmological astrophysical signals, such as thermal dust, CO, and low-frequency foregrounds.
Its results are used in two different ways in the 2018 hybrid likelihood.
\begin{enumerate}

\item The samples from the Bayesian exploration are reused to build a foreground-marginalised, large-scale temperature-only likelihood approximation, as is described in \citet{planck2014-a13}. This forms the large-scale $TT$ part of the hybrid likelihood, as in 2013. 

\item The \commander{} foreground-cleaned temperature map is used with the LFI large-scale polarization maps to build the $TE$ part of the large-scale alternative polarized likelihood. The map is also used to build a $TE$-based likelihood approximation with the HFI data, but its statistical characterization is shown to be too poor to build a large-scale $TE$ likelihood.

\end{enumerate}

While the \commander{} methodology has not changed significantly since the 2015 release, we have modified our choice of data cuts and accordingly the model. 
In order to produce a robust and conservative product for the 2018 release, we removed the dependency on external data, namely, the \WMAP\ and Haslam 408-MHz data sets \citep{bennett2012,haslam1982}.  While the HFI data processing has been greatly improved in terms of the number of systematic effects that are resolved on large scales, it no longer produces individual bolometer maps. In 2015, we used the slightly different bandpasses of the different individual bolometers and external data to constrain a more complex data model (in temperature). Because of the focus on polarization systematics, this is not possible with the 2018 data. 
For this reason, the usable sky fraction for \commander{} has shrunk from $94\,\%$ to $86\,\%$. Nevertheless, large-scale agreement between the different foreground-cleaned maps has improved compared to 2015, 
and in particular for the \SMICA\ map used for some specific applications (such as lensing or higher-order moment estimation).

\subsubsection{Large-scale HFI polarization and the \simall{} likelihood}

Following the work presented in \citet{planck2014-a10}, and building on the
improvements described in Sects.~\ref{sec:improvements},
\ref{sec:simulations}, and \ref{sec:mapimprovements}, the level of residual
systematics present in the large-scale HFI polarization data is now low enough
that the 100-GHz and 143-GHz maps can be used to build a large-scale $EE$
likelihood.
This likelihood allows for a roughly
$6\,\sigma$ determination of the reionization parameter, with
$\tau = 0.0506 \pm 0.0086$, using only the low-$\ell$ HFI polarized data
along with the \commander{} $TT$ large-scale likelihood, and fitting jointly
for $\tau$ and the amplitude of scalar fluctuations. 

\begin{figure*}[htpb]
\begin{center}
\resizebox{\textwidth}{!}{\includegraphics{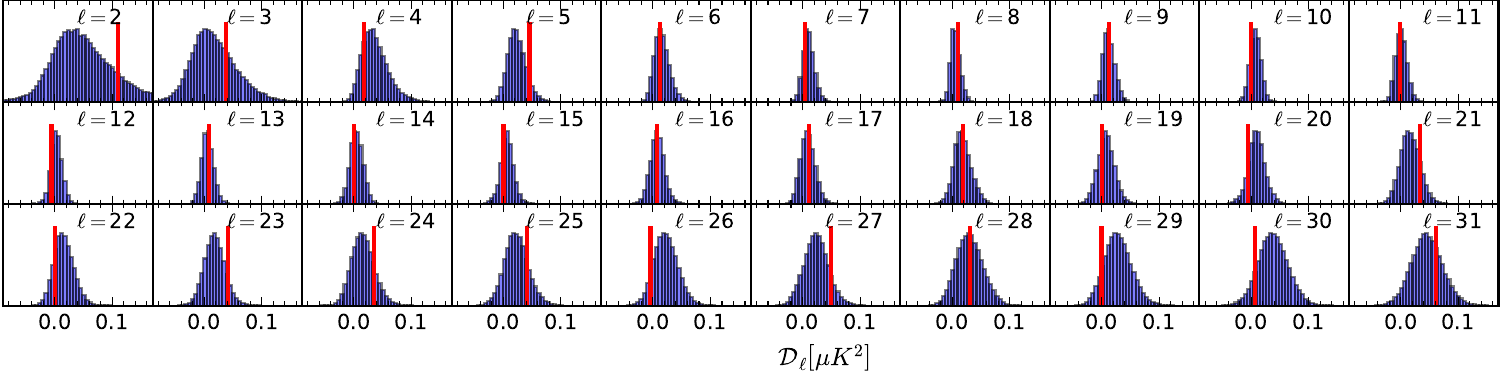}}
\end{center}
\caption{Empirical distributions for the $100\times143$\,GHz $EE$ cross-spectra, for every multipole up to $\ell = 31$. The distributions in blue are derived from 300 noise-plus-systematics simulations, combined with 1000 signal realizations with $\tau=0.05$. The power actually observed in HFI data is indicated by the red lines, showing that, at large scales, these data are well described by the end-to-end simulations. Equally consistent plots for $BB$, $TE$, $TB$, and $EB$, as well as all the corresponding PTEs, are given in \cite{planck2016-l03}.}
\label{fig:EE_lowell_consistency_datasims}
\end{figure*}

The large-scale HFI polarization likelihood is based on comparison between the cross-spectrum of the foreground-cleaned 100-GHz and 143-GHz polarization maps and very detailed, end-to-end simulations of the HFI data. Due to our inability to accurately account for ADC non-linearity, our modelling of the pixel covariance matrix is not sufficient and prevents us from using a more classical pixel-based likelihood approximation, as we do for LFI. 
In more detail, cleaned CMB maps at 100\,GHz and 143\,GHz are obtained by fitting for the dust and synchrotron contamination (using the 353-GHz and 30-GHz maps, respectively, as templates). The maps are further masked to avoid the highly contaminated regions, retaining $50\,\%$ of the sky. To reduce the level of scatter and correlation induced by the mask, the power spectra are estimated using the quadratic-maximum-likelihood (QML) method. 
The likelihood is computed by forming the product of the probabilities of  each of the QML power-spectrum multipoles, ignoring $\ell$-to-$\ell$ correlations. 
This probability is estimated by counting the number of end-to-end simulations computed for different input cosmologies that fall close to the observed value. 
\citet{planck2016-l05} presents a very thorough validation of this method, 
exploring the variations of the likelihood when changing masks, foreground-cleaning methods, data cuts, using part \WMAP\ or LFI data, etc. 
To give a flavour of the robustness of the approximated likelihood, and the fidelity of our simulations, we display in Fig.~\ref{fig:EE_lowell_consistency_datasims} the distribution of QML synthetic spectra measured from our end-to-end simulations for an input $\tau=0.05$, and compare this with the observed $EE$ spectrum. 

The $TE$ spectrum measured in a similar way also shows decent statistical
agreement with our simulations, but fails some of our null tests
\citep{planck2016-l05}.
Furthermore, our simulation-based likelihood estimation makes it very
difficult to accurately take into account $TE\times EE$ and $TE \times TT$
correlations.
For $TE$ this is particularly important, to avoid double counting the
constraining power of the temperature and polarization maps; however, it is
much less of an issue for $EE$, which has a much lower correlation with $TT$.
For these reasons we do not include the estimated, low-$\ell$ $TE$
spectrum in the likelihood.
Similar work was performed with the $BB$ spectrum, but at the level of
sensitivity of the HFI data, it is compatible with a null spectrum.

\subsubsection{Large-scale LFI polarization and its likelihood use}

As we did in 2015, we produce a full TEB likelihood using a pixel-based
approach based on the \commander{} and LFI polarization maps.
Given the lower sensitivity of LFI, this likelihood approximation has
an overall lower constraining power on the reionization fraction than
the HFI-based one, with a roughly $3\,\sigma$ determination of
$\tau = 0.063\pm  0.020$. 
Nevertheless, the 2018 version of the LFI-based likelihood can be used
when probing the TEB correlations, which may be important for specific
cosmological tests.

This pixel-based, low-$\ell$ approximation was already used in 2015, but has been improved and made more robust.
Thanks to the improvements in the LFI data processing and simulation pipelines, the sky fraction retained for the cosmological analysis has been increased from $46\,\%$ to $66\,\%$, and the second and fourth sky surveys, which were excluded 
from the 2015 likelihood, are now included. Robustness of the likelihood approximation has been further tested on different sky fractions, as well as through comparison with \WMAP\ and HFI data.

\subsubsection{Small-scale temperature and polarization HFI likelihood}

The methodology of the small-scale temperature and polarization likelihood
approximation has not changed since 2015, and remains very close to that
of 2013.  We continue to describe the statistical properties of the data
with a Gaussian approximation. 
We are still using cross-half-mission spectra of the 100, 143, and 217-GHz
channel maps, masking the highly foreground-contaminated sky regions
(due to Galactic contamination and, in the case of temperature, point sources).
The masks have not changed since 2015.
We are also discarding some of the spectrum multipoles that have a low
signal-to-noise ratio or high foreground contamination.
Compared to 2015, we have improved both the data and their characterization
to a level where we can now lift the reservations we had in 2015 on the
usage of the polarized small-scale data ($TE$ and $EE$) for cosmology.

On the data side, as described in Sects.~\ref{sec:improvements} and \ref{sec:mapimprovements}, most of the effort has translated into a decrease of the level of systematics at large scales in polarization. This also has some impact on the small-scale polarization likelihood, the most important being a reduced level of noise in the 143-GHz $Q$ and $U$ maps (by about $12\,\%$).

\begin{figure*}[htpb]
\begin{center}
\resizebox{\textwidth}{0.59\textwidth}{\includegraphics{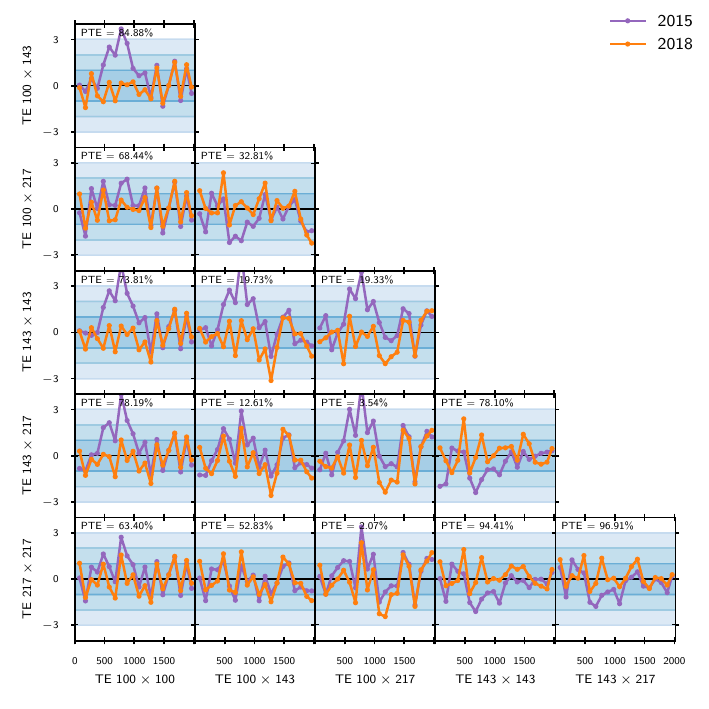}}
\resizebox{\textwidth}{0.59\textwidth}{\includegraphics{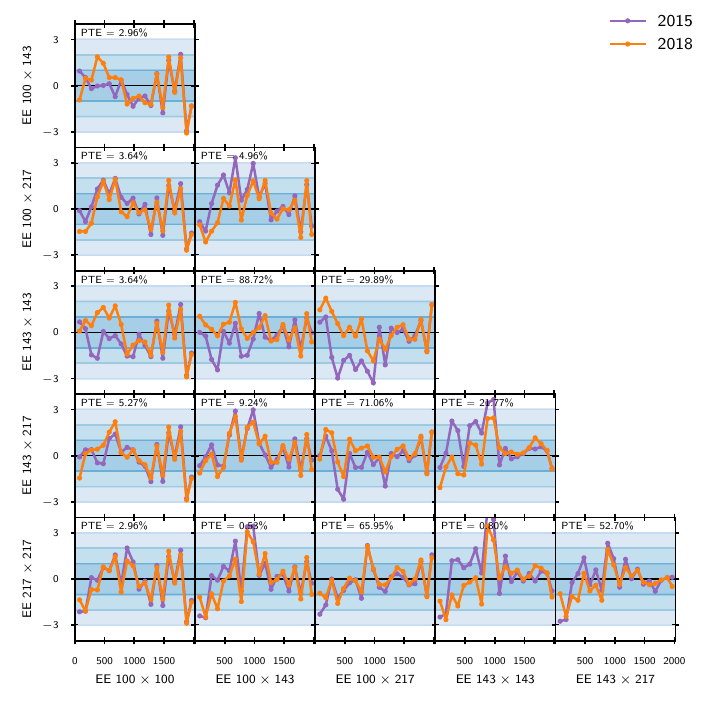}}
\end{center}
\caption{Inter-frequency null tests of CMB $TE$ and $EE$ power spectra. Each sub-panel shows the differences between two foreground-cleaned 
cross-spectra at different frequencies (horizontal minus vertical). We show the full spectra comparisons here, even though the likelihood discards
some of these data (according to the  multipole range). The two lines in each panel correspond to the 2015 data and nuisance model (purple) and the 2018 one (orange); for each case, foreground and nuisance cleaning is performed at the spectrum
level, as is done for the likelihood, using the best-fit nuisance
parameters from the baseline fit for each release.
The PTE values quoted in each sub-panel correspond
to the 2018 data (and nuisance model) for the full range
presented in the plot and with $\Delta\ell= 100$ flat binning.  There is 
impressive improvement in the 2015-to-2018 agreement in the inter-frequency spectra, in particular in $TE$, due in large part to the beam-leakage corrections.}
\label{fig:TE_consistency_triangle}
\end{figure*}

On the modelling side, the main improvements have been the correction of the so-called ``beam leakage,'' and a better determination of the polarization efficiencies of our detectors. These two refinements have a large effect on the consistency of the different $TE$ and $EE$ cross-spectra, as shown in Fig.~\ref{fig:TE_consistency_triangle}. Disagreement between the polarized cross-spectra in 2015 was the reason we did not recommend the use of the polarized data for cosmology applications.  With the new analysis, there is no longer such a
limitation. 

In detail, differences between the beams, gains, polarization efficiencies,
and polarization angles of the different data streams that enter the computation of a $Q$ or $U$ map are sources of temperature-to-polarization leakage.
In 2015 we could only evaluate those effects a posteriori, with a cosmology-dependent model. In 2018, following the methodology presented in \citet{2017A&A...598A..25H}, we can propagate the known characteristics (from measurements made on the ground) of the different detectors and compute beam-leakage templates for each cross-spectrum. We tested the fidelity of the templates against simulations and estimated their residual uncertainty. Correcting for beam leakage results in the large improvement of the $TE$ inter-frequency comparisons displayed in Fig.~\ref{fig:TE_consistency_triangle}. 

Temperature-to-polarization leakage corrections have a small effect on the $EE$ cross-spectra disagreement. Corrections to the estimated polarization efficiencies of the detectors are the source of the improvement displayed in the bottom panel of Fig.~\ref{fig:TE_consistency_triangle}.
Null tests performed on highly dust-contaminated regions in the high-frequency polarized channels (mainly 353\,GHz, but also 217\,GHz) have led us to revise upwards our previous estimate of the polarization-efficiency uncertainty by a factor of 5 to 10. 
Polarization-efficiency assessments performed using frequency-channel cross-spectra, with or without cosmology priors, as described in \citet{planck2016-l03} and \citet{planck2016-l05}, translate into percent-level corrections that need to be applied to the polarization efficiencies of the 100, 143, and 217-GHz channels. 

While the 100 and 217-GHz polarization-efficiency measurements are relatively stable, we find a $2\,\sigma$ discrepancy between the estimates performed at 143\,GHz, depending on whether the estimation is made on the $EE$ or $TE$ spectra. This difference is somewhat worrisome, since the 143-GHz channel dominates the cosmological constraints in polarization. At this time, we cannot tell whether this difference is a statistical fluctuation or a faint sign of residual systematics projecting onto the polarization efficiency estimates. 
We evaluate the effect of either enforcing the $EE$-based polarization efficiency estimation on $TE$ (the so-called ``map-based'' calibration model), which we retain for our baseline, or letting the $EE$ and $TE$
spectra have a different effective calibration (the so-called ``spectrum-based'' calibration model), which we use in an alternative likelihood implementation. The two different calibration models translate into $\leq0.5\,\sigma$ 
parameter shifts, which gives us an estimate of the level of possible residual systematics in the polarization analysis.

Numerous other improvement have also been applied to the high-$\ell$ likelihood. Beam corrections have been computed specifically for each of the  different masks used in temperature and polarization, and we have tightened our estimate of the beams and beam-leakage uncertainties, including effects that we neglected in previous analyses. We have significantly improved the quality of the residual Galactic contamination estimation and correction in the likelihood. Finally, we have also improved the estimation of the level of residual correlated noise in the spectra. We now include two, very small correlated-noise corrections, namely sub-pixel noise (due to the centroid of data samples falling in a pixel being different from the pixel centre) and a correlated-noise component in the auto-frequency $EE$ spectra that was observed in the high fidelity end-to-end simulations (see Sect.~\ref{sec:simulations}). 
All these refinements, while increasing the robustness of the likelihood approximation, have a much smaller impact than the beam-leakage and polarization-efficiency corrections.

With these improvements, the high-$\ell$ $TT$, $TE$, and $EE$ CMB power spectra are found to be in good agreement with each other in the context of a common $\Lambda$CDM model, as demonstrated by the conditional predictions displayed in Fig.~\ref{fig:cond_vs_coadd}.

\subsubsection{Lensing likelihood}

In 2018, the CMB power spectra (that already contain some information on
lensing) are complemented by the lensing power spectrum measured using the
reconstructed lensing-effect map (Sect.~\ref{sec:lensingspectra}).
In 2013 and 2015, a lensing-power-spectrum-based likelihood was already
provided, but it was only used for some specific cosmological applications.
It is now systematically added into the baseline hybrid likelihood mix.

The lensing estimation pipeline has been significantly improved compared to 2015.  Lensing maps are reconstructed from the \SMICA\ 2018 foreground-cleaned maps, using only a combination of the high-frequency maps. We now use inverse-noise weighting for polarization-only band powers to improve the S/N in reconstruction, a new mask to reduce point-source contamination, and a better model of the multiplicative bias.

The improved filters for polarization reconstruction allow us to perform a polarization-only lensing reconstruction, as a demonstration of consistency, and a cross-check on the paradigm. The robustness of the measurement pipeline has also been checked in numerous new ways, extending greatly the already quite thorough validation suite from 2013 and 2015. In particular, SZ and CIB leakage effects are checked, different Galactic masks are used to measure the impact of any residual Galactic contamination in the \SMICA{} maps, and alternative masks and data cuts (surveys, half missions, etc.) are used to check for any scanning-dependent feature in the lensing reconstruction maps. 

Thanks to our extensive validation suite, we have managed to increase the range of lensing multipoles usable for cosmological constraints, reducing the lower limit from $L\,{=}\,40$ to $L\,{=}\,8$. This helps to constrain some specific cosmological models.  Multipoles below this are adversely affected by a large and uncertain
mean-field correction \citep{planck2016-l08}. The upper limit, $L\,{=}\,400$, remains unchanged from our earlier releases, although data are provided to much smaller scales.

\section{HFI-LFI consistency} \label{sec:inst-consistency}

\begin{figure}[htbp]
\begin{center}
\resizebox{0.49\columnwidth}{!}{\includegraphics{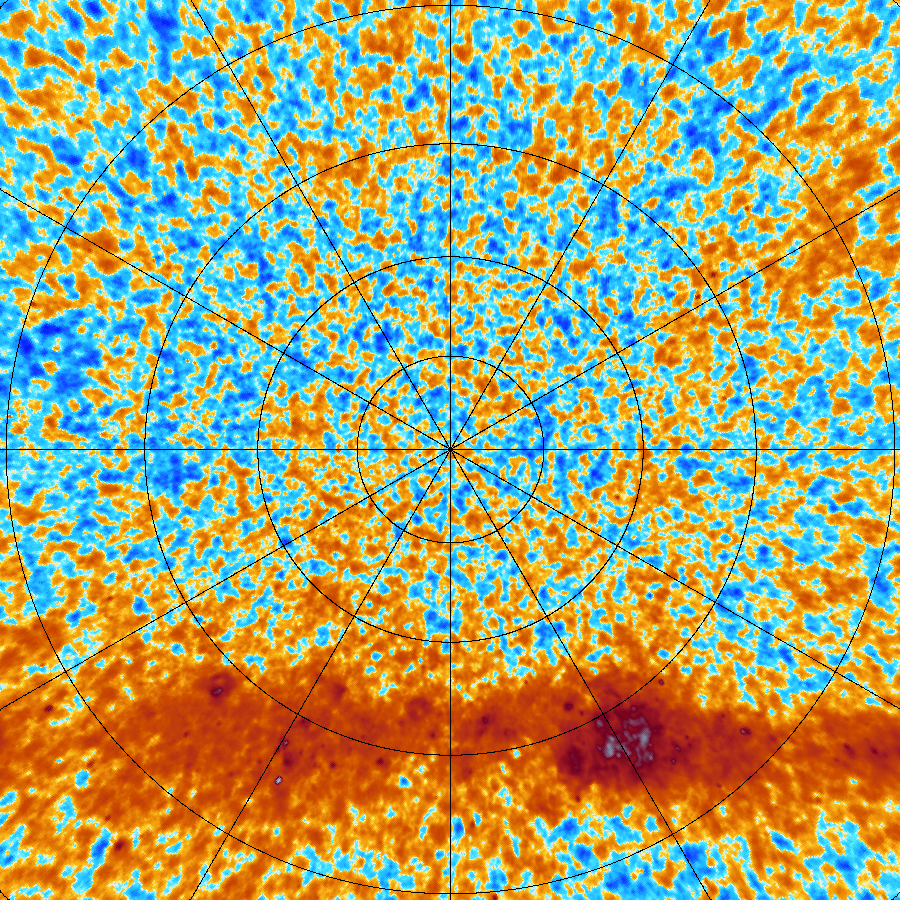}}
\resizebox{0.49\columnwidth}{!}{\includegraphics{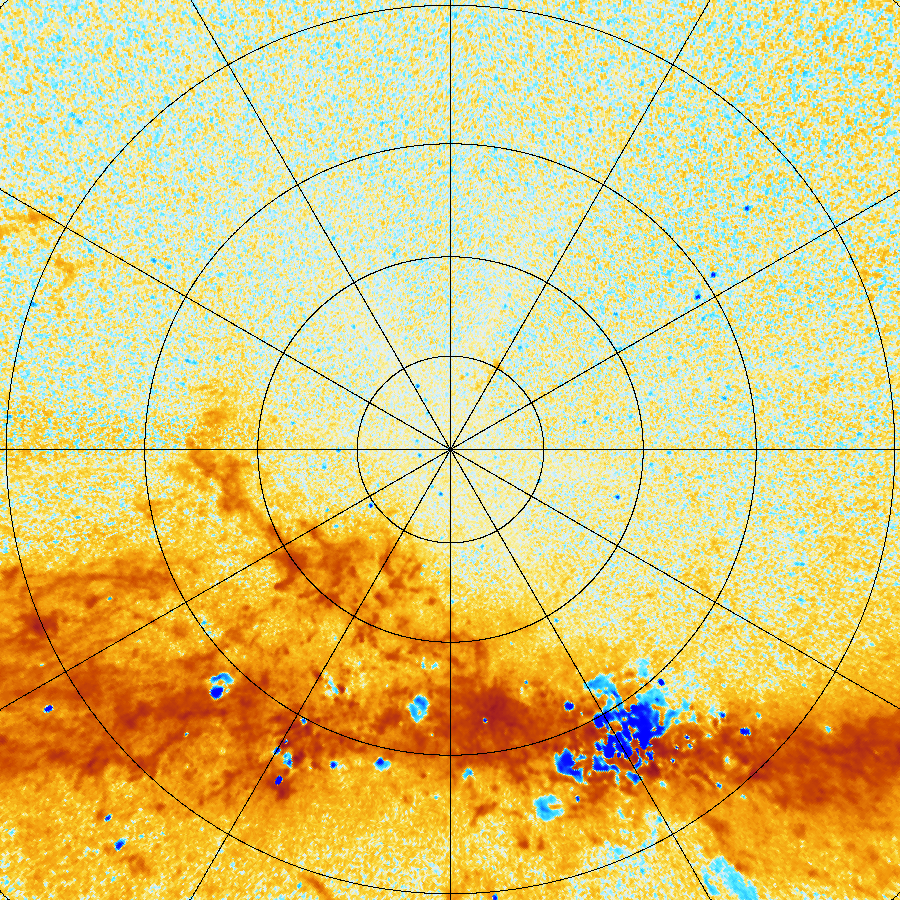}}
\resizebox{0.49\columnwidth}{!}{\includegraphics{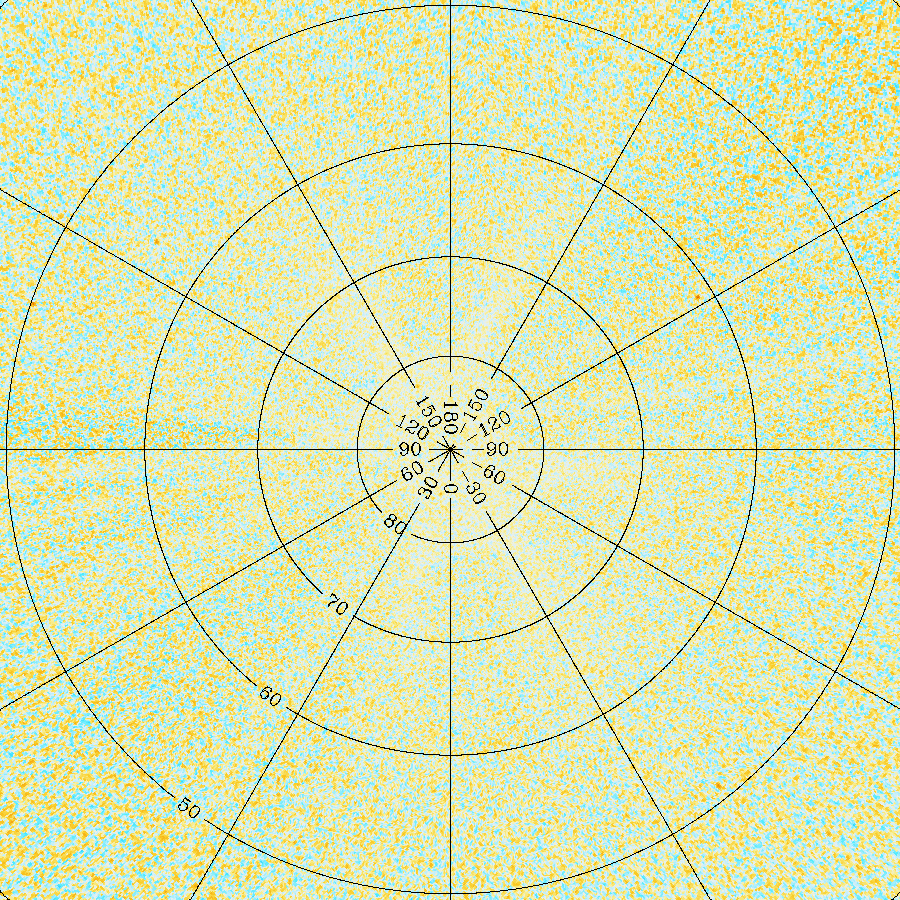}}
\resizebox{0.49\columnwidth}{!}{\includegraphics{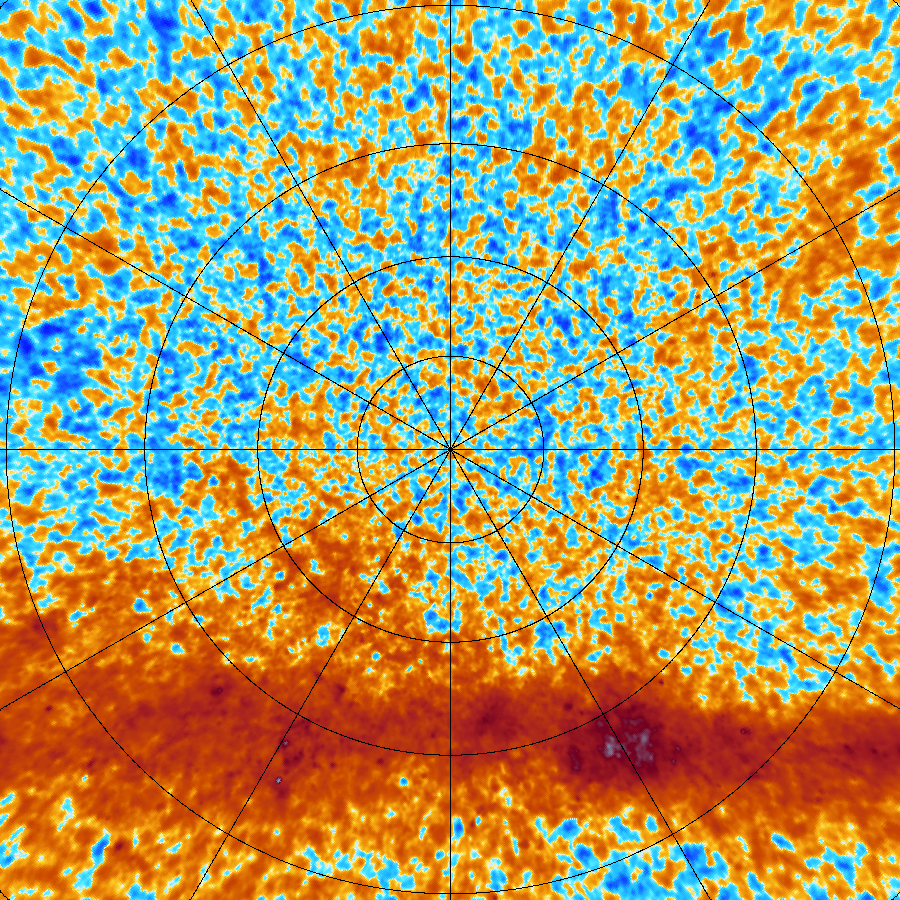}}
\resizebox{0.51\columnwidth}{!}{\includegraphics{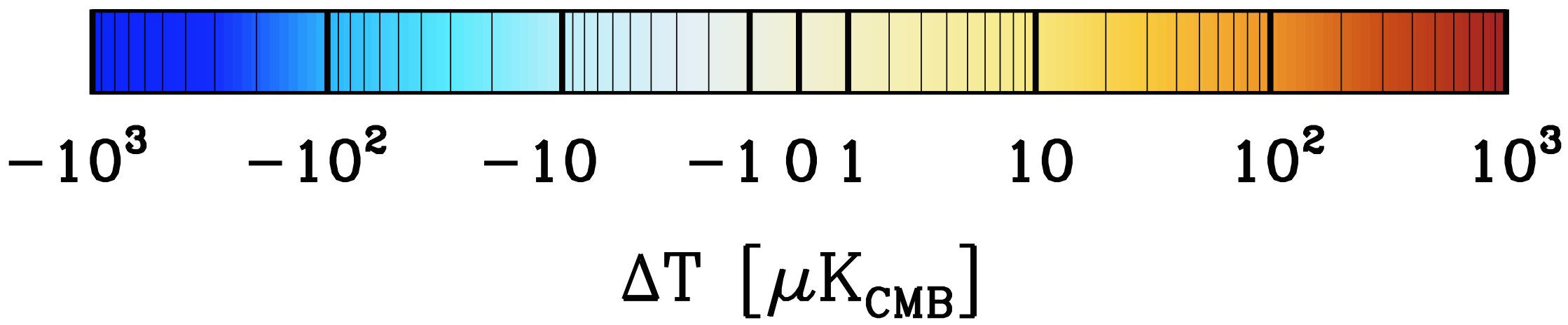}}
\resizebox{0.49\columnwidth}{!}{\includegraphics{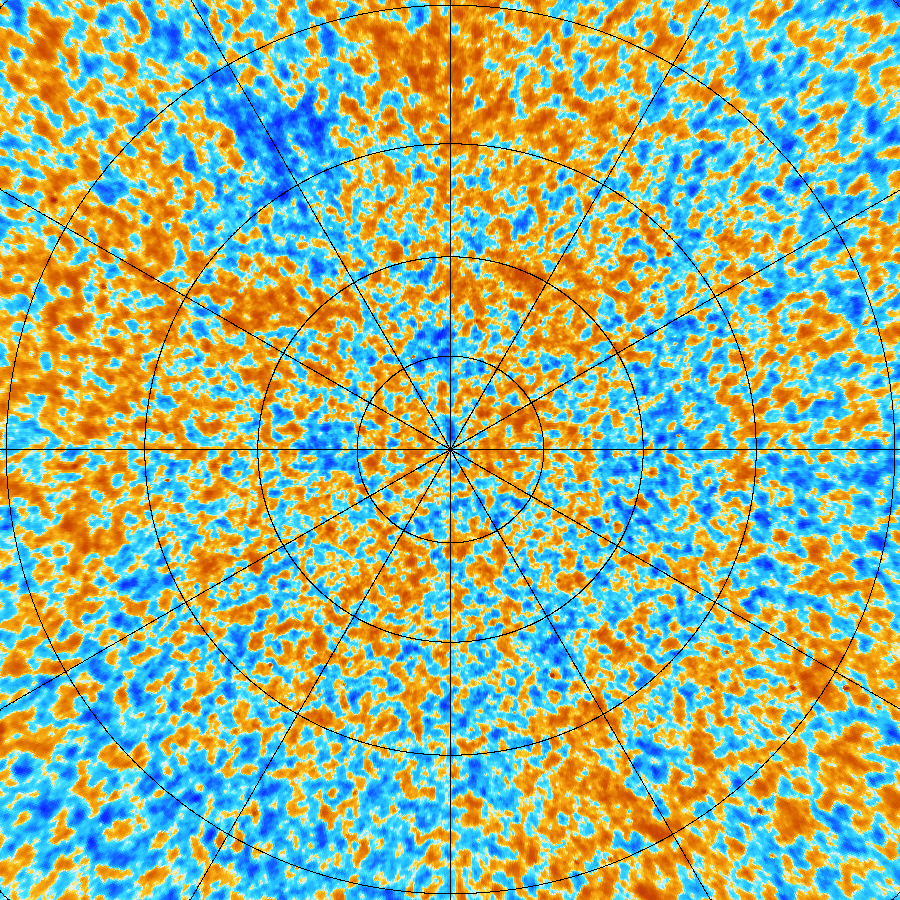}}
\resizebox{0.49\columnwidth}{!}{\includegraphics{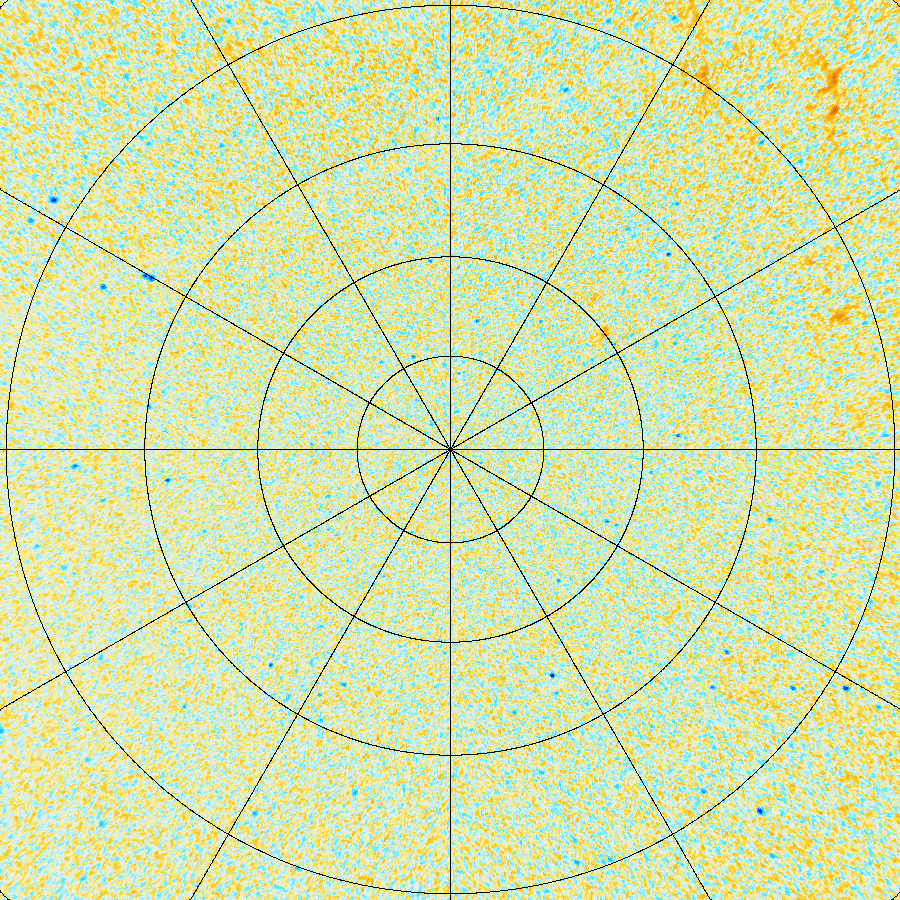}}
\resizebox{0.49\columnwidth}{!}{\includegraphics{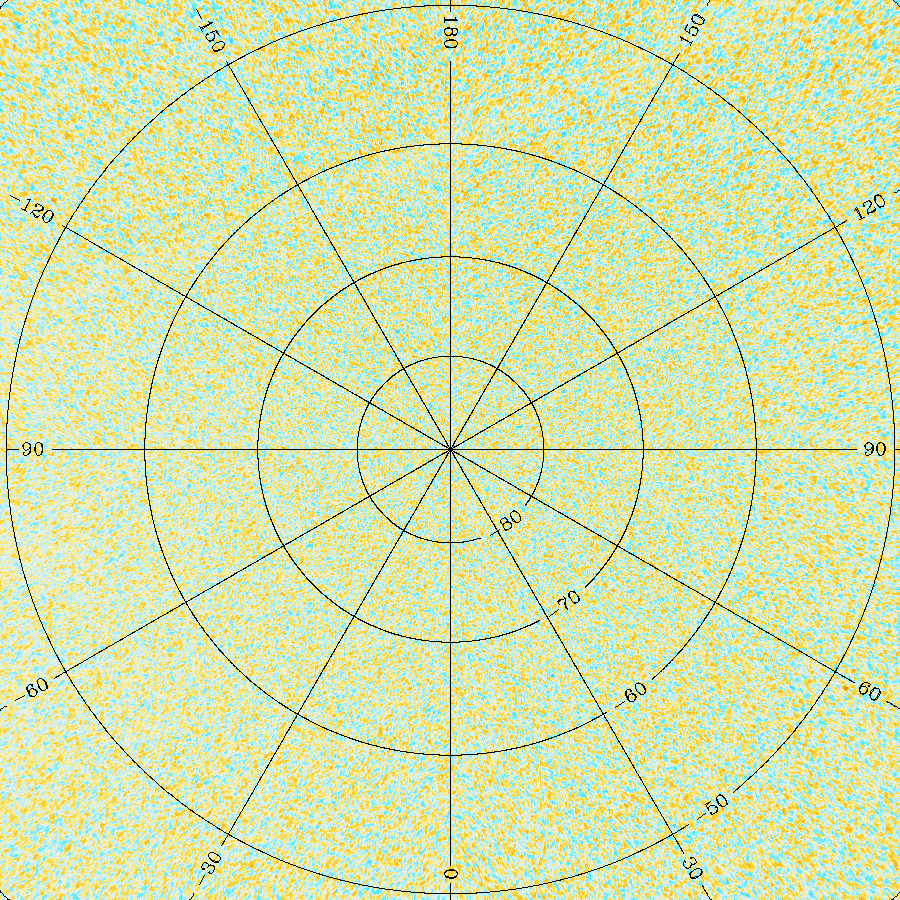}}
\resizebox{0.49\columnwidth}{!}{\includegraphics{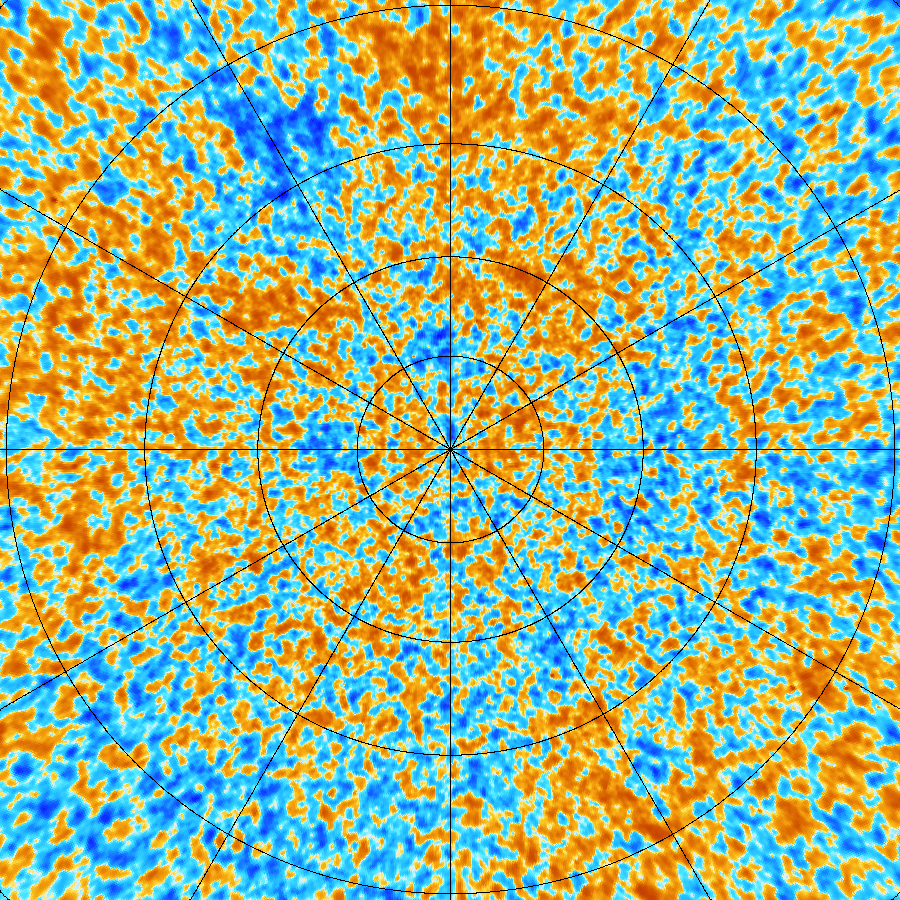}}
\end{center}
\caption{HFI and LFI maps and differences.
{\it Top}: North ecliptic pole region.
The four panels are: upper-left, 70\,GHz; lower-right, 100\,GHz; upper-right, $100\,{\rm GHz} - {\rm 70}\,{\rm GHz}$; and lower-left, difference of 70-GHz\ half-ring and 100-GHz\ half-mission difference maps. The frequency difference map demonstrates excellent visual consistency of the measured CMB anisotropies, and reveals diffuse foregrounds and point sources, specifically positive dust and CO emission and negative free-free and synchrotron emission. Note the large Cygnus region in the Galactic
plane.
{\it Bottom}: South Galactic pole region, with the same 4-panel layout as for
the top part.
We note the dust haze in the upper right part of the difference panel and the large, negative CMB fluctuation in the upper left of the 70- and 100-GHz frequency maps, called the Cold Spot. 
}
\label{fig:HFI-LFI-map}
\end{figure}

Having two instruments on-board \Planck\ offers the possibility of cross-checks between two renderings of the sky that are independent across detection technologies, processing pipelines, and to a large extent people, sharing only the satellite platform and scanning strategy. Such tests were performed in detail for the \Planck\ 2013 and 2015 data releases \citep{planck2013-p01a,planck2014-a01}. Here we show a subset of such tests for the \planck\ 2018 release.

Figure~\ref{fig:HFI-LFI-map} compares the LFI 70-GHz and the HFI 100-GHz maps in selected regions of the sky, when both are expressed in thermodynamic temperature units and smoothed to a common resolution of 15\arcmin. Frequency-difference (upper-right) maps demonstrate excellent consistency of the measured CMB anisotropies, and reveal diffuse foregrounds and point sources. The expected noise level is estimated (lower-left) by the difference of 70-GHz\ half-ring, and 100-GHz\ half-mission difference maps, each of which is a good noise estimate of the respective signal maps. 

The top four panels of Fig.~\ref{fig:HFI-LFI-map}
show an enlargement of the north ecliptic pole region, which was scanned by \Planck\ most frequently and is thus one of the least noisy parts of the sky. One can see in the difference map positive dust and CO emission, and negative free-free and synchrotron emission (because the lower frequency channel is subtracted from the higher frequency one). Note the large Cygnus region in the Galactic
plane. The four bottom panels are focused on the south Galactic pole region, with the same layout as for the top four panels. This a region with fairly reduced foreground emission; still, the haze of dust in the top right corner of the difference map is clearly visible. We note the large, negative CMB fluctuation in the upper left of the 70- and 100-GHz frequency maps, called the ``Cold Spot'' anomaly, which is rendered in the same way by LFI and HFI. Similar tests on the full sky, the entire equatorial, south ecliptic, and north Galactic pole regions, do not reveal any worrisome instrumental features.

\begin{figure*}[htbp]
\begin{center}
\resizebox{0.9\textwidth}{!}{\includegraphics{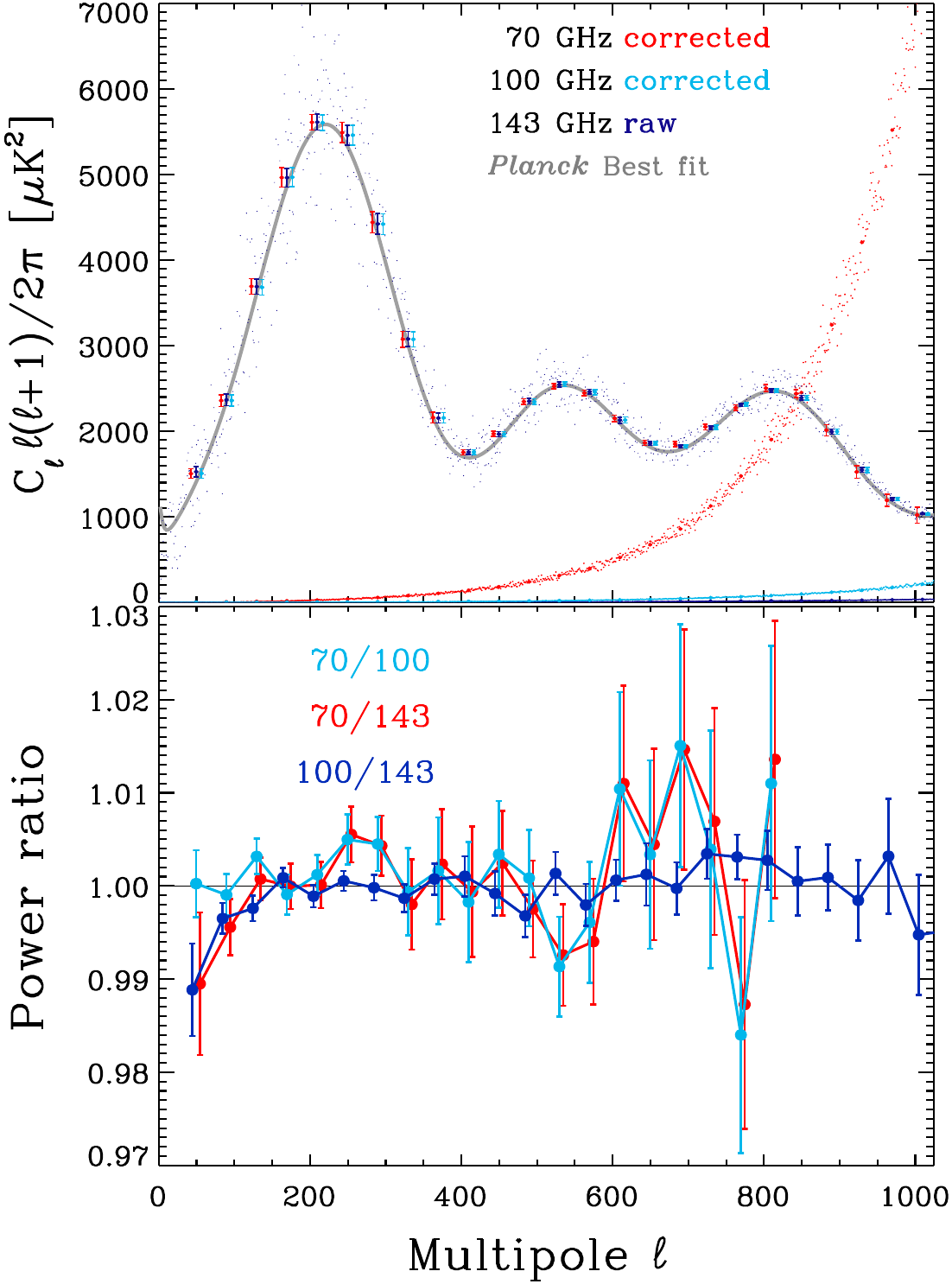}}
\end{center}
\caption{Power spectra of 70, 100, and 143-GHz maps evaluated after masking approximately 40\,\% of the sky. {\it Top}: Cross-spectra of half-ring maps at 70\,GHz and half-mission maps at 100 and 143\,GHz, as well as the spectra of the half-difference maps, which illustrate the noise content of the data. Binned 143-GHz raw cross-spectrum values indicate the bin centres; the other spectral values are spread in the bins for clarity (see further details in text). {\it Bottom}: Power ratios of the same spectra. The drop in the values of some ratios at $\ell < 150$ is due to uncorrected diffuse foreground emission that gets brighter at 143\,GHz outside the masked part of the Galaxy.  At $\ell > 150$ there is excellent consistency of all three spectra. One can see that at the top of the first acoustic peak near $\ell \simeq 220$ they are consistent at the level of a few $\times 10^{-3}$.
}
\label{fig:HFI-LFI-spectra}
\end{figure*}

One can make a more quantitative comparison by using power spectra in low foreground regions, masking 40\,\% of the sky by combining Galaxy and point-source frequency-specific masks. Figure~\ref{fig:HFI-LFI-spectra} compares binned cross-power spectra from the 70 and 100\,GHz channels.  The plotted noise spectra are the auto-spectra of the respective difference maps. Also shown is the raw spectrum at 143\,GHz, a channel whose noise is negligible at these angular scales (the noise spectrum is plotted, but it lies along the $x$-axis). We display the average power ($\propto (2\ell + 1) C_\ell$) in each bin, and show the error on the mean as an estimate of the binned power uncertainty (inclusive of cosmic variance within each bin). For 70 and 100\,GHz, the spectra are corrected for multiplicative calibration offsets with respect to the 143-GHz spectrum used as a fiducial for this check; the offsets are 0.997 for 70\,GHz, and 1.001 for 100\,GHz, very small corrections that indicate excellent calibration of \Planck\ frequency maps, combined with a small amount of residual power from Poisson-distributed, undetected point sources ($C^{70}_\ell \simeq 4.5\times 10^{-4}\mu\mathrm{K}^2$, and $C^{100}_\ell \simeq 1.75\times 10^{-4}\mu\mathrm{K}^2$). This simple procedure is enough to bring the three spectra into good agreement. The plot also shows the best-fit model as derived by \Planck\ multi-component likelihood fits with many nuisance parameters, including an optimal determination of calibration and various correction factors that become increasingly important at higher multipoles. In the $\ell$ range shown, the best fit is traced well by the 143-GHz raw spectrum. 

The bottom panel of Fig.~\ref{fig:HFI-LFI-spectra} shows the corresponding power ratio. The drop in the values of some ratios at $\ell < 150$ is due to uncorrected diffuse foreground emission that becomes brighter at 143\,GHz outside the masked region of the Galaxy. The nearby 70 and 100-GHz spectra indeed do not display such a drop. For the whole $\ell > 150$ range, there is excellent consistency of the 143 and 100-GHz spectra. The LFI 70-GHz spectrum becomes noisy at $\ell > 600$ and because of that we display it only up to 800. Nevertheless, we note the remarkable consistency (at the level of a few $\times 10^{-3}$) of all three spectra around the first acoustic peak near $\ell \simeq 220$.

Of course, such checks are too coarse to be directly useful for cosmology, which requires that we account for much smaller contributions than are visible by eye. Indeed, for analyses of isotropy and statistics of the CMB, one needs to resort to component separation and simulations, while for cosmological parameters one needs a likelihood analysis that directly accounts for the degeneracies between CMB and nuisance parameters of astrophysical and instrumental origin. However, the comparisons described here have the virtue of simplicity and provide a direct visual test of consistency.

\section{Blinding} \label{sec:blinding}

We end with an appendix addressing some of the principles that were
followed in the \Planck\ analysis.  In particular we address the extent
to which ``blinding'' \citep[see, e.g.,][]{Klein2005} was used in the
production of results from \Planck.
We include this discussion here because the question has often come
up, not least in the context of parameter tensions with other data sets.

The goal of blind analysis is the avoidance of biases and errors introduced by investigators.  The general principle is to shield relevant results from the view of investigators until analytical methods have been decided, implemented, debugged, and completed.  Various techniques, such as ``noising,'' ``biasing,'' ``cell scrambling,'' ``seeding,'' and ``item scrambling'' \citep[e.g.,][]{Maccoun2015} have proven to be useful in many situations.

For \Planck, and indeed all CMB experiments, the importance of the goal of blind analysis can hardly be overstated, and a quantitative demonstration is required that the goal has been met to a specified level.  The methods of analysis used must also satisfy another difficult requirement, that of extracting cosmological and astrophysical information from the data to a level of $10^{-6}$, $10^{-7}$, or even $10^{-8}$ of the input signal level.  The ultimate limits to this signal extraction are set by some
combination of instrument noise, noise in the sky signals themselves, the separation of signals from various sources (especially the CMB from Galactic and extragalactic foregrounds), and instrumental systematic errors, which in general are time-dependent and include transients.  All of these must be determined from the data themselves, in a process of disentanglement, identification, and mitigation, starting from the largest and most easily identifiable effects, and moving down to the smallest and
most degenerate.  Larger effects mask smaller effects, and combinations of effects may be particularly hard to recognize.  

An important tool for detecting systematic errors in astronomical measurements is redundancy in the observations themselves, that is, multiple observations of the same part of the sky.\footnote{Of course one must take into account the changing nature of the sky itself, whether from variable objects (e.g., essentially everything, but on varying timescales), moving objects (e.g., planets), or things that vary with location and direction (e.g., the zodiacal light).  Fortunately for CMB measurements, the CMB itself changes only on a cosmological timescale, and short-term changes in its characteristics that depend on the motion of the observer can be predicted with exquisite accuracy and used as a fundamental calibrator (in particular, the ``orbital dipole'').  \Planck\  incorporates observational redundancy on multiple timescales, from 1-min rotations of the spacecraft, with the spin axis fixed for many rotations, to the approximately 6-month repeat coverage of the sky (with the angle of attack on a given piece of sky alternating each time), to the exact 1-year repeat coverage of the sky.} 
Each of these redundancies provides null tests of the data, in other words, differences between two observations of the same sky that should be zero within the noise.  A common, and initially almost inevitable, cause of null-test failures is poor knowledge of the noise.  Other causes of null-test failures can sometimes be identified (e.g., Solar flares), and affected data removed from further processing.  The removal process can be specified with strict criteria that are applied without reference to their effect on final results (such as cosmological parameters), that is, ``blind''; this is always done in \Planck\ data processing.  

Null tests are necessary, but not sufficient, in revealing problems in the data.  For one thing, any systematic error that affects all of the data, such as overall calibration errors, has no effect on null tests.  Equally important, and much harder to address, are systematic effects that are too small or too distributed for detection in the timeline data, but that cause problems when concentrated in further processing.  These can be difficult to identify, especially in combination, although some  can be predicted from a priori understanding of the instruments and mission.  Nevertheless, an exhaustive search is impossible -- there are simply too many possibilities.  Instead, one must search all intermediate and even final data products (i.e., the parts of phase space where systematic errors really matter) for problems.  Such searches cannot be technically ``blind'', as indeed their value lies in the sensitivity of results to specific systematics.  In practice, however, they are effectively blind, for two reasons.  First, no one can look at a map of the sky or an angular power spectrum and know the values of cosmological parameters that will fit them.  Second, \Planck's results are complex, rather than just a few numbers, and with such complexity investigator bias is inherently less of a problem than with simple outcomes.  However, when apparent problems are found, by whatever method, the cause of the problem must be traced back to an instrumental or observational origin before corrective action can be taken with any sense of certainty.  

Still, the dangers of removing sky signal from the data along with some known
but partly degenerate systematic are real.  In the end, the most important
tool both for finding systematics and for demonstrating that the processing
of the data does not remove or bias the signals being investigated is
simulations.
While simulations including complex astrophysics and space-borne detectors
cannot approach the level of realism encountered in particle physics
experiments (which imposes fundamental limitations on how blinding can be
performed), they have progressed dramatically over the lifetime of \Planck.
As has been described in detail in many \Planck\ Collaboration papers,
simulations have been an essential tool in the analysis of \Planck\ data.

\end{document}